\documentclass[prx,twocolumn,superscriptaddress,longbibliography,prx]{revtex4-2}

\usepackage[]{fontenc}

\usepackage{graphicx}
\usepackage{bm}
\usepackage{mathtools} 

\usepackage[caption=false]{subfig}

\usepackage{hyperref}
\usepackage{chngcntr} 

\usepackage{multirow}
\usepackage{subeqnarray}

\usepackage{wasysym}

\usepackage[british]{babel}

\usepackage{amsmath}
\usepackage{amssymb}

\usepackage{framed}
\usepackage{soul}

\usepackage{pdfpages}

\usepackage{color}

\usepackage{alltt}

\usepackage{pstricks,pst-node,pst-tree,pst-grad,pst-text,graphics}

\usepackage{dcolumn}

\usepackage{url}

\usepackage{cases}

\usepackage{natbib}

\usepackage{ifthen}

\usepackage{afterpage}

\usepackage{comment}

\usepackage{supertabular}

\usepackage{xspace}

\usepackage{setspace}

\usepackage{textcomp}

\usepackage{tikz}
\usetikzlibrary{decorations.pathmorphing}
\usetikzlibrary{decorations.markings}
\usetikzlibrary{arrows,shapes,decorations,automata,backgrounds,petri}

\newcommand{\gpcomment}[1]{[GP: {\it #1}]}
\renewcommand{\gpcomment}[1]{}

\newcommand{\plaind}{\mathrm{d}}
\newcommand{\dint}[1]{\mathchoice{\!\plaind#1\,}{\!\plaind#1\,}{\!\plaind#1\,}{\!\plaind#1\,}}
\newcommand{\ddint}[1]{\ddintx{#1}{d}}
\newcommand{\ddintx}[2]{\mathchoice{\!\plaind^{#2}#1\,}{\!\plaind^{#2}#1\,}{\!\plaind^{#2}#1\,}{\!\plaind^{#2}#1\,}}

\newcommand{\dbar}{\text{\dj}}
\newcommand{\deltabar}{\delta\mkern-6mu\mathchar'26}
\newcommand{\dintbar}[1]{\mathchoice{\!\dbar#1\,}{\!\dbar#1\,}{\!\dbar#1\,}{\!\dbar#1\,}}

\newcommand{\Dint}[1]{\mathcal{D}\!#1\,}
\DeclareFontFamily{U}{wncy}{}
\DeclareFontShape{U}{wncy}{m}{n}{<->wncyr10}{}
\DeclareSymbolFont{mcy}{U}{wncy}{m}{n}
\DeclareMathSymbol{\sha}{\mathord}{mcy}{"58}

\newcommand{\MCLineVsMathEnv}[2]{\mathchoice{#1}{#2}{#2}{#2}}

\newcommand{\ddX}[1]{\MCLineVsMathEnv{\frac{\plaind}{\plaind #1}}{\plaind/\plaind #1}}
\newcommand{\ddt}{\ddX{t}}

\usepackage{dsfont}

\newcommand{\gpset}[1]{\mathds{#1}}

\newcommand{\canetset}[1]{{\mathchoice {\hbox{$\sf\textstyle #1\kern-0.4em #1$}}
{\hbox{$\sf\textstyle #1\kern-0.4em #1$}}
{\hbox{$\sf\scriptstyle #1\kern-0.3em #1$}}
{\hbox{$\sf\scriptscriptstyle #1\kern-0.2em #1$}}}}

\newcommand{\Nset}{\gpset{N}}
\newcommand{\Zset}{\gpset{Z}}

\DeclareMathOperator*{\SumInt}{%
\mathchoice%
  {\ooalign{$\displaystyle\sum$\cr\hidewidth$\displaystyle\int$\hidewidth\cr}}
  {\ooalign{\raisebox{.14\height}{\scalebox{.7}{$\textstyle\sum$}}\cr\hidewidth$\textstyle\int$\hidewidth\cr}}
  {\ooalign{\raisebox{.2\height}{\scalebox{.6}{$\scriptstyle\sum$}}\cr$\scriptstyle\int$\cr}}
  {\ooalign{\raisebox{.2\height}{\scalebox{.6}{$\scriptstyle\sum$}}\cr$\scriptstyle\int$\cr}}
}

\def\nbZ{{\mathchoice {\hbox{$\sf\textstyle Z\kern-0.4em Z$}}
{\hbox{$\sf\textstyle Z\kern-0.4em Z$}}
{\hbox{$\sf\scriptstyle Z\kern-0.3em Z$}}
{\hbox{$\sf\scriptscriptstyle Z\kern-0.2em Z$}}}}

\newcommand{\gpvec}[1]{\mathbf{#1}}

\newcommand{\zerovec}{\gpvec{0}}
\newcommand{\nullvec}{\zerovec}
\newcommand{\avec}{\gpvec{a}}
\newcommand{\bvec}{\gpvec{b}}

\newcommand{\evec}{\gpvec{e}}

\newcommand{\jvec}{\gpvec{j}}
\newcommand{\kvec}{\gpvec{k}}
\newcommand{\ellvec}{\gpvec{\ell}}
\newcommand{\mvec}{\gpvec{m}}
\newcommand{\nvec}{\gpvec{n}}

\newcommand{\rvec}{\gpvec{r}}

\newcommand{\uvec}{\gpvec{u}}
\newcommand{\vvec}{\gpvec{v}}
\newcommand{\wvec}{\gpvec{w}}
\newcommand{\xvec}{\gpvec{x}}
\newcommand{\yvec}{\gpvec{y}}
\newcommand{\zvec}{\gpvec{z}}

\newcommand{\nuvec}{\bm{\nu}}

\newcommand{\xivec}{\bm{\xi}}

\newcommand{\transpose}{\mathsf{T}}

\newcommand{\ident}{\mathbf{1}}

\newcommand*\laplace{\mathop{}\!\mathbin\bigtriangleup}

\renewcommand{\AC}{\mathcal{A}}
\newcommand{\BC}{\mathcal{B}}

\newcommand{\GC}{\mathcal{G}}
\newcommand{\HC}{\mathcal{H}}
\newcommand{\IC}{\mathcal{I}}

\newcommand{\KC}{\mathcal{K}}
\newcommand{\LC}{\mathcal{L}}
\newcommand{\MC}{\mathcal{M}}
\newcommand{\NC}{\mathcal{N}}
\newcommand{\OC}{\mathcal{O}}
\newcommand{\PC}{\mathcal{P}}

\newcommand{\WC}{\mathcal{W}}

\newcommand{\phitilde}{\tilde{\phi}}

\newcommand{\phidagger}{\phi^{\dagger}}

\newcommand{\half}{\mathchoice{\frac{1}{2}}{(1/2)}{\frac{1}{2}}{(1/2)}}

\newcommand{\Exp}[1]{\operatorname{exp}\left(#1\right)}
\renewcommand{\exp}[1]{\mathchoice%
{\mathrm{e}^{#1}}%
{\operatorname{exp}(#1)}
{\operatorname{exp}\left(#1\right)}%
{\operatorname{exp}\left(#1\right)}}

\newcommand{\sgn}{\operatorname{sgn}}

\newcommand{\elabel}[1]{\label{eq:#1}}
\newcommand{\eref}[1]{(\ref{eq:#1})}
\newcommand{\Eref}[1]{\mbox{Eq.~(\ref{eq:#1})}}
\newcommand{\Erefs}[1]{\mbox{Eqs.~(\ref{eq:#1})}}

\newcommand{\seclabel}[1]{\label{sec:#1}}
\newcommand{\sref}[1]{\ref{sec:#1}}
\newcommand{\Sref}[1]{Section~\ref{sec:#1}}

\newcommand{\Fref}[1]{Figure~\ref{fig:#1}}

\newcommand{\latin}[1]{{\it #1}}
\newcommand{\ie}{\latin{i.e.}\@\xspace}
\newcommand{\eg}{\latin{e.g.}\@\xspace}
\newcommand{\cf}{\latin{cf.}\@\xspace}
\newcommand{\etc}{\latin{etc.}\@\xspace}

\newcommand{\etal}{\latin{et~al}.\@\xspace}

\newlength \standardfigwidth
\DeclareMathAlphabet{\matheub}{U}{eur}{m}{n}

\newcommand{\gcomment}[1]{\PackageError{commentCommand}{Don't use comments in production LaTeX}{}}
\newcommand{\rcomment}[1]{\PackageError{commentCommand}{Don't use comments in production LaTeX}{}}
\DeclareMathOperator{\diag}{diag}

\usepackage{tikz}
\usetikzlibrary{decorations.pathmorphing}
\usetikzlibrary{decorations.markings}
\usetikzlibrary{arrows,shapes,decorations,automata,backgrounds,petri}
\usetikzlibrary{calc}
\usetikzlibrary{patterns}
\usetikzlibrary{decorations.text}
\tikzset{
OLDpotStyle/.style={black,dash dot},
potStyle/.style={black, thick,
dash pattern={on 2.5pt off 1pt on 0.75pt off 1pt}},
xxtsubstrate/.style={decorate, 
line width=1pt,
draw=olive, 
decoration=snake, 
segment amplitude=0.75mm, 
line after snake=0.25mm,
line before snake=0.25mm
},
tsubstrate/.style={decorate, 
line width=1pt,
draw=olive, 
decoration=snake, 
segment amplitude=0.5mm, 
segment length=5pt,
segment amplitude=0.2mm, 
line after snake=1mm,
line before snake=1mm
},
tAsubstrate/.style={decorate, 
thick,
draw=olive, 
decoration=snake, 
segment amplitude=0.5mm, 
segment length=5pt,
segment amplitude=0.2mm, 
line after snake=1mm,
line before snake=1mm
},
Bsubstrate/.style={decorate, 
line width=1pt,
draw=orange, 
decoration=snake,
segment length=5pt,
segment aspect=0,
segment amplitude=0.5mm, 
line after snake=0mm,
line before snake=0mm
},
substrate/.style={decorate, 
line width=1pt,
draw=orange,
decoration=snake, 
segment length=5pt,
segment amplitude=0.5mm, 
line after snake=0.5mm,
line before snake=0.5mm
},
ABoldActivity/.style={line width=1pt,double,draw=red},
BoldActivity/.style={line width=1pt,double,draw=red,postaction={decorate},
decoration={markings,mark=at position .5 with
{\arrow[draw=red]{>}}}},
activity/.style={very thick,draw=red,postaction={decorate},
decoration={markings,mark=at position .5 with
{\arrow[draw=red]{>}}}},
tactivity/.style={thick,draw=red,postaction={decorate},
decoration={markings,mark=at position .5 with
{\arrow[draw=red]{>}}}},
tEPSactivity/.style={thick,draw=red,postaction={decorate},
decoration={markings,mark=at position .55 with
{\arrow[draw=red]{>}}}},
tAactivity/.style={thick,draw=red},
Aactivity/.style={very thick,draw=red},
tSactivity/.style={thick,draw=red,postaction={decorate},
decoration={markings,mark=at position .7 with
{\arrow[draw=red]{>}}}},
Sactivity/.style={very thick,draw=red,postaction={decorate},
decoration={markings,mark=at position .7 with
{\arrow[draw=red]{>}}}},
polarity/.style={decorate, 
line width=1pt,
draw=red,
decoration={markings,mark=between positions 0 and 1 step 1.1mm with {\draw[red,thick]  (0,0) circle (0.03)} },
segment length=5pt,
segment amplitude=0.5mm, 
},
Bpolarity/.style={decorate, 
line width=1.5pt,
draw=red,
segment length=5pt,
segment amplitude=0.5mm, 
},
density/.style={ 
line width=1.5pt,
draw=black,
densely dashed,
segment length=5pt,
segment amplitude=0.5mm, 
},
}

\newcommand{\spave}[1]{\overline{#1}}

\newcommand{\ave}[2][]{\mathchoice%
{\left\langle #2 \right\rangle_{#1}}%
{\langle #2\rangle_{#1}}
{\langle #2\rangle_{#1}}%
{\langle #2\rangle_{#1}}}

\newcommand{\ket}[1]{\left|#1\right\rangle}

\makeatletter
\newcommand{\creatX}[2][]{a^{\@ifempty{#1}{\dagger}{\dagger\,#1}}\@ifempty{#2}{}{(#2)}}
\makeatother
\newcommand{\creatDS}{\tilde{a}}
\makeatletter
\newcommand{\creatDSX}[2][]{\@ifempty{#1}{\creatDS}{\creatDS^{#1}}\@ifempty{#2}{}{(#2)}}
\makeatother

\makeatletter
\newcommand{\annihX}[2][]{a\@ifempty{#1}{}{^{#1}}\@ifempty{#2}{}{(#2)}}

\newcommand{\tableofappendixcontents}{\@starttoc{toa}}
\newcommand{\l@toact}[2]{\@dottedtocline{1}{1.5em}{5.5em}{#1}{#2}}
\newcommand{\l@toactspace}[2]{\ \hfil}
\makeatother

\newcommand{\APref}[1]{\mbox{App.~\ref{sec:#1}}}
\newcommand{\SMref}[1]{\mbox{Suppl.~\ref{sec:#1}}}
\newcommand{\imag}{\mathring{\imath}}

\renewcommand{\exp}[1]{\mathchoice%
{e^{#1}}%
{\operatorname{exp}(#1)}%
{\operatorname{exp}(#1)}%
{\operatorname{exp}(#1)}}

\usepackage{xspace}

\renewcommand{\st}[1]{}

\newcommand{\entropyProduction}{\dot{S}_{\text{int}}}
\newcommand{\entropyProductionDensity}{\dot{\sigma}}

\newcommand{\Deltat}{{\Delta t}}
\newcommand{\Deltax}{{\Delta x}}

\newcommand{\transMatLin}{\MC}
\newcommand{\transMatNonLin}{\BC}
\newcommand{\PertOp}{\hat{\GC}}

\newcommand{\Transition}{\WC} 
\newcommand{\TransitionRate}{\mathring{\Transition}}

\newcommand{\density}{\rho}
\newcommand{\densityTilde}{\tilde{\rho}}

\makeatletter
\newcommand{\HMdensity}[2]{\rho_{#2}^{\@ifempty{#1}{}{(#1)}}}
\makeatother

\newcommand{\DPop}{\hat{\AC}}

\usepackage{textcomp}

\newcommand{\action}{\AC}

\newcommand{\extPot}{\Upsilon}
\newcommand{\pairPot}{U}
\newcommand{\drift}{w}
\newcommand{\diffusion}{D}
\newcommand{\vecDrift}{\wvec}

\newcommand{\hot}{\text{h.o.t.}}
\newcommand{\fdot}{\dot{f}}
\newcommand{\gdot}{\dot{g}}
\newcommand{\hdot}{\dot{h}}
\newcommand{\Bf}{\overline{f}}
\newcommand{\Bg}{\overline{g}}
\newcommand{\Bh}{\overline{h}}

\newcommand{\FPeqn}{FPE~\eref{FPeqn_main}\@\xspace}
\newcommand{\perturbative}{\text{pert}}
\newcommand{\FPprobability}{\rho}
\newcommand{\FPop}{\LC} 
\newcommand{\FPophat}{\hat{\LC}}

\newcommand{\Op}{\bm{\mathsf{K}}}
\newcommand{\Ln}{\operatorname{\bm{\mathsf{Ln}}}}
\newcommand{\order}[1]{\OC\left(#1\right)}

\newcommand{\genVertex}[1]{\draw[red,fill=red] (#1) circle (0.75ex);}
\newcommand{\tgenVertex}[1]{\draw[red,fill=red] (#1) circle (0.5ex);}

\newcommand{\fullBlobb}[2]{\tikz[baseline=-2.5pt]{\draw[tAactivity] (0.3,0) -- (-0.3,0);%
\node at (0.3,0) [above] {$#1$};%
\node at (-0.3,0) [above] {$#2$};%
\genVertex{0,0}}}

\newcommand{\tfullBlobb}[2]{\tikz[baseline=-2.5pt]{\draw[tAactivity] (0.2,0) -- (-0.2,0);%
\node at (0.15,0) [above] {$#1$};%
\node at (-0.15,0) [above] {$#2$};%
\tgenVertex{0,0}}}
\newcommand{\tfullDoubleBlobb}[2]{\tikz[baseline=-2.5pt]{\draw[tAactivity] (0.2,0) -- (-0.65,0);%
\node at (0.15,0) [above] {$#1$};%
\node at (-0.6,0) [above] {$#2$};%
\tgenVertex{0,0}%
\tgenVertex{-0.45,0}%
}}

\newcommand{\tbarePropagator}[2]{\tikz[baseline=-2.5pt]{
\node at (0.5,0) [above] {$#1$};
\node at (-0.5,0) [above] {$#2$};
\draw[tAactivity] (0.5,0) -- (-0.5,0);
}}

\newcommand{\tbarePropagatorL}[2]{\tikz[baseline=-2.5pt]{
\node at (0.7,0) [above] {$#1$};
\node at (-0.7,0) [above] {$#2$};
\draw[tAactivity] (0.7,0) -- (-0.7,0);
}}
\newcommand{\tbarePropagatorLS}[2]{\tikz[baseline=-2.5pt]{
\node at (0.7,0) [above] {${\scriptstyle #1}$};
\node at (-0.7,0) [above] {${\scriptstyle #2}$};
\draw[tAactivity] (0.7,0) -- (-0.7,0);
}}
\newcommand{\tblobbedPropagator}[2]{\tikz[baseline=-2.5pt]{
\node at (0.7,0) [above] {$#1$};
\node at (-0.7,0) [above] {$#2$};
\draw[tAactivity] (0.7,0) -- (-0.7,0);
\tgenVertex{0,0}
}}
\newcommand{\tblobbedPropagatorS}[2]{\tikz[baseline=-2.5pt]{
\node at (0.7,0) [above] {${\scriptstyle #1}$};
\node at (-0.7,0) [above] {${\scriptstyle #2}$};
\draw[tAactivity] (0.7,0) -- (-0.7,0);
\tgenVertex{0,0}
}}
\newcommand{\tblobbedDashedPropagator}[2]{\tikz[baseline=-2.5pt]{
\node at (0.7,0) [above] {$#1$};
\node at (-0.7,0) [above] {$#2$};
\draw[thin,red] (-0.2,-0.1) -- (-0.2,0.1);
\draw[tAactivity] (0.7,0) -- (-0.7,0);
\tgenVertex{0,0}
\node at (0,-0.1) [below] {$\vecDrift$};
}}
\newcommand{\blobbedDashedPotPropagator}[2]{\tikz[baseline=8pt]{
\begin{scope}[yshift=0.0cm]
\node at (\propWidth,0.4) [above] {$#1$};
\node at (-\propWidth,0.4) [above] {$#2$};
\draw[tAactivity] (\propWidth,0.4) -- (-\propWidth,0.4);
\draw[black,potStyle] (0,-0.2) -- (0,0.4);
\draw[black,thin] (-0.1,0.2) -- (0.1,0.2);
\draw[red,thin] (-0.2,0.3) -- (-0.2,0.5);
\tgenVertex{0,0.4}
\draw[black,fill=white] (0,-0.2) circle (3pt);
\node at (0,-0.2) [right,xshift=2pt] {$-\extPot'$};
\end{scope}
}
}
\newcommand{\blobbedDashedPotPropagatorText}[2]{\tikz[baseline=8pt]{
\begin{scope}[yshift=0.0cm]
\node at (\propWidth,0.4) [above] {$#1$};
\node at (-\propWidth,0.4) [above] {$#2$};
\draw[tAactivity] (\propWidth,0.4) -- (-\propWidth,0.4);
\draw[black,potStyle] (0,0.1) -- (0,0.4);
\draw[black,thin] (-0.1,0.25) -- (0.1,0.25);
\draw[red,thin] (-0.2,0.3) -- (-0.2,0.5);
\tgenVertex{0,0.4}
\draw[black,fill=white] (0,-0.00) circle (3pt);
\node at (0,0.1) [right,xshift=2pt] {$-\extPot'$};
\end{scope}
}
}

\newcommand{\tDblobbedPropagator}[2]{\tikz[baseline=-2.5pt]{
\node at (0.7,0) [above] {$#1$};
\node at (-1.4,0) [above] {$#2$};
\draw[tAactivity] (0.7,0) -- (-1.4,0);
\tgenVertex{0,0}
\tgenVertex{-0.7,0}
}}

\usepackage{color}
\definecolor{darkgreen}{rgb}{0,0.6,0}
\definecolor{darkblue}{rgb}{0,0,0.6}
\definecolor{darkred}{rgb}{0.6,0,0}
\definecolor{darkpurple}{rgb}{0.5,0,0.5}
\hypersetup{
bookmarksopen=true,
pdftitle="",
pdfauthor="", 
pdftoolbar=false, 
pdfstartview={FitH},		
pdfmenubar=true,			
pdfhighlight=/O,			
colorlinks=true,			
urlcolor=darkblue,
citecolor=darkblue,		
linkcolor=darkblue}

\makeatletter
\newcommand{\customlabel}[2]{%
   \protected@write \@auxout {}{\string \newlabel {#1}{{#2}{\thepage}{#2}{#1}{}} }%
   \hypertarget{#1}{\hspace{0pt}}
}
\makeatother

\makeatletter
\AtBeginDocument{\let\LS@rot\@undefined}
\makeatother

\begin{document}
\customlabel{eq:M_state_exact_asympotic_result}{S-I.2}
\customlabel{eq:N2_final_entropy_production}{S-II.30}
\customlabel{eq:entropy_production_final}{S-I.37}
\customlabel{eq:pair_pot_effect2}{S-II.23}
\customlabel{eq:pair_vertex1}{S-II.18}
\customlabel{eq:pair_vertex2}{S-II.19}
\customlabel{eq:trawler_A1v1}{S-II.14}
\customlabel{eq:trawler_A2v1}{S-II.16}
\customlabel{sec:HarmonicTrawlers}{S-II}
\customlabel{sec:HarmonicTrawlers_simple_physics}{S-II.1}
\customlabel{sec:app_ring}{S-I}
\customlabel{sec:model_description}{S-I.1}


 \newsavebox{\tempbox}
 \newcommand{\textboxL}[1]
{\savebox{\tempbox}{#1}
 \ifdim\wd\tempbox<4cm\relax
   \makebox[5cm]{\usebox{\tempbox}}%
 \else
   \parbox{5cm}{\centering #1}
 \fi}
 \newcommand{\textbox}[1]
{\savebox{\tempbox}{#1}
 \ifdim\wd\tempbox<4cm\relax
   \makebox[4cm]{\usebox{\tempbox}}%
 \else
   \parbox{4cm}{\raggedright #1}%
 \fi}

  \newcommand{\roadmap}{\begin{tikzpicture}
\tikzstyle{every node}=[font=\small]
\draw[draw=none,fill=gray,opacity=0.2] (2.8,6.6) rectangle (-2.8,-1.3);
\node[align=center] at (0,6) {\textboxL{\uppercase{Entropy production of active particle systems}}};
\node[draw] at (0,5) {\textbox{\sref{introduction} Introduction}};
\node[draw] at (0,4) {\textbox{\sref{DP_from_FP_main} Field theory from FPE}};
\node[draw] at (0,3) {\textbox{\sref{entropy_production_main} Entropy production}};
\node[draw] at (0,1.2) {\textbox{
{\hspace{25pt} \sref{examples_main} Examples}\newline Markov chain \newline Drift-diffusion \newline Harmonic trawlers \newline Interacting ABPs}};
\node[draw] at (0,-0.55) {\textbox{\sref{summary_outlook} Conclusions}};
\node[] at (-6,6) {\uppercase{Derivations}};
\node[draw] at (-6,4.58) {\textboxL{\SMref{app_ring} Failure of continuous particle number approximation: 
biased hopping on a ring}};
\node[draw] at (-6,3.58) {\textboxL{\APref{DP_plus_EP} ME/FPE to FT to EPR}};
\node[draw] at (-6,2.78) {\textboxL{\APref{appendixWhichDiagramsContribute} Short-time scaling of diagrams}};
\node[draw] at (-6,1.8) {\textboxL{\APref{drift_diffusion_on_ring} Drift-diffusion in external potential}};
\node[draw] at (-6,0.82) {\textboxL{\SMref{HarmonicTrawlers} Harmonic pair-interaction}};
\node[draw] at (-6,0.04) {\textboxL{\APref{MultipleParticles} Multiple particles}};
\node[align=center] at (5.5,6) {\textboxL{\uppercase{Applications and key results}}};
\node[draw] at (5.5,4) {\textbox{Action \eref{def_action0}}};
\node[draw] at (5.5,2.78) {\textbox{
Definition EPR \eref{def_entropyProduction}, kernel \eref{transition_from_action}, logarithm \eref{Ln_from_propagators}, EPR \eref{entropyProduction_for_pairPot} and \eref{entropyProduction_interacting_indistinguishable_example},
 and local EPR \eref{interacting_EPR_numerics}}};
\node[draw] at (5.5,0.97) {\textbox{
EPR \eref{EntropyProductionMarkovChain} \newline
EPR \eref{entropyProduction_driftdiffusion_main}, \eref{local_entropy_GWissel} and \eref{local_entropy_Gw} \newline
EPR \eref{trawler_final} \newline
Local EPR \eref{ABP_EPR_succinct}}};
\draw [-stealth,line width=2pt](0,4.78) -- +(0,-0.45);
\draw [-stealth,line width=2pt](0,3.75) -- +(0,-0.45);
\draw [-stealth,line width=2pt](0,2.75) -- +(0,-0.45);
\draw [-stealth,line width=2pt](0,0.17) -- +(0,-0.45);
\draw [-stealth,thick](-3.4,5) -- +(1.2,0);
\draw [-stealth,thick](-3.4,3.6) -- +(20:1.2);
\draw [-stealth,thick](-3.4,3.6) -- +(-20:1.2);
\draw [-stealth,thick](-3.4,3) -- +(1.2,0);
\draw [-stealth,thick](-3.4,1.7) -- +(-17:1.2);
\draw [-stealth,thick](-3.4,0.8) -- +(0:1.2);
\draw [-stealth,thick](-3.4,0.2) -- +(20:1.2);
\draw [-stealth,thick](-3.4,0.2) -- +(8:1.2);
\draw [-stealth,thick](2.1,4) -- +(0:1.2);
\draw [-stealth,thick](2.1,3) -- +(0:1.2);
\draw [-stealth,thick](2.1,1.6) -- +(0:1.2);
\draw [-stealth,thick](2.1,1.2) -- +(0:1.2);
\draw [-stealth,thick](2.1,0.8) -- +(0:1.2);
\draw [-stealth,thick](2.1,0.4) -- +(0:1.2);
\end{tikzpicture}}

\newcommand{\titleText}{Field theories of active particle systems and their entropy production}
\title{\titleText}

\author{Gunnar Pruessner}
\email{g.pruessner@imperial.ac.uk}
\affiliation{%
Department of Mathematics
and Centre of Complexity Science, 
Imperial College London, London SW7 2AZ, United Kingdom}%
\author{Rosalba Garcia-Millan}%
 \email{rosalba.garcia\_millan@kcl.ac.uk}
\affiliation{%
Department of Mathematics, King's College London, London WC2R 2LS, United Kingdom\\
DAMTP, Centre for Mathematical Sciences, University of Cambridge, Cambridge CB3 0WA, United Kingdom\\
St John's College, University of Cambridge, Cambridge CB2 1TP, United Kingdom}%
\affiliation{%
Department of Mathematics
and Centre of Complexity Science, 
Imperial College London, London SW7 2AZ, United Kingdom}%

\date{\today}

\begin{abstract}
Active particles that translate chemical energy into self-propulsion
can maintain a far-from-equilibrium steady state and perform work. 
The entropy production measures how far from equilibrium such a particle 
system operates and serves as a proxy for the work performed.
Field theory
offers a promising route to calculating entropy production,
as it allows for many interacting particles
to be considered simultaneously.
Approximate field theories obtained by coarse-graining or smoothing that draw on additive noise
can capture densities and correlations 
well, but they generally ignore the microscopic particle nature of the constituents, thereby producing spurious results for the
entropy production.
As an alternative we demonstrate how to use Doi-Peliti field theories, which capture the microscopic
dynamics, including reactions and interactions with external and pair potentials. 
Such field theories are in principle exact, while offering a systematic 
approximation scheme, in the form of diagrammatics.
We demonstrate how to construct them from 
a Fokker-Planck equation (FPE) and show how 
to calculate entropy production of active matter from first principles.
This framework is easily extended to include interaction. We use it to derive exact, compact and efficient general expressions for the entropy production for a vast range of interacting 
conserved
particle systems. 
These expressions 
are independent of the underlying field theory and
can be interpreted as the spatial average of the \emph{local} entropy production.
They are readily applicable to numerical and experimental data.
In general, 
the entropy production due to
any pair interaction draws at most on the three point, equal time density; and an $n$-point interaction on the $(2n-1)$-point density.
We illustrate the technique in a number of exact, tractable examples, including
some with pair-interaction 
as well as in a system of many interacting
Active Brownian Particles.
\end{abstract}

                              
\maketitle

\newcommand{\longtodo}[1]{\todo[inline,size=\tiny]{#1}}

\section{Introduction}
\seclabel{introduction}
Active matter has been the focus of much research in statistical mechanics 
and biophysics over the past decade, because of 
many surprising theoretical features
\cite{TonerTu:1995,Cates:2012,JuelicherGrillSalbreux:2018,MandalLiebchenLoewen:2019}, 
the rich phenomenology \cite{CatesTailleur:2015,LiebchenLevis:2017}
and a plethora of applications \cite{DiLeonardoETAL:2010,XiETAL:2019,PietzonkaETAL:2019}. 
At the heart of active matter lies 
the conversion of chemical fuel into mechanical work, 
often in the form of 
self-propulsion, which leads to sustained non-equilibrium behaviour
that is distinctly different from that of relaxing equilibrium thermodynamic 
systems \cite{Cates:2012}. How different, is quantified by the entropy
production, which also quantifies the work performed. If we want to harvest and utilise this work, we need to 
quantify
and control the system at the level it is observed and in the degrees of freedom that can be manipulated, rather than at a coarse-grained or 
smoothed
level. 
The problem is illustrated by a team of horses observed from high above, when they 
may look almost like a droplet squeezing through a pore as they push past obstacles. At this level of description it may be difficult to distinguish a forward from a backward movie of the scene. Zooming in on the individual animal, however, reveals the details of their movement \cite{Muybridge:1878} and thus the difference between forward and backward immediately. If the horses are to be hitched to a plough, this is the level of observation needed to asses their utility. Assessing smoothed quasi-horses \cite{Mattuck:1992} does not help.

Field theory has been the work-horse of statistical mechanics for many
decades \cite{DombGreenLebowitz:1972},
because it allows for an efficient calculation and a systematic approximation of universal and non-universal
observables in many-particle systems by means of a powerful machinery,
that can be cast in an elegant, physically meaningful language in the 
form of diagrams.
To apply this framework to active particle systems, effective field
theories have been proposed, that use the continuously varying local
particle density as the relevant degree of freedom. However, the entropy
production of an approximating field theory is not necessarily a good
approximation of the microscopic entropy production of the actual
particle system. An exact, fully microscopic framework to calculate the
entropy production systematically in active many-particle systems
remains a theoretical challenge. In recent years, several exact
results have been found \cite{Gaspard:2004}, although those are limited to linear
interaction forces \cite{LoosKlapp:2020,GodrecheLuck:2019},
or cases where the full time-dependent
particle probability density is known \cite{CocconiGarcia-MillanETAL:2020}.

\begin{figure*}
\roadmap
\caption{\label{fig:roadmap}
Road map
of the contents of this paper (central column),
appendices and supplements
where to find derivations (left column),
and applications of our framework and key 
equations
(right column).
Arrows indicate information flow and suggest
possible additional reading.
Abbreviations:
entropy production rate (EPR),
field theory (FT),
master equation (ME),
Fokker-Planck equation (FPE), and
Active Brownian Particles (ABPs).}
\end{figure*}

The entropy production crucially depends on the degrees of freedom used
to describe the system state. Coarse-graining by integrating out
degrees of freedom or by mapping sets of microstates to mesostates
generally underestimate the entropy production
\cite{RoldanETAL:2021, Esposito:2012, CocconiSalbreuxPruessner:2022, LoosPersonalCommunication:2021}.
In \cite{NardiniETAL:2017} the particle
dynamics has instead been approximated by recasting it as a continuously
varying density subject to a Langevin equation of motion with additive
noise. This approach captures much of the physics well, notably predicting
that most of the entropy is produced at interfaces between dense and
dilute phases \cite{NardiniETAL:2017,MartinETAL:2021,FodorETAL:2022}. Yet, it does not
provide a lower bound of the entropy production, as it replaces the
countably many particle degrees of freedom by the uncountably many of a
density in space. 
\SMref{app_ring} illustrates this in a simple, tractable example
wrongly, but quite possibly generically displaying a \emph{divergent} entropy production in the framework of \cite{NardiniETAL:2017}.

Doi-Peliti (DP) field theories 
\cite{Doi:1976,Doi:1976b,Peliti:1985} on the other hand,
retain the particle nature of the constituent
degrees of freedom, but can be cumbersome to derive,
normally
requiring discretisation and an explicit derivation of a master
equation. Instead, we demonstrate how 
the bilinear part 
a Doi-Peliti action for multiple pair-interacting particles can be
determined using the Fokker-Planck operator.
Interactions through external and pair potentials, as well as reactions can be added
by virtue of the same Poissonian ``superposition principle'', that
allows a master equation to account for concurrent processes by adding corresponding
gain and loss terms.
We further show how the
ensuing perturbation theory and its diagrammatics can be used to derive the entropy
production 
of conserved particle systems exactly, which turns out to draw entirely on first order terms, no matter how strong the interaction or how many particles interact simultaneously. These first order terms are  contained in the
bare propagator and the
lowest order perturbative vertices, weighted by certain correlation functions.
The diagrammatics of a field theory provides the small number of terms needed to calculate entropy production \emph{exactly}.
Perturbative field theory is used here only as to systematically determine exact expressions. 
Since 
only the lowest order 
corrections in the perturbative expansion are needed, 
our results do not rely on 
the perturbative expansion being convergent.
Our procedure results in very general formulae that need as system-specific input the details of the interaction potentials and a few low-order correlation functions. In the simplest case of non-interacting particles, the latter reduce to the one-point density, so that the entropy production becomes a spatial average of a local property. In general, if the interaction allows for up to $n$ particles interacting simultaneously, only the $(2n-1)$-point equal-time correlation function needs to be known, effectively quantifying where and how frequently such interactions take place.
For numerical and experimental data a simple estimator is derived that  that does not draw on any correlation function.
We thus introduce a generic scheme to derive tractable expressions for the entropy production of complex many-particle systems on the basis of their microscopic, stochastic equation of motion.
We illustrate the technique in a number of examples.

The general strategy in the following is to construct the field theory and derive the entropy production starting from the simplest possible setup of non-interacting particles,
then allowing for perturbative terms,
then allowing for multiple particles, 
then allowing for particle interactions. 
To keep the derivation tractable, we start with trivial non-interacting systems, which may make the framework look excessive and overly complicated, but only as to develop a clearer view of the steps necessary to treat interacting many-particle systems.
Expressions are derived explicitly for non-interacting particle systems as well as pair-interacting ones, and are easily generalised to higher order interactions. 
Perturbation theory is used only in so far as to provide us with first order terms, because these terms are all that is needed to calculate the entropy production. The underlying method is comparable to using the first order term of a Taylor series to determine the derivative of a function at the point it is an expansion about. DP field theory and its diagrammatics are a particularly convenient way to read off the required terms.
It serves solely the purpose of providing a tractable scheme to systematically identify the terms in the Fokker-Planck equation needed for the entropy production rate.
\emph{The formalism described in the present work is exact and applies universally to all stochastic particle dynamics that can be cast in a Fokker-Planck equation.} 
We confirm the validity of the formalism by means of stochastic particle systems that possess an exact solution, and in \Sref{examples_main} we apply it to a highly non-trivial but not exactly solvable setup, using numerical validation.

The present work brings to bear the power of field theory to the field of active matter, while retaining particle entity, by calculating entropy production of the relevant degrees of freedom using diagrammatics and avoiding approximations altogether. Details of our derivations can be found in the appendices 
and
the Supplemental Material.
\Fref{roadmap} shows a road map of 
the work and suggests possible reading paths.
We list the key results according to the structure of the article:
\begin{itemize} 

    \item[]\hspace{-20pt}\emph{\Sref{DP_from_FP_main}}: We show how a
    Doi-Peliti field theory is readily derived from a Fokker-Planck
    Equation, in particular \Eref{def_action0_cont} from
    \Eref{FPeqn_main} 
    in the free case and \Eref{interacting_action_example} from \Eref{interacting_FPE_example} in the case of pair-interaction (also \APref{DP_plus_EP}).
    \item[]\hspace{-20pt}\emph{\Sref{entropy_production_main}}: We introduce the framework to calculate entropy production, proceeding from the definition \Eref{def_entropyProduction} via \Eref{def_entropyProduction_elegant} to the diagrammatics of \Erefs{transition_from_action} and \eref{Ln_from_propagators} (also \APref{entropy_production} and \APref{appendixWhichDiagramsContribute}).
    \item[]\hspace{-20pt}\emph{\Sref{interaction_main}}: We include interaction, determining the relevant diagrams in \Eref{def_trans_multi_diag}, which immediately simplify to produce general expressions for $N$ pair-interacting indistinguishable particles such as \Eref{entropyProduction_for_pairPot} (also \APref{MultipleParticles}). A corresponding numerical scheme is readily derived as  \Eref{interacting_EPR_numerics}. 
    We find that the entropy production of pair-interacting particles draws at most on the $3$-point density \cite{LynnETAL:2022,ZhangGarcia-Millan:2023}.
    \item[]\hspace{-20pt}\emph{\Sref{examples_main}}: We give concrete examples: a Markov chain,  drift-diffusion of a single particle (also \APref{drift_diffusion_on_ring}), \Eref{entropy_production_drift_diffusion_main}, 
    two distinct particles on a circle (also \SMref{HarmonicTrawlers}),
    and $N$ interacting
    Active Brownian Particles.
\end{itemize}
We conclude in \emph{\Sref{summary_outlook}} with a discussion, a summary of our results and an outlook.

\section{Field Theory from Fokker-Planck equation}
\seclabel{DP_from_FP_main}
An efficient way to characterise a many-particle system is in terms of
occupation numbers, which allows for, in principle, arbitrary particle
numbers and species without having to change the parameterisation, as opposed to a description in terms of the
individual particle degrees of freedom. 
Doi-Peliti 
field theories provide a framework that readily caters for the spatio-temporal
evolution of \emph{particles} in terms of occupation numbers, in contrast to, say, the response field formalism 
\cite{MartinSiggiaRose:1973,Janssen:1976,DeDominicis:1976,Taeuber:2014}
which 
requires further non-linearities
in the form of Dean's equation \cite{Dean:1996,BotheETAL:2023}. As the derivation of a DP field theory
from a master equation can
be cumbersome in particular in the presence of external fields \cite{TaeuberHowardVollmayr-Lee:2005,Cardy:2008,BaishVollmayr-Lee:2024}, we demonstrate in \APref{DP_plus_EP} 
how 
a DP action is readily read off from a Fokker-Planck equation (FPE) for free particles, \APref{DP_field_theory_derivation}, and more generally with pair-interaction in \APref{FPE_inter}. 
This constitutes a much more explicit derivation than originally introduced
by Doi \cite{Doi:1976} and cuts short any tedious rewriting of particle dynamics and interactions in terms of chemical reactions \cite{Peliti:1985,Cardy:2008}.
The resulting correspondence between DP action and FPE is similar to that of the Martin-Siggia-Rose action and a Langevin equation
\cite{MartinSiggiaRose:1973,Janssen:1976,DeDominicis:1976,Taeuber:2014}. Any continuum limit
that has to be
taken in a lattice-based master equation to derive the 
continuum FPE can equivalently be applied in the field theory.
In other words, if the FPE of a density $\FPprobability(\yvec,t)$ reads
\begin{equation}\elabel{FPeqn_main}
    \partial_t \FPprobability(\yvec,t) = 
    \SumInt_{\xvec}
    \FPop_{\yvec,\xvec} \FPprobability(\xvec,t)
\end{equation}
with Fokker-Planck kernel $\FPop_{\yvec,\xvec}$,
then the DP action reads
\begin{equation}\elabel{def_action0}
    \action_0 = 
\int \dint{t}
    \SumInt_{\xvec,\yvec}  \phitilde(\yvec,t) (\FPop_{\yvec,\xvec} - \delta(\yvec-\xvec)\partial_t) \phi(\xvec,t)
\end{equation}
with annihilator field $\phi(\xvec,t)$ and Doi-shifted creator field \cite{Peliti:1985,Cardy:2008} $\phitilde(\yvec,t)$. 
The notation $\SumInt_{\xvec}$ indicates
sum or integral over the dummy variable $\xvec$, 
for discrete or continuous degree of freedom $\xvec$,
respectively.
The simple relationship between \Erefs{FPeqn_main} and \eref{def_action0} is the first key-result of the present work, with the derivations relegated to \APref{interaction_main}. 
Because \Eref{FPeqn_main} describes a non-interacting particle, the resulting action is bilinear, \ie Gaussian in the field $\phi(\xvec,t)$ and its conjugate $\phitilde(\yvec,t)$.

Henceforth, until \Sref{interaction_main}, we focus on non-interacting theories to demonstrate how a DP action is constructed from an FPE and how it determines the entropy production. It is demonstrated in \APref{FPE_inter} that a DP action is equally derived from an interacting FPE, which results in additional terms in the action, within a field-theory usually treated perturbatively. In \Sref{interaction_main} we show how interaction terms affect the entropy production, yielding exact results despite  drawing only on the lowest order terms.

Based on the field-theoretic action observables are calculated in the path-integral
\cite{TaeuberHowardVollmayr-Lee:2005,Taeuber:2014}
\begin{equation}\elabel{def_ave0}
    \ave[0]{\bullet} = \int\Dint{\phi}\Dint{\phitilde} \bullet \exp{\action_0} \ .
\end{equation}
For continuous degrees of freedom, 
the kernel in \Eref{FPeqn_main} is usually written as 
$\FPop_{\yvec,\xvec} = \FPophat^{\dagger}_{\xvec} \delta(\xvec-\yvec)=\FPophat_{\yvec} \delta(\xvec-\yvec)$ 
so that
$\SumInt_{\xvec}\FPop_{\yvec,\xvec} \FPprobability(\xvec,t) = \FPophat_{\yvec}\FPprobability(\yvec,t)$
with FP operator 
$\FPophat_{\yvec}$ and $\FPophat^\dagger_{\yvec}$ its
adjoint. In this case 
the action simplifies to
\begin{equation}\elabel{def_action0_cont}
\action_0 = 
\int \dint{t}
    \int \ddint{y}  \phitilde(\yvec,t) (\FPophat_\yvec - \partial_t) \phi(\yvec,t)
.
\end{equation}
The bare propagator 
\begin{equation}\elabel{bare_prop_example}
\ave[0]{\phi(\yvec,t')\phitilde(\xvec,t)}
\corresponds
\tbarePropagator{\xvec,t}{\yvec,t'}
\end{equation}
of the action
\Eref{def_action0} is the Green function of the \FPeqn, \APref{DP_plus_EP}, and thus solves
$\partial_{t'} \ave[0]{\phi(\yvec,t')\phitilde(\xvec,t)} = \FPophat_{\yvec}
\ave[0]{\phi(\yvec,t')\phitilde(\xvec,t)}$ with $\lim_{t'\downarrow t}
\ave[0]{\phi(\yvec,t')\phitilde(\xvec,t)} = \delta(\yvec-\xvec)$.
We use the symbol $\corresponds$ in \Eref{bare_prop_example} and henceforth to denote 
correspondence
between a mathematical expression and its representation in terms of
Feynman diagrams.

The action \Eref{def_action0} has by construction the same form as the
action obtained by formally applying
the Martin-Siggia-Rose-trick \cite{MartinSiggiaRose:1973,Janssen:1976,DeDominicis:1976,Taeuber:2014} to the FPE, despite the absence of a noise term. However, the DP field theory retains the
particle nature of the constituent degrees of freedom without the need
of additional terms, like Dean's
\cite{Dean:1996,BotheETAL:2023}.
As a small price, a DP field theory is endowed with a commutator relation that needs to 
be consulted every time an observable is constructed
from operators.
As a consequence, 
unlike in an effective Langevin-equation on the density, the annihilator
field $\phi$ of
a DP field theory is not a particle density \cite{Cardy:2008}, and the action is not the particle density probability functional. 
Recasting the
DP action as the Martin-Siggia-Rose action
of a Langevin equation on the field $\phi$ can produce unexpected features, such as imaginary noise \cite{HowardTaeuber:1997,LefevreBiroli:2007}.
Such an interpretation ignores 
that DP fields are introduced as complex conjugates, but are independent in Martin-Siggia-Rose. It generally
involves steps that are ``purely
formal" \cite{BenitezETAL:2016,Cardy:2008} rather than mathematically rigorous. That has lead to the common misunderstanding of DP field theory ``suffering from imaginary noise", despite being addressed and clarified in works such as \cite{HowardTaeuber:1997,LefevreBiroli:2007,Cardy:2008,BenitezETAL:2016,CorrealesETAL:2022}.
In a sense, the fields of a DP field theory are proxies, such that after expressing a desired observable in terms of fields according to the operators, the expectation of these fields is identical to that of the observable.

Drawing on the wealth of knowledge and intuition available for the construction 
of master equations, it is easy to incorporate 
into a DP field theory
a wide range of terms, including 
reactions, transmutations, interactions, 
pair-potentials or external potentials, as the field theory's action inherits the additivity of
concurrent Poisson processes in a master equation.
Some terms can be incorporated into the FP
operator, others have to be treated perturbatively. 
Pair-interaction terms as derived from
the FPE in \APref{FPE_inter} are usually included in the perturbative part of the action.
Henceforth, we will assume that the full action
\begin{equation}\elabel{def_action_full}
    \action = \action_0 + \action_{\perturbative}
\end{equation}
may contain perturbative terms such that expectations are calculated by
expanding the exponential on the right hand side of
\begin{equation}\elabel{perturbative_expansion}
    \ave{\bullet} 
    = \int\Dint{\phi}\Dint{\phitilde} \bullet \exp{\action} 
    = \ave{\ \bullet 
    \ \Exp{\action_{\perturbative}}}_0 \ ,
\end{equation}
and taking expectations as in \Eref{def_ave0}.
Even without interaction, $\action_{\perturbative}$ may absorb terms of $\FPop$ that are not readily
integrated in \Eref{def_ave0}, so that the solution of the \FPeqn becomes in fact
a perturbation theory. This is illustrated in 
\APref{drift_diffusion_on_ring} for drift-diffusion in an arbitrary,
periodic potential,
in \cite{Garcia-MillanPruessner:2021}
for
Run-and-Tumble particles in a harmonic potential and in
\cite{RobertsPruessner:2022} for boundary tumbling.

\section{Entropy production}
\seclabel{entropy_production_main}
In the present framework, the entropy production can be elegantly expressed in terms
of the bare propagators and the perturbative part of the action. 
At this point, we have not yet allowed for interaction terms, so for the time being, the perturbative terms to be considered are bilinear.
We show in the following that the field-theoretic observables needed to calculate the entropy production, are those to lowest order in the perturbation expansion, implying that convergence of the expansion is not required.
We will derive the entropy production first for a single particle before generalising step-by-step
to multiple 
conserved, interacting
particles.

Following the scheme by Gaspard \citep{Gaspard:2004}
to calculate entropy production in \emph{Markovian} systems,
we draw on the propagator $\ave{\phi(\yvec,t')\phitilde(\xvec,t)}$ as
the probability (density) for a particle to transition from $\xvec$ at
time $t$ to $\yvec$ at time $t'$.
The internal entropy production of an 
evolving degree of freedom may then be written as a functional of the instantaneous
probability (density) $\density(\xvec)$ to find it in state $\xvec$, 
namely
\begin{equation}\elabel{def_entropyProduction}
    \entropyProduction[\density] =
    \SumInt_{\xvec,\yvec}
      \density(\xvec) \Op_{\yvec,\xvec} \left\{ \Ln_{\yvec,\xvec} +
      \ln\left(
      \frac{\density(\xvec)}{\density(\yvec)}
      \right)
      \right\}
\end{equation}
with 
\begin{equation}\elabel{def_Op}
\Op_{\yvec,\xvec} = \lim_{t'\downarrow t} \frac{\plaind}{\plaind t'} \ave{\phi(\yvec,t')\phitilde(\xvec,t)}
\end{equation}
and 
\begin{equation}\elabel{def_Ln}
\Ln_{\yvec,\xvec} = \lim_{t'\downarrow t}
\ln\left(\frac{\ave{\phi(\yvec,t')\phitilde(\xvec,t)}}{\ave{\phi(\xvec,t')\phitilde(\yvec,t)}}\right)
\end{equation}
as we show in \APref{entropy_production}. 
In the presence of only a single particle, the probability density to find it somewhere is identical to the particle number density and we will refer to both simply as the \emph{density}.
\Eref{def_entropyProduction} is the starting
point for the derivation of the entropy production from a DP
field theory.
Much of what follows focuses on how to extract $\Op_{\yvec,\xvec}$
and $\Ln_{\yvec,\xvec}$
from the action.

As the field-theory correctly shows,
\APref{entropy_production},
if the states
$\yvec,\xvec$ are
discrete and the process is a simple Markov chain, $\Op_{\yvec,\xvec}$ reduces to the Markov (rate) matrix 
$\TransitionRate_{\yvec\xvec}$
of the
process of transitioning from $\xvec$ to $\yvec$, and $\Ln_{\yvec,\xvec}$
is the logarithm of ratios of these rates, 
\begin{equation}\elabel{Ln_discrete_case}
\Ln_{\yvec,\xvec} = \ln\left(
\frac{\TransitionRate_{\yvec\xvec}}{\TransitionRate_{\xvec\yvec}}
\right) \ ,
\end{equation}
\Erefs{Kn_from_TransitionRate} and \eref{Ln_from_TransitionRate}.
If the states $\xvec,\yvec$ are continuous, then $\Op_{\yvec,\xvec}$
can be cast as a kernel, 
which in the absence of a perturbative contribution to the action is
identical to the FP kernel, $\Op_{\yvec,\xvec}=\FPop_{\yvec,\xvec}$, 
given that the propagator is the Green function of the FPE, \APref{entropy_production}. Integrating by parts then gives
\begin{equation}\elabel{def_entropyProduction_elegant}
    \entropyProduction[\density] = 
    \SumInt_{\xvec,\yvec}\density(\xvec)\delta(\yvec-\xvec)
      \FPophat^\dagger_{\yvec} \left\{ \Ln_{\yvec,\xvec} + 
\ln\left(
      \frac{\density(\xvec)}{\density(\yvec)}
      \right)
\right\} \ .
\end{equation}

In principle, the density $\density(\xvec)$ to use in \Erefs{def_entropyProduction} and \eref{def_entropyProduction_elegant}
is given by the propagation from the initial state up until time $t$,
in which case it becomes an explicit function of $t$
\begin{equation}
    \density(\xvec;t)
    =\ave{\phi(\xvec,t)\phitilde(\xvec_0,t_0)} \ .
\end{equation}
In general, this density might be well approximated by an effective theory, that omits the microscopic details entering into the entropy production via \Erefs{def_Op} and \eref{def_Ln}.

The entropy production \Erefs{def_entropyProduction} and \eref{def_entropyProduction_elegant} simplifies further if $\density(\xvec)$ is stationary,
in a single particle system
\begin{equation}
    \density(\xvec) = \SumInt_{\yvec} \ave{\phi(\xvec,t')\phitilde(\yvec,t)} \density(\yvec)
\end{equation}
for any $t'-t>0$, 
in which case $\ln(\density(\xvec)/\density(\yvec))$ disappears from
\Eref{def_entropyProduction} and the expression reduces to that of the negative of the external entropy production
\cite{CocconiGarcia-MillanETAL:2020}.
In that case,
\Eref{def_entropyProduction} may be interpreted as the spatial average $\entropyProduction=\spave{\entropyProductionDensity(\xvec)}=\SumInt_{\xvec} \density(\xvec) \entropyProductionDensity(\xvec) $ of the \emph{local} entropy production
\begin{equation}\elabel{def_local_entropy}
    \entropyProductionDensity(\xvec) = \SumInt_{\yvec} \Op_{\yvec,\xvec} \Ln_{\yvec,\xvec} \ ,
\end{equation}
which derives from the dynamics only and is independent of the density. We will discuss the formalism in this form in greater detail below, after introducing interaction.

A priori, the full propagator is needed in \Erefs{def_Op} and
\eref{def_Ln}. However, as it turns out, provided the 
process is time-homogeneous,
generally in the discrete case
as well as in continuous perturbation theories about a Gaussian (details
in \APref{drift_diffusion_on_ring}), 
$\Op_{\yvec,\xvec}$ and $\Ln_{\yvec,\xvec}$ draw only on the bare propagator and
possibly on the first order perturbative term.
As detailed in
\APref{entropy_production}, 
the key argument for this simplification
is that 
the propagator
$\ave{\phi(\yvec,t')\phitilde(\xvec,t)}$
only ever enters in the limit $t'\downarrow t$,
either in the form of an explicit derivative, \Eref{def_Op},
or in the form of a ratio,
\Eref{def_Ln}, which may also draw on the derivative via L'H{\^o}pital.
The propagator therefore needs to be determined only to first order in
small $t'-t$. 
If the full propagator
$\ave{\phi(\yvec,t')\phitilde(\xvec,t)}$ is given by a perturbative
expansion of the action \Eref{perturbative_expansion}, diagrammatically
written as
\begin{widetext}
\begin{equation}\elabel{propagator_expansion}
\ave{\phi(\yvec,t')\phitilde(\xvec,t)} \corresponds
\tbarePropagator{\xvec,t}{\yvec,t'} + 
\tblobbedPropagator{\xvec,t}{\yvec,t'} +
\tDblobbedPropagator{\xvec,t}{\yvec,t'} + \ldots\ ,
\end{equation}
\end{widetext}
in principle every order in the perturbation theory might contribute to
$\ave{\phi(\yvec,t')\phitilde(\xvec,t)}$ to first order in $t'-t$.
As outlined in the following,
closer inspection, however, reveals a simple relationship, namely that
\emph{the $n$th order in $t'-t$ is fully given by the first $n+1$ diagrams} on the
right hand side of \Eref{propagator_expansion}. In continuum field theories,
this needs careful analysis, but it holds for perturbation theories about drift-diffusion, \APref{drift_diffusion_on_ring}, where the highest order derivative in $\Op_{\yvec,\xvec}$ is a second and $\Ln_{\yvec,\xvec}$, necessarily odd in $\yvec-\xvec$, therefore does not need to be known beyond second order.

Leaving the details to \APref{entropy_production}
and \APref{appendixWhichDiagramsContribute},
we proceed by demonstrating that the first time derivative of the
second order contribution 
\tDblobbedPropagator{}{}
on the right hand side of
\Eref{propagator_expansion}
vanishes at $t'=t$. This follows from differentiating with respect to
$t'$ the inverse Fourier-transform of 
\tDblobbedPropagator{}{},
which for
time-homogeneous processes has the form
\begin{equation}\elabel{derivative_integral}
    \dot{I}(t'-t)=\int \dintbar{\omega'}
    \frac{-\imag \omega' \exp{-\imag \omega' (t'-t)} C}
    {\prod_{j=1}^{3} (-\imag \omega' + p_j)} \ ,
\end{equation}
where the three propagators are $(-\imag \omega' + p_j)^{-1}$
with $j=1,2,3$ and $C$ denotes the couplings. 
The poles $-\imag p_j$ may be repeated, which does not
affect the argument. Crucially, all poles are situated in the
\emph{lower} half-plane, which is required by causality of each bare propagator
entering in \Eref{derivative_integral}. After taking $t'\to t$ the
contour can be closed in the \emph{upper} half-plane, as the integrand $\propto 1/\omega'^2$ decays
fast enough. It follows that $\dot{I}(0)$, \Eref{derivative_integral},  vanishes.
As shown in \APref{general_and_limit}, it further follows that $\lim_{t'\downarrow t}\dot{I}(t'-t)=0$.
These arguments easily generalise to higher derivatives and
correspondingly higher orders. Consequently, only the first two diagrams on the right hand side of \Eref{propagator_expansion} contribute to the propagator to first order in $t'-t$.

The argument above draws on the structure of the diagrams where bare
propagators connect ``blobs''. 
The diagrams 
in the propagators of \Erefs{def_Op} and \eref{def_Ln}
that end up 
contributing, contain at most one such blob. How the blob enters into $\Op_{\yvec,\xvec}$ and $\Ln_{\yvec,\xvec}$ is explained in the following.
The blobs can contain tadpole-like
loops only in the presence of source terms, such as \Eref{tadpole_diagram}. If such source terms are
absent, \emph{the blobs are merely the vertices of the perturbative part
of the action}. If the phenomenon studied is not time-homogeneous,
$\omega$ might have sinks and sources and the structure of the integrals
representing contributions to the propagator are no longer of the form
\Eref{derivative_integral}.

With these provisos in place, the kernel in \Eref{def_Op} reduces to the bare propagator plus the first order correction,
\begin{equation}\elabel{transition_from_action}
    \Op_{\yvec,\xvec}\corresponds
    \FPop_{\yvec,\xvec} + \fullBlobb{\xvec}{\yvec}
    \ ,
\end{equation}
where $\FPop$ refers to the non-perturbative part of the action \Eref{def_action0},
and $\tfullBlobb{}{}$ to the first order, one-particle 
irreducible, amputated contributions due to the perturbative part of the
action, the ``blob''. It may contain perturbative contributions due to the single-particle Fokker-Planck operator, or due to additional processes, such as interactions with external fields and reactions.
In field-theoretic terms, $\FPop_{\yvec,\xvec}$ is the inverse bare propagator 
evaluated at $\omega=0$ and $\tfullBlobb{}{}$ is a contribution to the ``self-energy''. 
In stochastic particle systems, 
$\FPop_{\yvec,\xvec}=(D\partial_y^2-w\partial_y)\delta(y-x)$ 
would be drift-diffusion (\APref{drift_diffusion_on_ring}) and 
$\tfullBlobb{}{}=-r$ an additional extinction with rate $r$, 
as the simplest possible perturbation, which would, of course, normally be dealt with in closed form.
In other systems, 
$\tfullBlobb{}{}$ may represent transmutation or an
external potential. Below it is extended to
pair interaction and particle collisions.
Generally, no higher orders, such as
$\tfullDoubleBlobb{}{}$,
or any loops carrying $\omega$, such as the middle term in \Eref{Npropagators_oneLoop},
enter (\APref{appendixWhichDiagramsContribute}). 
While maybe unsurprising as far as the kernel $\Op_{\yvec,\xvec}$ is concerned, 
this simplification to one blob carries through to the logarithm
on the basis of \Erefs{Ln_discrete_case} 
and \eref{Ln_from_propagator}
if states are discrete, and by an expansion of the form
\begin{widetext}
\begin{equation}\elabel{Ln_from_propagators}
\Ln_{\yvec,\xvec} \corresponds
\lim_{t'\downarrow t}\left\{
\ln
\left(
\frac{
\tbarePropagator{\xvec,t}{\yvec,t'}
}{
\tbarePropagator{\yvec,t}{\xvec,t'}
}
\right)
+
\frac
{\tblobbedPropagator{\xvec,t}{\yvec,t'}}
{\tbarePropagator{\xvec,t}{\yvec,t'}}
-
\frac
{\tblobbedPropagator{\yvec,t}{\xvec,t'}}
{\tbarePropagator{\yvec,t}{\xvec,t'}}
\right\}
\ ,
\end{equation}
\end{widetext}
in the continuum, \Eref{Ln_for_continuous} and similarly \APref{drift_diffusion_on_ring}, 
\Eref{Ln_from_drift_diffusion}.

In summary, what is needed 
to calculate the entropy production
\Eref{def_entropyProduction} 
of a single degree of freedom
is: (a) the density $\density(\xvec;t)$, which at stationarity may be well
approximated by an effective
theory,
and (b) the \emph{microscopic} 
action \Eref{def_action_full} to construct kernel $\Op_{\yvec,\xvec}$,
via \Eref{transition_from_action},
and logarithm $\Ln_{\yvec,\xvec}$ via \Eref{Ln_from_propagators} using at most one blob.

\subsection{Many conserved particles}
\seclabel{entropy_production_multiple_main}
In the presence of $N>1$ distinguishable particles,
\Eref{def_entropyProduction} remains
in principle
valid if $\xvec,\yvec$ are understood
to encapsulate all $N$ particle coordinates at once, with the 
density in \Eref{def_entropyProduction} replaced by the joint density $\density(\xvec_1,\ldots,\xvec_N)$
and the 
propagator
in \Erefs{def_Op} and \eref{def_Ln} replaced by the joint propagator
$\bigl\langle\phi_1(\yvec_1,t')\linebreak[1]\phi_2(\yvec_2,t')\linebreak[1]\ldots\linebreak[1]\phi_N(\yvec_N,t')\linebreak[1]\phitilde_1(\xvec_1,t)\linebreak[1]\phitilde_2(\xvec_2,t)\linebreak[1]\ldots\linebreak[1]\phitilde_N(\xvec_N,t)\bigr\rangle$,
where the indices of the fields refer to
\emph{distinguishable} particle species.
Each field $\phi_i(\yvec_i,t')$ has the effect of an indicator function probing for the presence of particle $i$.
Without interaction, the overall entropy production is the sum of the individual entropy productions, \Eref{entropyProductionDensity_independent_distinguishable_final_sum}.
If particles are \emph{indistinguishable}, dropping the indices
generally results in $N!$ as many terms from permutations of the fields,
as well as the joint particle number density $\HMdensity{N}{}(\xvec_1,\ldots,\xvec_N)$ 
at stationarity 
being $N!$ times that of equivalent
distinguishable particles.
Yet, for distinct $\xvec_i$ the density $\HMdensity{N}{}(\xvec_1,\ldots,\xvec_N)$ is clearly still a \emph{probability} density, provided the probability or phase space is adjusted to account for distinct arrangements of indistinguishable particles. This adjustment affects the sum or integral in \Eref{def_entropyProduction} and reflects that occupation numbers are the degrees of
freedom, not particle positions \cite{CocconiGarcia-MillanETAL:2020,ZhangGarcia-Millan:2023}.
In the case of sparse
occupation, where every site is occupied by
at most one particle, a condition usually met in continuum space,
this adjustment can be done by 
means of the Gibbs factor \cite{Sethna:2006}, 
which amounts to
dividing the corresponding phase space of distinguishable
particles by $N!$,
\begin{widetext}
\begin{equation}\elabel{entropyProduction_multipleParticles}
    \entropyProduction^{(N)}[\density] =
    \frac{1}{(N!)^2}
    \int\ddint{x_1}\ldots\ddint{x_N}
    \int\ddint{y_1}\ldots\ddint{y_N}
    \HMdensity{N}{}
    (\xvec_1,\ldots,\xvec_N)
    \Op^{(N)}_{\yvec_1,\ldots,\yvec_N,\xvec_1,\ldots,\xvec_N} 
    \Ln^{(N)}_{\yvec_1,\ldots,\yvec_N,\xvec_1,\ldots,\xvec_N} 
\end{equation}
\end{widetext}
with $N$-particle kernel 
$\Op^{(N)}_{\yvec_1,\ldots,\xvec_N}$
and logarithm 
$\Ln^{(N)}_{\yvec_1,\ldots,\xvec_N}$
defined by using the joint propagator 
$\bigl\langle\phi(\yvec_1,t')\linebreak[1]\ldots\linebreak[1]\phitilde(\xvec_N,t)\bigr\rangle$
on the right of \Erefs{def_Op} and \eref{def_Ln}.
As $\xvec_1,\ldots,\xvec_N$ and $\yvec_1,\ldots,\yvec_N$ are only dummy variables, the Gibbs-factor precisely cancels the multiplicity of the terms mentioned above, further discussed in \APref{N_independent_indistinguishable_particles}. 
Again, at sparse occupation, the annihilator fields $\phi(\yvec_i,t')$ can be seen as indicators of the presence of, in the case of indistinguishability, \emph{any} particle at $\yvec_i$. However, the assumption of sparse occupation is needed here only as to simplify the discussion. Doi-Peliti field theory, whose joint propagators give the expectation of the factorials of non-zero occupation numbers, is constructed precisely as to adjust the Gibbs factor by the correct multiplicity in the case of multiple occupation. 
Reassuringly, we find again, \Eref{entropy_production_indistinguishableN_as_density_independent3}, that the entropy production of $N$ indistinguishable particles is linear in $N$. These basic insights are the preliminaries to the complexities posed by interaction to be discussed in the next section.

The diagrams contributing to the joint propagator are generally disconnected, say
$
\tikz[baseline=-2.5pt]{
\draw[tAactivity] (0.5,0.12) -- (-0.5,0.12);
\draw[tAactivity] (0.5,0.0) -- (-0.5,0.0);
\draw[tAactivity] (0.5,-0.12) -- (-0.5,-0.12);
}
$
and may, 
in principle, involve any number of vertices, such as
$\tfullBlobb{}{}$, say 
$
\tikz[baseline=-2.5pt]{
\tgenVertex{0,0.2}
\draw[tAactivity] (0.5,0.2) -- (-0.5,0.2);
\draw[tAactivity] (0.5,0.03) -- (-0.5,0.03);
\draw[tAactivity] (0.5,-0.12) -- (-0.5,-0.12);
}
$
or
$
\tikz[baseline=-2.5pt]{
\tgenVertex{0,0.2}
\draw[tAactivity] (0.5,0.2) -- (-0.5,0.2);
\tgenVertex{0,0.0}
\draw[tAactivity] (0.5,0.0) -- (-0.5,0.0);
\draw[tAactivity] (0.5,-0.15) -- (-0.5,-0.15);
}
$. 
However, as detailed in \APref{appendixWhichDiagramsContribute},
the argument that 
reduces 
to at most one vertex
contributions 
to a single particle propagator,
similarly applies to multiple particle propagators, so that any blob inside a joint propagator raises the order of $t'-t$ by one.
Any contribution to the joint propagator
$\bigl\langle\phi(\yvec_1,t)\linebreak[1]\ldots\linebreak[1]\phitilde(\xvec_N,t)\bigr\rangle$
in the joint kernel $\Op_{\yvec_1,\ldots,\xvec_N}$ or the joint logarithm $\Ln_{\yvec_1,\ldots,\xvec_N}$
therefore 
contains at most one vertex.
The set of diagrams to be considered can be reduced further with an argument best made after allowing for interaction.

\subsection{Interaction}
\seclabel{interaction_main}
In the presence of interaction, the joint propagator contains contributions of the form 
$
\tikz[baseline=-2.5pt]{
\tgenVertex{0,0}
\draw[tAactivity] (0.5,0.1) -- (0,0) -- (-0.5,0.1);
\draw[tAactivity] (0.5,-0.1) -- (0,0) -- (-0.5,-0.1);
}
$. 
Each such vertex is also of order $t'-t$,
\APref{appendixWhichDiagramsContribute}. If particles are conserved, each vertex must 
have at least as many incoming legs as outgoing
ones.
At this stage, the joint propagator entering the $N$ particle kernel 
$\Op^{(N)}$
and logarithm 
$\Ln^{(N)}$
is of the form
\newcommand{\propWidth}{0.5}
\begin{widetext}
\begin{multline}\elabel{def_trans_multi_diag}
\ave{
\phi(\yvec_1,t')
\ldots\phi(\yvec_N,t')\phitilde(\xvec_1,t)
\ldots\phitilde(\xvec_N,t)
}\\
\corresponds
\tikz[baseline=-2.5pt]{
\begin{scope}[yshift=0.3cm]
\node at (\propWidth,0) [right] {$\xvec_1,t$};
\node at (-\propWidth,0) [left] {$\yvec_1,t'$};
\draw[tAactivity] (\propWidth,0) -- (-\propWidth,0);
\end{scope}
\begin{scope}[yshift=0.0cm]
\node at (\propWidth,0) [right] {$\xvec_2,t$};
\node at (-\propWidth,0) [left] {$\yvec_2,t'$};
\draw[tAactivity] (\propWidth,0) -- (-\propWidth,0);
\end{scope}
\node at (0,-0.2) {$\vdots$};
\begin{scope}[yshift=-0.65cm]
\node at (\propWidth,0) [right] {$\xvec_N,t$};
\node at (-\propWidth,0) [left] {$\yvec_N,t'$};
\draw[tAactivity] (\propWidth,0) -- (-\propWidth,0);
\end{scope}
}
+\text{perm.}+
\tikz[baseline=-2.5pt]{
\begin{scope}[yshift=0.3cm]
\tgenVertex{0,0}
\node at (\propWidth,0) [right] {$\xvec_1,t$};
\node at (-\propWidth,0) [left] {$\yvec_1,t'$};
\draw[tAactivity] (\propWidth,0) -- (-\propWidth,0);
\end{scope}
\begin{scope}[yshift=0.0cm]
\node at (\propWidth,0) [right] {$\xvec_2,t$};
\node at (-\propWidth,0) [left] {$\yvec_2,t'$};
\draw[tAactivity] (\propWidth,0) -- (-\propWidth,0);
\end{scope}
\node at (0,-0.2) {$\vdots$};
\begin{scope}[yshift=-0.65cm]
\node at (\propWidth,0) [right] {$\xvec_N,t$};
\node at (-\propWidth,0) [left] {$\yvec_N,t'$};
\draw[tAactivity] (\propWidth,0) -- (-\propWidth,0);
\end{scope}
}
+\text{perm.}+
\tikz[baseline=-2.5pt]{
\begin{scope}[yshift=0.3cm]
\tgenVertex{0,-0.15}
\node at (\propWidth,0) [right] {$\xvec_1,t$};
\node at (-\propWidth,0) [left] {$\yvec_1,t'$};
\draw[tAactivity] (\propWidth,0) -- (0,-0.15) -- (-\propWidth,0);
\node at (\propWidth,-0.3) [right] {$\xvec_2,t$};
\node at (-\propWidth,-0.3) [left] {$\yvec_2,t'$};
\draw[tAactivity] (\propWidth,-0.3) -- (0,-0.15) -- (-\propWidth,-0.3);
\end{scope}
\node at (0,-0.2) {$\vdots$};
\begin{scope}[yshift=-0.65cm]
\node at (\propWidth,0) [right] {$\xvec_N,t$};
\node at (-\propWidth,0) [left] {$\yvec_N,t'$};
\draw[tAactivity] (\propWidth,0) -- (-\propWidth,0);
\end{scope}
}
+\text{perm.}
+\order{(t'-t)^2}
\ ,
\end{multline}
\end{widetext}
each with all distinct permutations of incoming and outgoing particle coordinates, $\xvec_i$ and $\yvec_i$ respectively, as indicated by ``$\text{perm.}$''. What does \emph{not}
enter (\APref{appendixWhichDiagramsContribute}) are terms 
involving more than one vertex, such as
\renewcommand{\propWidth}{0.7}
\begin{equation}\elabel{Npropagators_oneLoop}
\tikz[baseline=-2.5pt]{
\begin{scope}[yshift=0.5cm]
\tgenVertex{0,0}
\draw[tAactivity] (\propWidth,0) -- (-\propWidth,0);
\end{scope}
\begin{scope}[yshift=0.2cm]
\tgenVertex{0,0}
\draw[tAactivity] (\propWidth,0) -- (-\propWidth,0);
\end{scope}
\begin{scope}[yshift=0.00cm]
\draw[tAactivity] (\propWidth,0) -- (-\propWidth,0);
\end{scope}
\node at (0,-0.22) {$\vdots$};
\begin{scope}[yshift=-0.65cm]
\draw[tAactivity] (\propWidth,0) -- (-\propWidth,0);
\end{scope}
}
\,\text{ or }\,
\tikz[baseline=-2.5pt]{
\begin{scope}[yshift=0.15cm]
\tgenVertex{0.35,0.15}
\tgenVertex{-0.35,0.15}
\draw[tAactivity] (\propWidth,0.3) -- (0.35,0.15) to[out=-135,in=-45] (-0.35,0.15) -- (-\propWidth,0.3);
\draw[tAactivity] (\propWidth,0.0) -- (0.35,0.15) to[out=135,in=45] (-0.35,0.15) -- (-\propWidth,0.0);
\end{scope}
\begin{scope}[yshift=0.00cm]
\draw[tAactivity] (\propWidth,0) -- (-\propWidth,0);
\end{scope}
\node at (0,-0.2) {$\vdots$};
\begin{scope}[yshift=-0.65cm]
\draw[tAactivity] (\propWidth,0) -- (-\propWidth,0);
\end{scope}
}
\,\text{ or }\,
\tikz[baseline=-6pt]{
\begin{scope}[yshift=0.00cm]
\tgenVertex{-0.35,0.2}
\tgenVertex{0.35,-0.2}
\draw[tAactivity] (\propWidth,0.2) -- (-0.35,0.2) -- (-\propWidth,0.2);
\draw[tAactivity] (\propWidth,-0.2) -- (0.35,-0.2) -- (-0.35,0.2)  -- (-\propWidth,0.0);
\draw[tAactivity] (\propWidth,0.0) -- (0.35,-0.2)  -- (-\propWidth,-0.2);
\end{scope}
\node at (0,-0.35) {$\vdots$};
\draw[tAactivity] (\propWidth,-0.8) -- (-\propWidth,-0.8);
}
\ .
\end{equation}
Even with the restriction to a single blob, \Eref{def_trans_multi_diag} contains many diagrams, seemingly involving many permutations of many initial and final coordinates. Similarly, the $N$-point equal time density 
$\HMdensity{N}{}(\xvec_1,\ldots,\xvec_N)$ 
is needed in \Eref{entropyProduction_multipleParticles}, which would be an arduous task to determine.
However, because every bare propagator degenerates into a $\delta$-function as $t'\downarrow t$, they simplify the expression for the entropy production considerably. 
As discussed in \APref{MultipleParticles} any bare propagator featuring together with, \ie multiplying, a blobbed diagram, effectively drops out in the limit $t'\downarrow t$. As the bare propagators drop away, so does the need for higher order densities. As a result
the entropy production of a system whose ``largest blob" has $n$ incoming and $n$ outgoing legs can be calculated on the basis of the $(2n-1)$-point joint density, restricting a hierarchy of terms to $2n-1$ rather than $N$ \cite{LynnETAL:2022}.

For example, 
$N$ indistinguishable particles with self-propulsion speed $\vecDrift$, diffusion $\diffusion$ and pair-interaction, $n=2$,  via an even potential $\pairPot$ have entropy production, 
\begin{widetext}
\begin{subequations}
\elabel{entropyProduction_for_pairPot}
\begin{align}
\entropyProduction^{(N)}[\HMdensity{N}{}] 
=&
\elabel{entropyProduction_for_pairPot1}
\int \ddint{x_1} \HMdensity{N}{1}(\xvec_1) \left\{\frac{\vecDrift^2}{\diffusion}\right\}\\
&+
\elabel{entropyProduction_for_pairPot2}
\int 
\ddint{x_{1,2}}
\HMdensity{N}{2}(\xvec_1,\xvec_2)
\left\{
\frac{1}{\diffusion} 
\left(\nabla\pairPot(\xvec_1-\xvec_2)\right)^2
-\laplace
\pairPot(\xvec_1-\xvec_2) 
\right\}\\
&+
\elabel{entropyProduction_for_pairPot3}
\int \ddint{x_{1,2,3}} \HMdensity{N}{3}(\xvec_1,\xvec_2,\xvec_3) 
\left\{
\frac{1}{\diffusion}
\nabla\pairPot(\xvec_1-\xvec_2)
\cdot
\nabla\pairPot(\xvec_1-\xvec_3)
\right\} \ ,
\end{align}
\end{subequations}
\end{widetext}
which is \Eref{entropyProduction_interacting_indistinguishable_example} with external potential $\extPot\equiv0$ and further simplified by using that the pair potential $\pairPot$ is even. It
demonstrably vanishes in the absence of a drift $\vecDrift$, as shown in \APref{no_EPR_without_drift}. 
\Eref{entropyProduction_for_pairPot} and \eref{entropyProduction_interacting_indistinguishable_example} are \emph{exact} results, assuming pair interaction being the highest order interaction.
\gpcomment{Keep your eyes pealed for $\density$ being called or not a density. It's always a particle number density, but it's a probability density only with the right probability space.}
The densities $\HMdensity{N}{n}(\xvec_1,\ldots,\xvec_n)$ thus denote the density of $n$ \emph{distinct} particles at positions $\xvec_1,\ldots\xvec_n$, normalising to $N!/(N-n)!$. Each term in curly brackets in \Eref{entropyProduction_for_pairPot} can be cast as a local entropy production, depending on one, two or three coordinates.
\Eref{entropyProduction_for_pairPot1} is the entropy production or work due to self-propulsion by individual particles, $N\vecDrift^2/\diffusion$,
\Eref{entropyProduction_for_pairPot2} is the work due to two particles excerting equal and opposite forces on each other and
\Eref{entropyProduction_for_pairPot3} is the work performed by one particle in the potential of another particle as it is being pushed or pulled by a third particle.

The integrals in \Eref{entropyProduction_for_pairPot} can be carried out and rendered as expectations,
\begin{widetext}
\begin{align}
\elabel{entropyProduction_for_pairPot_expectation}
\entropyProduction^{(N)}
&= N \frac{\drift^2}{\diffusion} 
+ N(N-1) \ave{
\frac{1}{\diffusion} 
\big(\nabla\pairPot(\xvec_1-\xvec_2)\big)^2
-\laplace
\pairPot(\xvec_1-\xvec_2) 
} \\
&\quad + 
N(N-1)(N-2)
\ave{
\frac{1}{\diffusion}
\nabla\pairPot(\xvec_1-\xvec_2)
\cdot
\nabla\pairPot(\xvec_1-\xvec_3)
} \nonumber\ ,
\end{align}
where each expectation is taken over all pairs of distinct particles, all triplets of distinct particles \etc, so that, for example, $\ave{(\xvec_1-\xvec_2)^2}$ would represent the average mean squared distance between any two particles. Using that $\nabla\pairPot$ is odd, one can write \Eref{entropyProduction_for_pairPot_expectation} as 
\begin{equation}
\elabel{entropyProduction_for_pairPot_expectation_rewrite}
\entropyProduction^{(N)}
= 
\sum_{i=1}^N \ave{\frac{1}{\diffusion} \big(
\vecDrift-\sum_{\substack{j=1\\ j\ne i}}^N \nabla\pairPot(\xvec_i-\xvec_j) \big)^2}
-
\sum_{i=1}^N \sum_{\substack{j=1\\ j\ne i}}^N \bigg\langle\!\!\laplace \pairPot(\xvec_i-\xvec_j)\bigg\rangle \ ,
\end{equation}
where the expectation is the ensemble average.


In the form of \Eref{entropyProduction_for_pairPot}, entropy production in an experiment or simulation can be estimated efficiently by using $Q$ samples $q=1,2,\ldots,Q$ of $N$ particle coordinates $\xvec^{(q)}_{i}$ with $i=1,\ldots,N$, 
\begin{equation}\elabel{interacting_EPR_numerics}
\entropyProduction=\frac{1}{Q}\sum_{q=1}^Q
\left[
\sum_{i_1=1}^N  
\entropyProductionDensity^{(1)}_1\left(\xvec^{(q)}_{i_1}\right)
+
\sum_{\substack{i_1,i_2=1\\ i_1\ne i_2}}^N 
\entropyProductionDensity^{(2)}_2\left(\xvec^{(q)}_{i_1},\xvec^{(q)}_{i_2}\right)
+
\sum_{\substack{i_1,i_2,i_3=1\\ i_1\ne i_2\ne i_3\ne i_1}}^N 
\entropyProductionDensity^{(3)}_3\left(\xvec^{(q)}_{i_1},\xvec^{(q)}_{i_2},\xvec^{(q)}_{i_3}\right)
\right]\ .
\end{equation}
\end{widetext}
with the $\entropyProductionDensity^{(i)}_i$ with $i=1,2,3$ given by the three pairs of curly brackets in \Eref{entropyProduction_for_pairPot} and generally in \Eref{def_entropyProductionDensities_indistinguishable}. 
Explicit expressions are shown in \Erefs{ABP_EPR_succinct} and \eref{entropyProduction_interacting_indistinguishable_example_numerics}, \APref{N_interacting_indistinguishable_particle_extPot}.
If all interactions are harmonic, at most two-point correlation functions $\ave{\xvec_i\cdot\xvec_j}$ are needed to calculate the entropy production.
All sums run over distinct particle indices, so that for example
$(1/Q)\sum_q 
\sum_{\substack{i_1,i_2=1\\ i_1\ne i_2}}^N 
\delta(\xvec_1-\xvec^{(q)}_{i_1})\delta(\xvec_2-\xvec^{(q)}_{i_2})$ estimates $\HMdensity{N}{2}(\xvec_1,\xvec_2)$.
Entropy production in a particle system with interaction can thus be estimated on the basis  of ``snapshots" and the microscopic action, without the need of introducing a new measure \cite{RoETAL:2021,TociuETAL:2020}. In the case of $n$-particle interaction, it generally draws on equal-time $(2n-1)$-point densities and the time-evolution terms given by the action. Neither the full $N$-point density nor the $2N$-point two-time correlation function are needed, which is what \Eref{entropyProduction_multipleParticles} suggests.

If the number of particles is not
fixed but becomes itself a degree of freedom, the phase space integrated or 
summed over in 
\Eref{def_entropyProduction} needs to be 
adjusted. This case is beyond the scope of
the present work.

\section{Examples}
\seclabel{examples_main}
In the following we illustrate the methods introduced above by calculating the
entropy production of 
1) a continuous time Markov chain,
2) a drift-diffusion Brownian particle on a torus with potential,
3) two drift-diffusion particles on a circle interacting via a 
harmonic pair potential,
and
4) interacting Active Brownian Particles.
The first three examples serve also as a sanity check, because the entropy production can be calculated straight-forwardly by physical reasoning from first principles and we demonstrate how it is reproduced by the present field-theoretic framework. The last example illustrates its power as there are few alternative schemes \cite{RoldanETAL:2021}, none of which, to our knowledge, based on field theory.
%
The Sekimoto-scheme 
\cite{Sekimoto:2010}
we compare our results to, in particular, requires access to \emph{consecutive} frames, which is not required in our scheme and is not easily available when analysing experimental data.

\paragraph*{Continuous time Markov chain.}
The single-particle master equation of a continuous time Markov chain is \Eref{FPeqn_main}
with $\FPop_{\yvec\xvec}$ the Markov-matrix $\transMatLin_{\yvec\xvec}$ for transitions from
discrete state
$\xvec$ to $\yvec$.
Following standard procedure \cite{TaeuberHowardVollmayr-Lee:2005},
\APref{DP_plus_EP},
the
action of the resulting field theory is \Eref{def_action0},
\begin{equation}
    \action_0 = \sum_{\xvec\yvec} \dint{t}  \phitilde(\yvec,t) (\transMatLin_{\yvec\xvec} - \delta_{\yvec,\xvec}\partial_t) \phi(\xvec,t)
    \ .
\end{equation}
From \Erefs{def_entropyProduction}, 
\eref{Ln_discrete_case}
and 
\eref{transition_from_action} in the absence of a perturbative term,
the entropy production immediately follows,
\begin{equation}\elabel{EntropyProductionMarkovChain}
    \entropyProduction[\density] = 
    \sum_{\xvec,\yvec}
      \transMatLin_{\yvec\xvec}\density(\xvec) 
      \ln\left(
      \frac
      {\transMatLin_{\yvec\xvec}\density(\xvec)}
      {\transMatLin_{\xvec\yvec}\density(\yvec)}
      \right) \ ,
\end{equation}
\APref{EP_from_propagators}, consistent with \cite{Gaspard:2004,CocconiGarcia-MillanETAL:2020}.
The contributions to first order in the perturbative vertex in
\Erefs{transition_from_action}, \eref{Op_in_diagrams} and \eref{Ln_from_propagator}
ensure that this expression for the entropy production
does not change even when some 
contributions to
$\transMatLin$
are moved to the perturbative part of the action, $\transMatNonLin$ in \Erefs{Op_in_diagrams} and \eref{Ln_from_propagator}.

\paragraph*{Drift-diffusion.}
This process is a paradigmatic example of a continuous space process as many
other, more complicated ones, 
in particular many active matter models \cite{ZhangPruessner:2022,ZhangETAL:2024,ZhenPruessner:2022}
can be studied as a perturbation of it.
The continuity of the degree of freedom means that a transform is needed
to render the process local in a new variable, here the Fourier-mode $\kvec$, so
that the path-integral \Eref{perturbative_expansion} can be performed. However, as detailed in
\APref{drift_diffusion_on_ring}, the transform
can in principle spoil the relationship between the number of blobs in the diagram and its order in $t'-t$ as discussed around \Eref{derivative_integral}. 
It turns out, \APref{Fourier_transformation}, 
in particular after \Erefs{first_order_bauble_t_x_final} and \eref{first_order_bauble_t_x_final2_fullPot_all},
that in the case of drift-diffusion processes, the number of blobs determines the leading order in the distance $\yvec-\xvec$ of any contribution finite in the limit $t'\downarrow t$, \eg \Erefs{first_order_bauble_t_x_final} and \eref{trick3}, in fact preserving \Erefs{transition_from_action} and \eref{Ln_from_propagators}.

To be general, we allow for an external potential, but to render drift diffusion stationary even without an external potential, we restrict it to a $d$-dimensional torus with circumference $L$.
As detailed in \APref{drift_diffusion_on_ring}, 
the drift can be captured either exactly or perturbatively, while a general
external potential has to be treated perturbatively.
The FPE of a particle diffusing with constant $D$ 
and drifting with velocity $\vecDrift$ on a torus with periodic
external potential $\extPot(\yvec)$ is
\Eref{FPeqn_main} with 
$\FPophat_\yvec=D\nabla_{\yvec}^2 + \nabla_{\yvec} (- \vecDrift +\extPot'(\yvec))$, where 
the operators act on everything to the right and
$\extPot'=\nabla\extPot$ denotes the gradient of the potential.
The propagator to first order is (\APref{drift_diffusion_on_ring},
\Erefs{phiphitilde_in_diagrams_all})
\begin{widetext}
\begin{align}\elabel{entropy_production_drift_diffusion_main}
\ave{\phi(\yvec,t')\phitilde(\xvec,t)}
= &
\frac{\theta(t'-t)\exp{-\frac{(\yvec-\xvec)^2}{4D(t'-t)}}}{(4\pi D(t'-t))^{d/2}} 
\left(
1 + (\yvec-\xvec)\cdot \frac{\vecDrift-\nabla\extPot(\xvec)}{2 D} + \order{(\yvec-\xvec)^2}
\right) \nonumber\\
\corresponds &
\tbarePropagator{\xvec,t}{\yvec,t'} 
+
\tblobbedDashedPropagator{\xvec,t}{\yvec,t'}
+
\blobbedDashedPotPropagator{\xvec,t}{\yvec,t'}+
\order{(t'-t)^2} \ ,
\end{align}
\end{widetext}
so that with \Eref{Ln_from_propagators}
\begin{multline}
\Ln_{\yvec,\xvec}=\frac{\yvec-\xvec}{2D} \cdot \big[2\vecDrift -\extPot'(\xvec) - \extPot'(\yvec)\big] \\
+
\order{(\yvec-\xvec)^3} \ ,
\end{multline}
\Erefs{Ln_drift_diffusion_Wissel} and \eref{Ln_drift}.
The diagrams \tblobbedDashedPropagator{}{}
and \blobbedDashedPotPropagatorText{}{} correspond to the first
order corrections to the propagator due to  drift
$\vecDrift$ and external potential $\extPot$, respectively, with dashes indicating spatial gradients.
Using in \Eref{def_entropyProduction} $\Ln_{\yvec,\xvec}$
and
$\Op_{\yvec,\xvec}=\FPophat_\yvec\delta(\yvec-\xvec)$, where the operator acts only on the Dirac $\delta$-function,
produces 
\begin{multline}\elabel{entropyProduction_driftdiffusion_main}
\entropyProduction[\density] = \int_0^L \ddint{x}
      \Biggl\{ \density(\xvec) \left(
      \frac{(\vecDrift-\extPot'(\xvec))^2}{\diffusion} - \laplace \extPot(\xvec)\right)\\
      + D \frac{(\density'(\xvec))^2}{\density(\xvec)} + \extPot'(\xvec)
      \density'(x) \Biggr\}
\end{multline}
with the last two terms that involve $\density'(\xvec)=\nabla_{\xvec} \density(\xvec)$ cancelling at stationarity, 
$\nullvec=\partial_t\density=D\density''-\partial_x(w-\extPot')\density$,
and the first two terms involving the potential cancelling at vanishing current $0=\jvec=-D\density'+(\vecDrift-\extPot')\density$. 
After the first cancellation, \Eref{entropyProduction_driftdiffusion_main} is consistent with the stationary
$\entropyProductionDensity$ in 
\Eref{local_entropy_Gw}, given that $\entropyProduction=\int\ddint{x} \density \entropyProductionDensity$, as well as with \cite{CocconiGarcia-MillanETAL:2020}.


\paragraph*{Harmonic trawlers.}
A free particle with diffusion constant $\diffusion$, drifting with velocity $w$ on a circle without
external potential produces entropy with rate $w^2/D$
\cite{CocconiGarcia-MillanETAL:2020}.
Entropy production being extensive, 
without interaction two identical particles
produce twice as much entropy. If they have different drift velocities
$w_1$ and $w_2$ the total entropy production is $(w_1^2+w_2^2)/D$. If
they are coupled by an attractive (binding) pair-potential, 
as if coupled by a spring,
they behave like a single particle
drifting with velocity $(w_1+w_2)/2$ and diffusing with constant $D/2$, 
so that the overall entropy production is $(w_1+w_2)^2/(2D)$. If
$w_1=w_2$, then the entropy production is identical to that of free
particles, but if $w_1\ne w_2$, the pair potential becomes ``visible''.

While easily derived using physical arguments, determining this expression
perturbatively from a field theory that is ``oblivious" to such physical intricacies is a non-trivial task and a good
litmus test for the power of the scheme presented in this work. 
As detailed in \SMref{HarmonicTrawlers}, the entropy production 
is indeed correctly reproduced, drawing in particular explicitly on
\Eref{def_trans_multi_diag}. The process is generalised to arbitrary attractive  pair-potentials in \APref{generalised_trawlers} and further qualified in \APref{no_EPR_without_drift} where it is confirmed that in the present framework arbitrarily many identical pair-interacting particles do not produce entropy without drift.


\begin{figure*}
\subfloat[Particle configuration with colour-coded ``instantaneous entropy production.''\label{fig:instantEPR}]{
\includegraphics[width=0.44\textwidth]{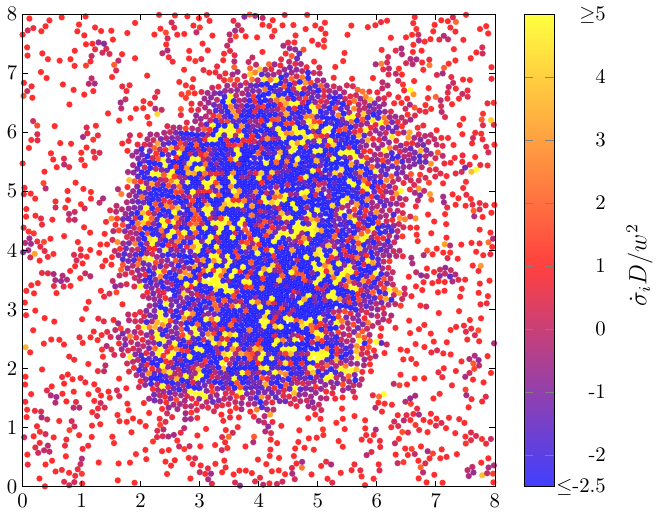}}
\subfloat[Entropy production rate per particle averaged over an increasing
time period]
{\label{fig:EPR_Sekimoto_vs_here}
\resizebox{0.54\textwidth}{!}{\begin{tikzpicture}
\node at (0,0) {\includegraphics[]{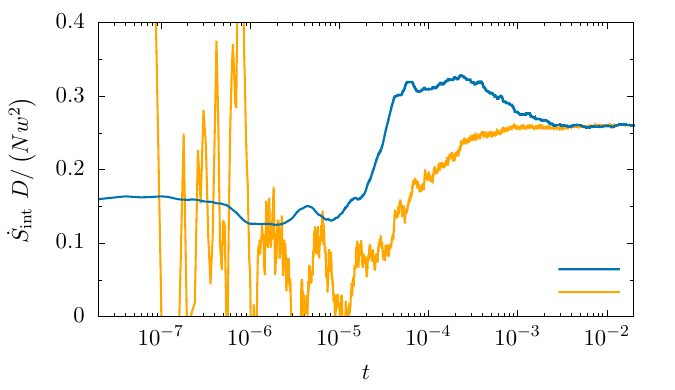}};
\node at (3.75,-1.25) [left] {\Eref{ABP_EPR_succinct}};
\node at (3.75,-1.6) [left] {Sekimoto's scheme};
\end{tikzpicture}
}}
\caption{
Illustration of \Eref{ABP_EPR_succinct} which allows for an instantaneous,
per-particle entropy production of pair-interacting ABPs. In all graphs the entropy production is
expressed in units of 
$\drift^2/\diffusion$, the entropy production of a single, free ABP.
The system parameters
are: 
linear size $L=8$,
translational diffusion constant $\diffusion=0.1$,
rotational diffusion constant $\diffusion_r=5$,
self-propulsion speed $\drift=40$,
WCA potential \cite{WeeksChandlerAndersen:1971} $\pairPot(r)=\epsilon+4\epsilon
[({R}/r)^{12}-({R}/r)^6]$ 
for $r\le2^{1/6}{R}$
with $\epsilon=0.1$ and ${R}=0.1$,
particle number $N=4000$, 
such that the packing fraction based on ${R}$ is $\phi=0.490874\dots$,
similar to the parameters used in the left panel of Fig.~5 in \cite{RednerHaganBaskaran:2013}.
The particle system was evolved using an Euler-Maruyama scheme with time steps $\Delta t = 8\cdot10^{-6}$ until a condensate formed and then with $\Delta t = 2\cdot10^{-8}$ to extract the time-series of the entropy production as shown.
\protect\subref{fig:instantEPR} Snapshot of a configuration with the colourcoding
indicating the ``instantaneous, per-particle entropy production'',
$\entropyProductionDensity_i
=\frac{1}{\diffusion}
\left(
\vecDrift_i-\sum_{j}\nabla\pairPot(\xvec_{i}-\xvec_{j})
\right)^2
- 
\sum_{j}
\laplace\pairPot(\xvec_{i}-\xvec_{j})$.
\protect\subref{fig:EPR_Sekimoto_vs_here} Entropy production rate per
particle,
$\entropyProduction/N$, 
estimated by averaging over all particles and increasingly long
times as shown on the log-scaled $x$-axis, on the one hand based on \Eref{ABP_EPR_succinct} and on the
other hand based on Sekimoto's scheme \cite{Sekimoto:2010,RoldanETAL:2021}.
The time ranges from the numerical discretisation-interval $t\sim10^{-8}$ until convergence around $t\sim10^{-2}$.
Both methods display fluctuations, with our scheme, being based on individual snapshots, showing less strong fluctuations compared to the one drawing on \emph{consecutive} snapshots needed in the Stratonovich product in Sekimoto's scheme.
}
\end{figure*}
\paragraph*{
Interacting Active Brownian particles.}
To illustrate our framework in a setting with very many
interacting particles that produce entropy we need to consider a system which does not admit
an exact solution like those above. In the following, we calculate the entropy production of Active Brownian
Particles (ABPs)
\cite{HowseETAL:2007,ZhangPruessner:2022,ZhangETAL:2024} interacting
through a purely repulsive WCA potential
\cite{WeeksChandlerAndersen:1971} and undergoing motility-induced phase separation in the
stationary state \cite{RednerHaganBaskaran:2013}.
%

The overdamped Langevin equation governing the particles' velocities is
$\dot{\xvec}_i = \vecDrift_i(t) - \sum_{j\neq i}\nabla U(\xvec_i-\xvec_j) + \sqrt{2\diffusion}\xivec_i(t)$
with self-propulsion 
$\vecDrift_i(t) = \drift\evec(\theta_i(t))$.
Here $\drift$ denotes the self-propulsion velocity and
$\evec(\theta)=(\cos\theta,\sin\theta)^\transpose$ 
the director, which
depends on the orientation angle $\theta$, 
evolving with stochastic dynamics $\dot{\theta}_i = \sqrt{2D_r}\eta_i(t)$, where 
$\eta_i$ is a
unit Gaussian noise 
$\ave{\eta_i(t)}=0$, 
$\ave{\eta_i(t)\eta_i(t')}=\delta(t-t')$,
and $D_r$ is the rotational diffusion constant.
Particles are further subject to a 
WCA pair-potential,
$\pairPot(r)=\epsilon+4\epsilon
[(R/r)^{12}-(R/r)^6]$ 
for $r\le2^{1/6}R$ and $\pairPot(r)=0$ otherwise,
parameterised by interaction energy $\epsilon$ and characteristic radius ${R}$. Particles diffuse with diffusion constant $\diffusion$, so that $\xi$ is a 
unit Gaussian noise with $\ave{\xi_i(t)}=0$, 
$\ave{\xi_i(t)\xi_j(t')}=\delta_{i,j}\delta(t-t')$.
We consider $N$ particles in a box of length $L$
with periodic boundary conditions.

For ABPs, we need to modify the local entropy production $\entropyProductionDensity^{(2)}_2$ entering into \Eref{interacting_EPR_numerics} 
because the self-propulsion velocity $\vecDrift$ as showing in
\Eref{entropyProduction_for_pairPot1} 
and as cancelled from
\Eref{entropyProduction_for_pairPot2} 
becomes a function of the particle index, so that the
cross-terms of \Eref{entropyProductionDensity2_example_final},
\[
\frac{2}{\diffusion}
\pairPot'(\xvec_1-\xvec_2)\cdot
\big(
- 
\vecDrift_1
\big)
+
\frac{2}{\diffusion}
\pairPot'(\xvec_2-\xvec_1)\cdot
\big(
- 
\vecDrift_2
\big)
\]
no longer
vanish when summed over,
as happens when $\vecDrift_1=\vecDrift_2$ and the pair-potential $\pairPot$ 
is even, resulting in \Eref{entropyProduction_for_pairPot}, as discussed after \Eref{entropyProduction_interacting_indistinguishable_example}.
The entropy production of pair-interacting ABPs works out as a simple
generalisation of
\Erefs{entropyProduction_interacting_indistinguishable_example} 
and
\eref{entropyProduction_interacting_indistinguishable_example_numerics}
without
external potential, $\extPot\equiv0$,
\begin{widetext}
\begin{multline}
\elabel{ABP_EPR_long}
\entropyProduction=\frac{1}{Q}\sum_{q=1}^Q
\Bigg[
\sum_{i_1=1}^N\frac{\vecDrift^2_{i_1}}{\diffusion}
+
\sum_{\substack{i_1,i_2=1\\ i_1\ne i_2}}^N 
\frac{2} {\diffusion}  
\nabla\pairPot(\xvec^{(q)}_{i_1}-\xvec^{(q)}_{i_2})\cdot
\left\{
- 
\vecDrift_{i_1}
\right\}
+
\frac{1}{\diffusion} 
\big(\nabla \pairPot(\xvec^{(q)}_{i_1}-\xvec^{(q)}_{i_2})\big)^2
-
\laplace\pairPot(\xvec^{(q)}_1-\xvec^{(q)}_2) \\
+
\sum_{\substack{i_1,i_2,i_3=1\\ i_1\ne i_2\ne i_3\ne i_1}}^N 
\frac{1}{\diffusion}
\nabla\pairPot(\xvec^{(q)}_{i_1}-\xvec^{(q)}_{i_2})
\cdot
\nabla\pairPot(\xvec^{(q)}_{i_1}-\xvec^{(q)}_{i_3})
\Bigg] \ .
\end{multline}
Similar to \Eref{entropyProduction_for_pairPot_expectation_rewrite},
this can be succinctly summarised as
\begin{equation}
\elabel{ABP_EPR_succinct}
\entropyProduction=\frac{1}{Q}\sum_{q=1}^Q
\Bigg[
\sum_{i=1}^N\frac{1}{\diffusion}
\bigg(
\vecDrift_i-\sum_{j.nn.i}\nabla\pairPot(\xvec^{(q)}_{i}-\xvec^{(q)}_{j})
\bigg)^2
- 
\sum_{i=1}^N \sum_{j.nn.i}
\laplace\pairPot(\xvec^{(q)}_{i}-\xvec^{(q)}_{j})
\Bigg] \ ,
\end{equation}
\end{widetext}
consistent with \cite[Eq.~(4.27)]{Sekimoto:2010},
where the ``nearest neighbour sums'' $\sum_{j.nn.i}$ sum over all
particles $j$
except $j=i$ 
within the range of the potential, outside of which both
$\nabla\pairPot(\xvec_{i}-\xvec_{j})$ and
$\laplace\pairPot(\xvec_{i}-\xvec_{j})$ vanish.
Structurally similar to \Eref{entropyProduction_driftdiffusion_main}, 
\Eref{ABP_EPR_succinct} suggests
that the effective speed
$\drift_{\text{eff.}\,i}=|\vecDrift_i-\sum_{j.nn.i}\nabla\pairPot(\xvec_{i}-\xvec_{j})|$ of the ABPs gives rise to all
positive entropy production in the typical form
$\drift_{\text{eff.}\,i}^2/\diffusion$.
The Laplacian of the potential is non-negative 
everywhere given the convexity of
the pair potential. 
Although not universally found \cite{FodorETAL:2016,KetaETAL:2021},
it can equally be obtained in an expansion of the gradient of the potential in its correlator with the velocity subject to Stratonovich noise.

\Erefs{ABP_EPR_long} and \eref{ABP_EPR_succinct} and more generally
\Eref{interacting_EPR_numerics} allow for an
\emph{instantaneous entropy
production of particle $i$}, namely
$\drift_{\text{eff.}\,i}^2/\diffusion-\sum_{j.nn.i}
\laplace\pairPot(\xvec_{i}-\xvec_{j})$.
This is what is shown in colour-coded fashion for ABPs interacting via a
WCA pair potential in
\Fref{instantEPR}.
As expected, freely moving particles produce entropy around
$\drift^2/\diffusion$ (red), whereas the phase-separated, dense condensate has
different regions. Around its perimeter, the entropy production is quite
uniformly large, as particles can move almost freely but at
comparatively high
density. Inside the condensate, particles interact, as many of
their close neighbours are located within the interaction range of the
WCA potential $2^{1/6}R$. Many arrested particles locally produce negative
entropy (blue), whereas some particles move under very strong forces in
steep potentials (yellow). The heterogeneity in the entropy production is reminiscent of that of
the pressure \cite[][Fig.~1]{RednerHaganBaskaran:2013}, but markedly
different to the homogeneous phase in Fig.~6 of \cite{MartinETAL:2021}
showing active Ornstein-Uhlenbeck particles in an external potential.

To demonstrate the validity of \Eref{ABP_EPR_succinct}
%
%
%
%
%
and by way of a sanity
check, \Fref{EPR_Sekimoto_vs_here} shows 
a comparison of the entropy production
estimated using \Eref{ABP_EPR_succinct} and Sekimoto's framework
\cite{Sekimoto:2010}, specifically using Eqs.~(7) and (E3) in \cite{RoldanETAL:2021}
with $k_B\mu_i=1$. 
As the stationary evolution is observed over a
brief period,
both procedures arrive at virtually the same estimate towards the end of
the observation interval shown,
albeit
the estimate based on Sekimoto's framework displays much stronger
fluctuations at very short times, as it incorporates the noise.
%
%
%



\section{Discussion, summary and outlook}
\seclabel{summary_outlook}
Above we have demonstrated how to construct a Doi-Peliti field theory,
\Erefs{def_action0}, \eref{def_ave0} and \eref{def_action0_cont},
from the Fokker-Planck or master equation \eref{FPeqn_main} governing single
particle dynamics, without having to resort to explicit discretisation.
The resulting expression for the entropy production
of conserved particle systems,
\Eref{def_entropyProduction_elegant}, is of a particularly simple form, indicating that entropy production can be interpreted as a mean of a \emph{local} expression \Eref{def_local_entropy}.
Additional processes, reactions and interaction, can be added to the action and, if
necessary, treated perturbatively, \Erefs{def_action_full} and
\eref{perturbative_expansion}. Expressing the entropy production in
terms of propagators, \Erefs{def_entropyProduction}, \eref{def_Op} and
\eref{def_Ln}, it turns out that 
the perturbative contributions enter only to first order,
\Erefs{transition_from_action} and
\eref{Ln_from_propagators}, because each such perturbation introduces a
term of order $t'-t$, \Eref{derivative_integral}. 
Higher order contributions from the perturbative terms produce higher powers of $t'-t$ which do not enter into the entropy production. Even when some phenomena are treated perturbatively the resulting expressions for the entropy production are exact and we do not need
to assume convergence of the perturbation expansion.
Loops enter into the entropy production only in the presence of external sources (tadpole-like diagrams).

Treating interaction perturbatively, the results are generalised to many interacting particles, \APref{MultipleParticles}, with significant simplifications taking place as diagrams with more than one blob do not enter, \eg \Eref{def_trans_multi_diag} (also \APref{appendixWhichDiagramsContribute}), and disconnected diagrams that simplify as bare propagators turn into $\delta$-functions. The resulting stationary entropy production, \Eref{entropyProduction_for_pairPot}, can again be understood as a spatial average involving equal-time densities. 
If interactions involve at most $n$ particles at once, the highest order density needed is $2n-1$.
Because of this structure, it can be used to estimate the entropy production in experimental and numerical systems, \Eref{interacting_EPR_numerics}, as well as on the basis of effective theories.

While the results are derived by means of a DP field theory, 
they apply universally.
Results such as \Erefs{entropyProduction_for_pairPot}, \eref{entropyProduction_interacting_indistinguishable_example} or \eref{entropyProduction_driftdiffusion_main} are exact and can be extended to include higher order interactions or even reactions. They can be used to answer vital questions
in applied and theoretical active matter that have previously 
been studied using approximative schemes \cite{TociuETAL:2020,RoETAL:2021}, 
such as the energy dissipation 
in hair-cell bundles \cite{RoldanETAL:2021}, in  
neuronal responses to visual stimuli \cite{LynnETAL:2022},
or
in Kramer's model \cite{Neri:2022}.

The general recipe to calculate entropy production in any system is thus to determine the basic or ``bare" characteristics, such as the self-propulsion speed or the pair-potential, as well as the $2n-1$ densities, which are used as the weight in an integral like \Eref{entropyProduction_for_pairPot}. Extensions to the formulae derived in the present work are a matter of inserting the new blobs into the field theory.
The present scheme then allows the systematic calculation of the entropy
production based on the microscopic dynamics of the process, while
retaining the particle nature of the degrees of freedom.

Calculating field theoretically the entropy production of particle systems has been
attempted before, 
notably by Nardini \etal \cite{NardiniETAL:2017}. Their approach is based on an
effective dynamics, given by Active Model B \cite{WittkowskiETAL:2014}, 
that describes the particle density as a continuous function in space  by means
of a Langevin equation with additive noise in order to smoothen or coarse-grain the dynamics. However, \emph{particle
systems} necessarily require multiplicative noise, for example in the
form of Dean's equation \cite{Dean:1996,BotheETAL:2023}, to allow the density to faithfully capture the dynamics of \emph{particles}. 
If not endowed with a mechanism to maintain the particle nature of the degrees of freedom, recasting the dynamics in terms of an unconstrained density constitutes a massive increase of the available phase space rather than a form of coarse graining.
There is no reason to assume that the entropy production of such an effective
field theory is an approximation of the microscopic entropy production.
We are not aware of an example of a \emph{particle system}, whose entropy production is correctly captured by an effective field theory
based on a 
Langevin equation on the continuously varying particle density with additive noise 
\cite{NardiniETAL:2017}.
In fact, attempting to use 
such an
approximation of a most basic, exactly solvable process and using
the coarse-graining scheme in \cite{NardiniETAL:2017}, produces 
a spurious dependence on the size of the state space
and a lack of
extensivity in the particle number, \SMref{app_ring}, 
\Eref{entropy_production_final},
while the present field theory trivially produces the exact expression for the microscopic entropy production, 
\Eref{M_state_exact_asympotic_result} in
\SMref{model_description}. 
We argue that the observable of entropy production needs to be constructed from the
microscopic dynamics, which is partially integrated out or ``blurred" in effective theories of the particle density. These generally capture correlations effectively and efficiently, but they do so at the expense of smoothing the microscopic details that give rise to entropy production, as they change the description of the dynamics from one in terms of particles to one in terms of space. However, the expression for the entropy production needs to be determined from the microscopics of the particle dynamics, even when eventually calculated from $(2n-1)$-point densities. 
Effective theories may contain the necessary information to determine these densities, but not to construct the functional for the entropy production in the first place.

Coarse graining is known to distort entropy production in a highly non-trivial way \cite{CocconiSalbreuxPruessner:2022}. In the continuum limit, just like in \SMref{app_ring}, divergences are to be expected.
This may be circumvented by the introduction of an
abritrary UV cutoff \cite{AlstonCocconiBertrand:2023,SuchanekKroyLoos:2023},
which renders the entropy production just as arbitrary.
In general, there is no relationship between the entropy production of an effective theory based on smooth particle densities and the particle dynamics it approximates, 
as the entropy production in the smoothened theory depends heavily on the choices made for the approximation. That said, we do not question the use an Onsager-Machlup functional to calculate the entropy production \cite{NardiniETAL:2017} of a process whose field is the degree of freedom.


In future research we may want to exploit further the general expressions for the entropy production of multiple interacting particles, such as \Eref{entropyProduction_for_pairPot} and those derived in \APref{MultipleParticles}. One may ask, in particular, for 
bounds
on the entropy production by an ensemble of interacting particles and the shape of the pair-interaction potential to maximise it.
The present framework can also be extended to the grand canonical ensemble, where particles are created and annihilated, as they branch and coagulate. 
The grand challenge, however, is to extend the present framework to non-Markovian systems, as to calculate the entropy production in systems where not all degrees of freedom are known, such as the orientation-integrated entropy production of Run-and-Tumble particles in a harmonic potential \cite{Garcia-MillanPruessner:2021}.

\begin{acknowledgments}
We would like to thank the many people with whom we discussed some aspects of the present work at some point:
Tal Agranov,
Marco Baiesi,
Ignacio Bordeu,
Michael Cates,
Luca Cocconi,
{\'E}tienne Fodor,
Sarah Loos,
Cesare Nardini,
Johannes Pausch, 
Patrick Pietzonka,
Guillaume Salbreux,
Elsen Tjhung,
Benjamin Walter, 
Fr{\'e}d{\'e}ric van Wijland,
Ziluo Zhang 
and
Zigan Zhen
for many enlightening discussions.

RG-M was supported in part by the European Research Council under the EU’s Horizon 2020 Programme (Grant number 740269).
RG-M acknowledges support from a St John’s College Research Fellowship, University of Cambridge.
\end{acknowledgments}

\bibliography{larticles,lbooks}
\bibliographystyle{apsrev4-2}

\newpage

\onecolumngrid
\appendix

\section*{appendices}
\tableofappendixcontents




\newcommand{\Xtoasection}[2]{%
  \section{#2}
  \addcontentsline{toa}{section}{\csname the#1\endcsname{} #2}
}
\newcommand{\toasection}[1]{%
  \section{#1}
  \addcontentsline{toa}{section}{\csname thesection\endcsname{} #1}
}
\newcommand{\toasubsection}[1]{%
  \subsection{#1}
  \addcontentsline{toa}{subsection}{\csname thesubsection\endcsname{} #1}
}
\newcommand{\toasubsubsection}[1]{%
  \subsubsection{#1}
  \addcontentsline{toa}{subsubsection}{\csname thesubsubsection\endcsname{} #1}
}
\newcommand{\toasubsubsubsection}[1]{%
  \subsubsection*{#1}
}

\toasection{From Master and Fokker-Planck Equation to Field Theory to Entropy Production}
\seclabel{DP_plus_EP}

\paragraph*{Abstract}
In this section, we derive a Doi-Peliti field theory 
with arbitrarily many particles
from 
the parameterisation of a single particle
master equation with discrete states. 
\Eref{time_evolution_operator_op_only} shows that the transition matrix of the master \Eref{rho_i_ME_matrix_rewrite} is identically the transition matrix in the action of the field theory.
It is further shown that any continuum limit that is taken in the master equation in order to obtain a Fokker-Planck equation (FPE), can equivalently be performed in the action of the field theory. As a result, we find a direct mapping from an FPE to a Doi-Peliti action, \Erefs{rho_i_ME_matrix_rewrite_kernel} and \eref{def_harmonic_action}. Rather than taking therefore the canonical \cite{Peliti:1985,TaeuberHowardVollmayr-Lee:2005,Cardy:2008,Taeuber:2014}, but cumbersome route from FPE to discretised master equation to discrete action to continuum action, we make explicit \cite{Doi:1976} a direct and very simple route from FPE to action in \APref{DP_field_theory_derivation}. 
In \APref{FPE_inter} we generalise this derivation to many, pair-interacting
particles.
In \APref{the_propagator} we construct the bare propagator of a Markov chain,
\Eref{orig_propagator_as_an_inverse}  in $\omega$ and \Eref{bare_propagator_realtime} in direct time. Allowing for a perturbative term in the action leads in principle to infinitely many additional terms in the propagagator, but crucially only a single correction in the short-time derivative \Eref{summary_propagator_first_order_summary_deri}, also \APref{appendixWhichDiagramsContribute}.
In \APref{entropy_production} we construct a field-theoretic formulation of entropy production by retracing
the basic reasoning by Gaspard's formulation \cite{Gaspard:2004} of the (internal) entropy production (rate) of a Markovian system, introducing in particular 
the kernel $\Op_{\yvec\xvec}$, \Eref{def_Op_app},
the logarithm $\Ln_{\yvec\xvec}$, \Eref{def_Ln_app},
and 
the local entropy production $\entropyProductionDensity$, \Eref{def_entropyProductionDensity_app}. In \APref{EP_from_propagators} kernel and logarithm are expressed in terms of diagrams and thus in terms of the propagator, \Eref{Op_and_Ln_from_diagrams}, and ultimately the action and the master equation, \Erefs{Op_in_diagrams} and \eref{Ln_from_propagator}. In \APref{EP_cont} this result is extended to continuous degrees of freedom, \Erefs{Op_continuum_Fpop_plus} and \eref{Ln_for_continuous}.

\toasubsection{Master equation for a single particle}
In the following $\transMatLin_{\yvec\xvec}$ for $\yvec\ne\xvec$ 
are non-negative rates for the transitioning of a particle from state $\xvec$ to state $\yvec$. The object
$\transMatLin$ may be thought of as a ``hopping matrix''.
A particle is then being lost from state $\yvec$ by a hop with 
rate 
$\sum_{\xvec,\xvec\ne\yvec} \transMatLin_{\xvec\yvec}$.
With suitable definition of $\transMatLin_{\yvec\yvec}$,
\Eref{def_Fjj},
$\transMatLin_{\yvec\xvec}$ has the form of a usual (conservative) Markov-matrix in continuous time.

If $\density(\xvec,t)$ is the probability to find an \emph{individual} particle  at time $t$ in state $\xvec$, which 
might be interpreted as a position in space, then a single degree of freedom evolves according to the master equation of a continuous time Markov chain,
\begin{equation}\elabel{rho_i_ME_orig}
\dot{\density}(\yvec,t) = 
\sum_{\substack{\xvec\\ \xvec\ne\yvec}} 
\big( \transMatLin_{\yvec\xvec}\density(\xvec,t) - \transMatLin_{\xvec\yvec} \density(\yvec,t) \big)  \ ,
\end{equation}
which is the usual Markovian evolution. The sum in \Eref{rho_i_ME_orig} runs over all states $\xvec$, excluding $\xvec=\yvec$.

To cater for the needs of the field theory, we need to break conservation of the Markovian evolution. We therefore amend \Eref{rho_i_ME_orig} by a term representing spontaneous extinction of a particle in state $\yvec$ with rate $r_\yvec$, 
\begin{equation}\elabel{rho_i_ME_mass}
\dot{\density}(\yvec,t) = 
- r_\yvec \density(\yvec,t) 
+
\sum_{\substack{\xvec\\ \xvec\ne\yvec}} 
\big( \transMatLin_{\yvec\xvec}\density(\xvec,t) - \transMatLin_{\xvec\yvec} \density(\yvec,t) \big) 
\ .
\end{equation}
The mass $r_\yvec>0$ is necessary to make the Doi-Peliti field theory causal. In the present work, it is a mere technicality and will be taken to $0^+$ whenever convenient.

Introducing the additional definition
\begin{equation}\elabel{def_Fjj}
\transMatLin_{\yvec\yvec}=-
\sum_{\substack{\xvec\\ \xvec\ne \yvec}} 
\transMatLin_{\xvec\yvec} \ ,
\end{equation}
allows us to rewrite the master equation in terms of a single rate matrix or \emph{kernel}
$-r_\yvec \delta_{\yvec,\xvec}+\transMatLin_{\yvec\xvec}$, so that
\begin{equation}\elabel{rho_i_ME_matrix_rewrite}
\dot{\density}(\yvec,t) = 
\sum_{\xvec}
\density(\xvec,t) 
\left[ -r_\yvec \delta_{\yvec,\xvec} + \transMatLin_{\yvec\xvec} \right]
\ ,
\end{equation}
using the Kronecker $\delta$-function $\delta_{\yvec,\xvec}$.

\toasubsubsection{Continuum limit}
\seclabel{continuum_limit}
It is instructive to consider the example of discretised drift-diffusion in one dimension,
\begin{equation}\elabel{transMatDiffusion}
\transMatLin_{b a}=
h_r \big(\delta_{a+1,b} - \delta_{a,b}\big)
+
h_l \big(\delta_{a-1,b} - \delta_{a,b}\big)
\end{equation}
with $h_l$ and $h_r$ the rates of hopping left and right respectively, with $a$ and $b$ numbering the position of origin and destination on a ring. 
The definition \Eref{transMatDiffusion} obeys \Eref{def_Fjj}.
In this case, the continuum limit of the master equation can be taken by introducing the parameterisation 
$\diffusion=(h_r+h_l)\Deltax^2 /2$ and $\drift=(h_r-h_l)\Deltax$, so that
$h_r=\diffusion/\Deltax^2+\drift/(2\Deltax)$ and 
$h_l=\diffusion/\Deltax^2-\drift/(2\Deltax)$. 
The rates $h_r$ and $h_l$ are bound
to be positive for any $\diffusion>0$ and 
sufficiently small $\Deltax$.
The master \Eref{rho_i_ME_matrix_rewrite} with the rate matrix $\transMatLin_{b a}$ given in \Eref{transMatDiffusion} can now be rewritten as
\begin{equation}\elabel{ME_diffusion_example}
\dot{\density}(b,t) = 
- r_b \density(b,t)
+
\diffusion \frac{\density(b+1,t)-2\density(b,t)+\density(b-1,t)}{\Deltax^2}
- \drift \frac{\density(b+1,t)-\density(b-1,t)}{2\Deltax}
\end{equation}
which turns into
\begin{equation}\elabel{FPE_diffusion_example}
\dot{\densityTilde}(\yvec,t) = \LC_\yvec \densityTilde(\yvec,t)
\quad\text{ with }\quad
\FPophat_\yvec = 
- r_\yvec
+
\diffusion \partial_\yvec^2 - \drift \partial_\yvec \ ,
\end{equation}
after introducing 
$\densityTilde(\yvec=b\Deltax,t)=\density(b,t)/\Deltax$ and
taking the continuum limit $\Deltax\to0$ while maintaining $\int\dint{y}\densityTilde(y)=1$. 

The continuum limit at fixed $\diffusion,\drift$ may be taken directly on the rates $\transMatLin_{b a}$
in \Eref{transMatDiffusion},
\begin{equation}
    \lim_{\Deltax\to0}
    \frac{h_r}{\Deltax} \big(\delta_{a+1,b} - \delta_{a,b}\big)
+
\frac{h_l}{\Deltax} \big(\delta_{a-1,b} - \delta_{a,b}\big)
=
\diffusion \delta''(\yvec-\xvec)-\drift \delta'(\yvec-\xvec)
\end{equation}
using
\begin{subequations}
\begin{align}\elabel{delta_dash_as_a_limit}
    \lim_{\Deltax\to0} \Deltax^{-2} 
    \big(\delta_{a+1,b} - \delta_{a-1,b}\big) &=-2
    \partial_\yvec
    \delta(\yvec-\xvec) = -2 \delta'(\yvec-\xvec)\\
    \lim_{\Deltax\to0} \Deltax^{-3} 
    \big(\delta_{a+1,b} - 2\delta_{a,b} + \delta_{a-1,b}\big) &=
    \partial^2_\yvec
    \delta(\yvec-\xvec) = \delta''(\yvec-\xvec)
\end{align}
\end{subequations}
with 
$\yvec=b\Deltax$ 
and
$\xvec=a\Deltax$ 
as well as $\delta(\yvec-\xvec) = \lim_{\Deltax\to0} \Deltax^{-1} \delta_{a,b}$, the Dirac $\delta$-function defined in terms of the Kronecker $\delta$-function. The kernel to be used in \Eref{rho_i_ME_matrix_rewrite} thus becomes a rate density,
\begin{subequations}
\elabel{continuum_limit_both}
\begin{align}
\elabel{continuum_limit_nointeraction}
\lim_{\Deltax\to0}
\Deltax^{-1} \left[ -r_b \delta_{b,a} + \transMatLin_{b a} \right]
= &
\left[ -r_\yvec \delta(\yvec-\xvec) + \diffusion \delta''(\yvec-\xvec) - \drift \delta'(\yvec-\xvec)\right]
\\
= &
\FPophat_y \delta(y-x)
=
\FPop_{yx}
\elabel{continuum_limit_kernel}
\end{align}
\end{subequations}
to be used in the continuum limit of \Eref{rho_i_ME_matrix_rewrite}, which now produces an integral,
\begin{equation}\elabel{rho_i_ME_matrix_rewrite_kernel}
    \dot{\density}(y,t) = 
\int\dint{x} 
\FPop_{yx} \density(x,t)
\ ,
\end{equation}
that turns into the usual FPE
\begin{equation}\elabel{FPE_standard}
    \dot{\density}(y,t) = 
    \big( \diffusion \partial_y^2 - \drift \partial_y - r\big) \density(y,t) \ ,
\end{equation}
using \Eref{continuum_limit_kernel}.
The procedure above is readily generalised to higher dimensions. To summarise this section, a master equation of the form \Eref{rho_i_ME_matrix_rewrite} can be turned into an FPE like \Erefs{rho_i_ME_matrix_rewrite_kernel} or \eref{FPE_standard} via a suitable continuum limit. \Erefs{rho_i_ME_mass}, \eref{rho_i_ME_matrix_rewrite} and later \eref{rho_i_ME_matrix_rewrite_kernel} form the basis of the action to be determined in the following.

\toasubsection{Doi-Peliti field theory}
\seclabel{DP_field_theory_derivation}
To build a Doi-Peliti field theory on the basis of the hopping matrix $\transMatLin$  and the extinction rates $r_\yvec$  that parameterise the master \Eref{rho_i_ME_mass}, we need to introduce the
probability 
$\PC\left( \{n\}; t \right)$
of occupation numbers $\{n\}=\{n_1,\ldots,n_\yvec,\ldots\}$, which quantify the number of particles on each site $\yvec$. Each of these particles is concurrently subject to a Poissonian change of state,
\begin{align}
\ddt{} \PC\left( \{n\}; t \right)
= &
\sum_{\yvec} 
\Big\{ 
(n_\yvec+1) r_\yvec 
\PC\left( \{\ldots, n_\yvec+1, \ldots\}; t \right)
-
n_\yvec r_\yvec
\PC\left( \{n\}; t \right)
\Big\}\nonumber\\
\elabel{Pdot}
& +
\sum_{\yvec}\sum_{\substack{\xvec\\ \xvec\ne\yvec}}
\Big\{
(n_\xvec+1) \transMatLin_{\yvec\xvec} \PC\left( \{\ldots, n_\xvec+1, \ldots, n_\yvec-1, \ldots\}; t \right)
- n_\yvec \transMatLin_{\xvec\yvec} \PC\left( \{n\}; t \right) 
\Big\} \ .
\end{align}
The second index $\xvec$ in the double-sum cannot take the value $\xvec=\yvec$, as otherwise the configuration $\{\ldots, n_\xvec+1, \ldots, n_\yvec-1, \ldots\}$ is ill-defined. That $\transMatLin_{\yvec\xvec}$ features on the right only with 
$\PC$ whose 
arguments $\{n\}$ have identical $\sum_i n_i$ 
is an expression of the conservation of particles in the dynamics parameterised by $\transMatLin_{\yvec\xvec}$. 

The master \Eref{Pdot} on the basis of occupation numbers differs crucially from the master \Eref{rho_i_ME_mass} on the basis of the state of a particle, in that the former
tracks many particles simultaneously, while
explicitly preserving the particle nature of the degrees of freedom, whereas the latter captures only the one-point density, \ie strictly the probability to find a single particle at a particular point. However, 
there is nothing in \Eref{rho_i_ME_mass} that forces $\xvec$ to be the sole degree of freedom of a \emph{particle} and $\density(\xvec,t)$ to be its probability. In fact, $\density(\xvec,t)$ may equally be an arbitrarily divisible quantity such as heat and \Eref{rho_i_ME_mass} its evolution. \Eref{rho_i_ME_mass} correctly describes the evolution of the one-point density of a \emph{particle}, but it makes no \emph{demand} on the particle nature and contains no information about higher correlation functions. In \Eref{Pdot}, on the other hand, the occupation numbers are strictly particle \emph{counts}, \ie non-negative integers. In order to arrive at \Eref{Pdot} from \Eref{rho_i_ME_mass} we have to demand that 
\Eref{rho_i_ME_mass} describes the probabilistic evolution of a single particle and 
\Eref{Pdot} 
the corresponding independent evolution of many of them.
And yet, because \Eref{Pdot} 
draws on the same transition matrix as \Eref{rho_i_ME_mass}, 
we will be able to take in \Eref{Pdot} the same continuum limit that turned \Eref{rho_i_ME_mass} into \eref{FPE_diffusion_example}.

We proceed by casting \Eref{Pdot} in a Doi-Peliti action following a well-established but somewhat cumbersome procedure \cite{Doi:1976,Doi:1976b,Peliti:1985,Taeuber:2014,TaeuberHowardVollmayr-Lee:2005,Cardy:2008}.
The temporal evolution of the weighted sum over Fock states 
$\ket{\{n\}}$ ,
\begin{equation}\elabel{weighted_sum_Fock}
\ket{\psi}(t) = \sum_{\{n\}} \PC\left( \{n\}; t \right) \ket{\{n\}}
\end{equation}
can be written in terms of ladder operators $\creatX{}=1+\creatDSX{}$ and $\annihX{}$ as
\elabel{time_evolution_operator}
\begin{equation}
\ddt \ket{\psi}(t)
= \DPop_0  \ket{\psi}(t)
\end{equation}
with a 
Gaussian
time-evolution operator as simple as
\begin{equation}\elabel{time_evolution_operator_op_only}
\DPop_0 
=
-
\sum_{\yvec} 
r_\yvec \creatDSX{\yvec}\annihX{\yvec} 
+
\sum_{\yvec}\sum_{\substack{\xvec\\ \xvec\ne\yvec}}
 \transMatLin_{\yvec\xvec}
\Big\{
\creatDSX{\yvec}-\creatDSX{\xvec}
\Big\}
\annihX{\xvec} \ .
\end{equation}
The term $\creatDSX{\yvec}-\creatDSX{\xvec}$ indicates a conservative particle transition from state $\xvec$ to state $\yvec$
parameterised by $\transMatLin_{\yvec\xvec}$, whereas 
$r_\yvec\creatDSX{\yvec} \annihX{\yvec}$ in \Eref{time_evolution_operator_op_only} is the signature of spontaneous particle
extinction from state $\yvec$ with rate $r_\yvec$. 

Using \Eref{def_Fjj} to 
rewrite \Eref{time_evolution_operator_op_only} again as
\begin{equation}
\elabel{time_evolution_operator_op_only_rewrite}
  \DPop_0 
=
\sum_{\yvec}
\creatDSX{\yvec}
\sum_\xvec
\annihX{\xvec} 
\bigg[
-
r_\yvec \delta_{\yvec,\xvec}
+
\transMatLin_{\yvec\xvec}
\bigg]
\end{equation}
reveals how closely the time evolution operator $\DPop_0$ of the Fock-space 
is related to the master \Eref{rho_i_ME_matrix_rewrite}, as the 
square bracketed rate matrix
$[-r_\yvec \delta_{\yvec,\xvec}
+
\transMatLin_{\yvec\xvec}
]$
in  \Eref{time_evolution_operator_op_only_rewrite} is the same as the one \Eref{continuum_limit_both}, which showed
that its continuum limit is the kernel $\FPop_{yx}$.

Proceeding along the canonical path \cite{Peliti:1985,TaeuberHowardVollmayr-Lee:2005,Cardy:2008,Taeuber:2014} turns \Eref{time_evolution_operator_op_only_rewrite} into the harmonic and therefore Gaussian action
\begin{equation}\elabel{def_harmonic_action_discrete}
    \action_0 = 
\int \dint{t} \sum_\yvec
\phitilde(\yvec,t)
\sum_\xvec 
\bigg[
-
\partial_t \delta_{\yvec,\xvec}
-
r_\yvec \delta_{\yvec,\xvec}
+
\transMatLin_{\yvec\xvec}
\bigg]
\phi(\xvec,t) \ .
\end{equation}
Comparing again to the original master \Erefs{rho_i_ME_mass} and \eref{rho_i_ME_matrix_rewrite} shows their simple relationship to the action. 
Upon taking the continuum limit, just as in \Eref{rho_i_ME_matrix_rewrite_kernel}, the sum over $\xvec$ turns into an integral.
To turn the sum over $\yvec$ into an integral, the product of the fields $\phitilde\phi$ is to be rescaled to a density. This is a trivial operation, as the fields are dummy variables,
\begin{equation}\elabel{def_harmonic_action}
    \action_0 = 
\int \dint{t} \int \ddint{x}\ddint{y}
\phitilde(\yvec,t)
\bigg[
-
\partial_t \delta(\yvec-\xvec)
+
\FPop_{yx}
\bigg]
\phi(\xvec,t) \ .
\end{equation}
Again, \Eref{def_harmonic_action} bears a striking resemblance to the master \Eref{rho_i_ME_matrix_rewrite_kernel}, a result closely related to Doi's original work \cite{Doi:1976}.
The relationship between FP operator and DP action is very much like that of a Langevin equation and its Martin-Siggia-Rose action.
Of course, \Eref{def_harmonic_action} simplifies significantly as some of the integrals can easily be carried out in the presence of $\delta$-functions.

After turning observables into fields, expectations on the basis of the harmonic action are calculated as
\begin{equation}\elabel{def_ave_0}
\ave[0]{\bullet} = 
\int \Dint{\phi}\Dint{\phitilde} 
\bullet \Exp{\action_0}  \ .
\end{equation}
In a DP field theory all terms in the action arise from a master or Fokker-Planck equation. More complicated ones, in particular those that describe interaction and reaction, are generally not bilinear and therefore need to be dealt with perturbatively. Yet, they are simply \emph{added} to the action, just as they are added to the master equation, being concurrent Poisson processes.
The full action $\action$ is then a sum of the harmonic part $\action_0$, 
whose path integral can be taken, and a perturbative part $\action_\perturbative$.

After turning observables into fields, expectations are now calculated as
\begin{equation}\elabel{perturbation_theory}
\ave{\bullet} = 
\int \Dint{\phi}\Dint{\phitilde} 
\bullet \Exp{\action} =
\int \Dint{\phi}\Dint{\phitilde} 
\big( \bullet \Exp{\action_\perturbative} \big) \Exp{\action_0} =
\ave[0]{ \bullet \Exp{\action_\perturbative} }
\end{equation}
with full action $\action=\action_0+\action_\perturbative$ and calculated perturbatively by expanding in powers of $\action_\perturbative$.

It is tempting to interpret $\phi(\xvec,t)$ as a particle density with corresponding units and $\phitilde(\yvec,t)$ as an auxiliary field like the one used in the response field formalism \cite{Taeuber:2014}. In fact, \Eref{def_harmonic_action} looks very much like the Martin-Siggia-Rose-Janssen-De Dominicis "trick" \cite{MartinSiggiaRose:1973,Janssen:1976,DeDominicis:1976,Taeuber:2014} applied to the FP \Eref{rho_i_ME_matrix_rewrite_kernel}, but without a noise source, given that the FPE is not a Langevin equation and thus does not carry a noise. There are, however, two crucial differences between Doi-Peliti field theories and reponse field field theories: Firstly, 
in Doi-Peliti field theories the fields are complex conjugates of each other abd conjugates of operators that obey a commutation relation. 
The operator formalism 
guarantees that the particle nature of the particles is maintained.
The fields are not densities. Consequently, observables are not simply fields $\phi$. 
Rather,
any observable has to be constructed on the basis of operators. That commutator produces additional terms that spoil any apparent interpretation of $\phi$ as the density. If $\phi$ were a \emph{particle density} and $\exp{\action}$ its statistical weight, the path integral over all \emph{allowed} density histories would have to be constrained to those paths that correspond to sums of $\delta$-functions. Secondly, observables in a Doi-Peliti field theory generally need to be initialised explicitly, with $\phi^\dagger=1+\phitilde$ "generating" a particle. This "auxiliary field" is not the response of the system to an external perturbation. The difference between response field and Doi-Peliti formalism is further illustrated and discussed in \cite{BotheETAL:2023}.

The Doi-Peliti formalism provides us with an action $\action$, a path integral and a commutator that allows us to construct desired observables which can be calculated as an expectation with $\exp{\action}$ as the apparent weight. The formalism may be seen as a recipe to replace a difficult calculation of observables in a particle system by an easier one in terms of continuously varying, unconstrained fields.
But because $\phi(\xvec,t)$ is not the particle density and the path integral not an integral over allowed paths,
$\int\Dint{\phitilde} \exp{\action[\phi,\phitilde]}$ is not the weight of a particular density history. This is the reason why the approach in \cite{NardiniETAL:2017} does not apply to Doi-Peliti field theories.

%
%
%
%
\toasubsection{
Alternative derivation suitable for interaction}
\seclabel{FPE_inter}
The above results, \Erefs{def_harmonic_action_discrete} and \eref{def_harmonic_action}, can be re-derived in a more general form, which
allows the inclusion of interaction terms in the Fokker-Planck
operator. To simplify the argument we assume a finite number $M$ of particle states for
now, so that the FPE becomes a master equation. We will
refer to the particle states as ``positions'' and
to the set of positions of all particles as a ``position state'', to be
distinguished from the particle number state, which is the set of
particle counts at all positions.
If
$\density(\yvec_1,\yvec_2,\ldots,\yvec_N;t)$ is the joint probability
to find particle $i=1,2,\ldots,N$ at positions
$\yvec_i\in\{1,2,\dots,M\}$ then it may evolve
according to
\begin{subequations}
\elabel{rho_interacting_FPE}
\begin{align}\elabel{rho_interacting_FPE_non_inter}
&\dot{\density}(\yvec_1,\yvec_2,\ldots,\yvec_N;t) = 
\sum_{k=1}^N
\sum_{\substack{\xvec_k=1\\ \xvec_k\ne\yvec_k}}^M
\big( 
\transMatLin_{\yvec_k\xvec_k}
\density(\yvec_1,\ldots,\yvec_{k-1},\xvec_k,\yvec_{k+1},\ldots,\yvec_N;t) 
- \
\transMatLin_{\xvec_k\yvec_k} 
\density(\yvec_1,\ldots,\yvec_N;t) \big)  \\
&\qquad +
\sum_{k=1}^N
\sum_{\substack{\xvec_k=1\\ \xvec_k\ne\yvec_k}}^M
\bigg( 
\sum_{\substack{\ell=1\\ \ell\ne k}}^N\transMatLin^{(\yvec_\ell)}_{\yvec_k\xvec_k}
\density(\yvec_1,\ldots,\yvec_{k-1},\xvec_k,\yvec_{k+1},\ldots,\yvec_N;t) 
- \
\sum_{\substack{\ell=1\\ \ell\ne k}}^N\transMatLin^{(\yvec_\ell)}_{\xvec_k\yvec_k}
\density(\yvec_1,\ldots,\yvec_N;t) \bigg) \ ,
\elabel{rho_interacting_FPE_inter}
\end{align}
\end{subequations}
where the first line corresponds to the single particle evolution
\Eref{rho_i_ME_orig} written out for multiple particles and the second line 
represents a generalised form of pair interaction, such that the transition
rate $\transMatLin^{(\yvec_\ell)}_{\yvec_k\xvec_k}$ from position $\xvec_k$ to
position $\yvec_k$ depends on the position $\yvec_\ell$ of particle $\ell$,
which has to be different from particle $k$ to avoid ``autocatalysis''. This
captures, in the continuum limit, the effect of a pair potential
$\pairPot(\yvec_k-\yvec_\ell)$, resulting in a Fokker-Planck operator
$\nabla_{\yvec_k}\cdot(\nabla_{\yvec_k}\pairPot(\yvec_k-\yvec_\ell))$. Both loss terms on the right-hand side of \Eref{rho_interacting_FPE} can be
removed by defining the reaction kernels $\transMatLin_{\yvec_k\xvec_k}$
and $\transMatLin^{(\yvec_\ell)}_{\yvec_k\xvec_k}$ suitably, as done for the former in
\Eref{def_Fjj} and for the latter, the interaction term, in \Eref{def_Fjj_extended}
below.

In the following, we assume indistinguishable particles, which is not
only more general, but also more relevant. The extension
to distinguishable particles is straight-forward. In case of
indistinguishability the probability (density) of a certain 
position state is
invariant under permutations of the particle label. This is the origin of
the factorial in the relation between the probability density of a 
position state used here
$\density(\yvec_1,\ldots;t)$ and the superscripted particle number density
$\HMdensity{N}{}(\yvec_1,\ldots,\yvec_N;t)=N!\density(\yvec_1,\ldots,\yvec_N;t)$
used in \APref{N_indistinguishable_particles} and further characterised
in \Erefs{marginalisation_densityN_i_indistinguishable}-\eref{marginalisation_densityN_consequences}.

We return to 
the joint probability $\PC\left(\{n\};t\right)$,
which
is the probability at time $t$ of a particular occupation number state,
$\{n\}=\{n_1,n_2,\ldots,n_M\}$, with $n_\yvec$ denoting the number
of particles at position $\yvec\in\{1,\ldots,M\}$, so that
$N=n_1+\ldots+n_M$. It
can be determined from
$\density(\yvec_1,\ldots;t)$ by summing over all position states
$\{\yvec_1,\ldots,\yvec_N\}$
that are compatible with the desired occupation numbers.
There are exactly $N!/\big(\prod_{\xvec=1}^M n_\xvec!\big)$ such compatible position states,
namely all permutations of the $N$ particle labels except those that leave
the position state unchanged, because they amount to permutations of
particles at the same position. As $\density(\yvec_1,\ldots;t)$ is
invariant under permutations of its arguments, we can make a convenient
choice,
\begin{equation}\elabel{PC_from_rho}
\PC\left(\{n_1,\ldots,n_M\};t\right)=
\density\left(\{\yvec_1,\ldots,\yvec_N\}_{\{n\}};t\right) N! \prod_{\uvec=1}^M
\frac{1}{n_\uvec!}
\end{equation}
where $\{\yvec_1,\ldots,\yvec_N\}_{\{n\}}$ denotes the simple, unique choice
\begin{equation}\elabel{convenient_state}
\{\yvec_1,\ldots,\yvec_N\}_{\{n\}} =
\{
\underbrace{1,\ldots,1}_{n_1\,\text{times}},
\underbrace{2,\ldots,2}_{n_2\,\text{times}},
\ldots
\underbrace{M,\ldots,M}_{n_M\,\text{times}}
\} \ ,
\end{equation}
considering $\density$ a function of an ordered set.
The time derivative of
$\PC\left(\{n_1,\ldots,n_M\};t\right)$ is determined by using \Eref{rho_interacting_FPE} in
\Eref{PC_from_rho}. For clarity, we first proceed without the
interaction term \Eref{rho_interacting_FPE_inter}, rewriting the master equation
\Eref{rho_interacting_FPE}
by summing over distinct positions $\zvec$ rather than particles,
\begin{multline}\elabel{rho_noninteracting_FPE_rewritten}
\dot{\density}\left(\{\yvec_1,\ldots,\yvec_N\}_{\{n\}};t\right) =
\sum_{\zvec=1}^M n_\zvec
\sum_{\substack{\xvec=1\\ \xvec\ne\zvec}}^M
\bigg( 
  \transMatLin_{\zvec\xvec}
  \density\left(\{\yvec_1,\ldots,\yvec_N\}_{\{\ldots,n_{\xvec}+1,\ldots,n_{\zvec}-1,\ldots\}};t\right) 
  \ - \ \transMatLin_{\xvec\zvec} 
\density(\{\yvec_1,\ldots,\yvec_N\}_{\{n\}};t) 
\bigg)  
\end{multline}
where
$\{\yvec_1,\ldots,\yvec_N\}_{\{\ldots,n_{\xvec}+1,\ldots,n_{\zvec}-1,\ldots\}}$
denotes the convenient position state \Eref{convenient_state} with one more particle at
position $\xvec$ and one less at $\zvec$, corresponding to the positive
gain term
in \Eref{rho_interacting_FPE_non_inter}, where particle
$k$ jumps from position $\xvec_k=\xvec$ to $\yvec_k=\zvec$. The factor $n_\zvec$ in
\Eref{rho_noninteracting_FPE_rewritten} 
multiplies equally gain and loss terms, because any two $\yvec_k$ in the
summand of $\sum_{k=1}^N$ in \Eref{rho_interacting_FPE_non_inter} that
happen to be equal, say
$\yvec_{k_1}=\yvec_{k_2}=\zvec$, contribute the same, so that $n_\zvec$ is
simply a counter of how many times $\yvec_{k}=\zvec$. 


With the help of \Eref{rho_noninteracting_FPE_rewritten} the time
derivative of $\PC\left(\{n\};t\right)$ defined in \Eref{PC_from_rho} can be written as
\begin{equation}
\ddt{} \PC\left(\{n\};t\right) = 
N! \left( \prod_{\uvec=1}^M \frac{1}{n_\uvec!} \right)
\sum_{\zvec=1}^M n_\zvec
\sum_{\substack{\xvec=1\\ \xvec\ne\zvec}}^M
\big( 
\transMatLin_{\zvec\xvec}
\density\left(\{\yvec_1,\ldots,\yvec_N\}_{\{\ldots,n_{\xvec}+1,\ldots,n_{\zvec}-1,\ldots\}};t\right) 
- \
\transMatLin_{\xvec\zvec} 
\density(\{\yvec_1,\ldots,\yvec_N\}_{\{n\}};t) \big)  
\ .
\end{equation}
Of the two terms on the right-hand side of
\Eref{rho_noninteracting_FPE_rewritten}, the negative loss term immediately
simplifies with the help of \Eref{PC_from_rho}, whereas the first term
needs some adjustments of the preceding factorials, resulting in
\begin{equation}\elabel{particle_num_ME}
\ddt{} \PC\left(\{n\};t\right) = 
\sum_{\zvec=1}^M 
\sum_{\substack{\xvec=1\\ \xvec\ne\zvec}}^M
\big( 
(n_{\xvec}+1)
\transMatLin_{\zvec\xvec}
\PC\left(\{\ldots,n_{\xvec}+1,\ldots,n_{\zvec}-1,\ldots\};t\right) 
- 
n_\zvec
\transMatLin_{\xvec\zvec} 
\PC\left(\{n\};t\right)
\big)  
\ .
\end{equation}
This is the standard form of the master equation for the occupation number
states, \Eref{Pdot}, that immediately translates into an action, \Eref{def_harmonic_action_discrete}. 
So far, we have thus simply rederived the results of
\APref{DP_field_theory_derivation}.

The derivation of \Eref{particle_num_ME} above is founded on the master 
\Eref{rho_noninteracting_FPE_rewritten} in the absence of
the interaction term $\transMatLin^{(\yvec_\ell)}_{\xvec_k\yvec_k}$ in
\Eref{rho_interacting_FPE_inter}. To account for it now, in particular the sum
$\sum_\ell$ in \Eref{rho_interacting_FPE_inter} over the
interacting particles at $\yvec_\ell$ has to be carried out, which in the gain
term is achieved by
\begin{equation}\elabel{transMatInter_simplified1}
\sum_{\substack{\ell=1\\ \ell\ne k}}^N\transMatLin^{(\yvec_\ell)}_{\yvec_k\xvec_k} 
=
\sum_{\vvec=1}^M (n_\vvec - \delta_{\vvec,\yvec_k}) \transMatLin^{(\vvec)}_{\yvec_k\xvec_k} 
\end{equation}
where the Kronecker $\delta$-function accounts for the condition
$\ell\ne k$ in the left sum.
The sum in the loss term can similarly be carried out,
\begin{equation}\elabel{transMatInter_simplified2}
\sum_{\substack{\ell=1\\ \ell\ne k}}^N\transMatLin^{(\yvec_\ell)}_{\xvec_k\yvec_k} 
=
\sum_{\vvec=1}^M (n_\vvec - \delta_{\vvec,\yvec_k}) \transMatLin^{(\vvec)}_{\xvec_k\yvec_k} 
\ ,
\end{equation}
which is \emph{not} \Eref{transMatInter_simplified1} with  $\yvec_k$ and $\xvec_k$ interchanged, because $\delta_{\vvec,\yvec_k}$ depends on $\yvec_k$ in both \Erefs{transMatInter_simplified1} and \eref{transMatInter_simplified2}.
Together with \Eref{particle_num_ME} this  
results in the master equation
\begin{align}\elabel{master_equation_particle_numbers_with_interaction}
\nonumber
\ddt{} \PC\left(\{n\};t\right) = 
\sum_{\zvec=1}^M 
\sum_{\substack{\xvec=1\\ \xvec\ne\zvec}}^M
\bigg\{ &
(n_{\xvec}+1)
  \big[\transMatLin_{\zvec\xvec} 
  + 
  \sum_{\vvec=1}^M (n_\vvec-\delta_{\vvec,\zvec}) \transMatLin^{(\vvec)}_{\zvec\xvec}
  \big]
\PC\left(\{\ldots,n_{\xvec}+1,\ldots,n_{\zvec}-1,\ldots\};t\right) 
\\
&- 
n_\zvec
  \big[\transMatLin_{\xvec\zvec} 
  +
  \sum_{\vvec=1}^M (n_\vvec-\delta_{\vvec,\zvec})
  \transMatLin^{(\vvec)}_{\xvec\zvec}
  \big]
\PC\left(\{n\};t\right)
\bigg\}  
\end{align}
on the occupation number states. The loss term can be included in the gain
term 
as $\xvec=\zvec$
after swapping dummy variables $\xvec$ and $\zvec$ in the summation over
the losses using \Eref{def_Fjj} and using it further
for each $\vvec$ separately,
\begin{equation}\elabel{def_Fjj_extended}
\transMatLin^{(\vvec)}_{\xvec\xvec}=
-\sum_{\substack{\zvec\\ \zvec\ne \xvec}}
\transMatLin^{(\vvec)}_{\zvec\yvec} \ .
\end{equation}
The master \Eref{master_equation_particle_numbers_with_interaction} can
be translated to an action via ladder operators, like \Eref{def_harmonic_action_discrete} was
derived from \Eref{Pdot}.
When introducing the ladder operators, 
the Kronecker
$\delta$-functions in
\Eref{master_equation_particle_numbers_with_interaction} 
are accounted for by the commutation rules of the operators, finally resulting in the action
\begin{equation}\elabel{def_interaction_action_discrete}
    \action = 
\int \dint{t} \sum_\yvec
\phitilde(\yvec,t)
\sum_\xvec 
\bigg[
-
\partial_t \delta_{\yvec,\xvec}
+
\transMatLin_{\yvec\xvec}
+
\sum_{\vvec}
\phidagger(\vvec,t)\phi(\vvec,t)
\transMatLin^{(\vvec)}_{\yvec\xvec}
\bigg]
\phi(\xvec,t) \ ,
\end{equation}
as an extension to \Eref{def_harmonic_action_discrete} beyond a purely harmonic,
Gaussian action. \Erefs{rho_interacting_FPE} and
\eref{def_interaction_action_discrete} are the key insights in the
present section, namely a one-to-one correspondence
between the interacting master \Eref{rho_interacting_FPE} and the action
\Eref{def_interaction_action_discrete} resulting from it, with the sum
over $\ell$ of the interaction partners in
\Eref{rho_interacting_FPE_inter} replaced by the particle
number (density) $\phidagger(\vvec,t)\phi(\vvec,t)$ at $\vvec$ in \Eref{def_interaction_action_discrete}.
We complete this section by taking the continuum limit in the case of
pair-interaction, which covers a vast range of FPEs.

\newcommand{\force}{h_F}

As an example for $\transMatLin^{(\vvec)}_{\yvec\xvec}$
in one dimension one may choose the pair-interaction modelled along the
lines of \Eref{transMatDiffusion},
\begin{equation}\elabel{def_tranMatLin_inter}
\transMatLin^{(c)}_{ba} = 
\half \force(a-c) \big(\delta_{a+1,b} - \delta_{a,b}\big)
-
\half \force(a-c) \big(\delta_{a-1,b} - \delta_{a,b}\big)
\ ,
\end{equation}
where the presence of a particle at 
$c$
facilitates the hopping of
another particle from 
$a$
to 
$b$
by a rate
$\force(a-c)$ that has the effect of a force in discrete space. It depends on the distance $a-c$ and results in a small
bias in the hopping rates in \Eref{transMatDiffusion},
not to
exceed $h_r$ or $h_l$, so that
$h_r+\force(a-c)>0$ and $h_l-\force(a-c)>0$. 
The Kronecker $\delta$-functions in \Eref{def_tranMatLin_inter} guarantee that 
$\transMatLin^{(c)}_{ba}$
obeys \Eref{def_Fjj_extended}.

To take the continuum limit, the rate $\force(a-c)$ is changed to
$-\pairPot'(\xvec-\vvec)/\Deltax$, so that with
\Eref{delta_dash_as_a_limit}
\begin{equation}\elabel{continuum_limit_interaction}
\lim_{\Deltax\to0} \Deltax^{-1} \transMatLin^{(c)}_{ba} 
= - \pairPot'(\xvec-\vvec) \partial_\xvec \delta(\xvec-\yvec) 
= \pairPot'(\xvec-\vvec) \partial_\yvec \delta(\yvec-\xvec) 
\end{equation}
at constant $\xvec=a\Deltax$ as well as $\yvec=b\Deltax$ and $\vvec=c\Deltax$ and
therefore $\pairPot'(\xvec-\vvec)$, thought to be the gradient of a
pair-potential. \Eref{continuum_limit_interaction} is easily generalised to higher
dimensions, with the restriction of the rate to be essentially a gradient only as to arrive at the standard form of an action with an external potential.
Using \Eref{continuum_limit_interaction} in the FP
\Eref{rho_interacting_FPE} together with \Eref{continuum_limit_nointeraction}
for the non-interacting rate matrix with finite, constant extinction rate
$r$,  produces the standard interacting
FPE
\begin{equation}\elabel{interacting_FPE_example}
\dot{\density}(\yvec_1,\ldots,\yvec_N;t) = 
\sum_{k=1}^N
\big(\diffusion \nabla_{\yvec_k}^2 - \drift\cdot\nabla_{\yvec_k} - r\big)
\density(\yvec_1,\ldots,\yvec_N;t)
+
\sum_{k=1}^N
\nabla_{\yvec_k}
\cdot
\bigg(
\sum_{\substack{\ell=1\\ \ell\ne k}}^N
\pairPot'(\yvec_k-\yvec_\ell)\density(\yvec_1,\ldots,\yvec_N;t)
\bigg)
\end{equation}
for drift-diffusion particles which interact by a pair-potential
$\pairPot$, where $\pairPot'(\yvec-\vvec)$ denotes $\nabla_\yvec \pairPot(\yvec-\vvec)$. 

The corresponding action in the continuum generalised to
$d$ dimensions with integrals over Dirac $\delta$-functions readily carried
out is
\begin{equation}\elabel{interacting_action_example}
    \action = 
\int \dint{t} \int \ddint{y}
\phitilde(\yvec,t)
(-\partial_t+\diffusion \nabla^2_\yvec - \drift\cdot\nabla_\yvec -r)
\phi(\yvec,t)\\
+ \int \dint{t} \int\ddint{y} 
 \phitilde(\yvec,t) \nabla_\yvec \cdot
\bigg(
\int \ddint{v}
\phidagger(\vvec,t)\phi(\vvec,t)
\pairPot'(\yvec-\vvec) 
\phi(\yvec,t)
\bigg)
\end{equation}
with the first double integral given by 
$\action_0$ in \Erefs{def_harmonic_action} or
\eref{action_drift_diffusion_appendix}. The interaction part
corresponds to that obtained by physical reasoning in 
\Erefs{trawler_A1v1}
and \eref{trawler_A2v1}. The term $\phidagger(\vvec,t)\phi(\vvec,t)$ is
easily
expressed in terms of Doi-shifted fields and has its origin in the particle
number at the point $\vvec$ of interaction.

\Erefs{interacting_FPE_example} and \eref{interacting_action_example}
demonstrate in a concrete example how immediate the correspondence is
between the FPE and the action. 
If the FP \Eref{interacting_FPE_example} applies to all $N\in\Nset$ simultaneously, then no
additional assumption is needed to derive the fully equivalent action
\Eref{interacting_action_example}, unlike the derivation of the master
\Eref{Pdot} for multiple, non-interacting
particles from that of a single particle \Eref{rho_i_ME_mass}, as discussed
before
\Eref{weighted_sum_Fock}, which needs the assumption of concurrent,
independent evolution.

When casting an FPE in an action, to obtain the
harmonic, Gaussian part, the sum over the
individual particles is replaced by an integral over all space and the
operator is sandwiched in creation and annihilation fields as if the FPE was subjected to the Martin-Siggia-Rose procedure
\cite{MartinSiggiaRose:1973,Janssen:1976,DeDominicis:1976,Taeuber:2014},
even when the DP annihilation field is not a simple particle
density and observables have to be constructed by recourse to
non-commuting operators. To cast the interaction part of the FPE into an DP
action, in addition to the steps above, the sum over all \emph{distinct}
interaction partners, $\sum_{\ell\ne k}$, is replaced by a suitable
``particle count'', $\int \ddint{v} \phidagger(\vvec,t)\phi(\vvec,t)$.
Both, FPE and DP field-theory, are an exact representation of the microscopic dynamics,
but the DP
framework readily provides a diagrammatic language and a systematic
perturbation theory to express any $n$-point density. In
\APref{entropy_production} we show that the only terms
needed in the construction of the entropy production are
the first order terms of the perturbative expansion. Before we do that, we extract and characterised the propagator.




\toasubsection{The propagator}\seclabel{the_propagator}
Using the canonical procedure \cite{Peliti:1985,TaeuberHowardVollmayr-Lee:2005,Cardy:2008,Taeuber:2014},
in the following we will derive some properties of the propagator 
$\ave{\phi(\yvec,t')\phitilde(\xvec,t)}$, which, strictly, is the expected particle number in discrete state $\yvec$ at time $t'$, given a single particle was initially placed in discrete state $\xvec$ at time $t$. In this case, because there is only one particle, the expected number of particles at $\xvec$ is identical to the probability that the particle is at $\xvec$. We determine first the bare propagator $\ave[0]{\phi(\yvec,t')\phitilde(\xvec,t)}$
for the discrete state action \Eref{def_harmonic_action_discrete} and then the full propagator perturbatively for an additional 
perturbative action 
\begin{equation}\elabel{gen_pert_action}
    \action_\perturbative = 
\int \dint{t} \sum_{\yvec,\xvec}
\phitilde(\yvec,t)
\transMatNonLin_{\yvec\xvec} 
\phi(\xvec,t) \ ,
\end{equation}
using \Eref{perturbation_theory}.
The perturbative expansion of the propagator will feed into an \emph{exact} expression for the entropy production in \APref{entropy_production}. That \Eref{gen_pert_action} is bilinear might look like a significant loss of generality, yet what matters below is not the precise form of the action, but the expansion of the propagator that results from it. We shall therefore consider $\transMatNonLin_{\yvec\xvec}$ as a generic higher order correction to the propagator. As qualified further below, we need to make certain assumptions on the time-dependence of $\transMatNonLin_{\yvec\xvec}$. For now, we may think of it as having no time-dependence. Given the conservative nature of the dynamics and general time-homogeneity this is not a strong restriction. In the continuum, a suitable perturbation might be self-propulsion or a potential, in discrete state space, the perturbation could be transitions beyond those convenient for the harmonic part.

The discreteness of the state space considered thus far also seems to reduce generality. This is indeed an important constraint, which will require careful resolution in \APref{drift_diffusion_on_ring}, in particular \APref{Fourier_transformation}. Even when we are able to determine the action of a continuous state process from its FPE, \Eref{def_harmonic_action} from \Eref{rho_i_ME_matrix_rewrite_kernel}, 
and derive an expression for the entropy production in the final \Sref{EP_cont},
in the following we will focus entirely on discrete states and leave the generalisation of the arguments for later.

\toasubsubsection{The bare propagator}
\seclabel{bare_propagator_diffusion}
The bare propagator $\ave[0]{\phi(\yvec,t')\phitilde(\xvec,t)}$ is most easily calculated after Fourier-transforming the fields,
\begin{equation}
    \phi(\yvec,t)=\int\dintbar{\omega}\exp{-\imag\omega t}\phi(\yvec,\omega)
\quad\text{ and }\quad
    \phitilde(\xvec,t')=\int\dintbar{\omega'}\exp{-\imag\omega' t'}\phi(\xvec,\omega')
\end{equation}
with $\dintbar{\omega}=\dint{\omega}/(2\pi)$, so that the harmonic part of the action \Eref{def_harmonic_action_discrete} becomes
\begin{equation}\elabel{def_harmonic_action_discrete_omega}
    \action_0 = 
\int \dintbar{\omega} \sum_{\yvec,\xvec}
\phitilde(\yvec,-\omega)
\bigg[
\imag\omega
\delta_{\yvec,\xvec}
-
r_\yvec \delta_{\yvec,\xvec}
+
\transMatLin_{\yvec\xvec}
\bigg]
\phi(\xvec,\omega)
\end{equation}
and correspondingly 
\begin{equation}\elabel{gen_pert_action_omega}
    \action_\perturbative = 
\int \dintbar{\omega} 
\sum_{\yvec,\xvec}
\phitilde(\yvec,-\omega)
\transMatNonLin_{\yvec\xvec} 
\phi(\xvec,\omega) \ .
\end{equation}
The bare propagator is 
\begin{equation}\elabel{orig_propagator_as_an_inverse}
    \tbarePropagator{\xvec,\omega}{\yvec,\omega'}
\corresponds
\ave[0]{\phi(\yvec,\omega')\phitilde(\xvec,\omega)}
=
\deltabar(\omega'+\omega)
\left(
    \left[-\imag \omega \ident +\diag(\rvec) - \transMatLin\right]^{-1}
\right)_{\yvec\xvec} \ ,
\end{equation}
derived, if necessary, using a transformation that diagonalises $\transMatLin$. Using that $\transMatLin$ is a Markov matrix and $\Re(r_y)>0$, this may be transformed into direct time
\begin{equation}\elabel{bare_propagator_realtime}
\tbarePropagator{\xvec,t}{\yvec,t'}
\corresponds
\ave[0]{\phi(\yvec,t')\phitilde(\xvec,t)}
= 
\theta(t'-t)
\Big(\Exp{(t'-t)\big[\transMatLin-\diag(\rvec)\big]}\Big)_{\yvec\xvec}
\ .
\end{equation}
\Eref{bare_propagator_realtime} implies that $\ave[0]{\phi(\yvec,t')\phitilde(\xvec,t)}$ solves the non-interacting master \Eref{rho_i_ME_matrix_rewrite} for $t'>t$, as
\begin{equation}\elabel{propagator_limit_is_delta}
\lim_{t'\downarrow t} \ave[0]{\phi(\yvec,t')\phitilde(\xvec,t)}
=
\delta_{\yvec,\xvec}
\quad\text{with}\quad
\lim_{t'\downarrow t} \ave[0]{\phi(\yvec,t')\phitilde(\xvec,t)}
=
\delta(\yvec-\xvec)
\quad\text{in the continuum}
\end{equation}
and thus
\begin{equation}\elabel{propagator_deri}
    \partial_{t'}
    \ave[0]{\phi(\yvec,t')\phitilde(\xvec,t)}
    =
    \delta_{\yvec,\xvec}\delta(t'-t) + 
    \sum_\zvec \left(\transMatLin_{\yvec\zvec} - r_\yvec \delta_{\yvec,\zvec}\right)
    \ave[0]{\phi(\zvec,t')\phitilde(\xvec,t)}
    \ ,
\end{equation}
where the term $\delta_{\yvec,\xvec}\delta(t'-t)$ is due to the derivative of the Heaviside $\theta$-function and  $\delta(t'-t)\Exp{(t'-t)(\transMatLin-\diag(\rvec))}=\ident\delta(t'-t)$.
After taking the continuum limit, the bare propagator will correspondingly be the Green function of a non-interacting FPE, say \Eref{FPE_standard}.

\toasubsubsection{Perturbative expansion of the full propagator}
\seclabel{pert_exp_full_prop}
The full propagator $\ave{\phi(\yvec,t')\phitilde(\xvec,t)}$ acquires corrections from the perturbative part of the action \Eref{gen_pert_action}, so that, \Eref{propagator_expansion},
\begin{align}
\tbarePropagator{\xvec,t}{\yvec,t'} & +
\tblobbedPropagator{\xvec,t}{\yvec,t'} +
\tDblobbedPropagator{\xvec,t}{\yvec,t'} + \ldots\,
\corresponds
\ave{\phi(\yvec,t')\phitilde(\xvec,t)} \nonumber\\
=&
\ave[0]{\phi(\yvec,t')\phitilde(\xvec,t)}
+
\int_{-\infty}^\infty \dint{s} \sum_{\avec,\bvec}
\ave[0]{\phi(\yvec,t')\phitilde(\bvec,s)}
\transMatNonLin_{\bvec\avec}
\ave[0]{\phi(\avec,s)\phitilde(\xvec,t)}
+
\ldots
\elabel{propagator_expansion_app}
\end{align}
The bare propagator is stated in \Eref{bare_propagator_realtime} and the first order correction is easily determined explicitly,
\begin{equation}\elabel{first_order_explicitly}
        \tblobbedPropagator{\xvec,t}{\yvec,t'} 
\corresponds
\int_{t}^{t'} \dint{s} \sum_{\avec,\bvec}
\left(\Exp{(t'-s)\transMatLin-\diag(\rvec)}\right)_{\yvec\bvec}
\transMatNonLin_{\bvec\avec}
\left(\Exp{(s-t)\transMatLin-\diag(\rvec)}\right)_{\avec\xvec} \ .
\end{equation}
This is generally not trivial to evaluate, because the matrix exponentials and $\transMatNonLin$ generally do not commute. Yet, \Eref{first_order_explicitly}
clearly vanishes as $t'\downarrow t$. The derivative of \Eref{first_order_explicitly} with respect to $t'$ produces two terms, one from the differentiation of the integrand and one from the differentiation of the integration limits. In the limit $t'\downarrow t$ only the latter contributes, as the integral vanishes for $t'\downarrow t$, so that
\begin{equation}\elabel{deri_first_order_correction_final}
\lim_{t'\downarrow t}
        \partial_{t'} 
        \tblobbedPropagator{\xvec,t}{\yvec,t'} 
\corresponds
\sum_{\avec,\bvec}
\delta_{\yvec,\bvec}
\transMatNonLin_{\bvec\avec}
\delta_{\avec,\xvec} =
\transMatNonLin_{\yvec\xvec}
\ .
\end{equation}
The diagrammatics in terms of perturbative ``blobs" is further discussed in \APref{appendixWhichDiagramsContribute}. Based on these arguments, or by direct evaluation of the convolutions using \Eref{bare_propagator_realtime}, one can show that a term to $n$th order in the perturbation vanishes like $(t'-t)^n$.

In summary, 
\begin{subequations}
\elabel{summary_propagator_first_order_summary}
\begin{align}
\elabel{summary_propagator_first_order_summary_nonDeri}
    \lim_{t'\downarrow t} 
    \ave{\phi(\yvec,t')\phitilde(\xvec,t)}
    &=
    \delta_{\yvec,\xvec}\\
    \elabel{summary_propagator_first_order_summary_deri}
    \lim_{t'\downarrow t} 
    \partial_{t'} \ave{\phi(\yvec,t')\phitilde(\xvec,t)}
    &=
    \transMatLin_{\yvec\xvec} - r_\yvec \delta_{\yvec,\xvec}
    +
    \transMatNonLin_{\yvec\xvec}
\end{align}
\end{subequations}
When we discuss entropy production in the following, we will drop the mass term $r_\yvec \delta_{\yvec,\xvec}$, as in the present work we treat only conservative dynamics.

\toasubsection{Entropy production}\seclabel{entropy_production}
To calculate the entropy production we cannot follow \cite{NardiniETAL:2017} and attempt to derive a "path density" in the form $\PC([\phi])\propto\int\Dint{\phitilde}\exp{\action}$, firstly because this integral generally cannot be sensibly performed as $\phitilde$ is introduced as the complex conjugate of $\phi$, and secondly because $\phi(\xvec,t)$ is not a particle density, but rather the conjugate of the annihilation operator. The quantity $\PC([\phi])$ therefore does not have the meaning of a probability density of a particular history of particle movements.

We will now use the propagator as characterised in \Eref{summary_propagator_first_order_summary} to calculate the entropy production of the continuous time Markov chain 
\Eref{rho_i_ME_mass} with the generic interaction \Eref{gen_pert_action} added.
The \emph{internal entropy production} (rate) by a single particle, whose sole degree of
freedom is its state $\xvec$, is generally given by \cite{Gaspard:2004}
\begin{equation}
\elabel{def_entropyProduction_Suppl_initial}
\entropyProduction[\density] = 
\lim_{\Deltat\downarrow0}
\frac{1}{2\Deltat}
\sum_{\yvec\xvec}
\Big\{
\density(\xvec)\Transition_{\yvec\xvec}(\Deltat)
-
\density(\yvec)\Transition_{\xvec\yvec}(\Deltat)
\Big\}
\ln
\left(
\frac
{\density(\xvec)\Transition_{\yvec\xvec}(\Deltat)}
{\density(\yvec)\Transition_{\xvec\yvec}(\Deltat)}
\right)
\end{equation}
where we define $0\ln(0/0)=0$ to make the expression well-defined even when some transition rates $\Transition_{\yvec\xvec}$ vanish. The external entropy production is closely related and identical to the negative of the internal entropy production at stationarity \cite{Gaspard:2004,CocconiGarcia-MillanETAL:2020}.
The \emph{functional}
$\entropyProduction[\density]$ is the rate of entropy production 
by the system given $\density(\xvec)$ as the probability  of finding the particle in state $\xvec$. Compared to 
\Eref{rho_i_ME_mass} we have dropped the time dependence of $\density(\xvec)$ to emphasise that in the expression above we consider the density as given.

Further,
$\Transition_{\yvec\xvec}(\Deltat)$ denotes the probability of the particle transitioning from state $\xvec$ to state $\yvec$ 
over the course of time $\Deltat$. 
With 
$\lim_{\Deltat\to0}\Transition_{\yvec\xvec}(\Deltat)=\delta_{\yvec,\xvec}$
and
\begin{equation}\elabel{Transition_dot}
\TransitionRate_{\yvec\xvec}
=
\lim_{\Deltat\downarrow0}
\ddX{\Deltat} \Transition_{\yvec\xvec}(\Deltat)
= \lim_{\Deltat\downarrow0}
\frac{\Transition_{\yvec\xvec}(\Deltat) - \Transition_{\yvec\xvec}(0)}{\Deltat}
\ .
\end{equation}
we have 
\begin{equation}\elabel{simplified_curly_bracket}
\lim_{\Deltat\downarrow0}\frac{1}{\Deltat}
\Big\{
\density(\xvec)\Transition_{\yvec\xvec}(\Deltat)
-
\density(\yvec)\Transition_{\xvec\yvec}(\Deltat)
\Big\}
=
\density(\xvec)\TransitionRate_{\yvec\xvec}
-
\density(\yvec)\TransitionRate_{\xvec\yvec} \ .
\end{equation}

Given we are studying a continuous time Markov chain, $\TransitionRate_{\yvec\xvec}$ is a rate matrix, so that \cite{Gaspard:2004}
\begin{equation}\elabel{transition_expansion}
    \Transition_{\yvec\xvec}(\Deltat) = \delta_{\yvec,\xvec}
    + \Deltat \TransitionRate_{\yvec\xvec}
    + \order{\Deltat^2} \ ,
\end{equation}
with $\TransitionRate_{\yvec\xvec}\ge0$ for $\yvec\ne\xvec$ and $\TransitionRate_{\yvec\yvec}<0$, accounting for the loss of any state $\yvec$ into all other accessible states, as normally implemented by definition of a Markovian rate matrix, \Eref{def_Fjj}.
In the Markov chain introduced at the beginning of the present appendix, the rate matrix $\TransitionRate$ of \Eref{Transition_dot} \emph{is} in fact the Markov matrix $\transMatLin$ of the master \Eref{rho_i_ME_mass} with \Eref{def_Fjj}, \ie $\TransitionRate=\transMatLin$. We will keep the notation separate to allow for $\TransitionRate$ to acquire corrections beyond $\transMatLin$ due to perturbations.

Below, we will demonstrate that the transition rate matrix $\TransitionRate$ plays the r{\^o}le of a kernel. Indeed, in the continuum, \APref{EP_cont}, it can be written as the Fokker-Planck operator acting on a Dirac $\delta$-function. To this end, we introduce separately
\begin{equation}\elabel{def_Op_app}
    \Op_{\yvec\xvec} = \lim_{\Deltat\downarrow0}
\ddX{\Deltat} \Transition_{\yvec\xvec}(\Deltat)
\end{equation}
even when in the present case of a Markov chain we simply have that $\Op=\TransitionRate$, \Eref{Transition_dot}. This term is the focus of much of this work.

Using \Eref{simplified_curly_bracket} in \eref{def_entropyProduction_Suppl_initial}, the entropy production (rate) is
\begin{align}
\elabel{def_entropyProduction_Suppl_step1}
\entropyProduction[\density] = &
\half
\sum_{\yvec\xvec}
\Big\{
\density(\xvec)\TransitionRate_{\yvec\xvec}
-
\density(\yvec)\TransitionRate_{\xvec\yvec}
\Big\}
\lim_{\Deltat\downarrow0}
\ln
\left(
\frac
{\density(\xvec)\Transition_{\yvec\xvec}(\Deltat)}
{\density(\yvec)\Transition_{\xvec\yvec}(\Deltat)}
\right) \nonumber\\
= &
\sum_{\yvec\xvec}
\density(\xvec)\Op_{\yvec\xvec}
\lim_{\Deltat\downarrow0}
\ln
\left(
\frac
{\density(\xvec)\Transition_{\yvec\xvec}(\Deltat)}
{\density(\yvec)\Transition_{\xvec\yvec}(\Deltat)}
\right)
\end{align}
assuming that both limits exist and defining now also $0\ln(0)=0$.

The logarithm vanishes for $\yvec=\xvec$ and we shall therefore proceed assuming $\yvec\ne\xvec$. It may be considered to be comprised of two terms: The first one, $\ln(\density(\xvec)/\density(\yvec))$, contains only the density $\density$ and is independent of the time $\Deltat$. The contribution from this term to the entropy production vanishes when $\density(\xvec)$ is stationary. The second logarithmic term in \Eref{def_entropyProduction_Suppl_step1} we define as
\begin{equation}\elabel{def_Ln_app}
    \Ln_{\yvec\xvec} =
    \lim_{\Deltat\downarrow0}
\ln
\left(
\frac
{\Transition_{\yvec\xvec}(\Deltat)}
{\Transition_{\xvec\yvec}(\Deltat)}
\right)
\ .
\end{equation}
This term generally contributes at 
stationarity
and is the second term the present work focuses on. With definitions \Erefs{def_Op_app} and \eref{def_Ln_app}
we can write the entropy production as
\begin{equation}
\elabel{def_entropyProduction_Suppl_step2}
\entropyProduction[\density] = 
\sum_{\yvec\xvec}
\density(\xvec)\Op_{\yvec\xvec}
\left\{
\Ln_{\yvec\xvec}
+ 
\ln\left(
\frac
{\density(\xvec)}
{\density(\yvec)}
\right)
\right\}
=
\sum_{\xvec} 
\density(\xvec)
\entropyProductionDensity(\xvec)
+
\sum_{\xvec} 
\density(\xvec)
\Op_{\yvec\xvec}
\sum_\yvec 
\ln\left(
\frac
{\density(\xvec)}
{\density(\yvec)}
\right)
\end{equation}
where we have introduced the \emph{(stationary) local entropy production},
\begin{equation}\elabel{def_entropyProductionDensity_app}
    \entropyProductionDensity(\xvec) =
    \sum_\yvec
    \Op_{\yvec\xvec}
\Ln_{\yvec\xvec} \ .
\end{equation}
This notation is also a reminder that this expression for the entropy production goes back to Kullback and Leibler \cite{KullbackLeibler:1951}.

Focusing now on
a Markov chain, the kernel is simply the transition rate matrix,
\begin{equation}\elabel{Kn_from_TransitionRate}
\Op_{\yvec\xvec}=\TransitionRate_{\yvec\xvec} \ ,    
\end{equation}
\Eref{def_Op_app} and \eref{Transition_dot}. 
The logarithm term $\Ln_{\yvec,\xvec}$, \Eref{def_Ln_app}, obviously vanishes when $\yvec=\xvec$ and is otherwise easily determined using \Eref{transition_expansion} and L'H{\^o}pital's rule.
In principle, this requires higher order derivatives beyond $\TransitionRate_{\yvec\xvec}$, if it vanishes. However, in this case $\Op_{\yvec\xvec}$ in \Erefs{def_entropyProduction_Suppl_step2} and \eref{def_entropyProductionDensity_app} vanishes as well and we thus write
\begin{subnumcases}{\elabel{Ln_from_TransitionRate}\Ln_{\yvec\xvec}=}
    0 & for $\yvec=\xvec$ \elabel{Ln_yy_vanishes}\\
    0 & for $\yvec\ne\xvec$ and $\TransitionRate_{\yvec\xvec}=0$\\
    \ln
\left(
\frac
{\TransitionRate_{\yvec\xvec}}
{\TransitionRate_{\xvec\yvec}}
\right)
& otherwise \ ,
\end{subnumcases}
making use of $0\ln(0/0)=0=0\ln(0)$ in case $\TransitionRate_{\yvec\xvec}$ or both $\TransitionRate_{\yvec\xvec}$ and $\TransitionRate_{\yvec\xvec}$ vanish. Strictly, \Eref{Ln_from_TransitionRate} is thus the limit \Eref{def_Ln_app} only in case of $\yvec=\xvec$ or whenever $\Op_{\yvec\xvec}=\TransitionRate_{\yvec\xvec}$ does not vanish.

In the present section, we have determined expressions for the entropy production \emph{given} the transition rate matrix $\TransitionRate$. We proceed by showing how transition rate matrix and thus entropy production are determined by a field theory.

\toasubsubsection{Expressing the entropy production in terms of propagators}
\seclabel{EP_from_propagators}
Both $\Op$ and $\Ln$ are based on the transition probability $\Transition_{\yvec\xvec}(\Deltat)$, \Erefs{def_Op_app} and \eref{def_Ln_app}. In a field-theoretic description, the probability to be in state $\yvec$ having started from state $\xvec$ is given by 
the propagator, which here serves the purpose of an indicator function, indicating the presence of the particle in state $\yvec$ through the number of particles at $\yvec$. This approach of using the propagator as a transition probability (density) therefore requires the vacuum as the initial state. It is readily extended to multiple distinguishable particles (\APref{N_indistinguishable_particles}), and, using a Gibbs-factor, even to indistinguishable particles with multiple occupation provided that \emph{all} particles are being annihilated. 

The transition probability to be in $\yvec$ starting from $\xvec$ is thus
\Eref{propagator_expansion_app}
\begin{align}
    \Transition_{\yvec\xvec}(\Deltat)
    = &
    \ave{\phi(\yvec,t+\Deltat)\phitilde(\xvec,t)}\nonumber\\
    \corresponds &
    \tbarePropagatorL{\xvec,t}{\yvec,t+\Deltat} + 
\tblobbedPropagator{\xvec,t}{\yvec,t+\Deltat} +
\tDblobbedPropagator{\xvec,t}{\yvec,t+\Deltat} + \ldots\ ,
\end{align}
which is independent of $t$ due to time translational invariance. Using this expression in \Erefs{def_Op_app}, \eref{def_Ln_app} and \eref{summary_propagator_first_order_summary} with $r_\yvec\downarrow0$ gives
\begin{subequations}
\elabel{Op_and_Ln_from_diagrams}
\begin{align} 
    \Op_{\yvec\xvec} &= 
    \lim_{\Deltat\downarrow0}
\ddX{\Deltat} \ave{\phi(\yvec,t+\Deltat)\phitilde(\xvec,t)}
=
\transMatLin_{\yvec\xvec} 
    +
    \transMatNonLin_{\yvec\xvec}
\nonumber\\
\elabel{Op_in_diagrams}
&\corresponds 
\lim_{\Deltat\downarrow0}
\ddX{\Deltat} 
\left(
\tbarePropagatorL{\xvec,t}{\yvec,t+\Deltat} + 
\tblobbedPropagator{\xvec,t}{\yvec,t+\Deltat}
\right)
\\
\Ln_{\yvec\xvec} &= 
    \lim_{\Deltat\downarrow0}
\ln
\left(
\frac
{\ave{\phi(\yvec,t+\Deltat)\phitilde(\xvec,t)}}
{\ave{\phi(\xvec,t+\Deltat)\phitilde(\yvec,t)}}
\right)
\nonumber\\
\elabel{Ln_in_diagrams}
&\corresponds 
\lim_{\Deltat\downarrow0}
\ln
\left(
\frac
{\tbarePropagatorL{\xvec,t}{\yvec,t+\Deltat} + 
\tblobbedPropagator{\xvec,t}{\yvec,t+\Deltat}}
{\tbarePropagatorL{\yvec,t}{\xvec,t+\Deltat} + 
\tblobbedPropagator{\yvec,t}{\xvec,t+\Deltat}}
\right)
\end{align}
\end{subequations}
where the diagrams are shown only to first order in the perturbation, as higher orders, those $\propto \Deltat^2$ and higher, cannot possibly contribute, \APref{pert_exp_full_prop}.

\Eref{Op_in_diagrams} is explicitly the first order contribution in $\Deltat$ to the propagator and is determined immediately 
using \Eref{summary_propagator_first_order_summary_deri} with $r_\yvec\downarrow0$ to preserve the particle number, unity. In the logarithm, we might first use \Eref{summary_propagator_first_order_summary_nonDeri}, but that produces a meaningful result only for $\yvec=\xvec$, in which case indeed $\Ln_{\yvec\yvec}=0$, \Eref{Ln_yy_vanishes}. For $\yvec\ne\xvec$ we need to apply L'H{\^o}pital, so that with \Eref{summary_propagator_first_order_summary_deri} for $r_\yvec\downarrow0$ (conserved particle number),
\begin{equation}\elabel{Ln_from_propagator}
    \Ln_{\yvec\xvec} =
    \ln\left(
    \frac
    {\transMatLin_{\yvec\xvec} 
    +
    \transMatNonLin_{\yvec\xvec}}
    {\transMatLin_{\xvec\yvec} 
    +
    \transMatNonLin_{\xvec\yvec}}
    \right) \ .
\end{equation}

In summary, the entropy production of a continuous time Markov chain with density $\density(\xvec)$ given, stationary or not, is \Eref{def_entropyProduction_Suppl_step2} with kernel $\Op$ in \Eref{Op_in_diagrams} and $\Ln$ in \Eref{Ln_from_propagator}.

\toasubsubsection{Continuum Limit}
\seclabel{EP_cont}
As long as states are discrete and rates therefore finite (\APref{Fourier_transformation}) the logarithm $\Ln$ \Eref{Ln_from_propagator} is a function of the kernel $\Op$, \Eref{Op_in_diagrams}. In the continuum this simple relationship breaks down. To find the relevant expressions in the continuum, we return to the propagator in the continuum, replacing rate matrices \etc by their continuum counterparts. Much of the following is done in further detail in \APref{drift_diffusion_on_ring} and illustrated further in \APref{MultipleParticles}. Below we present only the basic argument.

For continuous states $\xvec,\yvec$ the probability $\density(\xvec)$ in \Eref{def_entropyProduction_Suppl_step2} is a \emph{density} which we denote by the same symbol $\density(\xvec)$. Similarly, the kernel $\Op_{\yvec,\xvec}$, which for discrete states is a rate, has units of a rate \emph{density} on $\yvec$, with $\xvec$ given. Correspondingly, the expression for the entropy production \Eref{def_entropyProduction_Suppl_step2} becomes the double integral \begin{equation}
\elabel{def_entropyProduction_Suppl_step2_integral}
\entropyProduction[\density] = 
\int\dint{\xvec}\dint{\yvec}
\density(\xvec)\Op_{\yvec\xvec}
\left\{
\Ln_{\yvec\xvec}
+ 
\ln\left(
\frac
{\density(\xvec)}
{\density(\yvec)}
\right)
\right\}
=
\int\dint{\xvec}
\density(\xvec)
\entropyProductionDensity(\xvec)
+
\int\dint{\xvec}
\density(\xvec)
\Op_{\yvec\xvec}
\int\dint{\yvec}
\ln\left(
\frac
{\density(\xvec)}
{\density(\yvec)}
\right)
\end{equation}
with \Eref{def_entropyProductionDensity_app} replaced by 
\begin{equation}\elabel{def_entropyProductionDensity_app_continuous}
    \entropyProductionDensity(\xvec) =
\int\dint{\yvec}
    \Op_{\yvec\xvec}
\Ln_{\yvec\xvec} \ .
\end{equation}
The continuum limit of the kernel is easy to determine using \Eref{Op_in_diagrams} with \Erefs{continuum_limit_kernel} and \eref{def_harmonic_action}, effectively replacing $\transMatLin_{\yvec\xvec} - r_\yvec \delta_{\yvec,\xvec}$ 
in \Eref{Op_in_diagrams} by $\FPop_{\yvec\xvec}$, \Eref{continuum_limit_both}, setting again $r_\yvec=0$ to preserve the particle number.

Using further the definition similar to \Eref{continuum_limit_kernel},
\begin{equation}
    \PertOp_{\yvec\xvec}=
    \lim_{\Deltax\to0}\frac{1}{\Deltax} \transMatNonLin_{\yvec\xvec}
\corresponds
\fullBlobb{\xvec}{\yvec}
\end{equation}
to capture the contribution from the perturbative part in the continuum limit of \Eref{Op_in_diagrams}, we obtain \Eref{transition_from_action},
\begin{equation}\elabel{Op_continuum_Fpop_plus}
    \Op_{\yvec\xvec}=\FPop_{\yvec\xvec}  + \PertOp_{\yvec\xvec} \corresponds \FPop_{\yvec\xvec}  + 
    \fullBlobb{\xvec}{\yvec} \ .
\end{equation}

In the continuum,
the kernel $\Op_{\yvec\xvec}$ thus turns into an operator acting on $\delta$-functions . Using the L'H{\^o}pital route, one might expect the same for the logarithm, but it is hard to see how such ill-defined objects can be evaluated as a ratio within the logarithm. Instead, we assume at this stage and later demonstrate explicitly, \APref{drift_diffusion_on_ring}, that the following approach is useful. We write
\begin{equation}\elabel{to_be_expanded_mess}
    \frac
    {\tbarePropagatorL{\xvec,t}{\yvec,t+\Deltat} + 
\tblobbedPropagator{\xvec,t}{\yvec,t+\Deltat}}
{\tbarePropagatorL{\yvec,t}{\xvec,t+\Deltat} + 
\tblobbedPropagator{\yvec,t}{\xvec,t+\Deltat}}
=
    \frac
    {\tbarePropagatorL{\xvec,t}{\yvec,t+\Deltat}}
{\tbarePropagatorL{\yvec,t}{\xvec,t+\Deltat}}
\quad
    \frac
    {
    1+\frac
        {\tblobbedPropagatorS{\xvec,t}{\yvec,t+\Deltat}}
        {\tbarePropagatorLS{\xvec,t}{\yvec,t+\Deltat}}
    }
    {
    1+\frac
        {\tblobbedPropagatorS{\yvec,t}{\xvec,t+\Deltat}}
        {\tbarePropagatorLS{\yvec,t}{\xvec,t+\Deltat}}
    } \ ,
\end{equation}
for which we will assume that 
\begin{equation*}
\frac
        {\tblobbedPropagatorS{\xvec,t}{\yvec,t+\Deltat}}
        {\tbarePropagatorLS{\xvec,t}{\yvec,t+\Deltat}}
\qquad\text{ and }\qquad
    \frac
        {\tblobbedPropagator{\yvec,t}{\xvec,t+\Deltat}}
        {\tbarePropagatorL{\yvec,t}{\xvec,t+\Deltat}}
\end{equation*}
appearing in the numerator and the denominator respectively of the rightmost fraction in \Eref{to_be_expanded_mess},
are small in a sense further discussed in \APref{drift_diffusion_on_ring}.
Taking the limit of the logarithm of the above expression gives the right hand side of \Eref{Ln_in_diagrams} in the continuum limit,
\begin{equation}\elabel{Ln_for_continuous}
    \Ln_{\yvec\xvec} \corresponds
    \lim_{\Deltat\downarrow0}
    \left\{
    \ln\left(
    \frac
    {\tbarePropagatorL{\xvec,t}{\yvec,t+\Deltat}}
{\tbarePropagatorL{\yvec,t}{\xvec,t+\Deltat}}
    \right)
    +
    \frac
        {\tblobbedPropagator{\xvec,t}{\yvec,t+\Deltat}}
        {\tbarePropagatorL{\xvec,t}{\yvec,t+\Deltat}}
-
    \frac
        {\tblobbedPropagator{\yvec,t}{\xvec,t+\Deltat}}
        {\tbarePropagatorL{\yvec,t}{\xvec,t+\Deltat}}
    \right\} \ ,
\end{equation}
as illustrated in \APref{drift_diffusion_on_ring}, in particular \Eref{Ln_from_drift_diffusion}.
The limit of the logarithm of the ratio of bare propagators is in general available, because the bare propagator is known explicitly. The ratio of the correction and the bare propagator can be expected to be finite, as the correction draws itself on the bare propagation.

The above continuum limit concludes the present appendix, \APref{DP_plus_EP}. It contains essentially all technical details of how to proceed from the a master equation such as \Eref{rho_i_ME_matrix_rewrite} or an FPE such as \Eref{FPE_diffusion_example} or \eref{rho_i_ME_matrix_rewrite_kernel} to an action \Erefs{def_harmonic_action_discrete} or \eref{def_harmonic_action}. Expanding the resulting propagator for short times, \Eref{summary_propagator_first_order_summary}, finally produces expressions for the entropy production, \Eref{def_entropyProduction_Suppl_step2} with \eref{Op_and_Ln_from_diagrams}, and in the continuum \Eref{def_entropyProduction_Suppl_step2_integral} with \eref{Op_continuum_Fpop_plus} and \eref{Ln_for_continuous}.


\toasection{Short-time scaling of diagrams}\seclabel{appendixWhichDiagramsContribute}

\makeatletter
\makeatother
\newcommand{\IntMultiProp}{I_{n}}
\newcommand{\IntCross}{I_{\mathsf{X}}}
\newcommand{\IntStar}{I_{\mathsf{*}}}
\newcommand{\IntBCross}{I_\mathsf{-X}}
\newcommand{\IntLoop}{I_\mathsf{>o<}}
\newcommand{\IntDotCross}{\dot{I}_{\mathsf{X}}}
\newcommand{\IntDotStar}{\dot{I}_{\mathsf{*}}}
\newcommand{\IntDotBCross}{\dot{I}_\mathsf{-X}}
\newcommand{\IntDotLoop}{\dot{I}_\mathsf{>o<}}

\paragraph*{Abstract}
In the following we will consider different types of diagrams that possibly contribute to the propagator up to first order in time $\Deltat=t'-t$,
which are the diagrams that contribute to the
entropy production.
\emph{The general rule emerging from the arguments below is that a diagram containing $m$ blobs decays at least as fast as $\Deltat^m$ in small $\Deltat$. To calculate the entropy production only diagrams with up to one blob are needed.}

Diagrams enter in the entropy production
either through the kernel $\Op_{\yvec,\xvec}$, \Eref{transition_from_action} or \Eref{Op_in_diagrams}, or the logarithmic term $\Ln_{\yvec,\xvec}$, \Eref{Ln_from_propagators} or \Eref{Ln_in_diagrams}. By construction, both of these terms draw only on the first order in the small time difference $\Deltat=t'-t$ between creation at time $t$ and annihilation at time $t'$. For the kernel, this is established by the limit $\lim_{t'\downarrow t}$ after differentiation taken in \Erefs{def_Op} and \eref{Op_in_diagrams}. Such an operation extracts the first order in $\Deltat$ only, reducing it to the single particle Fokker-Planck operator plus corrections due to interactions and reactions. 

For the logarithm, the reason why diagrams enter only to first order in $\Deltat$ is more subtle; although \Erefs{Ln_from_propagators} and \eref{Ln_in_diagrams} contain a limit similar to \Erefs{def_Op} and \eref{Op_in_diagrams}, \latin{a priori}, L'H{\^o}pital's rule might require much higher derivatives: The first order is needed if the zeroth vanishes, the second if the first and zeroth order both vanish and so on. However, if the first order vanishes, then the kernel $\Op_{\yvec,\xvec}$ vanishes too, and the contribution to the entropy production according to \Eref{def_entropyProduction} is nil (\APref{DP_plus_EP}, remark after \Eref{Ln_from_TransitionRate}). 

In the present \APref{appendixWhichDiagramsContribute}
we derive some general principles 
of the short-time scaling of diagrams
in systems with a single particle
before considering systems  with multiple particles in \APref{MultipleParticles}. 
In \APref{contribs_to_prop} we present the basic arguments why contributions to the propagator order by order in the perturbation, shown as a ``blob" in the diagram, are in fact also order by order in the time $\Deltat$ that passes between creation and annihilation, \ie between initialisation and measurement. 
The argument carries through to more complicated objects, such as 
star-like vertices, \APref{interaction_vertices},
although the notion of blobs needs to be clarified in the case of an interaction potential, \Eref{interaction_example}, and
joint propagators, \APref{joint_props}. The scaling of diagrams with internal loops follows the pattern above, \APref{internal_blobs}. We include a discussion about  branching and coagulation vertices in the context of particle-conserving interactions, \APref{branching_vertices}.
We complete this section with a power-counting argument
to show that a diagram with $m$ blobs is of order $\Deltat^{m}$, \APref{general_power_counting}.

\toasubsection{Contributions to the full propagator}\seclabel{contribs_to_prop}
\toasubsubsection{
Zeroth order in small 
\texorpdfstring{$\Deltat$}{Delta t}}
First, we determine which diagrams in the full
propagator contribute to zeroth order in small
$\Deltat$.
Even when ultimately we are interested only in terms linear in $\Deltat$, whatever diagram contributes to zeroth order might also contribute to linear and, in fact, any higher order, \eg \Eref{simple_propagator_Fourier}.
Starting with the simplest such diagrams, we consider first the bare propagator like \Eref{orig_propagator_as_an_inverse} (\APref{bare_propagator_diffusion})
\begin{equation}\elabel{simple_propagator_with_extra}
    \tbarePropagator{\kvec,\omega}{\kvec',\omega'}
\corresponds
\ave[0]{\phi(\kvec',\omega')\phitilde(\kvec,\omega)}
=
\deltabar(\omega+\omega')
\deltabar(\kvec+\kvec')
\left\{
\frac{1}{-\imag\omega'+p} 
+
\order{\omega'^{-2}}
\right\} \ ,
\end{equation}
where we assume the typical conservation of momentum in the propagator and allow for some implicit dependence of the pole $\omega'=-\imag p$ on the momentum, $p=p(\kvec)$. The momenta are a proxy for any state-dependence and we will not make use of either $\deltabar(\kvec+\kvec')$ or $p(\kvec)$. Poles may be repeated, but that does not matter in the following considerations.
For the following discussion, it is helpful to retain the $\deltabar(\omega+\omega')$ function, even when
it can be easily integrated.
It is an indicator of time-translational invariance, to be discussed further 
below in the present section.

All that matters in \Eref{simple_propagator_with_extra} for the following arguments is that the bare propagator decays \emph{at least} as fast as $(-\imag\omega'+p)^{-1}$ in large $\omega'$, as shown for example in the case of a Markov chain in \Eref{orig_propagator_as_an_inverse}. However, since it must implement the feature \eref{propagator_limit_is_delta}
\begin{equation}\elabel{propagator_limit_is_delta_again_discrete}
    \lim_{t'\downarrow t} 
    \ave[0]{\phi(\yvec,t')\phitilde(\xvec,t)}
=\delta_{\yvec,\xvec}
\end{equation}
for discrete states and, say
\begin{equation}\elabel{propagator_limit_is_delta_again}
    \lim_{t'\downarrow t} 
    \ave[0]{\phi(\kvec',t')\phitilde(\kvec,t)}
=\deltabar(\kvec+\kvec')
\end{equation}
for continuous states,
it is also clear that it cannot decay faster than $\omega^{-1}$. If it were to decay like, say, $((-\imag\omega' + p_1)(-\imag\omega' + p_2))^{-1}$, then 
\begin{equation}
    \lim_{t'\downarrow t}\int\dintbar{\omega'} \exp{-\imag\omega' (t'-t)} 
    \frac{1}{-\imag\omega' + p_1}\frac{1}{-\imag\omega' + p_2}
    =
    \int\dintbar{\omega'} 
    \frac{1}{-\imag\omega' + p_1}\frac{1}{-\imag\omega' + p_2}
    =0 \ ,
\end{equation}
as will be discussed in further detail below. 

As indicated in \Eref{simple_propagator_with_extra}, a bare propagator might thus have contributions that vanish in large $\omega'$ as fast as $\order{\omega'^{-2}}$ or even faster
\cite{BothePruessner:2021,ZhangPruessner:2022,Garcia-MillanPruessner:2021}, but it always has one contribution of the form $(-\imag\omega'+p)^{-1}$. Its inverse Fourier transform reads
\begin{equation}\elabel{simple_propagator_Fourier}
    \int\dintbar{\omega}\dintbar{\omega'} 
    \exp{-\imag(\omega t+\omega't')}
    \frac{\deltabar(\omega+\omega')}{-\imag \omega'+ p} =
    \theta(\Re(\Deltat\, p))\sgn(\Deltat)
    \exp{-\Deltat\, p} \quad\text{ where }\quad \Deltat=t'-t\ ,
\end{equation}
and $\Re(\Deltat\, p)$ is the real part of $\Deltat\, p$.
Causality, \ie that a particle's presence cannot be measured before it is created, is then enforced by demanding that the real-part of $p$ is positive, so that the Heaviside $\theta$-function in \Eref{simple_propagator_Fourier} vanishes for $\Deltat=t'-t<0$. It is therefore safe to assume that all poles $\omega'=-\imag p$ of all propagators are located in the lower half-plane. 

To simplify the following discussion, we shall henceforth assume 
\begin{equation}\elabel{simple_propagator}
        \tbarePropagator{\kvec,\omega}{\kvec',\omega'}
\corresponds
\ave[0]{\phi(\kvec',\omega')\phitilde(\kvec,\omega)}
=
\deltabar(\omega+\omega')
\deltabar(\kvec+\kvec')
\frac{1}{-\imag\omega'+p} 
\end{equation}
and thus ignore the terms $\order{\omega'^{-2}}$ in \Eref{simple_propagator_with_extra}.

Are there any other diagrams contributing to the full
propagator to zeroth order in $\Deltat$?
Corrections to the propagator due to the perturbative part of the action, \eg \Eref{propagator_expansion_app} (\APref{pert_exp_full_prop}), may be written as
\begin{equation}\elabel{propagator_first_order_corrected}
\tblobbedPropagator{\kvec,\omega}{\kvec',\omega'} 
\corresponds
\frac{\deltabar(\omega+\omega')\deltabar(\kvec+\kvec')}{-\imag \omega' + p_2}
\transMatNonLin_{21}
\frac{1}{-\imag \omega' + p_1}
\end{equation}
with $p_1$, $p_2$ real and positive, and generally dependent on $\kvec=-\kvec'$. The effect of the blob in the diagram is captured by $\transMatNonLin_{21}$.

If $\transMatNonLin_{21}$ is a function of $\omega'$, its $\omega'$ dependence has to be analysed in more detail: Firstly, it cannot introduce poles in $\omega'$ that are located in the upper half-plane, as that would break causality. Secondly, $\transMatNonLin_{21}$ may diverge in $\omega'$ but never as fast as $\omega'$ itself, as it would not be captured in a perturbation theory in general, so it is safe to assume that $\lim_{\omega'\to\infty}\transMatNonLin_{21}(\omega')/\omega'=0$.
Thirdly, $\transMatNonLin_{21}(t)$ might be dependent on absolute time $t$, as if subject to some external forcing \cite{PauschGarcia-MillanPruessner:2020}, which amounts to allowing for sinks and sources of $\omega'$ in diagrams, which then no longer carry a factor of $\deltabar(\omega+\omega')$. This generalisation of $\transMatNonLin_{21}$ \emph{does} indeed invalidate the following arguments, because breaking time-translational invariance means that not all propagators might have to "carry" the same $\omega'$. 
To keep what follows simple, we will, however, assume time translational invariance, as will be manifest by any contribution to the propagator being proportional to $\deltabar(\omega+\omega')$.

\toasubsubsection{
General considerations and limit 
\texorpdfstring{$\Deltat\to0$}{Delta t->0}}
\seclabel{general_and_limit}
To determine to what order in $\Deltat$ the diagram in \eref{propagator_first_order_corrected} contributes
to the full propagator, 
we need to calculate its inverse Fourier transform. 
We may consider more generally an integral similar to \Eref{derivative_integral}, consisting of $n$ 
propagators and $n-1$ blobs,
\begin{equation}\elabel{def_IntMultiProp}
    \IntMultiProp(\Deltat)=\int \dintbar{\omega'}
    \frac{\exp{-\imag \omega' \Deltat}}
    {\prod_{j=1}^{n} (-\imag \omega' + p_j)} 
    \Big(
    \transMatNonLin_{n\,n-1}(\omega') 
    \transMatNonLin_{n-1\,n-2}(\omega') 
    \cdots
    \transMatNonLin_{21}(\omega') 
    \Big)
    \ ,
\end{equation}
which corresponds to the contribution of a diagram involving $n$ propagators each carrying $\omega'$, which vanishes most slowly in $\omega'$. If the $\transMatNonLin$-terms are constant in $\omega'$, the integrand at $\Deltat=0$ decays like $\propto\omega'^{-n}$. If  
they diverge in $\omega'$, 
say $\transMatNonLin_{i i-1}(\omega')\propto\omega'^\mu$
the overall behaviour of the integrand of \Eref{def_IntMultiProp} is $\propto\omega'^{-n+\mu(n-1)}$, with $\mu<1$ as discussed above.
An example of $I_{1}(\Deltat)$ and $I_{2}(\Deltat)$ are
the Fourier transforms of \eref{simple_propagator} and \eref{propagator_first_order_corrected} 
respectively.

Given that all poles of the integrand in \Eref{def_IntMultiProp} are in the lower half-plane, repeated or not, $\IntMultiProp(\Deltat)$ generally vanishes at $\Deltat<0$, for the same reason as \Eref{simple_propagator_Fourier} vanishes. At $\Deltat=0$ and $n\ge2$, the contour can be closed by the ML lemma \cite{AblowitzFokas:2003} 
either in the lower or in the upper half-plane, as the integrand vanishes strictly faster than $\omega'^{-1}$ in large $\omega'$. 
Because all poles are in the lower half-plane, the integral thus vanishes at $\Deltat=0$ and $n\ge2$, meaning that a diagram like \Eref{propagator_first_order_corrected} does not contribute at $\Deltat=0$. Not much can be said for $\Deltat=0$ and $n=1$, because then the integral \Eref{def_IntMultiProp} is logarithmically divergent.

Strictly speaking, however, we are interested in the behaviour of diagrams in the \emph{limit} $\Deltat\downarrow0$, as required in \Erefs{def_Op} and \eref{def_Ln}, not \emph{at} $\Deltat=0$. To make this connection, we make the following observation: In the limit of $\Deltat\downarrow0$ or $\Deltat\uparrow0$, the effect of the exponential like the one in \Eref{def_IntMultiProp} is solely that it directs the closure of the auxiliary contour to determine any of the integrals by the ML lemma. Otherwise, the exponential $\exp{-\imag \omega' \Deltat}$ has no further effect, its contribution to the residue in the limit $\Deltat\to0$ from above or from below is always a factor $1$, as it converges to unity irrespective of the value of the pole in $\omega'$. 
Of course, that by itself does not mean that the value of the integral is the same in both limits, as is known, for example, from the Fourier-transform of a Heaviside $\theta$-function. However, 
if the integral exists at $\Deltat=0$ (not just its principle value) and if the auxiliary path can be taken in both half planes without contributing by the ML lemma, then the value of the integral at $\Deltat=0$ is independent of the orientation of the auxiliary path. The integral at $\Deltat=0$ provides the "glue" between the two limits.
In other words, if the integral at $\Deltat=0$ exists, then it must be identical to the integral in the limit $\Deltat\downarrow0$, but in fact also identical to the integral in the limit $\Deltat\uparrow0$. The latter vanishes by causality, which means that if the integral at $\Deltat=0$ exists, it vanishes as well and so does the one for $\Deltat\downarrow0$. In brief: Defining $\IntMultiProp^\pm=\lim_{\Deltat\to0^\pm}\IntMultiProp(\Deltat)$, \Eref{def_IntMultiProp}, then the existence of $\IntMultiProp(0)$ and its independence from the orientation of the auxiliary path guarantees $\IntMultiProp(0)=\IntMultiProp^+$ as well as $\IntMultiProp(0)=\IntMultiProp^-$, but 
$\IntMultiProp^-=0$, which thus implies $\IntMultiProp^+=0$. 
If $\IntMultiProp(0)$ exists, then $\IntMultiProp(\Deltat)$ is continuous at $\Deltat=0$ and vanishes there.

We can use the argument in the following form: \emph{If a diagram with $n\geq2$ bare propagators can be calculated for $\Deltat=0$, then it is identical to its limits $\Deltat\to0^{\pm}$ and necessarily vanishes at $\Delta=0$ as well as $\Delta\to0^{\pm}$.}
Therefore, contributions to the zeroth order in $\Deltat$
solely come from the bare propagator, \Erefs{propagator_limit_is_delta_again_discrete}, \eref{propagator_limit_is_delta_again} and \eref{simple_propagator_Fourier}.

\toasubsubsection{
First and higher orders
in small 
\texorpdfstring{$\Deltat$}{Delta t}}

Using the result in the preceding section,
we determine which other diagrams in the full 
propagator contribute to first order in small $\Deltat$.
All of the following reasoning is done in Fourier time, \ie frequencies $\omega$, because in Fourier time it comes down to mere power counting, although equivalent arguments can of course be made in direct time.

Considering the first derivative of $\IntMultiProp(\Deltat)$, \Eref{def_IntMultiProp}, with respect to $\Deltat$, differentiation brings down a factor of $-\imag\omega'$. 
If there are $n$ bare propagators carrying $\omega'$
and thus $n-1$ factors of perturbative $\transMatNonLin_{i i-1}(\omega')\propto\omega'^{\mu}$, the integrand vanishes as fast as $\omega'^{1-n+\mu(n-1)}$. By the same arguments as outlined above, this integral vanishes at $\Deltat=0$ provided $1-n+\mu(n-1)<-1$, \ie $2-\mu<n(1-\mu)$. For this to hold for all $n\ge3$ we would need to require henceforth $\mu<1/2$. 
To simplify what follows, we will assume, however, $\transMatNonLin(\omega')$ to be independent of $\omega'$,
that is $\mu=0$.

For example
\begin{equation}
    \frac{\plaind}{\plaind \Deltat}\IntMultiProp(\Deltat)=\int \dintbar{\omega'}
    \frac{-\imag \omega' \exp{-\imag \omega' \Deltat}}
    {\prod_{j=1}^{n} (-\imag \omega' + p_j)} 
        \prod_{i=1}^{n-1}
        \transMatNonLin_{i+1\,i}
    \ ,
\end{equation}
which vanishes by the ML lemma for $\Deltat=0$ and $n\ge3$, such that, say
\begin{equation}
\left.\partial_{t'}
\left(
\tikz[baseline=1pt]{
\begin{scope}[yshift=0.0cm]
\node at (0.5,0) [right] {$\kvec_1,t$};
\node at (-1.5,0) [left] {$\kvec'_1,t'$};
\draw[tAactivity] (0.5,0) -- (-1.5,0);
\tgenVertex{0,0}
\tgenVertex{-1,0}
\end{scope}
}
\right)
\right|_{\Deltat=0}
\corresponds
0 \ .
\end{equation}
Similar arguments apply to higher derivatives, but these are not relevant in the present work.  \emph{For $\transMatNonLin(\omega')$ constant in $\omega'$ it follows that a contribution to the propagator with $n$ legs and thus $n-1$ blobs behaves like $\IntMultiProp(\Deltat)\in\order{\Deltat^{n-1}}$ in small $\Deltat$.}
Therefore, of all contributions to the full propagator, 
the bare propagator \tbarePropagator{}{} contributes to zeroth (and higher) order in
small $\Deltat$, and the first-order correction
\tblobbedPropagator{}{} contributes  
to first (and higher) order in small $\Deltat$. These are the only two terms
in the full propagator that
we need to calculate the entropy production rate.

A tadpole diagram like 
\begin{equation}
\elabel{tadpole_diagram}
\tikz[baseline=3pt]{
\begin{scope}[yshift=0cm]
\tgenVertex{0,0}
\draw[tAactivity] (\propWidth,0) -- (-\propWidth,0);
\draw[tAactivity] (0,0) to [out=60,in=180] (\propWidth,\propWidth);
\draw[tAactivity] (0,0) to [out=30,in=00] (\propWidth,\propWidth);
\end{scope}
}
\in\order{\Deltat}
\end{equation}
does not contribute to order $\Deltat^0$, but does so to order $\Deltat$, for the same reasons as \Eref{propagator_first_order_corrected}. The additional loop does not carry $\omega'$ and is in fact independent of external frequencies and momenta. The loop provides a constant pre-factor and does not modify the inverse Fourier transform in any other way.
Tadpoles are a rather exotic type of diagram in Doi-Peliti field theories, as they require a source, which is, however, generally found in 
response field theories \cite{Taeuber:2014,WalterSalbreuxPruessner:2022:FieldTheory}.

In summary, corrections to the propagator of the form \Eref{propagator_first_order_corrected} with $n\ge2$ bare propagators carrying $\omega'$ vanish at $\Deltat=0$ and in the limit $\Deltat\downarrow0$ provided $\mu<1$. 
Similarly, their first derivatives vanish at $\Deltat=0$ 
and in the limit $\Deltat\downarrow0$
for all $n\ge3$ provided $\mu<1/2$.

\toasubsection{Interaction vertices}\seclabel{interaction_vertices}
Taking the Fourier transform of a star-like diagram
\begin{align}\elabel{general_whiskers}
            \tikz[baseline=-4pt]{
\begin{scope}[yshift=0.0cm]
\tgenVertex{0,0}
\node at (20:1cm) [right,yshift=0pt] {$t$};
\node at (0:1cm) [right,yshift=0pt] {$t$};
\node at (-40:1cm) [right,yshift=0pt] {$t$};
\node at (-20:-1cm) [left,yshift=0pt] {$t'$};
\node at (0:-1cm) [left,yshift=0pt] {$t'$};
\node at (40:-1cm) [left,yshift=0pt] {$t'$};
\draw[tAactivity,dotted] (-35:0.8cm) arc (-35:-5:0.8cm);
\draw[tAactivity,dotted] (35:-0.8cm) arc (35:5:-0.8cm);
\draw[tAactivity] (20:1cm) -- (0,0);
\draw[tAactivity] (0:1cm) -- (0,0);
\draw[tAactivity] (-40:1cm) -- (0,0);
\draw[tAactivity] (-20:-1cm) -- (0,0);
\draw[tAactivity] (0:-1cm) -- (0,0);
\draw[tAactivity] (40:-1cm) -- (0,0);
\end{scope}
}
    &\corresponds \IntStar(\Deltat) = 
    \int 
    \dintbar{\omega_{1,\ldots,n}}
    \dintbar{\omega'_{1,\ldots,n}}
    \exp{-\imag(\omega_1+\ldots+\omega_n)\Deltat}\\
    &\nonumber \times
    \deltabar(\omega_1+\ldots+\omega_n+\omega'_1+\ldots+\omega'_n)
    \left(\prod_{i=1}^n \frac{1}{-\imag\omega_i+p_i}\right)
    \left(\prod_{i=1}^n \frac{1}{\imag\omega'_i+p'_i}\right)
\end{align}
one can show that 
\begin{equation}\elabel{IntStar_explicit}
    \IntStar(\Deltat)=\theta(\Deltat) \frac{\exp{-\Deltat\sum_{i=1}^n p'_i}-\exp{-\Deltat\sum_{i=1}^n p_i}}{\sum_{i=1}^n p_i-\sum_{i=1}^n p'_i} \ ,
\end{equation}
so that $\IntStar(0)=\lim_{\Deltat\downarrow0}\IntStar(\Deltat)=\lim_{\Deltat\uparrow0}\IntStar(\Deltat)=0$, and
\begin{equation}
\elabel{scaling_star_like_diagram}
    \IntStar(\Deltat)  \corresponds
    \tikz[baseline=-4pt]{
\begin{scope}[yshift=0.0cm]
\tgenVertex{0,0}
\node at (20:1cm) [right,yshift=0pt] {$t$};
\node at (0:1cm) [right,yshift=0pt] {$t$};
\node at (-40:1cm) [right,yshift=0pt] {$t$};
\node at (-20:-1cm) [left,yshift=0pt] {$t'$};
\node at (0:-1cm) [left,yshift=0pt] {$t'$};
\node at (40:-1cm) [left,yshift=0pt] {$t'$};
\draw[tAactivity,dotted] (-35:0.8cm) arc (-35:-5:0.8cm);
\draw[tAactivity,dotted] (35:-0.8cm) arc (35:5:-0.8cm);
\draw[tAactivity] (20:1cm) -- (0,0);
\draw[tAactivity] (0:1cm) -- (0,0);
\draw[tAactivity] (-40:1cm) -- (0,0);
\draw[tAactivity] (-20:-1cm) -- (0,0);
\draw[tAactivity] (0:-1cm) -- (0,0);
\draw[tAactivity] (40:-1cm) -- (0,0);
\end{scope}
}
    \in \order{\Deltat} \ .
\end{equation}
The interaction vertices in \Erefs{pair_vertex1} and \eref{pair_vertex2} of the type
\begin{equation}\elabel{interaction_example}
        \tikz[baseline=0pt]{
\draw [draw=none,fill=red!25] (0,0) ellipse (0.35cm and 0.75cm);
\begin{scope}[yshift=0.0cm]
\draw[black,potStyle] (0,-0.4) -- (0,0.4);
\draw[black,thin] (-0.1,0.2) -- (0.1,0.2);
\draw[red,thin] (-0.2,0.3) -- (-0.2,0.5);
\tgenVertex{0,-0.4}
\tgenVertex{0,0.4}
\node at (\propWidth,-0.5) [right,yshift=0pt] {$\kvec_1,t$};
\node at (-\propWidth,-0.5) [left,yshift=0pt] {$\kvec'_1,t'$};
\draw[tAactivity] (\propWidth,-0.5) -- (0,-0.4) -- (-\propWidth,-0.5);
\node at (\propWidth,0.5) [right,yshift=0pt] {$\kvec_2,t$};
\node at (-\propWidth,0.5) [left,yshift=0pt] {$\kvec'_2,t'$};
\draw[tAactivity] (\propWidth,0.5) -- (0,0.4) -- (-\propWidth,0.5);
\end{scope}
}
\in\order{\Deltat}
\end{equation}
are in fact of the form $\IntStar(\Deltat)$ with $n=2$ even when \Eref{interaction_example} seems to contain \emph{two blobs}. However, the dash-dotted vertical line, which represents the interaction potential, is not a propagator $\propto\omega^{-1}$ and has no frequency dependence. In \Eref{interaction_example} we show the two blobs suggestively as if located inside a large, faintly drawn blob, which is the blob to count as one.

\toasubsection{Joint propagators}\seclabel{joint_props}
The same arguments as above apply to joint propagators, which are just products of diagrams, such as 
\begin{equation}
\tikz[baseline=3pt]{
\begin{scope}[xscale=0.7]
\begin{scope}[yshift=0.4cm]
\tgenVertex{0,0}
\node at (\propWidth,0) [right] {$\kvec_1,t$};
\node at (-\propWidth,0) [left] {$\kvec'_1,t'$};
\draw[tAactivity] (\propWidth,0) -- (-\propWidth,0);
\end{scope}
\begin{scope}[yshift=0.0cm]
\node at (\propWidth,0) [right] {$\kvec_2,t$};
\node at (-\propWidth,0) [left] {$\kvec'_2,t'$};
\draw[tAactivity] (\propWidth,0) -- (-\propWidth,0);
\end{scope}
\end{scope}
}
\in\order{\Deltat}
    \quad\text{ and }\quad
\tikz[baseline=3pt]{
\begin{scope}[xscale=0.7]
\begin{scope}[yshift=0.4cm]
\tgenVertex{0,0}
\node at (\propWidth,0) [right] {$\kvec_1,t$};
\node at (-\propWidth,0) [left] {$\kvec'_1,t'$};
\draw[tAactivity] (\propWidth,0) -- (-\propWidth,0);
\end{scope}
\begin{scope}[yshift=0.0cm]
\tgenVertex{0,0}
\node at (\propWidth,0) [right] {$\kvec_2,t$};
\node at (-\propWidth,0) [left] {$\kvec'_2,t'$};
\draw[tAactivity] (\propWidth,0) -- (-\propWidth,0);
\end{scope}
\end{scope}
}
\in \order{\Deltat^2}
    \quad\text{ and }\quad
\tikz[baseline=3pt]{
\begin{scope}[xscale=0.7]
\begin{scope}[yshift=0.4cm]
\tgenVertex{0.25,0}
\tgenVertex{-0.25,0}
\node at (\propWidth,0) [right] {$\kvec_1,t$};
\node at (-\propWidth,0) [left] {$\kvec'_1,t'$};
\draw[tAactivity] (\propWidth,0) -- (-\propWidth,0);
\end{scope}
\begin{scope}[yshift=0.0cm]
\tgenVertex{0,0}
\node at (\propWidth,0) [right] {$\kvec_2,t$};
\node at (-\propWidth,0) [left] {$\kvec'_2,t'$};
\draw[tAactivity] (\propWidth,0) -- (-\propWidth,0);
\end{scope}
\end{scope}
}
\in \order{\Deltat^3}
\end{equation}
in the sense that these diagrams vanish in small $\Deltat$ at least like $\Deltat$, $\Deltat^2$ and $\Deltat^3$ respectively.
Similarly,
\begin{equation}
    \tikz[baseline=-11pt]{
\begin{scope}[yshift=0.0cm]
\tgenVertex{0,-0.15}
\node at (\propWidth,0) [right,yshift=3pt] {$\kvec_1,t$};
\node at (-\propWidth,0) [left,yshift=3pt] {$\kvec'_1,t'$};
\draw[tAactivity] (\propWidth,0) -- (0,-0.15) -- (-\propWidth,0);
\node at (\propWidth,-0.3) [right,yshift=0pt] {$\kvec_2,t$};
\node at (-\propWidth,-0.3) [left,yshift=0pt] {$\kvec'_2,t'$};
\draw[tAactivity] (\propWidth,-0.3) -- (0,-0.15) -- (-\propWidth,-0.3);
\begin{scope}[yshift=-0.7cm]
\node at (\propWidth,0) [right] {$\kvec_3,t$};
\node at (-\propWidth,0) [left] {$\kvec'_3,t'$};
\draw[tAactivity] (\propWidth,0) -- (-\propWidth,0);
\end{scope}
\end{scope}
}
\in \order{\Deltat}
    \qquad\text{ and }\qquad
    \tikz[baseline=-11pt]{
\begin{scope}[yshift=0.0cm]
\tgenVertex{0,-0.15}
\node at (\propWidth,0) [right,yshift=3pt] {$\kvec_1,t$};
\node at (-\propWidth,0) [left,yshift=3pt] {$\kvec'_1,t'$};
\draw[tAactivity] (\propWidth,0) -- (0,-0.15) -- (-\propWidth,0);
\node at (\propWidth,-0.3) [right,yshift=0pt] {$\kvec_2,t$};
\node at (-\propWidth,-0.3) [left,yshift=0pt] {$\kvec'_2,t'$};
\draw[tAactivity] (\propWidth,-0.3) -- (0,-0.15) -- (-\propWidth,-0.3);
\begin{scope}[yshift=-0.7cm]
\tgenVertex{0,0}
\node at (\propWidth,0) [right] {$\kvec_3,t$};
\node at (-\propWidth,0) [left] {$\kvec'_3,t'$};
\draw[tAactivity] (\propWidth,0) -- (-\propWidth,0);
\end{scope}
\end{scope}
}
\in \order{\Deltat^2}
\end{equation}
where the scaling of the interaction vertex
$\IntStar(\Deltat)$ with $n=2$
is that of \Eref{scaling_star_like_diagram}.

\toasubsection{Internal blobs and loops}\seclabel{internal_blobs}
The order of more complicated diagrams such as
\begin{equation}
    \tikz[baseline=-7.5pt]{
\begin{scope}[yshift=0.0cm,xscale=1]
\tgenVertex{0,-0.15}
\tgenVertex{-0.4cm,-0.07}
\node at (0.75cm,0.1) [right,yshift=0pt] {$t$};
\node at (-0.75cm,0.1) [left,yshift=0pt] {$t'$};
\draw[tAactivity] (-0.75cm,0) -- (0,-0.15) -- (0.75cm,0);
\node at (0.75cm,-0.3) [right,yshift=0pt] {$t$};
\node at (-0.75cm,-0.3) [left,yshift=0pt] {$t'$};
\draw[tAactivity] (0.75cm,-0.3) -- (0,-0.15) -- (-0.75cm,-0.3);
\end{scope}
}
\in\order{\Deltat^2} \,, \quad
    \tikz[baseline=-4pt]{
\begin{scope}[yshift=0.0cm,xscale=1]
\tgenVertex{-0.4cm,0}
\tgenVertex{0.4cm,0}
\node at (0.9cm,0.2cm) [right,yshift=0pt] {$t$};
\node at (-0.9cm,0.2cm) [left,yshift=0pt] {$t'$};
\node at (0.9cm,-0.2cm) [right,yshift=0pt] {$t$};
\node at (-0.9cm,-0.2cm) [left,yshift=0pt] {$t'$};
\draw[tAactivity] (-0.9cm,0.2) -- (-0.4cm,0);
\draw[tAactivity] (-0.9cm,-0.2) -- (-0.4cm,0);
\draw[tAactivity] (0.9cm,0.2) -- (0.4cm,0);
\draw[tAactivity] (0.9cm,-0.2) -- (0.4cm,0);
\draw[tAactivity] (-0.4cm,0) to [out=60,in=120] (0.4cm,0);
\draw[tAactivity] (-0.4cm,0) to [out=-60,in=-120] (0.4cm,0);
\end{scope}
}
\in\order{\Deltat^2} \,\quad\text{ and }\quad
    \tikz[baseline=-4pt]{
\begin{scope}[yshift=0.0cm,xscale=1]
\tgenVertex{-0.4cm,0}
\tgenVertex{0.4cm,0}
\node at (0.9cm,0.35cm) [right,yshift=0pt] {$t$};
\node at (-0.9cm,0.35cm) [left,yshift=0pt] {$t'$};
\node at (0.9cm,-0.35cm) [right,yshift=0pt] {$t$};
\node at (-0.9cm,-0.35cm) [left,yshift=0pt] {$t'$};
\node at (0.9cm,0cm) [right,yshift=0pt] {$t$};
\node at (-0.9cm,0cm) [left,yshift=0pt] {$t'$};
\draw[tAactivity] (-0.9cm,0.3) -- (-0.4cm,0);
\draw[tAactivity] (-0.9cm,-0.3) -- (-0.4cm,0);
\draw[tAactivity] (-0.9cm,0) -- (-0.4cm,0);
\draw[tAactivity] (0.9cm,0.3) -- (0.4cm,0);
\draw[tAactivity] (0.9cm,-0.3) -- (0.4cm,0);
\draw[tAactivity] (0.9cm,0) -- (0.4cm,0);
\draw[tAactivity] (-0.4cm,0) to [out=60,in=120] (0.4cm,0);
\draw[tAactivity] (-0.4cm,0) to [out=-60,in=-120] (0.4cm,0);
\end{scope}
}
\in\order{\Deltat^2} \ ,
\end{equation}
can be determined by studying them as a variation of star-like diagram \Eref{general_whiskers}. The key insight is that any additional internal propagator adds a pole \emph{on the same half-plane} as they can be found in the star-like diagrams. 
Loops result in additional integrals, but do not change the general argument.

\toasubsection{Branching and Coagulation vertices}\seclabel{branching_vertices}
The propagators considered in the entropy production of $N$ particles, \Erefs{entropyProduction_multipleParticles} and \eref{def_trans_multi_diag} are in principle of the form
$\ave{\phi(\kvec'_1,\Deltat)\ldots\phi(\kvec'_N,\Deltat)\phidagger(\kvec_1,0)\ldots\phidagger(\kvec_N,0)}$, 
probing for particles at $N$ positions after time $\Deltat$. The daggered fields allow for some initial particles to be ignored. 
After the Doi-shift $\phidagger=1+\phitilde$,
the propagators are represented by possibly disconnected diagrams that have $N$ outgoing legs and \emph{at most} $N$ incoming legs.

Since here we consider only processes where the
total particle number is conserved, there is no diagram with more outgoing than incoming legs,
such as the branching diagram
\begin{equation}\elabel{branching_and_coagulation}
\tikz[baseline=-4pt]{
\begin{scope}[yshift=0.0cm,xscale=1]
\tgenVertex{0,0}
\node at (0:\propWidth) [right,yshift=0pt] {$t$};
\node at (150:\propWidth) [left,yshift=0pt] {$t'$};
\node at (-150:\propWidth) [left,yshift=0pt] {$t'$};
\draw[tAactivity] (0:\propWidth) -- (0,0);
\draw[tAactivity] (150:\propWidth) -- (0,0);
\draw[tAactivity] (-150:\propWidth) -- (0,0);
\end{scope}
} \ .
\end{equation}
To have $N$ outgoing legs it therefore takes \emph{at least} $N$ incoming legs. All contributions to
 $\ave{\phi(\kvec'_1,\Deltat)\ldots\phi(\kvec'_N,\Deltat)\phidagger(\kvec_1,0)\ldots\phidagger(\kvec_N,0)}$,
which in principle can contain diagrams with fewer than $N$ incoming legs, in the processes considered here therefore have
exactly $N$ incoming legs and $N$ outgoing legs.
This constraint, together with the absence of
branching vertices, implies that coagulation-like
vertices, which have more incoming legs than 
outgoing legs, such as
\begin{equation}
    \tikz[baseline=-4pt]{
\begin{scope}[yshift=0.0cm,xscale=1]
\tgenVertex{0,0}
\node at (30:\propWidth) [right,yshift=0pt] {$t$};
\node at (-30:\propWidth) [right,yshift=0pt] {$t$};
\node at (180:\propWidth) [left,yshift=0pt] {$t'$};
\draw[tAactivity] (30:\propWidth) -- (0,0);
\draw[tAactivity] (-30:\propWidth) -- (0,0);
\draw[tAactivity] (180:\propWidth) -- (0,0);
\end{scope}
}
\end{equation}
do not contribute to the propagators needed to
calculate the entropy production, even if they
are present in the Doi-Peliti field theory as a
result of particles interactions, \eg \Erefs{pair_vertex1} and \eref{pair_vertex2}.

\toasubsection{General power counting}\seclabel{general_power_counting}
We complete the present discussion with a power counting argument showing that any diagram containing $m$ blobs, or vertices, scales like $\Deltat^{m}$ in the short-time limit $\Deltat\to0$. This section is a generalisation of \APref{contribs_to_prop}, since it includes 
diagrams possibly involving internal loops.
As disconnected diagrams scale like their product, we restrict the discussion to connected diagrams. Those are made from propagators and blobs, so that any two blobs are connected by a propagator proportional to $\omega^{-1}$ (but \Eref{interaction_example}). 
Since here we consider \emph{vertices} with  as many legs coming in as coming out only, we can further restrict the discussion to \emph{diagrams} having as many legs coming in as come out. 
To simplify the discussion, we restrict ourselves to vertices constant n $\omega$.

A connected diagram with $n$ incoming and outgoing legs can be thought of as being made from $n$ disconnected bare propagators which are "tied together" by inserting $m$ vertices. Initially, the $n$ propagators each scale like $\omega'^{-1}\deltabar(\omega'+\omega)$. The insertion of an $\ell$-legged vertex, with $\ell$ incoming and $\ell$ outgoing legs, splits $\ell$ propagators, effectively creating $\ell$ "internal" ones and splicing them, where we may ignore any additional internal $\deltabar$-functions as having cancelled with internal integrals. These internal integrals are trivial, as opposed to the ones discussed below.
As the vertex is time-translational invariant, it will further introduce integrals over $\ell$ internal $\omega$ as well as a single $\deltabar$-function.

As each such $\ell_i$-legged vertex $i$ with $i=1,\ldots,m$ effectively introduces $\ell_i$ new propagators by splitting and splicing, the total count of such new propagators is $\LC=\sum_{i=1}^m\ell_i$. Each of those gives rise to an internal $\omega$-integral, so that there are $\IC=\LC$ such internal, non-trivial integrals. There are $m$ vertices inserted, each will give rise to a frequency conserving $\deltabar$-function, so that starting with $n$ such $\deltabar$-functions from the initially disconnected $n$ propagators, there is a total $\KC=n+m$ 
Dirac $\deltabar$-functions and a total of $\NC=n+\LC$ propagators.

The final diagram is obtained by carrying out the $\IC$ internal integrals, using up as many $\deltabar$-functions as possible, but at most $\KC-1$ as the overall diagram has a $\deltabar$-prefactor.
The remaining $\IC-(\KC-1)\ge0$ integrals are loops.
Starting with an integrand that goes like $\omega^{-\NC}$ in large $\omega$ and integrating $\IC$ times with the help of $\KC-1$ Dirac $\deltabar$-functions produces a
final diagram that scales like $\omega'^{-\NC+\IC-(\KC-1)}\deltabar(\omega'_1+\ldots)$. From the above 
\begin{equation}
    -\NC+\IC-(\KC-1)
    = - (n+\LC) +\LC - (n+m-1)
    =-2n-m+1 \ ,
\end{equation}
\ie the diagram behaves in large, external $\omega'$ like $\omega'^{-2n-m+1}\deltabar(\omega_1'+\ldots)$.
Carrying out the inverse Fourier transform over $2n$ external $\omega'$ using up the remaining $\deltabar$-function, thus produces an integral proportional to $\Deltat^m$. This is the desired scaling behaviour.



\toasection{Entropy production of drift-diffusion particles on a torus with
potential}\seclabel{drift_diffusion_on_ring}
\paragraph*{Abstract}

\newcommand{\Gw}{G_{\vecDrift}}
\newcommand{\GD}{G_{\diffusion}}
\newcommand{\GWissel}{G_{\text{\tiny Wi}}}

In this section we consider a drift-diffusion
particle with diffusion constant $\diffusion$ and drift $\drift$ on a
$d$-dimensional torus with circumference $L$ and external potential $\extPot(x)$. 
We calculate its entropy production
in three different ways, to show that different
perturbative expansions produce the same result as well as to highlight some
peculiarities of continuum theories.
It is a pretty straight forward exercise to 
calculate the entropy production from first principles \cite[Sec.~3.11]{CocconiGarcia-MillanETAL:2020}.
This is done in the following within the framework of the main text,
first by drawing directly on Wissel's short-time propagator 
\cite{Wissel:1979} in
\SMref{entProd_Wissel}, \Eref{local_entropy_GWissel}, and then
field-theoretically in two different
setups: In \APref{FT_about_drift-diffusion}, only the potential is dealt with perturbatively,
\Eref{local_entropy_Gw},
in  \APref{FT_about_diffusion}, both the potential and the drift are dealt with
perturbatively. In \APref{Fourier_transformation} we discuss some of the details of continuous space and the particular r{\^o}le of the Fourier-transform.

\vspace{20pt}

The Fokker-Planck equation for the present
setup is
\begin{equation}\elabel{FPE_drift_diffusion}
\partial_t \FPprobability(\yvec,t) = \FPophat_{\yvec} \FPprobability(\yvec,t)
\text{ with }
\FPophat_\yvec = 
\diffusion \nabla_y^2 - \nabla_y \cdot ( \vecDrift - \extPot'(\yvec))
=
\diffusion \nabla_y^2 - ( \vecDrift - \extPot'(\yvec))
\cdot \nabla_y + \extPot''(\yvec)
\end{equation}
where $\extPot'(\yvec)=\nabla\extPot(\yvec)$ and 
$\extPot''(\yvec)=\laplace\extPot(\yvec)$ are used to emphasise that a derivative acts only on $\extPot(\yvec)$, in contrast to the nabla in front of the bracket, $\nabla_y ( \vecDrift - \extPot(\yvec))$, which acts on everything to the right of it, just like the first term $\diffusion \nabla_y^2=\diffusion\laplace_\yvec$. 
After adding a small mass $r\downarrow0$ to maintain causality,
the action in real-space and direct time, \Eref{def_action0},
is
\begin{equation}\elabel{action_drift_diffusion_appendix}
\action
= \int \ddint{x}\ddint{y}\ddint{t} \phitilde(\yvec,t) ( - \delta(\yvec-\xvec)\partial_t + \FPop_{\yvec,\xvec} -r) \phi(\xvec,t)
= \int \ddint{x}\ddint{t} \phitilde(\xvec,t) ( - \partial_t + \FPophat_{\xvec} -r) \phi(\xvec,t)
\end{equation}
according to \Eref{def_harmonic_action}
as $\FPop_{\yvec,\xvec} = \FPophat^{\dagger}_{\xvec} \delta(\xvec-\yvec)$. In one dimension
the Fokker-Planck equation has a known stationary solution and the entropy
production can readily be calculated \cite{CocconiGarcia-MillanETAL:2020}.

\toasubsection{Entropy production from the short-time propagator}
\seclabel{entProd_Wissel}
In the present section we derive the entropy production on the basis of the
short-time propagator introduced by Wissel \cite{Wissel:1979}. This will serve as a reference for the following
sections. The short-time propagator $\GWissel(\xvec\to\yvec; t'-t)$ may be constructed by some very
basic physical reasoning, namely that ``the derivative of the potential
plays the same r{\^o}le as a drift''.
It is the probability 
density to transition from position $\xvec$ to $\yvec$ within time $\Deltat=t'-t$,
given by \cite{Wissel:1979}
\begin{equation}\elabel{def_GWissel} 
\GWissel(\xvec\to \yvec; t'-t)
    =
    \frac{\theta(t'-t)}{(4\pi \diffusion (t'-t))^{d/2}}\\
\Exp{-\frac{\Big(\yvec-\xvec-(\vecDrift-\extPot'(\xvec))(t'-t)\Big)^2}{4 \diffusion (t'-t)}}
\end{equation}
which, by inspection, solves the differential equation
\begin{equation}\elabel{Wissel_PDE}
\partial_{t'} \GWissel(\xvec\to \yvec; t'-t) = 
D \nabla_\yvec^2 \GWissel(\xvec\to \yvec; t'-t)
- (\vecDrift-\extPot'(\xvec)) \nabla_\yvec \GWissel(\xvec\to \yvec; t'-t) \ .
\end{equation}
What is missing in \Eref{Wissel_PDE} compared to \Eref{FPE_drift_diffusion} is the $\extPot''$-term. 
\Eref{def_GWissel} is therefore \emph{not} a solution of the FP \Eref{FPE_drift_diffusion}, but because $\lim_{t'\downarrow t} \GWissel(\xvec\to \yvec; t'-t) = \delta(\yvec-\xvec)$ and
\begin{equation}\elabel{actual_op_Wissel_identity}
\big(\vecDrift-\extPot'(\xvec)\big) \cdot \nabla_\yvec \delta(\yvec-\xvec) =
\nabla_\yvec \cdot \big( (\vecDrift-\extPot'(\xvec)) \delta(\yvec-\xvec) \big) =
\nabla_\yvec \cdot \big( (\vecDrift-\extPot'(\yvec)) \delta(\yvec-\xvec) \big) 
\end{equation}
\Eref{def_GWissel} produces the correct kernel,
\begin{equation}\elabel{Op_drift_diffusion_Wissel}
    \lim_{t'\downarrow t} \partial_{t'} \GWissel(\xvec\to \yvec; t'-t) = \FPophat_\yvec \delta(\yvec-\xvec)
\end{equation}
in other words, the full propagator $\ave{\phi(\yvec,t')\phitilde(\xvec,t)}$ is approximated to first order by the short-time propagator $\GWissel(\xvec\to \yvec; t'-t)$,
\begin{equation}\elabel{propagator_approximated_by_Wissel}
    \ave{\phi(\yvec,t')\phitilde(\xvec,t)}
    = \GWissel(\xvec\to \yvec; t'-t)
    \Big(1 + \order{(t'-t)^2}\Big) \ .
\end{equation}
The kernel $\Op_{\yvec,\xvec}$,
\Eref{def_Op},
calculated from the short-time propagator \Eref{def_GWissel} therefore reproduces correctly the Fokker-Planck kernel,
\begin{equation}
\Op_{\yvec,\xvec}  = \lim_{t'\downarrow t} \partial_{t'} \GWissel(\xvec\to \yvec; t'-t) = \FPophat_\yvec \delta(\yvec-\xvec)\ , 
\end{equation}
\Eref{Op_drift_diffusion_Wissel}.

As for the logarithm $\Ln_{\yvec,\xvec}$, \Eref{def_Ln}, using \Eref{propagator_approximated_by_Wissel} gives
\begin{equation}\elabel{Ln_drift_diffusion_Wissel}
\Ln_{\yvec,\xvec}=\lim_{t'\downarrow t} \ln\left(
      \frac
      {\ave{\phi(\yvec,t')\phitilde(\xvec,t)}}
      {\ave{\phi(\xvec,t')\phitilde(\yvec,t)}}
      \right)
      = 
\frac{(\yvec-\xvec)\bigg(2 \vecDrift-\extPot'(\xvec)-\extPot'(\yvec)\bigg)}{2\diffusion} \ .
\end{equation} 
by explicit use of \Eref{def_GWissel}
and thus the local entropy production \Eref{def_local_entropy}
\begin{equation}\elabel{local_entropy_GWissel}
    \entropyProductionDensity(\xvec)=
    \int \ddint{y} \Op_{\yvec,\xvec} \Ln_{\yvec,\xvec} =
    -\extPot''(\xvec) + \frac{( \vecDrift -
\extPot'(\xvec))^2}{\diffusion}  
\quad\text{so that}\quad
\entropyProduction[\density]=
\int\ddint{x} \density(\xvec) \entropyProductionDensity(\xvec) \ .
\end{equation}
\Eref{local_entropy_GWissel} is exact.
Away from stationarity, the logarithm $\ln(\density(\xvec)/\density(\yvec))$ needs to be added to 
$\Ln_{\yvec,\xvec}$ to capture all entropy production, 
\Eref{def_entropyProduction},
but this contribution is not considered in the present derivation.
The results above are very well known,
\eg \cite{Seifert:2012} or \cite[Sec.~3.11]{CocconiGarcia-MillanETAL:2020},
and are here retraced only to highlight 
which short-time details enter.

\toasubsection{Entropy production from a perturbation theory about drift diffusion}\seclabel{FT_about_drift-diffusion}
In the present section, we calculate the entropy production of a
drift-diffusion particle in an external potential in a perturbative
field theory about drift-diffusion. 
To this end, we split the action \Eref{action_drift_diffusion_appendix} into two terms,
$\action=\action_0+\action_\perturbative$ with
\begin{subequations}
\elabel{action_pert_over_drift}
\begin{equation}\elabel{action0_drift_diffusion}
\action_0 = 
\int \ddint{x} \int  \dint{t} \phitilde(\xvec,t) (D \nabla_\xvec^2 - \vecDrift \cdot \nabla_\xvec - \partial_t) \phi(x,t)
\end{equation}
and
\begin{equation}\elabel{action_pert_pot_only}
\action_\perturbative = 
- \int \ddint{x} \int \dint{t} \left(\extPot'(\xvec) \phi(\xvec,t) \right) \cdot \nabla_{\xvec} \phitilde(\xvec,t) \ ,
\end{equation}
\end{subequations}
so that any expectation of the full theory can be calculated along the lines of \Eref{perturbative_expansion}. The bare propagator 
\begin{align}\elabel{def_Gw}
    \Gw(\xvec\to \yvec; t'-t) 
    &=
     \frac{\theta(t'-t)}{(4\pi \diffusion (t'-t))^{d/2}}
\Exp{-\frac{\Big(\yvec-\xvec-\vecDrift(t'-t)\Big)^2}{4 \diffusion (t'-t)}}\\
\nonumber &\corresponds
\tikz[baseline=7.5pt]{
\begin{scope}[yshift=0.0cm]
\node at (\propWidth,0.4) [right,yshift=0pt] {$\xvec,t$};
\node at (-\propWidth,0.4) [left,yshift=0pt] {$\yvec,t'$};
\draw[tAactivity] (\propWidth,0.4) -- (-\propWidth,0.4);
\end{scope}
}
\end{align}
solves
\begin{equation}\elabel{Gw_PDE}
\partial_{t'} \Gw(\xvec\to \yvec; t'-t) 
=
\Big(\diffusion \nabla_{\yvec}^2 - \vecDrift\cdot\nabla_{\yvec}\Big)
\Gw(\xvec\to \yvec; t'-t) \ ,
\end{equation}
less vividly denoted by $g(y;x;t'-t)$ in \APref{simplified_notation}.
The bare propagator may be read off from $\action_0$ either in its present form or after Fourier transforming
\begin{equation}\elabel{def_Gw_k}
    \tikz[baseline=7.5pt]{
\begin{scope}[yshift=0.0cm]
\node at (\propWidth,0.4) [right,yshift=0pt] {$\kvec_\nvec,\omega$};
\node at (-\propWidth,0.4) [left,yshift=0pt] {$\kvec_\mvec,\omega'$};
\draw[tAactivity] (\propWidth,0.4) -- (-\propWidth,0.4);
\end{scope}
}
\corresponds
\frac{\deltabar(\omega'+\omega)L^d\delta_{\mvec+\nvec,0}}
{-\imag\omega'+\diffusion \kvec_\mvec^2+\imag\vecDrift\cdot\kvec_\mvec+r}
\ ,
\end{equation}
with discretised $\kvec_\nvec=2\pi\nvec/L$ and $\nvec\in\Zset^d$.
Although the Fourier-transform does not add anything crucial to the calculations to come, we discuss it here nevertheless, because of some puzzling implications.

\toasubsubsection{Fourier transformation}
\seclabel{Fourier_transformation}
To have a guaranteed stationary state, we need the present system to have a finite size of $L^d$ in the following. 
We therefore need to introduce a 
Fourier series representation for the spatial coordinates and a
Fourier transform for time
\begin{subequations}
\begin{align}
\phi(\xvec,t) & = 
\frac{1}{L^d}\sum_{\nvec\in\Zset^d} \exp{\imag \kvec_\nvec \cdot \xvec}
\int \dintbar{\omega} \exp{-\imag \omega t}
\phi_\nvec(\omega) \\
\phi_\nvec(\omega) & =
\int_{L^d}\ddint{x} \exp{-\imag \kvec_\nvec\cdot\xvec} 
\int \dint{t} \exp{\imag \omega t}
\phi(\xvec,t) \ .
\end{align}
\end{subequations}
The action \Eref{action_pert_over_drift} may then be written as
\begin{subequations}
\elabel{actions_Fouriered}
\begin{align}
\elabel{action_harm_Fouriered}
\action_0 &= 
\frac{1}{L^d} \sum_{\nvec\in\Zset^d}
\int \dintbar{\omega}
\phitilde_{-\nvec}(-\omega) 
(-\diffusion\kvec_\nvec^2 - \imag \vecDrift\cdot\kvec_\nvec -r+ \imag\omega) 
\phi_\nvec(\omega)\\
\elabel{action_pert_pot_only_again}
\action_\perturbative &=
\frac{1}{L^{3d}} \sum_{\nvec\mvec\ellvec}
\int \dintbar{\omega}
\phitilde_{\nvec}(-\omega) 
\kvec_\mvec\cdot\kvec_{\ellvec}\extPot_{\ellvec}
\phi_\mvec(\omega)
L^d \delta_{\nvec+\mvec+\ellvec,\nullvec} \ ,
\end{align}
\end{subequations}
with
\begin{equation}\elabel{extPot_Fourier_series}
\extPot(\xvec) =
\frac{1}{L^d}\sum_{\ellvec\in\Zset^d} \exp{\imag \kvec_\ellvec \cdot \xvec}
\extPot_\ellvec
\quad\text{ and }\quad
    \extPot_{\ellvec} = 
\int_{L^d}\ddint{x} \exp{-\imag \kvec_\ellvec\cdot\xvec} 
\extPot(\xvec) \ .
\end{equation}
The bare propagator then follows immediately, \Eref{def_Gw_k},
\begin{align}
\Gw(\kvec_\nvec\to \kvec_\mvec;\omega\to\omega') = &
\ave[0]{\phi(\kvec_\mvec,\omega')\phitilde(\kvec_\nvec,\omega)} \nonumber\\
=&
\frac{\deltabar(\omega'+\omega)L^{d}
\delta_{\mvec+\nvec,\nullvec}
}{-\imag\omega'+\diffusion \kvec_\mvec^2+\imag\vecDrift\cdot\kvec_\mvec+r}
\corresponds
\tikz[baseline=7.5pt]{
\begin{scope}[yshift=0.0cm]
\node at (\propWidth,0.4) [right,yshift=0pt] {$\kvec_\nvec,\omega$};
\node at (-\propWidth,0.4) [left,yshift=0pt] {$\kvec_\mvec,\omega'$};
\draw[tAactivity] (\propWidth,0.4) -- (-\propWidth,0.4);
\end{scope}
}
\ ,
\end{align}
with $\delta_{\mvec+\nvec,\nullvec}$ enforcing $\nvec=-\mvec$ and thereby momentum conservation.
The diagrammatic expansion produces corrections of the form
\begin{align}
    \ave{\phi(\kvec_\mvec,\omega')\phitilde(\kvec_\nvec,\omega)}
    \corresponds
    \tikz[baseline=0pt]{
\begin{scope}[yshift=0.0cm]
\node at (\propWidth,0.4) [right,yshift=0pt] {$\kvec_\nvec,\omega$};
\node at (-\propWidth,0.4) [left,yshift=0pt] {$\kvec_\mvec,\omega'$};
\draw[tAactivity] (\propWidth,0.4) -- (-\propWidth,0.4);
\draw[white,fill=white] (0,-0.3) circle (3pt);
\end{scope}
}
+
\tikz[baseline=0pt]{
\begin{scope}[yshift=0.0cm]
\node at (\propWidth,0.4) [right,yshift=0pt] {$\kvec_\nvec,\omega$};
\node at (-\propWidth,0.4) [left,yshift=0pt] {$\kvec_\mvec,\omega'$};
\draw[tAactivity] (\propWidth,0.4) -- (-\propWidth,0.4);
\draw[black,potStyle] (0,-0.2) -- (0,0.4);
\draw[black,thin] (-0.1,0.2) -- (0.1,0.2);
\draw[red,thin] (-0.2,0.3) -- (-0.2,0.5);
\tgenVertex{0,0.4}
\draw[black,fill=white] (0,-0.2) circle (3pt);
\end{scope}
}
+
\tikz[baseline=0pt]{
\begin{scope}[yshift=0.0cm]
\node at (-\propWidth,0.4) [left,yshift=0pt] {$\kvec_\mvec,\omega'$};
\draw[tAactivity] (\propWidth,0.4) -- (-\propWidth,0.4);
\draw[black,potStyle] (0,-0.2) -- (0,0.4);
\draw[black,thin] (-0.1,0.2) -- (0.1,0.2);
\draw[red,thin] (-0.2,0.3) -- (-0.2,0.5);
\tgenVertex{0,0.4}
\draw[black,fill=white] (0,-0.2) circle (3pt);
\end{scope}
\begin{scope}[xshift=0.8cm]
\node at (\propWidth,0.4) [right,yshift=0pt] {$\kvec_\nvec,\omega$};
\draw[tAactivity] (\propWidth,0.4) -- (-\propWidth,0.4);
\draw[black,potStyle] (0,-0.2) -- (0,0.4);
\draw[black,thin] (-0.1,0.2) -- (0.1,0.2);
\draw[red,thin] (-0.2,0.3) -- (-0.2,0.5);
\tgenVertex{0,0.4}
\draw[black,fill=white] (0,-0.2) circle (3pt);
\end{scope}
}
+
\ldots \ ,
\end{align}
where each ``bauble" represents the effect of the external potential that serves as a source for momentum, thereby breaking translational invariance \cite{ZhangETAL:2024}.

Apart from the technical subtleties of the full propagator not being a Gaussian but rather a Jacobi-theta function, which is of little interest in the following, allowing for a spatial Fourier series has deeper consequences. To see this, we determine the first order correction 
\begin{equation}\elabel{first_order_bauble_omega_k}
    \tikz[baseline=0pt]{
\begin{scope}[yshift=0.0cm]
\node at (\propWidth,0.4) [right,yshift=0pt] {$\kvec_\nvec,\omega$};
\node at (-\propWidth,0.4) [left,yshift=0pt] {$\kvec_\mvec,\omega'$};
\draw[tAactivity] (\propWidth,0.4) -- (-\propWidth,0.4);
\draw[black,potStyle] (0,-0.2) -- (0,0.4);
\draw[black,thin] (-0.1,0.2) -- (0.1,0.2);
\draw[red,thin] (-0.2,0.3) -- (-0.2,0.5);
\tgenVertex{0,0.4}
\draw[black,fill=white] (0,-0.2) circle (3pt);
\end{scope}
}
\corresponds
-
\frac{\deltabar(\omega'+\omega)}
{-\imag\omega'+\diffusion \kvec_\mvec^2+\imag\vecDrift\cdot\kvec_\mvec+r}
\kvec_\mvec\cdot\kvec_{\mvec+\nvec} \extPot_{\mvec+\nvec}
\frac{1}
{-\imag\omega'+\diffusion \kvec_\nvec^2-\imag\vecDrift\cdot\kvec_\nvec+r}
\end{equation}
for the spurious potential $\kvec_\ellvec\extPot_{\ellvec}=\imag\nuvec L^d \delta_{\ellvec,0}$, which has the same effect as an additional drift by $\nuvec$, as can be verified by direct evaluation in \Eref{action_pert_pot_only_again} and comparison to the corresponding term in \Eref{action_harm_Fouriered}. 
The inverse Fourier series \Eref{extPot_Fourier_series} to take $\extPot_\ell$ to $\extPot(\xvec)$ generally does not converge as $\extPot_\ell$ vanishes for all $\ell$ except $\ell=0$ where is diverges,
but that has no bearing on the arguments that follow.

After inversely Fourier transforming \Eref{first_order_bauble_omega_k} back to direct time, 
\begin{equation}\elabel{first_order_bauble_t_k}
    \tikz[baseline=0pt]{
\begin{scope}[yshift=0.0cm]
\node at (\propWidth,0.4) [right,yshift=0pt] {$\kvec_\nvec,t$};
\node at (-\propWidth,0.4) [left,yshift=0pt] {$\kvec_\mvec,t'$};
\draw[tAactivity] (\propWidth,0.4) -- (-\propWidth,0.4);
\draw[black,potStyle] (0,-0.2) -- (0,0.4);
\draw[black,thin] (-0.1,0.2) -- (0.1,0.2);
\draw[red,thin] (-0.2,0.3) -- (-0.2,0.5);
\tgenVertex{0,0.4}
\draw[black,fill=white] (0,-0.2) circle (3pt);
\end{scope}
}
\corresponds
-\imag (t'-t) \theta(t'-t) \kvec_\mvec\cdot\nuvec 
\Exp{-(t'-t) (\diffusion \kvec_\mvec^2+\imag\vecDrift\cdot\kvec_\mvec+r)}
\end{equation}
one can see explicitly the linear dependence on $t'-t$. This observation, of "each blob producing an order of $t'-t$", 
\APref{appendixWhichDiagramsContribute},
is what simplifies the calculation of the entropy production in the field-theoretic framework so dramatically.

The right-hand side of \Eref{first_order_bauble_t_k} can be recognised as the Fourier-transform in space of the gradient of $\Gw$,
\Erefs{def_Gw} and \eref{def_Gw_k},
which is immediately inverted to real space,
\begin{align}
    \tikz[baseline=0pt]{
\begin{scope}[yshift=0.0cm]
\node at (\propWidth,0.4) [right,yshift=0pt] {$\xvec,t$};
\node at (-\propWidth,0.4) [left,yshift=0pt] {$\yvec,t'$};
\draw[tAactivity] (\propWidth,0.4) -- (-\propWidth,0.4);
\draw[black,potStyle] (0,-0.2) -- (0,0.4);
\draw[black,thin] (-0.1,0.2) -- (0.1,0.2);
\draw[red,thin] (-0.2,0.3) -- (-0.2,0.5);
\tgenVertex{0,0.4}
\draw[black,fill=white] (0,-0.2) circle (3pt);
\end{scope}
}
&\corresponds
-(t'-t) 
\nuvec\cdot\nabla_\yvec
\Gw(\xvec\to \yvec; t'-t) \nonumber\\
&=
\elabel{first_order_bauble_t_x_final}
\nuvec\cdot
\frac{\yvec-\xvec-\vecDrift(t'-t)}{2\diffusion}
\Gw(\xvec\to \yvec; t'-t)
\end{align}
using \Eref{def_Gw}.

The key-difference between
\Erefs{first_order_bauble_t_k} and \eref{first_order_bauble_t_x_final} is the absence of an overall pre-factor $t'-t$ in the latter.
In a
perturbation theory of the propagator, terms that are of order $t'-t$ in
one representation of the degree of freedom, say $\kvec$, may no longer
seem to be of that order after a Fourier transform. However, the
right-hand side of \Eref{first_order_bauble_t_x_final} still
vanishes as $t'-t\downarrow0$ for any $\yvec-\xvec\ne\nullvec$ 
due to the exponential in $\Gw$, \Eref{def_Gw},
and indeed
it vanishes much faster than linearly in $t'-t>0$ for any such
$\yvec-\xvec\ne\nullvec$. For 
$t'-t<0$ it vanishes for any $\yvec-\xvec$ due to the Heaviside $\theta$-function in $\Gw$ and for
$\yvec-\xvec=\nullvec$ it vanishes linearly in $t'-t>0$ as the prefactor becomes $-\nuvec\cdot\vecDrift(t'-t)/(2\diffusion)$.

This phenomenon, that the order in $t'-t$ is changed by a transformation, is unique to continuous states and physically related to infinite rates being at play in the continuum limit, \APref{continuum_limit}. If all rates remain finite, as is generally the case for discrete states, it cannot occur, and neither does it happen when all "states" decouple, as is the case \emph{after} a Fourier-transform here.

\toasubsubsection{\texorpdfstring{$\Op$}{K} and \texorpdfstring{$\Ln$}{Ln} for drift diffusion in a perturbative potential}
\seclabel{Kn_Ln_for_drift_diffusion_in_pert_pot}
Following from the arguments in the main text and in \APref{appendixWhichDiagramsContribute}, we re-state the key-ingredients to calculate the entropy production. Firstly, the kernel $\Op_{\yvec,\xvec}$ can immediately be read off from the action or the Fokker-Planck operator, \Eref{FPE_drift_diffusion},
\begin{equation}\elabel{Kn_from_FP_drift_diffusion}
\Op_{\yvec,\xvec}
    = \FPophat_\yvec \delta(\yvec-\xvec) 
    = \Big(
\diffusion \nabla_y^2 - 
\nabla_y \cdot ( \vecDrift - \extPot'(\yvec)) 
\Big) \delta(\yvec-\xvec) 
    \ ,
\end{equation}
with $\nabla_y\cdot\extPot'\delta(\yvec-\xvec)$ intended to result in two terms by the product rule and with \Eref{actual_op_Wissel_identity} available to re-arrange the right-hand side. Even though the kernel is extracted easily, we will reproduce it below via the propagator to illustrate our scheme.
Secondly, the logarithm $\Ln_{\yvec,\xvec}$ is constructed from the propagator to first order. To this end, we state the first order correction \Eref{first_order_bauble_omega_k} in real space and direct time for arbitrary potentials using \Eref{action_pert_pot_only}
\begin{equation}\elabel{first_order_bauble_t_x_fullPot}
\tikz[baseline=0pt]{
\begin{scope}[yshift=0.0cm]
\node at (\propWidth,0.4) [right,yshift=0pt] {$\xvec,t$};
\node at (-\propWidth,0.4) [left,yshift=0pt] {$\yvec,t'$};
\draw[tAactivity] (\propWidth,0.4) -- (-\propWidth,0.4);
\draw[black,potStyle] (0,-0.2) -- (0,0.4);
\draw[black,thin] (-0.1,0.2) -- (0.1,0.2);
\draw[red,thin] (-0.2,0.3) -- (-0.2,0.5);
\tgenVertex{0,0.4}
\draw[black,fill=white] (0,-0.2) circle (3pt);
\end{scope}
}
\corresponds
-\int \ddint{z} \int \dint{s}
\Gw(\xvec\to\zvec,s-t) \extPot'(\zvec)\cdot\nabla_{\zvec} \Gw(\zvec\to\yvec,t'-s) \ ,
\end{equation}
which was previously stated in \Eref{first_order_bauble_t_k} only for the specific choice of the (spurious) potential that has the effect of a uniform drift.

To simplify \Eref{first_order_bauble_t_x_fullPot} by direct calculation, we draw on four "tricks": Firstly, 
\begin{equation}\elabel{trick1}
    \nabla_{\zvec} \Gw(\zvec\to\yvec,t'-s) = -\nabla_{\yvec} \Gw(\zvec\to\yvec,t'-s)
\end{equation}
so that the $\nabla_{\zvec}$ can be taken outside the integral in \Eref{first_order_bauble_t_x_fullPot}.
Secondly, we Taylor-expand $\extPot'(\zvec)$ about $(\xvec+\yvec)/2$,
\begin{equation}\elabel{trick2}
\nabla_\zvec \extPot(\zvec) = \extPot'(\zvec) =
  \extPot' \left(\frac{\yvec+\xvec}{2}\right)
+ \left( 
\frac{\zvec-\xvec}{2} - \frac{\yvec-\zvec}{2}
\right)\cdot\nabla \extPot' \left(\frac{\yvec+\xvec}{2}\right)
+ \ldots 
\end{equation}
so that parity in $\yvec-\xvec$ is readily determined, in contrast to, say, expanding about $\xvec$ or $\yvec$. \Eref{trick2} also allows us
to use, thirdly, 
\begin{equation}
\elabel{trick3}
  (\zvec-\xvec) \Gw(\xvec\to\zvec; s-t) 
= (s-t) \big( 2\diffusion \nabla_{\xvec} + \vecDrift \big) \Gw(\xvec\to\zvec; s-t) \ ,
\end{equation}
by inspection of \Eref{def_Gw}.
Finally, by the time-uniformity of the bare Markov process of drift-diffusion
\begin{equation}\elabel{trick4}
    \int \ddint{z} \Gw(\xvec\to\zvec,s-t) \Gw(\zvec\to\yvec,t'-s) = \theta(s-t) \theta(t'-s) \Gw(\xvec\to\yvec,t'-t) \ ,
\end{equation}
so that the spatial integral in \Eref{first_order_bauble_t_x_fullPot} can be carried out.

It turns out that of the expansion \Eref{trick2} only the first order is needed, 
\begin{subequations}
    \elabel{first_order_bauble_t_x_final2_fullPot_all}
\begin{align}
\elabel{first_order_bauble_t_x_final2_fullPot_1st}
    \tikz[baseline=0pt]{
\begin{scope}[yshift=0.0cm]
\node at (\propWidth,0.4) [right,yshift=0pt] {$\xvec,t$};
\node at (-\propWidth,0.4) [left,yshift=0pt] {$\yvec,t'$};
\draw[tAactivity] (\propWidth,0.4) -- (-\propWidth,0.4);
\draw[black,potStyle] (0,-0.2) -- (0,0.4);
\draw[black,thin] (-0.1,0.2) -- (0.1,0.2);
\draw[red,thin] (-0.2,0.3) -- (-0.2,0.5);
\tgenVertex{0,0.4}
\draw[black,fill=white] (0,-0.2) circle (3pt);
\end{scope}
}
&\corresponds
(t'-t) 
\nabla_\yvec\cdot\Big(
\extPot'\left(\frac{\xvec+\yvec}{2}\right)
\Gw(\xvec\to \yvec; t'-t) \Big) + \ldots \\
&=
\elabel{first_order_bauble_t_x_final2_fullPot_2nd}
-
\Gw(\xvec\to \yvec; t'-t)
\left(
\frac{\yvec-\xvec-\vecDrift(t'-t)}{2\diffusion}
\right)\cdot\extPot'\left(\frac{\xvec+\yvec}{2}\right)
+ \ldots
\end{align}
\end{subequations}
as higher order terms contribute neither to the kernel nor to the logarithm. 
In particular, the Laplacian of the external potential $\nabla_\yvec\cdot
\extPot'$ in \Eref{first_order_bauble_t_x_final2_fullPot_1st} is preceded by a factor $t'-t$ and thus vanishes from the logarithm as $t'\downarrow t$. As the logarithm is odd in $\yvec-\xvec$ by construction and the highest spatial derivative in the kernel is a second, the logarithm needs to be known only to linear order in $\yvec-\xvec$. Similarly, the kernel is a limit of a first derivative in time and thus draws only on terms linear in time, related to orders in space via \Eref{trick3}. 
The propagator may thus be written as
\begin{equation}\elabel{st_dd_prop}
    \ave{\phi(\yvec,t')\phitilde(\xvec,t)}=
    \Gw(\xvec\to \yvec; t'-t)
 + (t'-t) 
\nabla_\yvec\cdot\Big(
\extPot'\left(\frac{\xvec+\yvec}{2}\right)
\Gw(\xvec\to \yvec; t'-t) \Big) + \ldots
 \ .
\end{equation}
Applying \Eref{def_Op} and equally \Eref{transition_from_action} to \eref{st_dd_prop} reproduces the kernel \Eref{Kn_from_FP_drift_diffusion},
\begin{equation}\elabel{Kn_from_prop}
    \Op_{\yvec,\xvec} = 
    \Big(\diffusion \nabla_{\yvec}^2 - \vecDrift\cdot\nabla_{\yvec}\Big) 
    \delta(\yvec-\xvec)
+
\nabla_\yvec\cdot\left\{
\extPot'\left(\frac{\xvec+\yvec}{2}\right)
\delta(\yvec-\xvec) \right\}
\end{equation}
using \Eref{Gw_PDE} and $\lim_{t'\downarrow t} \Gw(\xvec\to \yvec; t'-t)=\delta(\yvec-\xvec)$. As $\extPot'((\xvec+\yvec)/2)\delta(\yvec-\xvec)=\extPot'(\yvec)\delta(\yvec-\xvec)=\extPot'(\xvec)\delta(\yvec-\xvec)$ the gradient of the potential can be taken outside the divergence, \Eref{actual_op_Wissel_identity}, but under an integral, this manipulation makes no difference.

The logarithm \Eref{def_Ln} is correspondingly
\begin{subequations}
\elabel{Ln_from_drift_diffusion}
\begin{align}
\Ln_{\yvec,\xvec} &\corresponds 
\lim_{t'\to t}
\ln\left(
\frac{
\tikz[baseline=0pt]{
\begin{scope}[yshift=0.0cm]
\node at (\propWidth,0.4) [right,yshift=0pt] {$\xvec,t$};
\node at (-\propWidth,0.4) [left,yshift=0pt] {$\yvec,t'$};
\draw[tAactivity] (\propWidth,0.4) -- (-\propWidth,0.4);
\end{scope}
}
+
\tikz[baseline=0pt]{
\begin{scope}[yshift=0.0cm]
\node at (\propWidth,0.4) [right,yshift=0pt] {$\xvec,t$};
\node at (-\propWidth,0.4) [left,yshift=0pt] {$\yvec,t'$};
\draw[tAactivity] (\propWidth,0.4) -- (-\propWidth,0.4);
\draw[black,potStyle] (0,-0.2) -- (0,0.4);
\draw[black,thin] (-0.1,0.2) -- (0.1,0.2);
\draw[red,thin] (-0.2,0.3) -- (-0.2,0.5);
\tgenVertex{0,0.4}
\draw[black,fill=white] (0,-0.2) circle (3pt);
\end{scope}
}
}{
\tikz[baseline=0pt]{
\begin{scope}[yshift=0.0cm]
\node at (\propWidth,0.4) [right,yshift=0pt] {$\yvec,t$};
\node at (-\propWidth,0.4) [left,yshift=0pt] {$\xvec,t'$};
\draw[tAactivity] (\propWidth,0.4) -- (-\propWidth,0.4);
\end{scope}
}
+
\tikz[baseline=0pt]{
\begin{scope}[yshift=0.0cm]
\node at (\propWidth,0.4) [right,yshift=0pt] {$\yvec,t$};
\node at (-\propWidth,0.4) [left,yshift=0pt] {$\xvec,t'$};
\draw[tAactivity] (\propWidth,0.4) -- (-\propWidth,0.4);
\draw[black,potStyle] (0,-0.2) -- (0,0.4);
\draw[black,thin] (-0.1,0.2) -- (0.1,0.2);
\draw[red,thin] (-0.2,0.3) -- (-0.2,0.5);
\tgenVertex{0,0.4}
\draw[black,fill=white] (0,-0.2) circle (3pt);
\end{scope}
}
}
\right)\\
\elabel{Ln_from_drift_diffusion_ln_pre_lim}
&\corresponds
\lim_{t'\to t}\left\{
\ln\left(
\frac{\Gw(\xvec\to \yvec; t'-t)}{\Gw(\yvec\to \xvec; t'-t)}
\right)
+
\ln\left(
\frac{1-
\left(
\frac{\yvec-\xvec-\vecDrift(t'-t)}{2\diffusion}
\right)\cdot\extPot'\left(\frac{\xvec+\yvec}{2}\right)
+ \ldots }
{1-
\left(
\frac{\xvec-\yvec-\vecDrift(t'-t)}{2\diffusion}
\right)\cdot\extPot'\left(\frac{\xvec+\yvec}{2}\right)
+\ldots }
\right)
\right\}
\\
&=
\frac{(\yvec-\xvec)\cdot\vecDrift}{\diffusion}
-
\frac{\yvec-\xvec}{\diffusion}
\cdot\extPot'\left(\frac{\xvec+\yvec}{2}\right)
+
\order{(\yvec-\xvec)^3}\ ,
\elabel{Ln_drift}
\end{align}
\end{subequations}
using \Eref{first_order_bauble_t_x_final2_fullPot_2nd} to arrive at \Eref{Ln_from_drift_diffusion_ln_pre_lim}.
This expression is identical to \Eref{Ln_drift_diffusion_Wissel} based on Wissel's short-time propagator if one allows for corrections of order $(\yvec-\xvec)^3$, where the expansion of $\extPot'$ being about $(\xvec+\yvec)/2$ becomes important. Together with \Eref{Kn_from_prop} this reproduces \Eref{local_entropy_GWissel}
\begin{equation}\elabel{local_entropy_Gw}
    \entropyProductionDensity(\xvec)=
    \int \ddint{y} \Op_{\yvec,\xvec} \Ln_{\yvec,\xvec} =
    -\extPot''(\xvec) + \frac{( \vecDrift -
\extPot'(\xvec))^2}{\diffusion}  
\ .
\end{equation}
This concludes the present derivation of the entropy production of a drift-diffusion particle using field theory only.

\toasubsection{Entropy production from a perturbation theory about diffusion}\seclabel{FT_about_diffusion}
We repeat the above derivation treating both potential and drift perturbatively. 
Expanding thus about pure diffusion,
the action \Eref{action_drift_diffusion_appendix} is split into two
terms, $\action=\action_0+\action_\perturbative$ with
\begin{subequations}
\elabel{action_pure_diffusion}
\begin{equation}\elabel{action0_pure_diffusion}
\action_0 = 
\int \ddint{x} \int  \dint{t} \phitilde(\xvec,t) (D \nabla_\xvec^2 - \partial_t) \phi(\xvec,t)
\end{equation}
and
\begin{equation}\elabel{actionPert_pure_diffusion}
\action_\perturbative = 
\int \ddint{x} \int \dint{t} \big( (\vecDrift - \extPot'(\xvec)) \phi(\xvec,t) \big) \cdot \nabla_\xvec \phitilde(\xvec,t)
\ ,
\end{equation}
\end{subequations}
where the drift $\vecDrift$ now features as a shift of the
force exerted by the potential $\extPot'$.
The bare propagator from \Eref{action0_pure_diffusion} is $\Gw$ of \Eref{def_Gw} with $\vecDrift=0$,
\begin{equation}
    \GD(\xvec\to \yvec; t'-t) 
    =
     \frac{\theta(t'-t)}{(4\pi \diffusion (t'-t))^{d/2}}
\Exp{-\frac{(\yvec-\xvec)^2}{4 \diffusion (t'-t)}}\\
\corresponds
\tikz[baseline=7.5pt]{
\begin{scope}[yshift=0.0cm]
\node at (\propWidth,0.4) [right,yshift=0pt] {$\xvec,t$};
\node at (-\propWidth,0.4) [left,yshift=0pt] {$\yvec,t'$};
\draw[tAactivity] (\propWidth,0.4) -- (-\propWidth,0.4);
\end{scope}
}
\end{equation}
and the two corrections from \Eref{actionPert_pure_diffusion} are \Eref{first_order_bauble_t_x_final2_fullPot_all} with $\extPot'$ replaced by $\extPot'-\vecDrift$,
\begin{subequations}
\elabel{phiphitilde_in_diagrams_all}
\begin{align}
\ave{\phi(\yvec,t')\phitilde(\xvec,t)}&\corresponds
\tbarePropagator{\xvec,t}{\yvec,t'} 
+
\tblobbedDashedPropagator{\xvec,t}{\yvec,t'}
+
\blobbedDashedPotPropagator{\xvec,t}{\yvec,t'}
+ \ldots\\
\elabel{first_order_bauble_t_x_pure_diff1}
&
\corresponds
\GD(\xvec\to \yvec; t'-t)
+
(t'-t) 
\nabla_\yvec\cdot\left(
\left[ \extPot'\left(\frac{\xvec+\yvec}{2}\right)
-\vecDrift \right]
\GD(\xvec\to \yvec; t'-t) \right) + \ldots \\
&
=
\elabel{first_order_bauble_t_x_pure_diff2}
\GD(\xvec\to \yvec; t'-t)
\left\{
1 -
\left(
\frac{\yvec-\xvec}{2\diffusion}
\right)\cdot\left[
\extPot'\left(\frac{\xvec+\yvec}{2}\right)
-
\vecDrift
\right]
\right\}
+ \ldots
\end{align}
\end{subequations}
As the total action \Eref{action_pure_diffusion} is identical to \Eref{action_pert_over_drift}, the kernel from the action of course is the same as \Eref{Kn_from_prop}, as confirmed by reading it off from the propagator in the form \Eref{first_order_bauble_t_x_pure_diff1},
\begin{equation}\elabel{Kn_from_prop_pure_diff}
    \Op_{\yvec,\xvec} = 
    \diffusion \nabla_{\yvec}^2 
    \delta(\yvec-\xvec)
+
\nabla_\yvec\cdot\left\{
\left[ 
\extPot'\left(\frac{\xvec+\yvec}{2}\right)
-\vecDrift\right]
\delta(\yvec-\xvec) \right\}\ .
\end{equation}
Similarly, the logarithmic term \Eref{Ln_from_drift_diffusion} is confirmed as
\begin{equation}
\Ln_{\yvec,\xvec} = 
-
\frac{\yvec-\xvec}{\diffusion}
\cdot\left[
\extPot'\left(\frac{\xvec+\yvec}{2}\right)
-
\vecDrift\right]
+
\order{(\yvec-\xvec)^3}
\end{equation}
obviously reproducing \Eref{local_entropy_Gw}.

This completes the present appendix. We have shown that the short-time propagator \Eref{def_GWissel} by Wissel \cite{Wissel:1979} used in \Erefs{def_Op}, \eref{def_Ln} and \eref{def_local_entropy} reproduces the entropy production \Eref{local_entropy_GWissel} in the literature \cite{CocconiGarcia-MillanETAL:2020}. We have further shown that a field-theoretic perturbation theory about drift-diffusion, \APref{FT_about_drift-diffusion}, or about pure diffusion, \APref{FT_about_diffusion}, equally reproduces these results. This is not a triviality, given the effect of spatial Fourier transform in a continuous state process, \cf \Erefs{first_order_bauble_t_k} and \eref{first_order_bauble_t_x_final}.



\toasection{Entropy production of multiple particles}\seclabel{MultipleParticles}

\newcommand{\ddtDpt}{\partial_{t'}}

\paragraph*{Abstract}
In the following we derive 
the entropy production of a system of $N$ 
conserved
particles,
\ie particles do not appear or disappear spontaneously. 
We treat distinguishable and indistinguishable particles separately.
In the case of distinguishable particles, the set of indexed particle coordinates describes a state fully. In the case of indistinguishable particles, all permutations of the indexed particle coordinates correspond to the same state. 
This ambiguity can be efficiently discounted by dividing the phase space by 
the Gibbs factor $N!$. 
We
build up our framework step-by-step: 
First for $N$ distinguishable and then also for indistinguishable particles, we derive the general principles in \APref{N_distinguishable_particles} and \ref{sec:N_indistinguishable_particles} respectively, before considering more concretely independent particles, \APref{N_independent_distinguishable_particles} and \ref{sec:N_independent_indistinguishable_particles}, before generalising to pair-interacting particles in \APref{N_interacting_distinguishable_particles} and \ref{sec:N_interacting_indistinguishable_particles}. 
We apply the present framework  to calculate the entropy production \Eref{trawler_final} of two pair-interacting, distinguishable particles in \APref{generalised_trawlers}, reproducing in a generalised form the ``trawler" system of \SMref{HarmonicTrawlers}.
We further apply this framework to calculate
the entropy production \Eref{entropyProduction_interacting_indistinguishable_example} of $N$ indistinguishable, pair-interacting particles in an external potential in \APref{N_interacting_indistinguishable_particle_extPot}, reproduced without external potential in \Eref{entropyProduction_for_pairPot}.
In the same section we show how to estimate the entropy production numerically from $Q$ samples of particle configurations at stationarity.

\toasubsection{\texorpdfstring{$N$}{N} distinguishable particles}
\seclabel{N_distinguishable_particles}
For  \emph{distinguishable} particles,
the starting point of the derivation is the entropy production of a single particle \Eref{def_entropyProduction}, with the particle coordinates $\xvec$ and $\yvec$ re-interpreted as those of multiple particles,
so that, say, components $(i-1)d+1$ to $id$ of $\xvec$ and $\yvec$ are the components of $\xvec_i$ and $\yvec_i$ of particle $i$ respectively.
The one-point probability or density $\density(\xvec)$ is then rewritten as the $N$-point probability or density $\HMdensity{N}{}(\xvec_1,\xvec_2,\ldots,\xvec_N)$ of $N$ \emph{distinguishable} particles. 
The constraint of being distinguishable comes about, because in \Eref{def_entropyProduction} each component of $\xvec$ and $\yvec$ refers to distinguishable spatial directions. This ``shortcut'' of deriving the expression for the entropy production of $N$ particles can therefore not be taken in the case of indistinguishable particles, which we treat separately
in \APref{N_indistinguishable_particles}.

In the field theory, distinguishability is implemented by having different species of particles, each represented by a pair of fields $\phi_i$ and $\phidagger_i$,  whereas indistinguisable particles belong to the same species, and are then represented by a single pair of fields $\phi$ and $\phidagger$.
The propagator of a single particle $\ave{\phi(\yvec,t')\phitilde(\xvec,t)}$ that used to make up the kernel $\Op_{\yvec,\xvec}$ and the log-term $\Ln_{\yvec,\xvec}$, \Erefs{def_Op} and \eref{def_Ln}, correspondingly is to be replaced by the joint propagator of all $N$ particle coordinates, $\bigl\langle\phi_1(\yvec_1,t')\linebreak[1]\phi_2(\yvec_2,t')\linebreak[1]\ldots\linebreak[1]\phi_N(\yvec_N,t')\linebreak[1]\phidagger_1(\xvec_1,t)\linebreak[1]\phidagger_2(\xvec_2,t)\linebreak[1]\ldots\linebreak[1]\phidagger_N(\xvec_N,t)\bigr\rangle$,
which contains the sum 
of all 
diagrams with $N$ incoming and $N$ outgoing
legs. 
The entropy production \Eref{def_entropyProduction} can then be written as a functional of the $N$-point density
$\HMdensity{N}{}(\xvec_1,\xvec_2,\ldots,\xvec_N)$ as
\begin{equation}\elabel{entropy_production_distinguishableN}
    \entropyProduction^{(N)}[\HMdensity{N}{}] =
    \SumInt_{\substack{\xvec_1,\ldots,\xvec_N \\ \yvec_1,\ldots,\yvec_N}}
      \HMdensity{N}{}(\xvec_1,\ldots,\xvec_N)
      \Op^{(N)}_{\yvec_1,\ldots,\yvec_N,\xvec_1,\ldots,\xvec_N} \left\{
      \Ln^{(N)}_{\yvec_1,\ldots,\yvec_N,\xvec_1,\ldots,\xvec_N} +
      \ln\left(
      \frac{\HMdensity{N}{}(\xvec_1,\ldots,\xvec_N)}{\HMdensity{N}{}(\yvec_1,\ldots,\yvec_N)}
      \right)
      \right\}\ ,
\end{equation}
where we allow for a sum over discrete states or an integral over continuous states,
with
\begin{equation}\elabel{Op_distinguishableN}
\Op^{(N)}_{\yvec_1,\ldots,\yvec_N,\xvec_1,\ldots,\xvec_N} = \lim_{t'\downarrow t} 
\ddtDpt
\ave{\phi_1(\yvec_1,t')\ldots\phi_N(\yvec_N,t')\phitilde_1(\xvec_1,t)\ldots\phitilde_N(\xvec_N,t)}
\end{equation}
and
\begin{equation}\elabel{Ln_distinguishableN}
\Ln^{(N)}_{\yvec_1,\ldots,\yvec_N,\xvec_1,\ldots,\xvec_N} = \lim_{t'\downarrow t}
\ln\left(\frac{\ave{\phi_1(\yvec_1,t')\ldots\phi_N(\yvec_N,t')\phitilde_1(\xvec_1,t)\ldots\phitilde_N(\xvec_N,t)}}{\ave{\phi_1(\xvec_1,t')\ldots\phi_N(\xvec_N,t')\phitilde_1(\yvec_1,t)\ldots\phitilde_N(\yvec_N,t)}}\right)
\ .
\end{equation}
The density $\HMdensity{N}{}(\xvec_1,\xvec_2,\ldots,\xvec_N)$ disappears from the
curly bracket in \Eref{entropy_production_distinguishableN}
at stationarity. 
Indeed, in the following, \emph{we  focus on the entropy production at
stationarity}, neglecting the term
\begin{equation}\elabel{entropyProductionNeglectedAtStationarity}
\Delta \entropyProduction^{(N)}[\HMdensity{N}{}] = 
    \SumInt_{\substack{\xvec_1,\ldots,\xvec_N \\ \yvec_1,\ldots,\yvec_N}}
      \HMdensity{N}{}(\xvec_1,\ldots,\xvec_N)
      \Op^{(N)}_{\yvec_1,\ldots,\yvec_N,\xvec_1,\ldots,\xvec_N} \left\{
      \ln\left(
      \frac{\HMdensity{N}{}(\xvec_1,\ldots,\xvec_N)}{\HMdensity{N}{}(\yvec_1,\ldots,\yvec_N)}
      \right)
      \right\}\ .
\end{equation}

In \Eref{entropy_production_distinguishableN}, the entropy production at stationarity is written as a functional of $\HMdensity{N}{}$, which may be ``supplied externally'', to emphasise that the entropy production can be thought of as a spatial average of the \emph{local entropy production}
\begin{equation}
\elabel{def_entropyProductionDensity}
\entropyProductionDensity^{(N)}(\xvec_1,\ldots,\xvec_N) = \SumInt_{\yvec_1,\ldots,\yvec_N} 
\Op^{(N)}_{\yvec_1,\ldots,\yvec_N,\xvec_1,\ldots,\xvec_N}
\Ln^{(N)}_{\yvec_1,\ldots,\yvec_N,\xvec_1,\ldots,\xvec_N} \ ,
\end{equation}
which is a function of $\xvec_1,\ldots\xvec_N$ only, so that
\begin{equation}\elabel{entropyProduction_as_spave}
    \entropyProduction^{(N)}[\HMdensity{N}{}] =
\SumInt_{\xvec_1,\ldots,\xvec_N} \HMdensity{N}{}(\xvec_1,\ldots,\xvec_N) \entropyProductionDensity^{(N)}(\xvec_1,\ldots,\xvec_N) 
\end{equation}
is a spatial mean.

The need to know the full
$\HMdensity{N}{}(\xvec_1,\ldots,\xvec_N)$ is generally a major
obstacle. If $N$ is large then little is generally known analytically about it in
an interacting system. 
Even numerical or experimental estimates of the $\HMdensity{N}{}$ are of
limited use, because often the statistics is poor. 
Below, this obstacle is overcome as it turns out that a theory with $n$-point interaction needs at most the $(2n-1)$-density and, under the assumption of short-rangedness, only the $n$-point density. 
In the field theory, the \emph{exact, stationary} $N$-point density $\HMdensity{N}{}$ is
\begin{align}
\elabel{density_is_propagator}
\HMdensity{N}{}(\xvec_1,\xvec_2,\ldots,\xvec_N) 
=
\lim_{t_{01},\ldots,t_{0N}\to-\infty}
\ave{\phi_1(\xvec_1,t)\ldots\phi_N(\xvec_N,t)\phitilde_1(\xvec_{01},t_{01})\ldots\phitilde_N(\xvec_{0N},t_{0N})} \ ,
\end{align}
independent of the initialisation $\xvec_{01},\ldots,\xvec_{0N}$ provided the system is ergodic. The limit of each $t_{0i}\to-\infty$ may be replaced by $t\to\infty$.

In principle, the propagator $\langle\phi_1\ldots\phitilde_N\rangle$ entering into 
the entropy production \Eref{entropy_production_distinguishableN} via
$\Op^{(N)}$ and $\Ln^{(N)}$, \Erefs{Op_distinguishableN} and
\eref{Ln_distinguishableN}
contains a plethora of terms. Without perturbative terms, however, it is simply the product of $N$ single-particle propagators,
\begin{align}\elabel{propagator_factorising}
\bigl\langle\phi_1(\yvec_1,t')\linebreak[1]\phi_2(\yvec_2,t')\linebreak[1]\ldots\linebreak[1]\phi_N(\yvec_N,t')\linebreak[1]\phidagger_1(\xvec_1,t)\linebreak[1]\phidagger_2(\xvec_2,t)\linebreak[1]\ldots\linebreak[1]\phidagger_N(\xvec_N,t)\bigr\rangle
&=
\prod^N_i \ave[0]{\phi_i(\yvec_i,t')\phitilde_i(\xvec_i,t)}
\nonumber\\&
\corresponds
\tikz[baseline=-2.5pt,scale=1.0]{
\begin{scope}[yshift=0.45cm]
\node at (\propWidth,0) [right] {$\xvec_1,t$};
\node at (-\propWidth,0) [left] {$\yvec_1,t'$};
\draw[tAactivity] (\propWidth,0) -- (-\propWidth,0) node[midway,above,yshift=-0.5mm] {$1$};
\end{scope}
\begin{scope}[yshift=0.0cm]
\node at (\propWidth,0) [right] {$\xvec_2,t$};
\node at (-\propWidth,0) [left] {$\yvec_2,t'$};
\draw[tAactivity] (\propWidth,0) -- (-\propWidth,0) node[midway,above,yshift=-0.5mm] {$2$};
\end{scope}
\node at (0,-0.4) {$\vdots$};
\node at (1,-0.4) [right] {$\vdots$};
\node at (-1,-0.4) [left] {$\vdots$};
\begin{scope}[yshift=-0.9cm]
\node at (\propWidth,0) [right] {$\xvec_N,t$};
\node at (-\propWidth,0) [left] {$\yvec_N,t'$};
\draw[tAactivity] (\propWidth,0) -- (-\propWidth,0) node[midway,below] {$N$};
\end{scope}
} \ ,
\end{align}
each propagator distinguishable by the particle species as indicated by the additional label on the line. To simplify the diagrammatics, we will omit many of the labels in the following.
Throughout this work, 
we are not considering multiple particles of the same of many species. We are also not considering any form of branching, such that all diagrams have the same number of incoming and outgoing legs, as discussed in \APref{branching_vertices}.

Next, allowing for perturbative terms, such as a single ``blob", 
$\tikz[baseline=-2.5pt]{\draw[tAactivity] (0.3,0) -- (-0.3,0); \tgenVertex{0,0};}$, for example, when particle drift or an external potential is implemented perturbatively, produces
\begin{multline}\elabel{joint_propagator_plus_first_order}
\bigl\langle\phi_1(\yvec_1,t')\linebreak[1]\phi_2(\yvec_2,t')\linebreak[1]\ldots\linebreak[1]\phi_N(\yvec_N,t')\linebreak[1]\phidagger_1(\xvec_1,t)\linebreak[1]\phidagger_2(\xvec_2,t)\linebreak[1]\ldots\linebreak[1]\phidagger_N(\xvec_N,t)\bigr\rangle\\
\corresponds
\tikz[baseline=-2.5pt,scale=1.0]{
\begin{scope}[yshift=0.45cm]
\node at (\propWidth,0) [right] {$\xvec_1$};
\node at (-\propWidth,0) [left] {$\yvec_1$};
\draw[tAactivity] (\propWidth,0) -- (-\propWidth,0) node[midway,above,yshift=-0.5mm] {$1$};
\end{scope}
\begin{scope}[yshift=0.0cm]
\node at (\propWidth,0) [right] {$\xvec_2$};
\node at (-\propWidth,0) [left] {$\yvec_2$};
\draw[tAactivity] (\propWidth,0) -- (-\propWidth,0) node[midway,above,yshift=-0.5mm] {$2$};
\end{scope}
\node at (0,-0.4) {$\vdots$};
\node at (1,-0.4) {$\vdots$};
\node at (-1,-0.4) {$\vdots$};
\begin{scope}[yshift=-0.9cm]
\node at (\propWidth,0) [right] {$\xvec_N$};
\node at (-\propWidth,0) [left] {$\yvec_N$};
\draw[tAactivity] (\propWidth,0) -- (-\propWidth,0) node[midway,below] {$N$};
\end{scope}
} 
+
\tikz[baseline=-2.5pt,scale=1.0]{
\begin{scope}[yshift=0.45cm]
\tgenVertex{0,0}
\node at (\propWidth,0) [right] {$\xvec_1$};
\node at (-\propWidth,0) [left] {$\yvec_1$};
\draw[tAactivity] (\propWidth,0) -- (-\propWidth,0) node[midway,above,yshift=-0.0mm] {$1$};
\end{scope}
\begin{scope}[yshift=0.0cm]
\node at (\propWidth,0) [right] {$\xvec_2$};
\node at (-\propWidth,0) [left] {$\yvec_2$};
\draw[tAactivity] (\propWidth,0) -- (-\propWidth,0) node[midway,above,yshift=-0.5mm] {$2$};
\end{scope}
\node at (0,-0.4) {$\vdots$};
\node at (1,-0.4) {$\vdots$};
\node at (-1,-0.4) {$\vdots$};
\begin{scope}[yshift=-0.9cm]
\node at (\propWidth,0) [right] {$\xvec_N$};
\node at (-\propWidth,0) [left] {$\yvec_N$};
\draw[tAactivity] (\propWidth,0) -- (-\propWidth,0) node[midway,below] {$N$};
\end{scope}
} 
+
\tikz[baseline=-2.5pt,scale=1.0]{
\begin{scope}[yshift=0.45cm]
\node at (\propWidth,0) [right] {$\xvec_1$};
\node at (-\propWidth,0) [left] {$\yvec_1$};
\draw[tAactivity] (\propWidth,0) -- (-\propWidth,0) node[midway,above,yshift=-0.5mm] {$1$};
\end{scope}
\begin{scope}[yshift=0.0cm]
\tgenVertex{0,0}
\node at (\propWidth,0) [right] {$\xvec_2$};
\node at (-\propWidth,0) [left] {$\yvec_2$};
\draw[tAactivity] (\propWidth,0) -- (-\propWidth,0) node[midway,above] {$2$};
\end{scope}
\node at (0,-0.4) {$\vdots$};
\node at (1,-0.4) {$\vdots$};
\node at (-1,-0.4) {$\vdots$};
\begin{scope}[yshift=-0.9cm]
\node at (\propWidth,0) [right] {$\xvec_N$};
\node at (-\propWidth,0) [left] {$\yvec_N$};
\draw[tAactivity] (\propWidth,0) -- (-\propWidth,0) node[midway,below] {$N$};
\end{scope}
} 
+ \ldots +
\tikz[baseline=-2.5pt,scale=1.0]{
\begin{scope}[yshift=0.45cm]
\node at (\propWidth,0) [right] {$\xvec_1$};
\node at (-\propWidth,0) [left] {$\yvec_1$};
\draw[tAactivity] (\propWidth,0) -- (-\propWidth,0) node[midway,above,yshift=-0.5mm] {$1$};
\end{scope}
\begin{scope}[yshift=0.0cm]
\node at (\propWidth,0) [right] {$\xvec_2$};
\node at (-\propWidth,0) [left] {$\yvec_2$};
\draw[tAactivity] (\propWidth,0) -- (-\propWidth,0) node[midway,above,yshift=-0.5mm] {$2$};
\end{scope}
\node at (0,-0.4) {$\vdots$};
\node at (1,-0.4) {$\vdots$};
\node at (-1,-0.4) {$\vdots$};
\begin{scope}[yshift=-0.9cm]
\tgenVertex{0,0}
\node at (\propWidth,0) [right] {$\xvec_N$};
\node at (-\propWidth,0) [left] {$\yvec_N$};
\draw[tAactivity] (\propWidth,0) -- (-\propWidth,0) node[midway,below] {$N$};
\end{scope}
} 
+ \hot
\end{multline}
On the right there is a single product of bare propagators without a blob, followed by $N$ terms consisting of a product of $N-1$ bare propagators and a single propagator with blob. Higher order terms with multiple bobs do not contribute, \APref{appendixWhichDiagramsContribute}.

\toasubsubsubsection{Simplified notation and example}
\seclabel{simplified_notation}
To facilitate the derivations in the following sections, we introduce a simplified notation and an example at this stage. Firstly, a plain, bare propagator of particle species $i$ shall be written as
\begin{equation}\elabel{def_g_i}
\tikz[baseline=-2.5pt,scale=1.0]{
\node at (\propWidth,0) [right] {$\xvec_i,t$};
\node at (-\propWidth,0) [left] {$\yvec_i,t'$};
\draw[tAactivity] (\propWidth,0) -- (-\propWidth,0) node[midway,above,yshift=-0.5mm] {$i$};
}
\corresponds
\ave[0]{\phi_i(\yvec_i,t')\phitilde_i(\xvec_i,t)} = g_i(\yvec_i;\xvec_i;t'-t)
= g_i
\end{equation}
with \Eref{propagator_limit_is_delta} 
\begin{equation}\elabel{g_i_limit}
\lim_{t'\downarrow t} 
\ave[0]{\phi_i(\yvec_i,t')\phitilde_i(\xvec_i,t)} = 
\lim_{t'\downarrow t}
g_i(\yvec_i;\xvec_i;t'-t) = \delta(\yvec_i-\xvec_i) = \delta_i
\end{equation}
where we have also introduced the shorthand $\delta_i$, whose gradient and higher order derivatives we will denote by dashes. 
We further denote the time derivative of $g_i$ in the limit of $t'\downarrow t$  by
\begin{equation}\elabel{def_gdot}
\lim_{t'\downarrow t} 
\ddtDpt \ave[0]{\phi_i(\yvec_i,t')\phitilde_i(\xvec_i,t)}=
\lim_{t'\downarrow t} 
\ddtDpt g_i(\yvec_i;\xvec_i;t'-t)=
\gdot_i(\yvec_i;\xvec_i)=
\gdot_i \ .
\end{equation}
The latter derives its properties from the Fokker-Planck operator, $\FPophat_{\yvec}$,
in \Eref{FPeqn_main}
\begin{equation}
\gdot_i=\FPophat_{\yvec_i}\delta(\yvec_i-\xvec_i) \ .
\end{equation}
The perturbative, generic transmutation-like terms, such as those with a single blob in 
\Eref{propagator_expansion_app}, will be denoted by 
\begin{equation}\elabel{def_f_i}
\tikz[baseline=5pt,scale=1.0]{
\begin{scope}[yshift=0.25cm]
\tgenVertex{0,0}
\node at (\propWidth,0) [right] {$\xvec_i,t$};
\node at (-\propWidth,0) [left] {$\yvec_i,t'$};
\draw[tAactivity] (\propWidth,0) -- (-\propWidth,0) node[midway,above,yshift=0.3mm] {$i$};
\end{scope}
}
\corresponds
f_i(\yvec_i;\xvec_i;t'-t)
= f_i \ .
\end{equation}
Such a term
may have a complicated dependence on $t'-t$, but is generally evaluated to first order in $t'-t$.
We denote its time derivative in the limit of $t'\downarrow t$  as
\begin{equation}\elabel{def_fdot}
\lim_{t'\downarrow t} 
\ddtDpt f_i(\yvec_i;\xvec_i;t'-t)=
\fdot_i(\yvec_i;\xvec_i)=
\fdot_i \ .
\end{equation}
The notation of $g_i$ and $f_i$ allows us to succinctly express the full propagator as
\begin{equation}
\ave{\phi_i(\yvec_i,t')\phitilde_i(\xvec_i,t)} = g_i + f_i + \order{(t'-t)^2} \ ,
\end{equation}
where $f_i$ generally vanishes linearly in $t'-t$, 
\APref{appendixWhichDiagramsContribute} and \ref{sec:drift_diffusion_on_ring},
so that
\begin{equation}\elabel{f_vanishes}
\lim_{t'\downarrow t} f_i = 0 \ .
\end{equation}
The time derivatives of the full propagators that we will need can be succinctly expressed as
\begin{equation}
\lim_{t'\downarrow t}\elabel{single_propagator_derivative}
\ddtDpt
\ave{\phi_i(\yvec_i,t')\phitilde_i(\xvec_i,t)} = \gdot_i + \fdot_i \ .
\end{equation}
Beyond the narrow definitions above, expanding the propagator can be rather dangerous. For example, it would be wrong to say that
$\ave{\phi_i(\yvec_i,t')\phitilde_i(\xvec_i,t)}=\delta(\yvec_i-\xvec_i)+(t'-t)(\gdot_i + \fdot_i) + \order{(t'-t)^2}$, because the $\delta$-function is truly absent from $\ave{\phi_i\phitilde_i}$ at $t'-t>0$.
Also, $\gdot_i$ and $\fdot_i$ are kernels, generally containing derivatives of $\delta$-functions, unsuitable, for example, to appear in the logarithm. There, we will need limits of the form $\lim_{t'\downarrow t} f_i/g_i$, such as \Eref{f_over_g_example}.
It is further useful to introduce a succinct notation for $g_i$ and $f_i$ with reversed arguments
\begin{subequations}\elabel{def_barred}
\begin{align}
\Bg_i&= g_i(\xvec_i;\yvec_i;t'-t)\\
\Bf_i&= f_i(\xvec_i;\yvec_i;t'-t) \ .
\end{align} 
\end{subequations}

A useful example of $g_i$ and $\gdot_i$ is drift-diffusion in $d$ dimension, \Eref{def_Gw},
\begin{equation}\elabel{example_g_i}
g_i=\frac{\theta(t'-t)}{(4\pi \diffusion_i (t'-t))^{d/2}} \Exp{-\frac{(\yvec_i-\xvec_i-\vecDrift_i(t'-t))^2}{4 \diffusion_i (t'-t)}}
\end{equation}
with drift velocity $\vecDrift_i$ and diffusion constant $\diffusion_i$ of particle species $i$, so that
\begin{equation}\elabel{gi_over_Bgi}
    \lim_{t'\downarrow t} \frac{g_i}{\Bg_i} = \Exp{\frac{\vecDrift_i\cdot(\yvec_i-\xvec_i)}{\diffusion_i}} \ .
\end{equation}
A propagator has generally the property \Eref{trick4}, 
\begin{equation}\elabel{propoagator_convolution}
g_i(\yvec_i;\xvec_i;t'-t) = \int \ddint{z_i} g_i(\yvec_i;\zvec_i;t'-s) g_i(\zvec_i;\xvec_i;s-t) 
\end{equation}
for any $s\in(t,t')$. For $s\notin(t,t')$ the integral vanishes, as each $g_i$ enforces causality via a Heaviside-$\theta$ function, \Eref{example_g_i}.
The bare propagator in \Eref{example_g_i} solves the \FPeqn for individual particle species $i$ with operator 
\begin{equation}
\FPophat^{(i)}_{\yvec_i}=
\diffusion_i\nabla_{\yvec_i}^2 - \vecDrift_i\cdot\nabla_{\yvec_i}\ ,
\end{equation}
and therefore 
\begin{equation}\elabel{gdot_example}
\gdot_i =
\left(\diffusion_i\nabla_{\yvec_i}^2 - \vecDrift_i\cdot\nabla_{\yvec_i}\right)
\delta(\yvec_i-\xvec_i)
=
\diffusion_i\delta''(\yvec_i-\xvec_i)
-
\vecDrift_i\cdot\delta'(\yvec_i-\xvec_i) \ .
\end{equation}
The perturbative term $f_i$ may be another source of drift, either constant or due to an external potential $\extPot_i(\xvec)$.
It is constructed via the convolution \Eref{first_order_bauble_t_x_fullPot}
\begin{align}
    \tikz[baseline=0pt]{
\begin{scope}[yshift=0.0cm]
\node at (\propWidth,0.4) [right,yshift=0pt] {$\xvec_i,t$};
\node at (-\propWidth,0.4) [left,yshift=0pt] {$\yvec_i,t'$};
\draw[tAactivity] (\propWidth,0.4) -- (-\propWidth,0.4);
\draw[potStyle] (0,-0.2) -- (0,0.4);
\draw[black,thin] (-0.1,0.2) -- (0.1,0.2);
\draw[red,thin] (-0.2,0.3) -- (-0.2,0.5);
\tgenVertex{0,0.4}
\draw[black,fill=white] (0,-0.2) circle (3pt) node[right] {$\extPot_i'$};
\end{scope}
}
&\corresponds f_i(\yvec_i;\xvec_i;t'-t) 
\nonumber\\
&=
(t'-t) 
\nabla_{\yvec_i}\cdot\Big(
\extPot_i'\left(\frac{\xvec_i+\yvec_i}{2}\right)
g_i(\yvec_i;\xvec_i;t'-t)
\Big) + \hot \ ,
\elabel{f_example}
\end{align}
where $\extPot_i'$ denotes the gradient of $\extPot_i$ with respect to its argument.
Of the two terms resulting from the nabla acting on the product to its right, only the differentiation of $g_i$ results in a term that eventually enters in the entropy production, as discussed after \Eref{first_order_bauble_t_x_final2_fullPot_all},  \APref{Kn_Ln_for_drift_diffusion_in_pert_pot}. Using 
\Eref{example_g_i}
explicitly, one finds in particular
the ratio $f_i/g_i$ that will be useful for the logarithm,
\begin{equation}\elabel{f_over_g_example}
\lim_{t'\downarrow t} \frac{f_i}{g_i}
=
-\frac{\yvec_i-\xvec_i}{2 \diffusion_i} \cdot \extPot'_i\left(\frac{\yvec_i+\xvec_i}{2}\right) \ ,
\end{equation}
so that $\delta_i \lim_{t'\downarrow t} f_i/g_i=0$. In general, we will make the weaker assumption
\begin{equation}\elabel{f_over_g_delta_limit}
\lim_{t'\downarrow t} \delta_i \frac{f_i}{g_i} =0 \ ,
\end{equation}
which might be taken most easily as the $\delta_i$ in front of $f_i/g_i$ can greatly simplify this ratio.
As for the kernel,
differentiating \Eref{f_example} with respect to $t'$, \Eref{def_fdot}, gives
\begin{equation}\elabel{fdot_example}
\fdot_i = 
\lim_{t'\downarrow t}
\nabla_{\yvec_i}\cdot\Big(
\extPot_i'\left(\frac{\xvec_i+\yvec_i}{2}\right)
g_i(\yvec_i;\xvec_i;t'-t)
\Big)=
\nabla_{\yvec_i}\cdot\left(
\extPot_i'\left(\frac{\xvec_i+\yvec_i}{2}\right)
\delta_i
\right)
=
\extPot_i'\left(\frac{\xvec_i+\yvec_i}{2}\right)
\cdot
\delta_i'
\ ,
\end{equation}
similar to \Eref{Kn_from_prop} further discussed thereafter.

Carrying on with the simplified notation, we also need to introduce the notation for pair interactions.
The structure of a pair potential term follows that of the external potential \Eref{f_example}, to leading order in $t'-t$,
\begin{align}
&\tikz[baseline=0pt]{
\begin{scope}[yshift=0.0cm]
\draw[potStyle] (0,-0.4) -- (0,0.4);
\draw[black,thin] (-0.1,0.2) -- (0.1,0.2);
\draw[red,thin] (-0.2,0.3) -- (-0.2,0.5);
\tgenVertex{0,-0.4}
\tgenVertex{0,0.4}
\node at (\propWidth,-0.5) [right,yshift=0pt] {$\xvec_j,t$};
\node at (-\propWidth,-0.5) [left,yshift=0pt] {$\yvec_j,t'$};
\draw[tAactivity] (\propWidth,-0.5) -- (0,-0.4) -- (-\propWidth,-0.5);
\node at (\propWidth,0.5) [right,yshift=0pt] {$\xvec_i,t$};
\node at (-\propWidth,0.5) [left,yshift=0pt] {$\yvec_i,t'$};
\draw[tAactivity] (\propWidth,0.5) -- (0,0.4) -- (-\propWidth,0.5);
\end{scope}
}
\corresponds h_{ij}(\yvec_i,\yvec_j;\xvec_i,\xvec_j; t'-t) \nonumber\\
&= - \int \dint{s} \ddint{z_i}\ddint{z_j}
g_i(\zvec_i; \xvec_i; s-t)  \left(\nabla_{\zvec_i} \pairPot_{ij}(\zvec_i-\zvec_j)\right)
\cdot
\left(\nabla_{\zvec_i} g_i(\yvec_i; \zvec_i; t'-s)\right) 
g_j(\zvec_j; \xvec_j; s-t) g_j(\yvec_j; \zvec_j; t'-s) \nonumber\\
&= (t'-t) \left(\nabla_{\xvec_i} \pairPot_{ij}(\xvec_i-\xvec_j)\right)\cdot\left(\nabla_{\yvec_i}g_i(\yvec_i; \xvec_i; t'-t)\right) g_j(\yvec_j; \xvec_j; t'-t) + \hot
\elabel{h_example}
\end{align}
where we have 
used "tricks" similar to \Erefs{trick1} to \eref{trick4}.
In brief, we may write $h_{ij}$ as \Eref{pair_pot_effect2}
\begin{equation}\elabel{h_example_compact}
h_{ij}
= h_{ij}(\yvec_i,\yvec_j;\xvec_i,\xvec_j; t'-t)= (t'-t)\pairPot'_{ij}(\xvec_i-\xvec_j)\cdot g_i'(\yvec_i; \xvec_i; t'-t)  g_j(\yvec_j; \xvec_j; t'-t) + \hot \ .
\end{equation}
and with \Eref{example_g_i} explicitly, the relevant limit is
\begin{equation}\elabel{h_over_gg}
\lim_{t'\downarrow t} \frac{h_{ij}}{g_ig_j} = 
-\frac{\yvec_i-\xvec_i}{2 \diffusion_i} \cdot \pairPot'_{ij}(\xvec_i-\xvec_j) \ .
\end{equation}

Just like $f_i$, the interaction $h_{ij}$ is only ever evaluated to first order in $t'-t$ and we may therefore be occasionally found sloppilly dropping higher order terms in their entirety.
We denote the limit of the time-derivative of $h_{ij}$ 
by
\begin{equation}\elabel{def_hdot}
\lim_{t'\downarrow t} \ddtDpt h_{ij}(\yvec_i,\yvec_j;\xvec_i,\xvec_j; t'-t)
= \hdot_{ij}(\yvec_i,\yvec_j;\xvec_i,\xvec_j)
=\hdot_{ij}
 \ ,
\end{equation}
and assume that it is $\delta$-like in $\yvec_j-\xvec_j$,
\begin{equation}
    \hdot_{ij}\propto\delta_j \ .
    \elabel{hderi_delta}
\end{equation}
For the example in \Eref{h_example_compact}, this means
\begin{equation}\elabel{hdot_example}
\hdot_{ij}=\pairPot'_{ij}(\xvec_i-\xvec_j)\cdot \delta_i' \delta_j\ .
\end{equation}
The interaction term evaluated with inverted arguments is denoted by
\begin{align}
\Bh_{ij}    &=  h_{ij}(\xvec_i,\xvec_j;\yvec_i,\yvec_j; t'-t)\ ,
\end{align}
corresponding to the notation introduced above. The properties of $h_{ij}$ are very similar to those of $f_i$, as it affects the motion of particle $i$, otherwise only evaluating the position of particle $j$.

\toasubsubsection{\texorpdfstring{$N$}{N} independent, distinguishable particles}
\seclabel{N_independent_distinguishable_particles}
When particles are independent, the propagator factorises, 
\begin{align}\nonumber
\bigl\langle\phi_1(\yvec_1,t')\linebreak[1]\ldots\linebreak[1]\phi_N(\yvec_N,t')\linebreak[1]\phidagger_1(\xvec_1,t)\linebreak[1]\ldots\linebreak[1]\phidagger_N(\xvec_N,t)\bigr\rangle = &
\prod^N_i \ave{\phi_i(\yvec_i,t')\phitilde_i(\xvec_i,t)}\\
= &
\prod^N_i g_i + \sum_i^N f_i \prod^N_{\substack{j\ne i}} g_j + \order{(t'-t)^2} \ , 
\elabel{Npropagators_distinguishable}
\end{align}
so that the kernel becomes,
using \Erefs{g_i_limit}, \eref{def_gdot}, \eref{def_fdot} and \eref{f_vanishes}, or simply \Erefs{propagator_factorising} and \eref{single_propagator_derivative},
\begin{align}\elabel{Op_distinguishable_independent}
  \Op^{(N)}_{\yvec_1,\ldots,\yvec_N,\xvec_1,\ldots,\xvec_N} = &
  \sum_{i=1}^N 
  \lim_{t'\downarrow t} 
  \left(
  \ddtDpt
  \ave{\phi_i(\yvec_i,t')\phitilde_i(\xvec_i,t)}
  \right)
  \prod_{j\ne i}^N
  \ave{\phi_j(\yvec_j,t')\phitilde_j(\xvec_j,t)} 
\nonumber \\ = &
\sum_i^N (\gdot_i + \fdot_i) \prod^N_{\substack{j\ne i}} \delta_j
\end{align}
and the logarithm of the ratio of the propagators,
\begin{equation}\elabel{Ln_distinguishable_independent}
\Ln^{(N)}_{\yvec_1,\ldots,\yvec_N,\xvec_1,\ldots,\xvec_N} = \lim_{t'\downarrow t}
\ln\left(
\frac
{\prod^N_i   g_i + \sum_i^N   f_i \prod^N_{\substack{j\ne i}}   g_j + \order{(t'-t)^2}}
{\prod^N_i \Bg_i + \sum_i^N \Bf_i \prod^N_{\substack{j\ne i}} \Bg_j + \order{(t'-t)^2}}
\right) \ ,
\end{equation}
where we have made use of the barred notation \Eref{def_barred}. 
There is no need to retain terms of order $(t'-t)^2$, because if the lower order terms vanish, so does the kernel and the entire logarithm does not contribute.
For a continuous variable, the logarithm is efficiently written as
\begin{equation}\elabel{Ln_distinguishable_independent_simplified}
\Ln^{(N)}_{\yvec_1,\ldots,\yvec_N,\xvec_1,\ldots,\xvec_N} = \lim_{t'\downarrow t}
\sum_i^N
\ln\left(
\frac
{g_i}
{\Bg_i}
\right)
+
\ln\left(
\frac
{1 + \sum_i^N   f_i/g_i  }
{1 + \sum_i^N \Bf_i/\Bg_i} 
\right) \ .
\end{equation}
If states are discrete, a slightly different approach is needed and $\Ln^{(N)}$ is best kept in the form \Eref{Ln_distinguishable_independent} as neither $g_i/\Bg_i$ nor $f_i/g_i$ might be well-defined in the limit $t'\downarrow t$, while $\delta_i$ itself evaluates to either $0$ or $1$ inside the logarithm. Henceforth, we will focus entirely on continuous states $\xvec_i$.

As the kernel $\Op^{(N)}$ is expected to be at most second order in spatial derivatives, \APref{Kn_Ln_for_drift_diffusion_in_pert_pot}, the logarithm $\Ln^{(N)}$, which is odd in $\yvec-\xvec$, can be expanded in small $\yvec-\xvec$, 
\begin{equation}\elabel{Ln_distinguishable_independent_simplified2}
\Ln^{(N)}_{\yvec_1,\ldots,\yvec_N,\xvec_1,\ldots,\xvec_N} = \lim_{t'\downarrow t}
\sum_i^N
\ln\left(
\frac
{g_i}
{\Bg_i}
\right)
+
\sum_i^N \left[ \frac{f_i}{g_i} - \frac{\Bf_i}{\Bg_i} \right]\ ,
\end{equation}
so that
 the local entropy production
\Eref{def_entropyProductionDensity} at stationarity,
is 
\begin{equation}\elabel{entropyProductionDensity_independent_distinguishable_initial}
\entropyProductionDensity^{(N)}(\xvec_1,\ldots,\xvec_N) 
= \int \ddint{y_{1,\ldots,N}}
\left\{
\sum_i^N (\gdot_i + \fdot_i) \prod^N_{\substack{j\ne i}} \delta_j
\right\}
\lim_{t'\downarrow t}
\left\{ 
\sum_k^N \left[ \ln\left(\frac{g_k}{\Bg_k}\right) + \frac{f_k}{g_k} - \frac{\Bf_k}{\Bg_k} \right]
\right\} \ .
\end{equation}
The two sums in this expression produce $N^2$ terms in total. Under the integral, the product $\prod^N_{\substack{j\ne i}} \delta_j$ forces $g_k/\Bg_k$ to converge to $1$ and $f_k/g_k$ to vanish for all $k\ne i$, \Eref{f_over_g_delta_limit}. 
Of the second sum, only the terms $k=i$ remain, so that
\begin{align}
\entropyProductionDensity^{(N)}(\xvec_1,\ldots,\xvec_N) 
= & \int \ddint{y_{1,\ldots,N}}
\sum_i^N (\gdot_i + \fdot_i) \left\{
\prod^N_{\substack{j\ne i}} \delta_j
\lim_{t'\downarrow t}
\left[ \ln\left(\frac{g_i}{\Bg_i}\right) + \frac{f_i}{g_i} - \frac{\Bf_i}{\Bg_i} \right]\right\}
\nonumber \\ = &
\sum_i^N \entropyProductionDensity^{(N)}_i(\xvec_i)
\elabel{entropyProductionDensity_independent_distinguishable_final_sum}
\end{align}
with
\begin{align}
\entropyProductionDensity^{(N)}_i(\xvec_i) = 
\int \ddint{y_i} (\gdot_i + \fdot_i) 
\lim_{t'\downarrow t}
\left[ \ln\left(\frac{g_i}{\Bg_i}\right) + \frac{f_i}{g_i} - \frac{\Bf_i}{\Bg_i} \right]
\ ,
\elabel{entropyProductionDensity_independent_distinguishable_final}
\end{align}
in fact independent of any other particles around, so that $\entropyProductionDensity^{(N)}_i(\xvec_i)=\entropyProductionDensity^{(1)}_i(\xvec_i)$ is the same for any $N\ge1$.
The local entropy production is therefore the sum of the local entropy production of each particle. Using this expression in \Eref{entropyProduction_as_spave},
\begin{equation}
\entropyProduction^{(N)}[\HMdensity{N}{}] = \int \ddint{x_{1,\ldots,N}} \HMdensity{N}{}(\xvec_1,\ldots,\xvec_N) \sum_i^N \entropyProductionDensity^{(1)}_i(\xvec_i)
\end{equation}
allows all integrals except the one over $\xvec_i$ to be carried out as a marginalisation,
\begin{equation}\elabel{marginalisation_densityN_i_distinguishable}
\HMdensity{N}{i}(\xvec_i) =  \int \ddint{x_{1\ldots i-1,i+1,\ldots,N}} \HMdensity{N}{}(\xvec_1,\ldots,\xvec_N)  \ ,
\end{equation}
so that $\HMdensity{N}{i}(\xvec_i)$ is the density of particle species $i$ at $\xvec_i$ and
\begin{align}
\entropyProduction^{(N)}[\HMdensity{N}{}]
=  \sum_i^N \int \ddint{x_i} \HMdensity{N}{i}(\xvec_i) \,\entropyProductionDensity^{(1)}_i(\xvec_i) 
\ ,
\elabel{entropyProduction_independent_distinguishable_final}
\end{align}
confirming the overall entropy production as the sum of single particle entropy productions, \ie confirming extensivity.

To illustrate \Eref{entropyProductionDensity_independent_distinguishable_final} with an example, 
\begin{align}
\entropyProductionDensity^{(1)}_i(\xvec_i) = &
\int \ddint{y_i} \left(\diffusion_i\delta''_i - \vecDrift_i\cdot\delta'_i + \extPot'_i\left(\frac{\yvec_i+\xvec_i}{2}\right) \cdot \delta'_i + \half \extPot''_i\left(\frac{\yvec_i+\xvec_i}{2}\right) \delta_i\right)\nonumber\\
 & \times
\left[ \frac{\vecDrift_i\cdot(\yvec_i-\xvec_i)}{\diffusion_i}- \frac{\yvec_i-\xvec_i}{\diffusion_i}\cdot  \extPot'_i\left(\frac{\yvec_i+\xvec_i}{2}\right)   \right] 
=
-\nabla^2 \extPot_i(\xvec_i) +\frac{\left(\nabla \extPot(\xvec_i) - \vecDrift_i\right)^2}{\diffusion_i}
\elabel{entropyProductionDensity_distinguishable_non-interacting_example}
\end{align}
using \Erefs{example_g_i}, \eref{gi_over_Bgi}, \eref{gdot_example}, \eref{f_over_g_example} and \eref{fdot_example}. 
This result is identical to \Eref{local_entropy_Gw}. 
Without drift, 
it is easy to show that the spatial average of
\Eref{entropyProductionDensity_distinguishable_non-interacting_example} 
vanishes for a Boltzmann density, $\HMdensity{N}{i}(\xvec_i) \propto \exp{-\extPot(\xvec_i)/\diffusion_i}$ so that the entropy production vanishes.

\Erefs{entropyProductionDensity_independent_distinguishable_final} and \eref{entropyProduction_independent_distinguishable_final} are the central results of this section. Adding interaction, as done in the next section,  results in more terms, but a remarkably similar structure.

\toasubsubsection{\texorpdfstring{$N$}{N} pairwise interacting, distinguishable particles}
\seclabel{N_interacting_distinguishable_particles}
In the presence of interaction, the $n$-point propagator acquires additional terms to order $t'-t$. Diagrammatically, such terms to be added to $\ave{\phi_1(\yvec_1,t')\ldots\phitilde_N(\xvec_N,t)}$, beyond those shown in \Eref{joint_propagator_plus_first_order}, are of the form \Eref{h_example},
\begin{align}\elabel{Npropagators_distinguishable_interaction_diagrams}
\ave{\phi_1(\yvec_1,t')\ldots\phitilde_N(\xvec_N,t)} 
\corresponds \ldots +
\tikz[baseline=-20pt]{
\begin{scope}[yshift=0.0cm]
\draw[potStyle] (0,-0.25) -- (0,0.25);
\draw[black,thin] (-0.1,0.1) -- (0.1,0.1);
\draw[red,thin] (-0.2,0.15) -- (-0.2,0.35);
\tgenVertex{0,-0.25}
\tgenVertex{0,0.25}
\node at (\propWidth,-0.3) [right,yshift=0pt] {$\xvec_2,t$};
\node at (-\propWidth,-0.3) [left,yshift=0pt] {$\yvec_2,t'$};
\draw[tAactivity] (\propWidth,-0.3) -- (0,-0.25) -- (-\propWidth,-0.3);
\node at (\propWidth,0.3) [right,yshift=0pt] {$\xvec_1,t$};
\node at (-\propWidth,0.3) [left,yshift=0pt] {$\yvec_1,t'$};
\draw[tAactivity] (\propWidth,0.3) -- (0,0.25) -- (-\propWidth,0.3);
\end{scope}
\begin{scope}[yshift=-0.8cm]
\node at (\propWidth,0) [right,yshift=0pt] {$\xvec_3,t$};
\node at (-\propWidth,0) [left,yshift=0pt] {$\yvec_3,t'$};
\draw[tAactivity] (\propWidth,0) -- (-\propWidth,0);
\node at (0,-0.45) {$\vdots$};
\node at (1,-0.45) [right] {$\vdots$};
\node at (-1,-0.45) [left] {$\vdots$};
\end{scope} 
\begin{scope}[yshift=-1.75cm]
\node at (\propWidth,0) [right,yshift=0pt] {$\xvec_N,t$};
\node at (-\propWidth,0) [left,yshift=0pt] {$\yvec_N,t'$};
\draw[tAactivity] (\propWidth,0) -- (-\propWidth,0);
\end{scope} 
}
+
\tikz[baseline=-20pt]{
\begin{scope}[yshift=0.0cm]
\draw[potStyle] (0,-0.25) -- (0,0.25);
\draw[black,thin] (-0.1,0.1) -- (0.1,0.1);
\draw[red,thin] (-0.2,0.15) -- (-0.2,0.35);
\tgenVertex{0,-0.25}
\tgenVertex{0,0.25}
\node at (\propWidth,-0.3) [right,yshift=0pt] {$\xvec_3,t$};
\node at (-\propWidth,-0.3) [left,yshift=0pt] {$\yvec_3,t'$};
\draw[tAactivity] (\propWidth,-0.3) -- (0,-0.25) -- (-\propWidth,-0.3);
\node at (\propWidth,0.3) [right,yshift=0pt] {$\xvec_1,t$};
\node at (-\propWidth,0.3) [left,yshift=0pt] {$\yvec_1,t'$};
\draw[tAactivity] (\propWidth,0.3) -- (0,0.25) -- (-\propWidth,0.3);
\end{scope}
\begin{scope}[yshift=-0.8cm]
\node at (\propWidth,0) [right,yshift=0pt] {$\xvec_2,t$};
\node at (-\propWidth,0) [left,yshift=0pt] {$\yvec_2,t'$};
\draw[tAactivity] (\propWidth,0) -- (-\propWidth,0);
\node at (0,-0.45) {$\vdots$};
\node at (1,-0.45) [right] {$\vdots$};
\node at (-1,-0.45) [left] {$\vdots$};
\end{scope} 
\begin{scope}[yshift=-1.75cm]
\node at (\propWidth,0) [right,yshift=0pt] {$\xvec_N,t$};
\node at (-\propWidth,0) [left,yshift=0pt] {$\yvec_N,t'$};
\draw[tAactivity] (\propWidth,0) -- (-\propWidth,0);
\end{scope} 
}
+
\tikz[baseline=-20pt]{
\begin{scope}[yshift=-0.5cm]
\draw[potStyle] (0,-0.25) -- (0,0.25);
\draw[black,thin] (-0.1,0.1) -- (0.1,0.1);
\draw[red,thin] (-0.2,0.15) -- (-0.2,0.35);
\tgenVertex{0,-0.25}
\tgenVertex{0,0.25}
\node at (\propWidth,-0.3) [right,yshift=0pt] {$\xvec_3,t$};
\node at (-\propWidth,-0.3) [left,yshift=0pt] {$\yvec_3,t'$};
\draw[tAactivity] (\propWidth,-0.3) -- (0,-0.25) -- (-\propWidth,-0.3);
\node at (\propWidth,0.3) [right,yshift=0pt] {$\xvec_2,t$};
\node at (-\propWidth,0.3) [left,yshift=0pt] {$\yvec_2,t'$};
\draw[tAactivity] (\propWidth,0.3) -- (0,0.25) -- (-\propWidth,0.3);
\node at (0,-0.75) {$\vdots$};
\node at (1,-0.75) [right] {$\vdots$};
\node at (-1,-0.75) [left] {$\vdots$};
\end{scope}
\begin{scope}[yshift=0.3cm]
\node at (\propWidth,0) [right,yshift=0pt] {$\xvec_1,t$};
\node at (-\propWidth,0) [left,yshift=0pt] {$\yvec_1,t'$};
\draw[tAactivity] (\propWidth,0) -- (-\propWidth,0);
\end{scope} 
\begin{scope}[yshift=-1.75cm]
\node at (\propWidth,0) [right,yshift=0pt] {$\xvec_N,t$};
\node at (-\propWidth,0) [left,yshift=0pt] {$\yvec_N,t'$};
\draw[tAactivity] (\propWidth,0) -- (-\propWidth,0);
\end{scope} 
}
+
\ldots
\end{align}
each particle potentially interacting with any other particle. The underlying vertex, \eref{h_example}, is not symmetric, as the force is exerted on a particle attached via a dashed, red leg by the particle attached via undashed legs. The force is mediated by the dash-dotted line and need not be symmetric, \ie might be non-reciprocal \cite{ZhangGarcia-Millan:2023}. There are therefore $N(N-1)$ such contributions.

With each of the interaction terms \Eref{Npropagators_distinguishable_interaction_diagrams} being of order $t'-t$, \Eref{h_example_compact}, they add to the propagator \Eref{Npropagators_distinguishable} in the form
\begin{equation}\elabel{Npropagators_distinguishable_with_interaction}
\ave{\phi_1(\yvec_1,t')\ldots\phitilde_N(\xvec_N,t)} 
= 
\prod^N_i g_i + \sum_i^N f_i \prod^N_{\substack{j\ne i}} g_j + \sum_i^N \sum_{j\ne i}^N  h_{ij} \prod_{k\notin\{i,j\}}   g_{k} + \order{(t'-t)^2} \ ,
\end{equation}
affecting both the kernel $\Op^{(N)}$, \Eref{Op_distinguishableN}, and the logarithm $\Ln^{(N)}$, \Eref{Ln_distinguishableN}, of the entropy production \Eref{entropy_production_distinguishableN}.

The effect of the interaction term $h_{ij}$ on the kernel is similar to that of $f_i$, \Eref{def_f_i}, as the time derivative of $h_{ij}$ in the limit $t'\downarrow t$ renders it a kernel on $\yvec_i,\xvec_i$. The coordinates $\yvec_j$ and $\xvec_j$ enter into the amplitude of the force, but otherwise, under the limit, enter merely into a $\delta$-function, so that, starting from \Eref{Op_distinguishable_independent}
\begin{equation}\elabel{Op_distinguishable_interacting}
  \Op^{(N)}_{\yvec_1,\ldots,\yvec_N,\xvec_1,\ldots,\xvec_N}
=
\sum_i^N (\gdot_i + \fdot_i) \prod^N_{\substack{j\ne i}} \delta_j
+
\sum_i^N \sum_{j\ne i}^N  \hdot_{ij} \prod_{k\notin\{i,j\}}^N  \delta_{k} \ .
\end{equation}
Because of the $\delta_j$-like nature of $\hdot_{ij}$, each term 
multiplied by it has $\yvec_j=\xvec_j$ enforced for all $j$ except $j=i$.

The logarithm $\Ln^{(N)}$ also acquires $N(N-1)$ new terms, conveniently written in the form
\begin{equation}\elabel{Ln_distinguishable_interacting_simplified}
\Ln^{(N)}_{\yvec_1,\ldots,\yvec_N,\xvec_1,\ldots,\xvec_N} = \lim_{t'\downarrow t}
\sum_i^N
\ln\left(
\frac
{g_i}
{\Bg_i}
\right)
+
\ln\left(
\frac
{1 + \sum_i^N   f_i/g_i  + \sum_i^N\sum_{j\ne i}^N h_{ij}/(g_i g_j)  }
{1 + \sum_i^N \Bf_i/\Bg_i+ \sum_i^N\sum_{j\ne i}^N \Bh_{ij}/(\Bg_i \Bg_j)  } 
\right) \ .
\end{equation}
following the steps from \Eref{Ln_distinguishable_independent} to \eref{Ln_distinguishable_independent_simplified}. Again, we expand the terms in the rightmost logarithm, so that
\begin{equation}\elabel{Ln_distinguishable_interacting_simplified2}
\Ln^{(N)}_{\yvec_1,\ldots,\yvec_N,\xvec_1,\ldots,\xvec_N} = \lim_{t'\downarrow t}
\sum_i^N
\ln\left(
\frac
{g_i}
{\Bg_i}
\right)
+
\sum_i^N \left[ \frac{f_i}{g_i} - \frac{\Bf_i}{\Bg_i} \right]
+
\sum_i^N \sum_{j\ne i}^N \left[ \frac{h_{ij}}{g_i g_j} - \frac{\Bh_{ij}}{\Bg_i\Bg_j} \right]\ ,
\end{equation}
producing an expression that is more efficiently analysed. Using \Eref{Op_distinguishable_interacting} for $\Op^{(N)}$
and \Eref{Ln_distinguishable_interacting_simplified2} for $\Ln^{(N)}$ in \Eref{def_entropyProductionDensity}, we have
\begin{align}
\entropyProductionDensity^{(N)}(\xvec_1,\ldots,\xvec_N) 
=& \int \ddint{y_{1,\ldots,N}}
\left\{
\sum_i^N (\gdot_i + \fdot_i) \prod^N_{\substack{j\ne i}} \delta_j
+
\sum_i^N \sum_{j\ne i}^N  \hdot_{ij} \!\!\prod_{k\notin\{i,j\}}\!\!\delta_{k} 
\right\}\nonumber\\
&\times
\lim_{t'\downarrow t}
\left\{
\sum_\ell^N
\ln\left(
\frac
{g_\ell}
{\Bg_\ell}
\right)
+
\sum_\ell^N \left[ \frac{f_\ell}{g_\ell} - \frac{\Bf_\ell}{\Bg_\ell} \right]
+
\sum_\ell^N \sum_{m\ne \ell}^N \left[ \frac{h_{\ell m}}{g_\ell g_m} - \frac{\Bh_{\ell m}}{\Bg_\ell\Bg_m} \right]
\right\} \ ,
\elabel{entropyProductionDensity_interacting_distinguishable_initial}
\end{align}
which contains all the terms of 
\Eref{entropyProductionDensity_independent_distinguishable_final_sum}, in addition to any terms involving $h_{ij}$, in particular
\begin{align}\nonumber
\entropyProductionDensity^{(N)}_{\text{term 1}}(\xvec_1,\ldots,\xvec_N)
= &
\int \ddint{y_{1,\ldots,N}}
\Bigg(
\sum_i^N \sum_{j\ne i}^N  \hdot_{ij} \prod_{k\notin\{i,j\}}  \delta_{k}
\Bigg)\\
&\times
\lim_{t'\downarrow t}
\Bigg\{
\sum_\ell^N
\ln\left(
\frac
{g_\ell}
{\Bg_\ell}
\right)
+
\sum_\ell^N \left[ \frac{f_\ell}{g_\ell} - \frac{\Bf_\ell}{\Bg_\ell} \right]
+
\sum_\ell^N \sum_{m\ne \ell}^N \left[ \frac{h_{\ell m}}{g_\ell g_m} - \frac{\Bh_{\ell m}}{\Bg_\ell\Bg_m} \right]
\Bigg\}
\elabel{entropyProductionDensity_interacting_distinguishable_term1}
\end{align}
and
\begin{equation}\elabel{entropyProductionDensity_interacting_distinguishable_term2}
\entropyProductionDensity^{(N)}_{\text{term 2}}(\xvec_1,\ldots,\xvec_N) 
= \int \ddint{y_{1,\ldots,N}}
\Bigg(
\sum_i^N (\gdot_i + \fdot_i) \prod^N_{\substack{j\ne i}} \delta_j
\Bigg)
\lim_{t'\downarrow t}
\left\{
\sum_\ell^N \sum_{m\ne \ell}^N \left[ \frac{h_{\ell m}}{g_\ell g_m} - \frac{\Bh_{\ell m}}{\Bg_\ell\Bg_m} \right]
\right\}\ ,
\end{equation}
so that with the simplifications from 
\Erefs{entropyProductionDensity_independent_distinguishable_final_sum} and \eref{entropyProductionDensity_independent_distinguishable_final}
\begin{equation}\elabel{entropyProduction_interacting_distinguishable_sum}
\entropyProductionDensity^{(N)}(\xvec_1,\ldots,\xvec_N)
=
\entropyProductionDensity^{(N)}_{\text{term 1}}(\xvec_1,\ldots,\xvec_N)
+
\entropyProductionDensity^{(N)}_{\text{term 2}}(\xvec_1,\ldots,\xvec_N)
+
\sum_i^N \entropyProductionDensity^{(1)}_i(\xvec_i)
\ .
\end{equation}
Analysing \Eref{entropyProductionDensity_interacting_distinguishable_term1} first, the term $\hdot_{ij}\prod_k\delta_k$ enforces $\yvec_\ell=\xvec_\ell$ for all $\ell$ except $\ell=i$, because $\hdot_{ij}$ is proportional to $\delta_j$. As a result, $\ln(g_\ell/\Bg_\ell)$ vanishes for all $\ell\ne i$, \Eref{gi_over_Bgi}, as do $f_\ell/g_\ell$ and $\Bf_\ell/\Bg_\ell$, \Eref{f_over_g_delta_limit}. 
For the same reason, the terms 
$h_{\ell m}/(g_\ell g_m)$ 
and
$\Bh_{\ell m}/(\Bg_\ell \Bg_m)$ 
vanish for $\ell\ne i$, \Eref{h_over_gg}.
The only terms in the curly brackets of \Eref{entropyProductionDensity_interacting_distinguishable_term1} that do not vanish by the $\hdot_{ij}\prod_k\delta_k$ pre-factor are therefore $\ell=i$.

The first set of terms \Eref{entropyProductionDensity_interacting_distinguishable_term1} thus simplify to
\begin{align}
\nonumber
\entropyProductionDensity^{(N)}_{\text{term 1}}(\xvec_1,\ldots,\xvec_N)
&=
\int \ddint{y_{1,\ldots,N}}
\sum_i^N \sum_{j\ne i}^N  \hdot_{ij} \prod_{k\notin\{i,j\}}  \delta_{k}
\lim_{t'\downarrow t}
\Bigg\{
\ln\left(
\frac
{g_i}
{\Bg_i}
\right)
+
\left[ \frac{f_i}{g_i} - \frac{\Bf_i}{\Bg_i} \right]
+
\sum_{m\ne i} \left[ \frac{h_{i m}}{g_i g_m} - \frac{\Bh_{i m}}{\Bg_i\Bg_m} \right]
\Bigg\}\\
&=
\sum_i^N \sum_{j\ne i}^N  
\int \ddint{y_{i,j}}
\hdot_{ij}
\left\{
\ln\left(
\frac
{g_i}
{\Bg_i}
\right)
+
\left[ \frac{f_i}{g_i} - \frac{\Bf_i}{\Bg_i} \right]
+
\left[ \frac{h_{i j}}{g_i g_j} - \frac{\Bh_{i j}}{\Bg_i\Bg_j} \right]
\right\}
\nonumber
\\& 
\elabel{entropyProductionDensity_interacting_distinguishable_term1_simplified2}
\qquad + 
\sum_i^N \sum_{j\ne i}^N \sum_{m\notin \{i,j\}} 
\int \ddint{y_{i,j,m}}
\hdot_{ij}\delta_m
\left[ \frac{h_{i m}}{g_i g_m} - \frac{\Bh_{i m}}{\Bg_i\Bg_m} \right]
\ ,
\end{align}
where the last term contributes only when particles are sufficiently densely packed on the scale of the potential range. This is because a term of the form
\[
\hdot_{ij} \left[ \frac{h_{i m}}{g_i g_m} - \frac{\Bh_{i m}}{\Bg_i\Bg_m} \right]
\]
needs $\xvec_j$ to be sufficiently close to $\xvec_i$ such that $\hdot_{ij}$ contributes, and $\xvec_m$ to be sufficiently close to $\xvec_i$ such that $h_{i m}$ and $\Bh_{i m}$ contribute. In other words, this term contributes only if three particles might be interacting simultaneously by pairwise interaction \cite{SuzukiETAL:2015,ChatterjeeGoldenfeld:2019,ZampetakiETAL:2021}, which may be exceedingly rare for short-ranged, repulsive potentials and low enough densities.

The second set of terms, \Eref{entropyProductionDensity_interacting_distinguishable_term2}, can be simplified similarly. Since $\yvec_\ell=\xvec_\ell$ for all $\ell\ne i$, all terms 
$h_{\ell m}/(g_\ell g_m)$ and $ \Bh_{\ell m}/(\Bg_\ell \Bg_m)$ 
vanish for all $\ell\ne i$,
\begin{align}
\entropyProductionDensity^{(N)}_{\text{term 2}}(\xvec_1,\ldots,\xvec_N) 
= & \int \ddint{y_{1,\ldots,N}}
\sum_i^N (\gdot_i + \fdot_i) \prod^N_{\substack{j\ne i}} \delta_j
\lim_{t'\downarrow t}
\left\{
\sum_{m\ne i}^N \left[ \frac{h_{i m}}{g_i g_m} - \frac{\Bh_{i m}}{\Bg_i\Bg_m} \right]
\right\} \nonumber\\
= & \sum_i^N \sum_{m\ne i}^N \int \ddint{y_{i,m}}
(\gdot_i + \fdot_i)\delta_m
\lim_{t'\downarrow t}
\left[ \frac{h_{i m}}{g_i g_m} - \frac{\Bh_{i m}}{\Bg_i\Bg_m} \right] \ .
\elabel{entropyProductionDensity_interacting_distinguishable_term2_simplified}
\end{align}
This term has a smaller contribution the quicker $h_{im}$ drops off when particle
$i$ and particle $m$ are far apart. This is because $(\gdot_i +
\fdot_i)\delta_m$ does not provide extra weight for $i$ and $m$ being
close to each other, which is the condition for 
$h_{im}$ and $\Bh_{im}$ in
the logarithmic term
$h_{im}/(g_ig_m)-\Bh_{im}/(\Bg_i\Bg_m)$ to contribute.

The local entropy production of interacting, distinguishable particles thus consists of three types of terms:
Firstly, $\entropyProductionDensity^{(1)}_i(\xvec_i)$, \Eref{entropyProductionDensity_independent_distinguishable_final}, collects all contributions due to $g_i$ and $f_i$ only, 
which are due to the free motion of particle $i$ and the effect
of any external potential on it.
Secondly, $\entropyProductionDensity^{(N)}_{ij}(\xvec_i,\xvec_j)=\entropyProductionDensity^{(2)}_{ij}(\xvec_i,\xvec_j)$, 
which contains those terms of $\entropyProductionDensity^{(N)}_{\text{term 1}}$ and  $\entropyProductionDensity^{(N)}_{\text{term 2}}$, that depend on the coordinates, $\xvec_i$ and $\xvec_j$ of only \emph{two distinct} particles $i$ and $j$, \Eref{entropyProductionDensity_interacting_distinguishable_term1_simplified2}
and \Eref{entropyProductionDensity_interacting_distinguishable_term2_simplified},
\begin{align}\elabel{entropyProductionDensity_interacting_distinguishable_ij}
\entropyProductionDensity^{(2)}_{ij}(\xvec_i,\xvec_j)=
\int \ddint{y_{i,j}}
\left(
\hdot_{ij}
\lim_{t'\downarrow t}
\left\{
\ln\left(
\frac
{g_i}
{\Bg_i}
\right)
+
\left[ \frac{f_i}{g_i} - \frac{\Bf_i}{\Bg_i} \right]
+
\left[ \frac{h_{i j}}{g_i g_j} - \frac{\Bh_{i j}}{\Bg_i\Bg_j} \right]
\right\}
+
(\gdot_i + \fdot_i)\delta_j
\lim_{t'\downarrow t}
\left\{ \frac{h_{i j}}{g_i g_j} - \frac{\Bh_{i j}}{\Bg_i\Bg_j} \right\}\right) \ ,
\end{align}
where the superscript of $\entropyProductionDensity^{(2)}_{ij}$ indicates the number of particles involved.
It gives the entropy
produced by particle $i$ due to its interaction with particle $j$.
Thirdly, a term that depends on three coordinates, the last term of \Eref{entropyProductionDensity_interacting_distinguishable_term1_simplified2} for one triplet $i,j,k$ of distinct particles, that contributes only for particle systems so dense that more than two particles might be interacting at once,
\begin{equation}\elabel{entropyProductionDensity_interacting_distinguishable_ijk}
\entropyProductionDensity^{(N)}_{ijk}(\xvec_i,\xvec_j,\xvec_k)=
\entropyProductionDensity^{(3)}_{ijk}(\xvec_i,\xvec_j,\xvec_k)=
\int \ddint{y_{i,j,k}}
\hdot_{ij}\delta_k
\lim_{t'\downarrow t}
\left[ \frac{h_{i k}}{g_i g_k} - \frac{\Bh_{i k}}{\Bg_i\Bg_k} \right] \ .
\end{equation}
This term gives the entropy produced by particle $i$ due to its 
interaction with particles $j$ and $k$ simultaneously. 
In general, the local entropy productions are not invariant under index permutations, as the order of indices determines the specific r{\^o}le each particle plays.
We calculate $\entropyProductionDensity^{(1)}_i$, $\entropyProductionDensity^{(2)}_{ij}$ and $\entropyProductionDensity^{(3)}_{ijk}$ explicitly in the
 examples studied in \APref{generalised_trawlers} and \ref{sec:N_interacting_indistinguishable_particle_extPot}.

By construction, an $n$-point vertex will result in a local entropy production depending on up to $2n-1$ locations, namely one location of the particle experiencing the displacement, $n-1$ locations of other particles interacting with it in the kernel and another $n-1$ of particles interacting with it in the logarithm. Under the assumption of short-rangedness, terms depending on more than $n$ locations may be neglected, by assuming that if $n$ particles happen to be close enough to interact, the probability of finding more than $n$ will be exceedingly low. We are \emph{not} making any such assumption in the present work.
The \emph{different} notion of sparseness in \APref{N_indistinguishable_particles} is introduced only to facilitate the discussion.

With these local entropy productions of $N$ pair-interacting, distinguishable particles in place, the overall entropy production at stationarity is
\begin{align}
\entropyProduction^{(N)}[\HMdensity{N}{}] = &
\sum_i^N \int \ddint{x_i} \HMdensity{N}{i}(\xvec_i) \entropyProductionDensity^{(1)}_i(\xvec_i)
+
\sum_i^N \sum_{j\ne i}^N \int \ddint{x_i} \ddint{x_j} \HMdensity{N}{ij}(\xvec_i,\xvec_j) \entropyProductionDensity^{(2)}_{ij}(\xvec_i,\xvec_j) \nonumber\\
&+
\sum_i^N \sum_{j\ne i}^N \sum_{k\notin \{i,j\}} \int \ddint{x_i} \ddint{x_j} \ddint{x_k} \HMdensity{N}{ijk}(\xvec_i,\xvec_j,\xvec_k) \entropyProductionDensity^{(3)}_{ijk}(\xvec_i,\xvec_j,\xvec_k)\ ,
\elabel{entropyProduction_interacting_distinguishable_final}
\end{align}
where we have introduced various marginalisations of the density, similar to \Eref{marginalisation_densityN_i_distinguishable}, 
\begin{subequations}
\elabel{marginalisation_densityN_i2orMore_distinguishable}
\begin{align}
\HMdensity{N}{ij}(\xvec_i,\xvec_j)          &= \int \!\prod_{\ell\notin\{i,j\}}^N\!
\ddint{x_\ell}\,
\HMdensity{N}{}(\xvec_1,\xvec_2,\ldots,\xvec_N)\\
\HMdensity{N}{ijk}(\xvec_i,\xvec_j,\xvec_k) &= \int \!\prod_{\ell\notin\{i,j,k\}}^N\!
\ddint{x_\ell}\,
\HMdensity{N}{}(\xvec_1,\xvec_2,\ldots,\xvec_N)
\ .
\end{align}
\end{subequations}
The two-point density $\HMdensity{N}{ij}(\xvec_i,\xvec_j)$ is the joint density of particle species $i$ at $\xvec_i$ and species $j$ at $\xvec_j$, and similarly for the three-point density $\HMdensity{N}{ijk}(\xvec_i,\xvec_j,\xvec_k)$.
These densities
are invariant under permutations of the indices, say
\begin{equation}
    \HMdensity{N}{ij}(\xvec_i,\xvec_j)=\HMdensity{N}{ji}(\xvec_j,\xvec_i)
    \quad\text{and}\quad
    \HMdensity{N}{ijk}(\xvec_i,\xvec_j,\xvec_k)=\HMdensity{N}{kij}(\xvec_k,\xvec_i,\xvec_j) \ .
\end{equation}

Because the local entropy production \Erefs{entropyProductionDensity_independent_distinguishable_final}, \eref{entropyProductionDensity_interacting_distinguishable_ij} and \eref{entropyProductionDensity_interacting_distinguishable_ijk} depends only on a very reduced set of coordinates, the others can be integrated out.
These marginalisations, \Erefs{marginalisation_densityN_i_distinguishable} and \eref{marginalisation_densityN_i2orMore_distinguishable}, are what makes the calculation of the entropy production feasible in practice. 
Rather than having to determine the full $N$-point density in, say, \Eref{entropy_production_distinguishableN}, \Eref{entropyProduction_interacting_distinguishable_final} reduces what needs to be measured to more easily accessible quantities. In the following, we rewrite it in a form that lends itself more naturally to experimental data.
Denoting by $\xvec^{(q)}_i$ the particle locations of species $i$ in measurement $q$ of $Q$ measurements, with the help of \Eref{entropyProduction_interacting_distinguishable_final} the entropy production may then be estimated by
\begin{equation}\elabel{entropyProduction_interacting_distinguishable_experimental}
\entropyProduction^{(N)} =
\frac{1}{Q} \sum_{q}^{Q} \left\{
\sum_i^N \entropyProductionDensity^{(1)}_i(\xvec^{(q)}_i)
+
\sum_i^N \sum_{j\ne i}^N \entropyProductionDensity^{(2)}_{ij}(\xvec^{(q)}_i,\xvec^{(q)}_j)
+
\sum_i^N \sum_{j\ne i}^N \sum_{k\notin \{i,j\}} \entropyProductionDensity^{(3)}_{ijk}(\xvec^{(q)}_i,\xvec^{(q)}_j,\xvec^{(q)}_k)
\right\} \ ,
\end{equation}
replacing, for example, $\HMdensity{N}{2}(\xvec_1,\xvec_2)$ by the experimental estimate $(1/Q)\sum_q^Q 
\sum_{\substack{i_1,i_2=1\\ i_1\ne i_2}}^N 
\delta(\xvec_1-\xvec^{(q)}_{i_1}) \delta(\xvec_2-\xvec^{(q)}_{i_2})$. 

For drift-diffusive particles in pair and external potentials, the expressions derived in this section are \emph{exact}. The qualification to drift-diffusion and potentials is necessary, only in so far as asumptions have been made about the properties of $g_i$, $f_i$ and $h_{ij}$
under various limits, such as \Erefs{g_i_limit}, \eref{f_vanishes}, \eref{f_over_g_delta_limit} and \eref{hderi_delta}.

This concludes the derivations for distinguishable particles, with the crucial results \Eref{entropyProduction_independent_distinguishable_final} drawing on \Eref{entropyProductionDensity_independent_distinguishable_final}, and \Eref{entropyProduction_interacting_distinguishable_final} drawing on \Erefs{entropyProductionDensity_independent_distinguishable_final}, \eref{entropyProductionDensity_interacting_distinguishable_ij} and \eref{entropyProductionDensity_interacting_distinguishable_ijk}. In the next example, 
we re-derive the results of \SMref{HarmonicTrawlers} using the present, general framework, and in \APref{N_indistinguishable_particles}, we extend this framework to indistinguishable particles.

\toasubsubsection{Example: Entropy production of two pair-interacting distinguishable drift-diffusion particles without external potential}
\seclabel{generalised_trawlers}
To illustrate the framework outlined in \APref{N_interacting_distinguishable_particles}, we use the example of two pair-interacting drift-diffusion particles on a circle of circumference $L$, which is calculated ``from first principles'' in \SMref{HarmonicTrawlers}, where it is found that the entropy production is $\entropyProduction=(\drift_1+\drift_2)^2/(2\diffusion)$, if the particles drift with velocities $\drift_1$ and $\drift_2$ respectively and both diffuse with diffusion contant $\diffusion$. Even when calculated in \SMref{HarmonicTrawlers} using an attractive harmonic potental, this result ought to be independent of the details of the  pair potential $\pairPot$, given the simple physical reasoning in \SMref{HarmonicTrawlers_simple_physics}, as we confirm in the following.

To calculate the entropy production on the basis of \Eref{entropyProduction_interacting_distinguishable_final}, we need the local entropy productions,
$\entropyProductionDensity^{(1)}_i(x_i)$, \Eref{entropyProductionDensity_independent_distinguishable_final},
and
$\entropyProductionDensity^{(2)}_{ij}(x_i,x_j)$, \Eref{entropyProductionDensity_interacting_distinguishable_ij},
but in the absence of a third particle, not $\entropyProductionDensity^{(3)}_{ijk}$. We further need
the obviously uniform one-point densities
\begin{align}
\HMdensity{2}{1}(x_1)=\HMdensity{2}{2}(x_2)=1/L
\end{align}
at stationarity, \SMref{HarmonicTrawlers}, and the two-point densities $\HMdensity{2}{12}$ and $\HMdensity{2}{21}$. As it turns out, these do not need to be known explicitly in terms of the interaction potential $\pairPot(x_1-x_2)$. Firstly, by translational invariance, the two-point densities factorise into a uniform distribution and a distribution of the distance $r=x_1-x_2$,
\begin{equation}\elabel{density_12_trawler_example}
\HMdensity{2}{12}(x_1,x_2)=
\HMdensity{2}{21}(x_2,x_1)=
\frac{1}{L}
\density_r(x_1-x_2) \ .
\end{equation}
Secondly, assuming Newton's third law, so that the force acting on one particle $\pairPot'_{1\,2}(x_1-x_2)=\pairPot'(x_1-x_2)$ is the negative of the force acting on the other particle, $\pairPot'_{2\,1}(x_2-x_1)=-\pairPot'(x_1-x_2)$ 
\cite{LoosKlapp:2020,Garcia-MillanZhang:2023}
and further assuming that the potential is even so that $\pairPot'$ is odd, we can write 
the equations of motion
\begin{subequations}\elabel{Langevin_x12}
\begin{align}
\dot{x}_1 &= \drift_1 - \pairPot'(x_1-x_2) + \xi_1(t) \\
\dot{x}_2 &= \drift_2 - \pairPot'(x_2-x_1) + \xi_2(t)
\end{align}
\end{subequations}
where $\pairPot(r)=kr^2/2$ in \SMref{HarmonicTrawlers}, but shall be left unspecified here. We can then, thirdly, determine the equation of motion of the distance $r$ because the right hand sides of \Eref{Langevin_x12} are solely a function of $r$,
\begin{equation}
    \dot{r}=\big(\drift_1-\drift_2-2\pairPot'(r)\big) + \xi_1(t)-\xi_2(t)\ ,
\end{equation}
so that $r$ diffuses with diffusion constant $2\diffusion$ and drifts with velocity $\drift_1-\drift_2-2\pairPot'(r)$, giving rise to a Fokker-Planck equation of the density $\density_r(r,t)$ of $r$,
\begin{equation}
    \dot{\density}_r = -\partial_r 
    \bigg(\big(\drift_1-\drift_2-2\pairPot'(r)\big)\density_r\bigg) + 2\diffusion\partial_r^2\density_r \ ,
\end{equation}
which determines the probability current $j_r$ via $\dot{\density}_r=-\diffusion\partial_r j_r$ up to a constant.
A simplifying assumption that allows simple physical reasoning to reproduce the results below, \SMref{HarmonicTrawlers_simple_physics}, is that one particle ends up towing the other, implying that the particle distance $r$ does not increase indefinitely. We thus demand
that $j_r$ vanishes at stationarity,
\begin{equation}\elabel{j_r_vanishes}
    0=-j_r=2\density_r'(r)+\frac{1}{\diffusion}\big(2\pairPot'(r)-(\drift_1-\drift_2) \big)\density_r(r) \ ,
\end{equation}
which, in the presence of drift $\drift_1-\drift_2$ implies that the potential $\pairPot(r)$ is binding.
The differential \Eref{j_r_vanishes} is all we  need to know about $\density_r(r)$ in the following.

To calculate the local entropy production $\entropyProductionDensity^{(1)}_i(x_i)$ on the basis of \Eref{entropyProductionDensity_independent_distinguishable_final}, we require $f_i$ and $g_i$. 
Without an external potential and with the drift being dealt with non-perturbatively, $f_i$ vanishes and $g_i$ is given by \Eref{example_g_i}, so that, from \Erefs{gi_over_Bgi} and \eref{gdot_example},
\begin{subequations}
\begin{align}
\gdot_i &= \diffusion \delta_i'' - \drift_i \delta_i' \ ,\\
\ln\left(\frac{g_i}{\Bg_i}\right)&=\frac{(y_i-x_i)\drift_i}{\diffusion} \ ,
\end{align}
\end{subequations}
where dashed $\delta$-functions are differentiated with respect to their argument, $\delta_i=\delta(y_i-x_i)$,
and therefore
\begin{equation}\elabel{entropyProductionDensity_1_example_final}
\entropyProductionDensity^{(1)}_i(x_i) = 
\int \dint{y_i} \left( \diffusion \delta_i'' - \drift_i \delta_i' \right)
\frac{(y_i-x_i)\drift_i}{\diffusion}
=
\frac{\drift_i^2}{\diffusion} \ .
\end{equation}

The interaction term is equally easily determined, \Erefs{h_example_compact} and \eref{hdot_example} give
\begin{equation}
\hdot_{ij}=\pairPot'(x_i-x_j) \delta_i' \delta_j 
\elabel{hdot_example_trawler_final}
\end{equation}
and by \Eref{trick3} 
\begin{equation}
h_{ij}=-\frac{y_i-x_i}{2\diffusion} g_i g_j \pairPot'(x_i-x_j) + \hot \ .
\elabel{h_example_trawler_final}
\end{equation}
Using \Erefs{hdot_example_trawler_final} and \eref{h_example_trawler_final} in the local entropy production in \Eref{entropyProductionDensity_interacting_distinguishable_ij},
\begin{align}
\entropyProductionDensity^{(2)}_{ij}(x_i,x_j)&=
\int \dint{y_{i,j}}
\left(
\pairPot'(x_i-x_j) \delta_i' \delta_j
\Bigg\{
\frac{(y_i-x_i)\drift_i}{\diffusion}
 +
\left[ -\frac{y_i-x_i}{2\diffusion} \pairPot'(x_i-x_j) + \frac{x_i-y_i}{2\diffusion} \pairPot'(y_i-y_j) \right]
\right\}\nonumber\\
&\nonumber\qquad 
+
(\diffusion \delta_i'' - \drift_i \delta_i')\delta_j
\left[ -\frac{y_i-x_i}{2\diffusion} \pairPot'(x_i-x_j) + \frac{x_i-y_i}{2\diffusion} \pairPot'(y_i-y_j) \right]\Bigg)\\
&=
-\frac{\pairPot'(x_i-x_j)}{\diffusion}\big(2\drift_i-\pairPot'(x_i-x_j)\big)
-\pairPot''(x_i-x_j) \ .
\elabel{entropyProductionDensity_2_example_final}
\end{align}

We proceed to calculate the entropy production by using $\entropyProductionDensity^{(1)}_i(x_i)$, \Eref{entropyProductionDensity_1_example_final},
and 
$\entropyProductionDensity^{(2)}_{ij}(x_i,x_j)$, \Eref{entropyProductionDensity_2_example_final}, 
in
\Eref{entropyProduction_interacting_distinguishable_final},
\begin{align}
\entropyProduction^{(2)}[\HMdensity{2}{}] =&
\int_0^L \dint{x_1} 
\HMdensity{2}{1}(x_1) \entropyProductionDensity^{(1)}_1(x_1)
+
\int_0^L \dint{x_2} 
\HMdensity{2}{2}(x_2) \entropyProductionDensity^{(1)}_2(x_2)
\nonumber\\&
+
\int_0^L \dint{x_1}\dint{x_2}
\left\{
\HMdensity{2}{12}(x_1,x_2) \entropyProductionDensity^{(2)}_{12}(x_1,x_2)
+
\HMdensity{2}{21}(x_2,x_1) \entropyProductionDensity^{(2)}_{21}(x_2,x_1)
\right\}
\elabel{entropyProduction_example_distinguishable_first_step}
\end{align}
Given that $\HMdensity{2}{1}(x_1)=\HMdensity{2}{2}(x_2)=1/L$ is constant, the first two integrals give simply 
\begin{equation}\elabel{entropyProduction_example_distinguishable_first_step_first_integral}
\int_0^L \dint{x_1}
\HMdensity{2}{1}(x_1) \entropyProductionDensity^{(1)}_1(x_1)
+
\int_0^L \dint{x_2}
\HMdensity{2}{2}(x_2) \entropyProductionDensity^{(1)}_2(x_2)
=
\frac{\drift_1^2}{\diffusion}+\frac{\drift_2^2}{\diffusion}
\ ,
\end{equation}
which is the entropy production of two independent drift-diffusion particles.
The remaining double integrals are 
\begin{align}
&\int_0^L \dint{x_1}\dint{x_2}
\left\{
\HMdensity{2}{12}(x_1,x_2) \entropyProductionDensity^{(2)}_{12}(x_1,x_2)
+
\HMdensity{2}{21}(x_2,x_1) \entropyProductionDensity^{(2)}_{21}(x_2,x_1)
\right\}\nonumber\\
&=
- 2
\int_0^L \dint{x_1}\dint{x_2}
\HMdensity{2}{12}(x_1,x_2) 
\left\{
 \frac{\pairPot'(x_1-x_2)}{\diffusion}\big(\drift_1-\drift_2-\pairPot'(x_1-x_2)\big)
+\pairPot''(x_1-x_2)\right\}
\end{align}
where we have used that $\pairPot'$ is odd and $\pairPot''$ is even. Inserting \Eref{density_12_trawler_example} and using $\density_r (\pairPot'^2/\diffusion - (\drift_1-\drift_2)\pairPot'/\diffusion - \pairPot'') = \diffusion \density''_r - (\drift_1-\drift_2)^2\density_r/(4\diffusion)$ from \Eref{j_r_vanishes} finally gives
\begin{equation}\elabel{entropyProduction_example_distinguishable_second_integral}
\int_0^L \dint{x_1}\dint{x_2}
\left\{
\HMdensity{2}{12}(x_1,x_2) \entropyProductionDensity^{(2)}_{12}(x_1,x_2)
+
\HMdensity{2}{21}(x_2,x_1) \entropyProductionDensity^{(2)}_{21}(x_2,x_1)
\right\}
=
- \frac{(\drift_1-\drift_2)^2}{2\diffusion} \ .
\end{equation}
The sum of \Erefs{entropyProduction_example_distinguishable_first_step_first_integral} and \eref{entropyProduction_example_distinguishable_second_integral} gives the total entropy production \Eref{entropyProduction_example_distinguishable_first_step},
\begin{equation}\elabel{trawler_final}
\entropyProduction^{(2)}[\HMdensity{2}{}] =
\frac{\drift_1^2}{\diffusion}+\frac{\drift_2^2}{\diffusion}
- \frac{(\drift_1-\drift_2)^2}{2\diffusion}
=
\frac{(\drift_1+\drift_2)^2}{2\diffusion} \ ,
\end{equation}
where the two-particle contributions $\entropyProductionDensity^{(2)}_{12}$ and $\entropyProductionDensity^{(2)}_{21}$ cancel some of the entropy generated by the free case.
\Eref{trawler_final} is
indeed identical to the result \Eref{N2_final_entropy_production} in \SMref{HarmonicTrawlers}. As opposed to the calculation there, the present result holds for all even, reciprocal \cite{LoosKlapp:2020,Garcia-MillanZhang:2023} interaction potentials as outlined before \Erefs{Langevin_x12}. As particles drag each other by attraction or push each other by repulsion, provided only the potential 
prevents a current in $r=x_1-x_2$ \Eref{j_r_vanishes}, the entropy production is independent of its details.

\toasubsection{\texorpdfstring{$N$}{N} indistinguishable particles}
\seclabel{N_indistinguishable_particles}
Assuming that no particle position is occupied more than once, the 
integral over the 
phase space occupied by $N$ indistinguishable
particles is correctly captured by the $N$-fold integral over the
particle coordinates, as if the particles were distinguishable, but
dividing by $N!$ to compensate for the $N!$-fold degeneracy and thus overcounting of
equivalent states. 
Using the Gibbs factor $1/N!$ to account for indistinguishability is 
obviously
allowable whenever multiple occupation of the same position has vanishing measure, an assumption which we refer to as \emph{sparse occupation}. Sparse occupation re-establishes distinguishability at equal times, so that indistinguishability needs to be accounted for only in transitions: Observing two particles at $\yvec_1,\yvec_2$ at one time and $\xvec_1,\xvec_2$ a moment later allows for the transitions 
$(\yvec_1,\yvec_2)\to(\xvec_1,\xvec_2)$ or $(\yvec_1,\yvec_2)\to(\xvec_2,\xvec_1)$.
Obviously, observables, such as the entropy production, must reflect that
particles are indistinguishable and thus must be invariant under
permutations of coordinates, 
but this apparent simplification is difficult to implement.

The difference between distinguishable and indistinguishable particles
becomes apparent already at the level of the $N$-point density
$\HMdensity{N}{N}(\xvec_1,\xvec_2,\ldots,\xvec_N)$,
which for indistinguishable particles is invariant under permutations of
the arguments and is normalised differently. While the $N$-point density
$\HMdensity{N}{}(\xvec_1,\xvec_2,\ldots,\xvec_N)$
of distinguishable particles equals the \emph{probability density} to find the different particles $i$ at their respective locations
$\xvec_i$, for indistinguishable particles it is the
\emph{number density of
any particle} at $\xvec_1$, \emph{any other} particle at $\xvec_2$
and so on. As long as all coordinates are distinct, there is no need to
re-introduce distinguishability in order to satisfy the requirement of
locating \emph{another} particle.
Further, when coordinates are distinct and can be ordered in a unique way, the particle number density can be interpreted as a probability density in the ordered probability space.

While the sparse occupation assumption simplifies the following derivation as it allows the interpretation of the particle number density as a probability density in the form needed, Doi-Peliti field theory does not require this assumption. In Doi-Peliti field theory the Gibbs factor alone renders all number densities effectively probability densities. This is because this field theory is based on factorial moments, which produce exactly the right correction factors to undo the overcompensation by the Gibbs-factor in the case of multiple occupation.

More generally, a joint propagator can be interpreted as the expectation of the product of indicator functions:
For distinguishable particles, the joint propagator is the expectation of an indicator function and therefore a probability density. 
For indistinguishable particles, the joint propagator is the expectation of an indicator function multiplied by the correct combination of factorials, 
as to compensate for the overcompensation done by the Gibbs factor.
These factorials are produced by the multiplicity of diagrams or, equivalently, by the factorial moments generated by the annihilator operators.

Similar to \Eref{density_is_propagator}, 
the density $\HMdensity{N}{}$ can be determined elegantly on the basis of the field theory with one pair of fields, $\phi$ and $\phidagger$. At stationarity, we introduce
\begin{equation}\elabel{density_indistinguishable_from_propagator}
\HMdensity{N}{n}(\xvec_1,\ldots,\xvec_n) 
=
\lim_{t_{01},\ldots,t_{0N}\to-\infty}
\ave{\phi(\xvec_1,t)\ldots\phi(\xvec_n,t)\phidagger(\xvec_{01},t_{01})\ldots\phidagger(\xvec_{0N},t_{0N})} \ ,
\end{equation}
as the $n$-point 
number density of $N$ particles of the same species.
Fixing $n-1$ particle coordinates and considering only the dependence of
the $n$-point density on $\xvec_n$, the latter might ``encounter'' any
of the ``undetermined, other'' $N-(n-1)$ particles in an integral, so that integrating over $\xvec_n$ produces 
\begin{equation}\elabel{marginalisation_densityN}
\int \ddint{x_n} \HMdensity{N}{n}(\xvec_1,\ldots,\xvec_n) 
=
\big( N-(n-1) \big)
\HMdensity{N}{n-1}(\xvec_1,\ldots,\xvec_{n-1}) \ .
\end{equation}
This marginalisation property applies even when the system is ``not ergodic'' with particles being trapped or not equilibrated, in which case both right and left-hand side of \Eref{marginalisation_densityN} depend also on the initial positions of all particles as well as time. 
\Eref{marginalisation_densityN}
is owed to the density accounting for
\emph{distinct particles}, 
implemented in
\Eref{density_indistinguishable_from_propagator}, which is constructed using $n$ annihilator operators. Each of those 
contribute with a local particle number count and then remove (annihilate) one particle locally, so that it cannot contribute towards further counts. 

Using \Eref{marginalisation_densityN} repeatedly gives
\begin{equation}\elabel{marginalisation_densityN_i_indistinguishable}
\HMdensity{N}{1}(\xvec_1) =
\frac{1}{(N-1)!} 
\int \ddint{x_{N,N-1,\ldots,2}}
\HMdensity{N}{N}(\xvec_1,\xvec_2,\ldots,\xvec_N) \ ,
\end{equation}
and generally
\begin{equation}
    \elabel{def_rho_N_n_indistinguishable}
    \HMdensity{N}{n}(\xvec_1,\ldots,\xvec_n)=\frac{1}{(N-n)!}
    \int\ddint{x_{N,N-1,\ldots,n+1}}
    \HMdensity{N}{N}(\xvec_1,\xvec_2,\ldots,\xvec_N)
\end{equation}
to be contrasted with the one-point
density of distinguishable particles,
\Eref{marginalisation_densityN_i_distinguishable}.

For the special case of $n=1$ in \Eref{marginalisation_densityN} we may define
$\HMdensity{N}{0}(\emptyset)=1$, so that
\begin{equation}\elabel{marginalisation_densityN_consequences}
\int \ddint{x} \HMdensity{N}{1}(\xvec) = N
\qquad\text{and}\qquad
\int \ddint{x_{N,\ldots,1}}
\HMdensity{N}{N}(\xvec_1,\ldots,\xvec_N) = N! \ .
\end{equation}
The integral over the phase space of occupation numbers then
suggestively produces
\begin{equation}
\frac{1}{N!} \int \ddint{x_{N,\ldots,1}}
\HMdensity{N}{N}(\xvec_1,\ldots,\xvec_N) = 1 \ .
\end{equation}

Propagators in a Doi-Peliti field theory,
designed for occupation-number
states,
naturally implement
indistinguishability.
Unless different species are specified in the form of different fields,
an expression such as \Eref{density_indistinguishable_from_propagator} produces diagrams of all possible
permutations of incoming and outgoing coordinates by virtue of Wick's
theorem. The propagators used in \Erefs{def_Op} and \eref{def_Ln} are
therefore naturally the transition probability densities of occupation
number states and the expression for the entropy production rate only needs to
account for the phase space being that of indistinguishable particles,
\begin{align}
    \entropyProduction^{(N)}[\HMdensity{N}{N}] =
    \frac{1}{(N!)^2}
    \int&\ddint{x_{1,\ldots,N}}\ddint{y_{1,\ldots,N}}
      \HMdensity{N}{N}(\xvec_1,\ldots,\xvec_N)
      \Op^{(N)}_{\yvec_1,\ldots,\yvec_N,\xvec_1,\ldots,\xvec_N} \nonumber\\
      &\times \left\{
      \Ln^{(N)}_{\yvec_1,\ldots,\yvec_N,\xvec_1,\ldots,\xvec_N} +
      \ln\left(
      \frac{\HMdensity{N}{N}(\xvec_1,\ldots,\xvec_N)}{\HMdensity{N}{N}(\yvec_1,\ldots,\yvec_N)}
      \right)
      \right\}
\elabel{entropy_production_indistinguishableN}
\end{align}
with $\Op^{(N)}$ and $\Ln^{(N)}$ given by the expressions for
indistinguishable particles corresponding to \Erefs{Op_distinguishableN}
and \eref{Ln_distinguishableN} respectively,
\begin{equation}\elabel{Op_indistinguishableN}
\Op^{(N)}_{\yvec_1,\ldots,\yvec_N,\xvec_1,\ldots,\xvec_N} = \lim_{t'\downarrow t} 
\ddtDpt
\ave{\phi(\yvec_1,t')\ldots\phi(\yvec_N,t')\phitilde(\xvec_1,t)\ldots\phitilde(\xvec_N,t)}
\end{equation}
and
\begin{equation}\elabel{Ln_indistinguishableN}
\Ln^{(N)}_{\yvec_1,\ldots,\yvec_N,\xvec_1,\ldots,\xvec_N} = \lim_{t'\downarrow t}
\ln\left(\frac{\ave{\phi(\yvec_1,t')\ldots\phi(\yvec_N,t')\phitilde(\xvec_1,t)\ldots\phitilde(\xvec_N,t)}}{\ave{\phi(\xvec_1,t')\ldots\phi(\xvec_N,t')\phitilde(\yvec_1,t)\ldots\phitilde(\yvec_N,t)}}\right)
\ .
\end{equation}
The joint propagator used here accounts for strictly \emph{distinct} particles as we characterise the transition probabilities of \emph{all} particles.

In keeping with \Erefs{def_entropyProductionDensity} and \eref{entropyProduction_as_spave}, we can write \Eref{entropy_production_indistinguishableN}
at stationarity as a weighted average,
\begin{equation}\elabel{entropy_production_indistinguishableN_as_density}
    \entropyProduction^{(N)}[\HMdensity{N}{N}] =
    \frac{1}{N!}
    \int \ddint{x_{1,\ldots,N}}
      \HMdensity{N}{N}(\xvec_1,\ldots,\xvec_N)
      \entropyProductionDensity^{(N)}(\xvec_1,\ldots,\xvec_N)
\end{equation}
with the local entropy production $\entropyProductionDensity^{(N)}$ for independent particles at stationarity defined as
\begin{equation}\elabel{def_entropyProductionDensity_indistinguishableN}
      \entropyProductionDensity^{(N)}(\xvec_1,\ldots,\xvec_N)=
      \frac{1}{N!}
      \int \ddint{y_{1,\ldots,N}}
      \Op^{(N)}_{\yvec_1,\ldots,\yvec_N,\xvec_1,\ldots,\xvec_N} 
      \Ln^{(N)}_{\yvec_1,\ldots,\yvec_N,\xvec_1,\ldots,\xvec_N} \ .
\end{equation}

To ease notation, in the following we use the notation of $g_i$, $f_i$, \etc, as introduced
in \Eref{def_g_i},
for example
\begin{equation}\elabel{g_the_same}
\ave[0]{\phi(\yvec_i,t')\phitilde(\xvec_i,t)} 
= g_i = g(\yvec_i;\xvec_i;t'-t)\ ,
\end{equation}
adopted for indistinguishable particles by dropping the index from the fields. However, all $g_i$ now are the \emph{same} function evaluated for different variables, namely $\yvec_i$ and $\xvec_i$, as suggested by the final $g(\yvec_i;\xvec_i;t'-t)$ in \Eref{g_the_same} not carrying an index $i$. The same applies to $f_i$, $h_{ij}$ and the corresponding functions with inverted arguments $\yvec_i$ and $\xvec_i$, for example
\begin{equation}\elabel{f_the_same}
\Bf_i = f(\xvec_i;\yvec_i;t'-t) \ .
\end{equation}

\toasubsubsection{\texorpdfstring{$N$}{N} independent, indistinguishable particles}
\seclabel{N_independent_indistinguishable_particles}
The $N$-particle joint propagator 
$\bigl\langle\phi(\yvec_1,t')\linebreak[1]\phi(\yvec_2,t')\linebreak[1]\ldots\linebreak[1]\phi(\yvec_N,t')\linebreak[1]\phitilde(\xvec_1,t)\linebreak[1]\phitilde(\xvec_2,t)\linebreak[1]\ldots\linebreak[1]\phitilde(\xvec_N,t)\bigr\rangle$
immediately factorises in the absence of interactions. However, rather
than resulting in a single product of $N$ propagators like
\Eref{propagator_factorising}, it is the sum of the $N!$ distinct
products of propagators, each accounting for a particular pairing of
Doi-shifted creator and annihilator fields, \APref{branching_vertices},
\begin{align}
\bigl\langle\phi(\yvec_1,t') & \phi(\yvec_2,t')\ldots\phi(\yvec_N,t')\phitilde(\xvec_1,t)\phitilde(\xvec_2,t)\ldots\phitilde(\xvec_N,t)\bigr\rangle \nonumber\\
= &
\ave{\phi(\yvec_1,t')\phitilde(\xvec_1,t)}
\ave{\phi(\yvec_2,t')\phitilde(\xvec_2,t)}
\ldots
\ave{\phi(\yvec_N,t')\phitilde(\xvec_N,t)} \nonumber\\
&+
\ave{\phi(\yvec_2,t')\phitilde(\xvec_1,t)}
\ave{\phi(\yvec_1,t')\phitilde(\xvec_2,t)}
\ldots
\ave{\phi(\yvec_N,t')\phitilde(\xvec_N,t)}
+
\ldots 
\elabel{permutation_propagator}
\end{align}

As a result of \Eref{permutation_propagator}, both $\Op^{(N)}$ and $\Ln^{(N)}$ contain $N!$ times as many
terms as in the case of distinguishable particles. As far as
$\Op^{(N)}$ is concerned, \Eref{Op_indistinguishableN}, this seems to barely complicate the expression
for the entropy production, because the permutation of the fields can be
undone by a permutation of the dummy variables of the integral it is sitting in, so that,
say $\yvec_i$ is paired with $\xvec_i$ in each and every of the
propagators appearing in $\Op^{(N)}$ according to \Eref{permutation_propagator}. 
As $\Ln^{(N)}$ and the logarithm of the joint density both are
invariant under permutations of the $\yvec_i$, 
and the joint density $\HMdensity{N}{N}(\xvec_1,\ldots,\xvec_N)$ in the pre-factor is not even
affected by such a permutation of the $\yvec_i$,
there are $N!$ such
permutations, all equal and therefore cancelling the factor of $1/N!$ in
\Eref{def_entropyProductionDensity_indistinguishableN}, so that 
\begin{equation}\elabel{def_entropyProductionDensity_indistinguishableN2}
\entropyProductionDensity^{(N)}(\xvec_1,\ldots,\xvec_N) =
\int \ddint{y_{1,\ldots,N}}
\left\{
\sum_{i=1}^N  (\gdot_i + \fdot_i)  \prod_{j\ne i}^N \delta_j
\right\}
\Ln^{(N)}_{\yvec_1,\ldots,\yvec_N,\xvec_1,\ldots,\xvec_N} \ ,
\end{equation}
similar to \Eref{Op_distinguishable_independent}.

Using \Eref{permutation_propagator} in \Eref{Ln_indistinguishableN} to calculate the logarithm, we have
\begin{align}
\Ln^{(N)}_{\yvec_1,\ldots,\yvec_N,\xvec_1,\ldots,\xvec_N} 
\nonumber &= 
\lim_{t'\downarrow t}
\Bigg\{
\ln\Big(
\ave{\phi(\yvec_1,t')\phitilde(\xvec_1,t)}
\ave{\phi(\yvec_2,t')\phitilde(\xvec_2,t)}
\ldots
\ave{\phi(\yvec_N,t')\phitilde(\xvec_N,t)}
\\\nonumber 
&\quad\qquad +
\ave{\phi(\yvec_2,t')\phitilde(\xvec_1,t)}
\ave{\phi(\yvec_1,t')\phitilde(\xvec_2,t)} 
\ldots
\ave{\phi(\yvec_N,t')\phitilde(\xvec_N,t)} 
+
\ldots
\Big)\\\nonumber 
&\quad -
\ln\Big(
\ave{\phi(\xvec_1,t')\phitilde(\yvec_1,t)}
\ave{\phi(\xvec_2,t')\phitilde(\yvec_2,t)}
\ldots
\ave{\phi(\xvec_N,t')\phitilde(\yvec_N,t)}
\\ 
&\quad\qquad+
\ave{\phi(\xvec_2,t')\phitilde(\yvec_1,t)}
\ave{\phi(\xvec_1,t')\phitilde(\yvec_2,t)} 
\ldots
\ave{\phi(\xvec_N,t')\phitilde(\yvec_N,t)} 
+
\ldots
\Big)
\Bigg\} \ .
\end{align}
We will use \Eref{g_i_limit}
in the form  
that any of the propagators $\ave{\phi(\yvec_i,t')\phitilde(\xvec_j,t)}$ in the limit of $t'\downarrow t$ will vanish if $i\ne j$ because 
$\yvec_i=\xvec_j$ for $i\ne j$ is not enforced by the $\delta$-functions in the kernel
and therefore $\yvec_i=\xvec_j$ has vanishing measure under the integral. As the kernel enforces $\yvec_i=\xvec_j$ only for $i=j$, 
under the integral the logarithm 
simplifies just like in \Eref{entropyProductionDensity_independent_distinguishable_final_sum}. 
Because all the $g_i$ and $f_i$ are the same function for indistinguishable particles just with different arguments $\yvec_i$ and $\xvec_i$,
the local entropy production of indistinguishable particles corresponding to
\Eref{entropyProductionDensity_independent_distinguishable_final_sum}
is invariant under permutations of the arguments $\xvec_1,\ldots,\xvec_N$
and the single-particle local entropy production $\entropyProductionDensity_i^{(N)}(\xvec_i)$ is the same for any particle $i$, \ie
$\entropyProductionDensity_i^{(N)}(\xvec_i)=\entropyProductionDensity_1^{(1)}(\xvec_i)$, \Eref{entropyProductionDensity_independent_distinguishable_final}.
Inserting $\entropyProductionDensity^{(N)}$ in \Eref{entropyProductionDensity_independent_distinguishable_final_sum}
into the entropy production \Eref{entropy_production_indistinguishableN_as_density}, and using
that $\entropyProductionDensity^{(N)}$ and
$\HMdensity{N}{N}$ are invariant
under permutations of $\xvec_1,\ldots,\xvec_N$, 
we obtain
\begin{equation}\elabel{entropy_production_indistinguishableN_as_density_independent2}
    \entropyProduction^{(N)}[\HMdensity{N}{N}] 
=
    \frac{N}{N!}
    \int \ddint{x_{1,\ldots,N}}
      \HMdensity{N}{N}(\xvec_1,\ldots,\xvec_N)
\entropyProductionDensity_1^{(1)}(\xvec_1)
\ .
\end{equation}
Using \Eref{marginalisation_densityN_i_indistinguishable} to marginalise $\HMdensity{N}{N}$ over $\xvec_2,\ldots,\xvec_N$ finally gives
\begin{equation}\elabel{entropy_production_indistinguishableN_as_density_independent2_final}
    \entropyProduction^{(N)}[\HMdensity{N}{N}]
=
    \int \ddint{x_1}
      \HMdensity{N}{1}(\xvec_1)
      \entropyProductionDensity_1^{(1)}(\xvec_1)
\ ,
\end{equation}
which is $N$ times the entropy production of a single particle, provided $\HMdensity{N}{1}(\xvec_1)=N\HMdensity{1}{1}(\xvec_1)$, consistent with the normalisation \Eref{marginalisation_densityN_consequences}. This is not always the case, in particular not when ``the system is not ergodic'' or not stationary, 
for example when particles are trapped or their position is not equilibrated,
so that the density is in fact a function of the initial positions of the particles.
If the density, however, obeys $\HMdensity{N}{1}(\xvec_1)=N\HMdensity{1}{1}(\xvec_1)$ we can write 
\begin{equation}\elabel{entropy_production_indistinguishableN_as_density_independent3}
    \entropyProduction^{(N)}[\HMdensity{N}{N}]
=
    N \int \ddint{x_1}
      \HMdensity{1}{1}(\xvec_1)
\entropyProductionDensity_1^{(1)}(\xvec_1) 
\ .
\end{equation}
For indistinguishable particles, we henceforth use the notation $\entropyProductionDensity_n^{(N)}(\xvec_1,\ldots,\xvec_n)$ for the local entropy production depending on $n$ locations in an $N$ particle system.

\toasubsubsection{\texorpdfstring{$N$}{N} pairwise interacting, indistinguishable particles}
\seclabel{N_interacting_indistinguishable_particles}
In the following, we generalise the result in 
\Eref{entropy_production_indistinguishableN_as_density_independent2_final} to
interacting, indistinguishable particles. 
In the case of interaction, neither density nor propagator factorise. However, just as in the discussion of interacting distinguishable particles, the propagator can still be expanded systematically, very much along the same lines as \Eref{Npropagators_distinguishable_interaction_diagrams}, with the added benefit of having to draw only on \emph{one} type of interaction,
\begin{equation}
h_{ij}=h(\yvec_i,\yvec_j;\xvec_i,\xvec_j; t'-t)  \ ,
\end{equation}
which is, similar to $g$ and $f$, \Erefs{g_the_same} and \eref{f_the_same}, the same function $h$ for any two particles with positions as indicated. Using the propagator in \Eref{Npropagators_distinguishable_with_interaction} as the starting point, we may write
\begin{align}\elabel{Npropagators_indistinguishable_with_interaction}
\ave{\phi(\yvec_1,t')\ldots\phitilde(\xvec_N,t)} 
= &
  \prod^N_i g(\yvec_i;\xvec_i;t'-t) 
+ \sum_i^N f(\yvec_i;\xvec_i;t'-t) \prod^N_{\substack{j\ne i}} g(\yvec_j;\xvec_j;t'-t)
\nonumber\\
&+ \sum_i^N \sum_{j\ne i}^N  h(\yvec_i,\yvec_j;\xvec_i,\xvec_j; t'-t) \prod_{k\notin\{i,j\}} g(\yvec_k;\xvec_k;t'-t) 
+ \text{perm.}
+ \order{(t'-t)^2} \ ,
\end{align}
where  "perm." refers to distinct permutations of the coordinates, as seen earlier in \Eref{permutation_propagator}. 
For example, the term $\prod_i g$ exists in $N!$ distinct permutations: One being the first term in \Eref{Npropagators_indistinguishable_with_interaction}, $\prod_ig_i$, and the remaining containing at least two terms such as $g(\yvec_1;\xvec_2;t'-t)$, that do not adhere to the pattern of the shorthand $g_i=g(\yvec_i,\xvec_i;t'-t)$. When $\xvec_i=\yvec_i$ is enforced for all $i$ except one, all these additional permutations essentially vanish under the integral, as discussed below.
The term $\sum_i f \prod_j g$ exists in $N(N!)$ distinct permutations, as 
$f(\yvec_i;\xvec_j;t'-t)$ exists in $N^2$ permutations and the remaining $\prod_j g$ in a further $(N-1)!$. The term involving $h$, correspondingly comes in $N(N-1)(N!)$ permutations. 

\Eref{Npropagators_indistinguishable_with_interaction}  enters the kernel with a time-derivative and a limit $t'\downarrow t$, producing $N(N!)$ distinct terms of the form $(\gdot+\fdot)\prod \delta$, as seen in the case without interaction, \Eref{def_entropyProductionDensity_indistinguishableN2}. Permuting the $\yvec_1,\ldots,\yvec_N$, so that every $\yvec_i$ is paired with $\xvec_i$ produces $N!$ times the same $N$ terms involving $(\gdot+\fdot)$ and $N(N-1)$ terms
involving $h$, specifically,
\begin{equation}\elabel{def_entropyProductionDensity_indistinguishable_interacting1}
\entropyProductionDensity^{(N)}(\xvec_1,\ldots,\xvec_N) 
=
\int \ddint{y_{1,\ldots,N}}
\left\{
\sum_{i=1}^N  (\gdot_i + \fdot_i)  
\prod_{j\ne i}^N
\delta_j
+
\sum_{i=1}^N  
\sum_{j\ne i}^N
\hdot_{ij} 
\prod_{k\notin \{i,j\}}^N
\delta_k
\right\}
\Ln^{(N)}_{\yvec_1,\ldots,\yvec_N,\xvec_1,\ldots,\xvec_N} \ ,
\end{equation}
similar to \Eref{def_entropyProductionDensity_indistinguishableN2}. 

As in the previous section, the logarithmic term is \emph{a priori} unaffected by any of the permutations, because being based on the propagator it is invariant under any permutations among the $\yvec_i$ and among the $\xvec_i$.
However, the same argument as in the previous section applies to all terms that the logarithm is comprised of, namely that in each one which demands $\yvec_i$ to be arbitrarily close to $\xvec_j$, this needs to be enforced by a $\delta$-function in the kernel, as it otherwise happens only with vanishing measure. All terms entering the logarithm make this demand in $N-1$ of $N$ pairs of $\yvec_i$ and $\xvec_j$, as $h(\yvec_k,\yvec_{\ell};\xvec_i,\xvec_j; t'-t)$ is $\delta$-like in $\yvec_{\ell}-\xvec_j$. 
What remains of the logarithm in \Eref{def_entropyProductionDensity_indistinguishable_interacting1} is therefore
\begin{equation}\elabel{def_entropyProductionDensity_indistinguishable_interacting1b}
\Ln^{(N)}_{\yvec_1,\ldots,\yvec_N,\xvec_1,\ldots,\xvec_N} =
\lim_{t'\downarrow t}
\ln\left(\frac
{\prod^N_\ell   g_\ell + \sum_\ell^N   f_\ell \prod^N_{\substack{m\ne \ell}}   g_m + \sum_\ell^N\sum_{m\ne \ell}^N   h_{\ell m} \prod_{n\notin\{\ell,m\}}^N   g_n + \ldots}
{\prod^N_\ell \Bg_\ell + \sum_\ell^N \Bf_\ell \prod^N_{\substack{m\ne \ell}} \Bg_m + \sum_\ell^N\sum_{m\ne \ell}^N \Bh_{\ell m} \prod_{n\notin\{\ell,m\}}^N \Bg_n + \ldots}
\right)  \ ,
\end{equation}
with the terms in $\ldots$ vanishing as some of the proximities are not enforced.
After dividing out $\prod^N_\ell   g_\ell/\Bg_\ell$ from the argument of the logarithm, the resulting expression
for $\entropyProductionDensity^{(N)}(\xvec_1,\ldots,\xvec_N)$ in \Eref{def_entropyProductionDensity_indistinguishableN} is identical to that for distinguishable particles, \Eref{entropyProductionDensity_interacting_distinguishable_initial}.
The factor of $1/N!$ in \Eref{entropy_production_indistinguishableN}, and the factorial factors produced by marginalisation, \Eref{marginalisation_densityN}, further simplify the total entropy production. Moreover, since the functions $g$, $f$ and $h$ are the same for all particles, the resulting expressions 
are not as cluttered with running indeces as their distinguishable counterparts, \eg \Erefs{entropyProductionDensity_interacting_distinguishable_ij} and \eref{entropyProductionDensity_interacting_distinguishable_ijk}.

Focussing firstly on the overall structure, using
$\entropyProductionDensity^{(N)}(\xvec_1,\ldots,\xvec_N)$ 
in \Eref{entropy_production_indistinguishableN_as_density}
with the same simplifications as carried out on
\Eref{entropyProductionDensity_interacting_distinguishable_initial} via \Eref{entropyProduction_interacting_distinguishable_sum} to \Eref{entropyProductionDensity_interacting_distinguishable_ijk}
in the case of distinguishable particles gives, for indistinguishable particles,
\begin{align}\elabel{entropy_production_indistinguishableN_with_interaction}
    \entropyProduction^{(N)}[\HMdensity{N}{N}] =
    \frac{1}{N!}
    \int \ddint{x_{1,\ldots,N}}
      \HMdensity{N}{N}(\xvec_1,\ldots,\xvec_N)
      \left\{
\sum_i^N \entropyProductionDensity^{(1)}_i(\xvec_i)
+
\sum_i^N \sum_{j\ne i}^N \entropyProductionDensity^{(2)}_{ij}(\xvec_i,\xvec_j)
+
\sum_i^N \sum_{j\ne i}^N \sum_{k\notin \{i,j\}} \entropyProductionDensity^{(3)}_{ijk}(\xvec_i,\xvec_j,\xvec_k)
\right\}
\end{align}
with the local entropy production 
$\entropyProductionDensity^{(1)}_i(\xvec_i)$,
$\entropyProductionDensity^{(2)}_{ij}(\xvec_i,\xvec_j)$
and
$\entropyProductionDensity^{(3)}_{ijk}(\xvec_i,\xvec_j,\xvec_k)$
as defined  in
\Erefs{entropyProductionDensity_independent_distinguishable_final},
\eref{entropyProductionDensity_interacting_distinguishable_ij},
and
\eref{entropyProductionDensity_interacting_distinguishable_ijk}
respectively.
\Eref{entropy_production_indistinguishableN_with_interaction} is
essentially \Eref{entropyProduction_interacting_distinguishable_final} but
with a different notion of the $N$-point density $\HMdensity{N}{N}$, which can be further simplified by
marginalisation, \Erefs{marginalisation_densityN}, \eref{marginalisation_densityN_i_indistinguishable} and 
\eref{def_rho_N_n_indistinguishable}.
Finally, because the functions $g_i$, $f_i$ and $h_{ij}$ depend on the particle index only in as far as the coordinates are concerned, the summations above can all be carried out and the local entropy productions  reduce to
\begin{subequations}
\elabel{def_entropyProductionDensities_indistinguishable}
\begin{align}
\elabel{def_entropyProductionDensities_indistinguishable_1}
\entropyProductionDensity^{(1)}_1(\xvec_1) &=
\int \ddint{y_1} (\gdot_1 + \fdot_1)
\lim_{t'\downarrow t}
\left[ \ln\left(\frac{g_1}{\Bg_1}\right) + \frac{f_1}{g_1} - \frac{\Bf_1}{\Bg_1} \right] \\
\elabel{def_entropyProductionDensities_indistinguishable_2}
\entropyProductionDensity^{(2)}_2(\xvec_1,\xvec_2) &=
\int \ddint{y_{1,2}}
\hdot_{12}
\lim_{t'\downarrow t}
\left\{
\ln\left(
\frac
{g_1}
{\Bg_1}
\right)
+
\left[ \frac{f_1}{g_1} - \frac{\Bf_1}{\Bg_1} \right]
+
\left[ \frac{h_{1 2}}{g_1 g_2} - \frac{\Bh_{1 2}}{\Bg_1\Bg_2} \right]
\right\}
+
(\gdot_1 + \fdot_1)\delta_2
\lim_{t'\downarrow t}
\left\{ \frac{h_{1 2}}{g_1 g_2} - \frac{\Bh_{1 2}}{\Bg_1\Bg_2} \right\} \\
\elabel{def_entropyProductionDensities_indistinguishable_3}
\entropyProductionDensity^{(3)}_{3}(\xvec_1,\xvec_2,\xvec_3)&=
\int \ddint{y_{1,2,3}}
\hdot_{12}\delta_3
\lim_{t'\downarrow t}
\left[ \frac{h_{1 3}}{g_1 g_3} - \frac{\Bh_{1 3}}{\Bg_1\Bg_3} \right] \ ,
\end{align}
\end{subequations}
using a slightly more suitable notation, 
where the subscript of $\entropyProductionDensity^{(N)}_n$ refers to the number of particles considered rather than the particle index, as in \Erefs{entropyProductionDensity_independent_distinguishable_final}, 
\eref{entropyProductionDensity_interacting_distinguishable_ij} 
and 
\eref{entropyProductionDensity_interacting_distinguishable_ijk}, 
and the superscript to the number of particles involved and thus to the minimal particle number.
With these definitions, the integrated entropy production is then
\begin{align}
\entropyProduction^{(N)}[\HMdensity{N}{}] = 
\int \ddint{x_1} \HMdensity{N}{1}(\xvec_1) \entropyProductionDensity^{(1)}_1(\xvec_1)
+
\int \ddint{x_{1,2}} \HMdensity{N}{2}(\xvec_1,\xvec_2) \entropyProductionDensity^{(2)}_2(\xvec_1,\xvec_2)
+
\int \ddint{x_{1,2,3}} \HMdensity{N}{3}(\xvec_1,\xvec_2,\xvec_3) \entropyProductionDensity^{(3)}_3(\xvec_1,\xvec_2,\xvec_3)\ ,
\elabel{entropyProduction_interacting_indistinguishable_final2}
\end{align}
neatly cancelling the factorial pre-factor.

\toasubsubsection{Example: Entropy production of \texorpdfstring{$N$}{N} pair-interacting indistinguishable particles in an external potential}
\seclabel{N_interacting_indistinguishable_particle_extPot}
The example of a drift-diffusion particle in an external potential has been introduced in \APref{simplified_notation}, in particular \Eref{entropyProductionDensity_distinguishable_non-interacting_example}: An example for 
$g$ is shown in \Eref{example_g_i}, $\gdot$ in \Eref{gdot_example},
$f$ in \Eref{f_example}, $\fdot$ in \Eref{fdot_example},
$h$ in \Eref{h_example_compact} and $\hdot$ in \Eref{hdot_example}. We will use those for 
$\entropyProductionDensity^{(1,2,3)}$, 
\Erefs{def_entropyProductionDensities_indistinguishable},
in \Erefs{entropyProduction_interacting_indistinguishable_final2}.
The local entropy production
$\entropyProductionDensity^{(1)}_1$ 
is given by \Erefs{entropyProductionDensity_distinguishable_non-interacting_example} with the same velocity and diffusion for all particles,
\begin{equation}
\elabel{entropyProductionDensity1_example_final}
\entropyProductionDensity^{(1)}_1(\xvec_1) =
-\extPot''(\xvec_1) + \frac{1}{\diffusion}\left(\vecDrift - \extPot'(\xvec_1)\right)^2
\ ,
\end{equation}
which is due to 
self-propulsion with velocity $\vecDrift$ in
the external potential $\extPot(\xvec)$, while 
$\entropyProductionDensity^{(2)}_2$ from \Eref{def_entropyProductionDensities_indistinguishable_2} is, 
using \Eref{fdot_example} and \eref{hdot_example_trawler_final},
\begin{equation}\elabel{entropyProductionDensity2_example_final}
\entropyProductionDensity^{(2)}_2(\xvec_1,\xvec_2) 
=
\frac{2} {\diffusion}  
\pairPot'(\xvec_1-\xvec_2)\cdot
\big(
\extPot'(\xvec_1)
- 
\vecDrift
\big)
+
\frac{1}{\diffusion} 
\pairPot'^2(\xvec_1-\xvec_2)
-
\pairPot''(\xvec_1-\xvec_2) 
\end{equation}
which originates from pair interactions, and equals \Eref{entropyProductionDensity_2_example_final} when $\extPot\equiv0$ and all drifts are the same.
Finally, $\entropyProductionDensity^{(3)}_3$ in \Eref{def_entropyProductionDensities_indistinguishable_3} is, 
using \Eref{h_example_trawler_final},
\begin{align}
\entropyProductionDensity^{(3)}_{3}(\xvec_1,\xvec_2,\xvec_3)
&=
\int \ddint{y_{1,2,3}}
\pairPot'(\xvec_1-\xvec_2)\cdot\delta_1' \delta_2 \delta_3
\left[
-\pairPot'(\xvec_1-\xvec_3) \cdot \frac{\yvec_1-\xvec_1}{2\diffusion} 
+\pairPot'(\yvec_1-\yvec_3) \cdot \frac{\xvec_1-\yvec_1}{2\diffusion}
\right] \nonumber\\
\elabel{entropyProductionDensity3_example_final}
&=
\frac{1}{\diffusion}
\pairPot'(\xvec_1-\xvec_2)
\cdot
\pairPot'(\xvec_1-\xvec_3) \ ,
\end{align}
showing that, for this choice of interactions, $\entropyProductionDensity^{(3)}_{3}$ has a distinctive r{\^o}le for particle $1$ while particles $2$ and $3$ play the same r{\^o}le.

Collecting all terms to construct the entropy production 
of $N$ pair-interacting, indistinguishable particles
according to \Eref{entropyProduction_interacting_indistinguishable_final2} 
from
$\entropyProductionDensity^{(1)}_1$ in 
\Eref{entropyProductionDensity1_example_final},
$\entropyProductionDensity^{(2)}_2$ in \Eref{entropyProductionDensity2_example_final}
and
$\entropyProductionDensity^{(3)}_3$ in \Eref{entropyProductionDensity3_example_final},
then gives
\begin{align}
\nonumber
\entropyProduction^{(N)}[\HMdensity{N}{}] 
=&
\int \ddint{x_1} \HMdensity{N}{1}(\xvec_1) 
\left( 
-\extPot''(\xvec_1) + \frac{1}{\diffusion}\left(\vecDrift - \extPot'(\xvec_1)\right)^2 
\right)\\
\nonumber
&+
\int 
\ddint{x_{1,2}}
\HMdensity{N}{2}(\xvec_1,\xvec_2)
\left(
\frac{2} {\diffusion}  
\pairPot'(\xvec_1-\xvec_2)\cdot
\left\{
\extPot'(\xvec_1)
- 
\vecDrift
\right\}
+
\frac{1}{\diffusion} 
\pairPot'^2(\xvec_1-\xvec_2)
-
\pairPot''(\xvec_1-\xvec_2) 
\right)\\
&+
\int \ddint{x_{1,2,3}} \HMdensity{N}{3}(\xvec_1,\xvec_2,\xvec_3) 
\left(
\frac{1}{\diffusion}
\pairPot'(\xvec_1-\xvec_2)
\cdot
\pairPot'(\xvec_1-\xvec_3)
\right) \ .
\elabel{entropyProduction_interacting_indistinguishable_example}
\end{align}
If $\pairPot$ is even, then the term $\HMdensity{N}{2}(\xvec_1,\xvec_2) \pairPot'(\xvec_1-\xvec_2)\cdot\vecDrift$ 
in the second line of \Eref{entropyProduction_interacting_indistinguishable_example} changes sign under exchange of the dummy variables $\xvec_1$ and $\xvec_2$ and thus drops out under integration. 
The corresponding term projecting on $\extPot'(\xvec_1)$ rather than $\vecDrift$ does not posses the same symmetry. \Eref{entropyProduction_interacting_indistinguishable_example} with external potential vanishing, $\extPot\equiv0$, and even pair potential, $\pairPot'(\xvec_1-\xvec_2)=-\pairPot'(\xvec_2-\xvec_1)$, is \Eref{entropyProduction_for_pairPot} in the main text.

\toasubsubsubsection{Sample-based entropy production}
Determining the $n$-point densities $\HMdensity{N}{1}$, $\HMdensity{N}{2}$ and $\HMdensity{N}{3}$ in \Eref{entropyProduction_interacting_indistinguishable_example}
by numerical or experimental estimates is generally demanding. 
In the following, we derive sample-based expressions similar to \Eref{entropyProduction_interacting_distinguishable_experimental} at the end of \APref{N_interacting_indistinguishable_particles}.
At stationarity, the densities are
\begin{subequations}
\begin{align}
\lim_{Q\to\infty}
\frac{1}{Q}\sum_{q=1}^Q \sum_{i_1=1}^N \delta(\xvec_1-\xvec^{(q)}_{i_1})
& = \HMdensity{N}{1}(\xvec_1)\\
\lim_{Q\to\infty}
\frac{1}{Q}\sum_{q=1}^Q \sum_{i_1=1}^N
\sum_{\substack{i_2=1\\ i_1\ne i_2}}^N
\delta(\xvec_1-\xvec^{(q)}_{i_1})\delta(\xvec_2-\xvec^{(q)}_{i_2})
& = \HMdensity{N}{2}(\xvec_1,\xvec_2)\\
\lim_{Q\to\infty}
\frac{1}{Q}\sum_{q=1}^Q \sum_{i_1=1}^N
\sum_{\substack{i_2=1\\ i_1\ne i_2}}^N
\sum_{\substack{i_3=1\\ i_3\notin \{i_1,i_2\}}}^N 
\delta(\xvec_1-\xvec^{(q)}_{i_1})\delta(\xvec_2-\xvec^{(q)}_{i_2})\delta(\xvec_3-\xvec^{(q)}_{i_3})
& = \HMdensity{N}{3}(\xvec_1,\xvec_2,\xvec_3) \ ,
\end{align}
\end{subequations}
based on samples $q=1,2,\ldots,Q$ of particle positions $\xvec^{(q)}_{i}$ indexed by $i=1,2,\ldots,N$. 
At finite sample size $Q$ the sums over Dirac $\delta$-functions serve as estimates of the $n$-point densities, as they are expectations of indicator
functions.
At stationarity, the sample-based entropy production on the basis of
\Eref{entropyProduction_interacting_indistinguishable_example}  is thus
\begin{equation}\elabel{entropyProduction_interacting_indistinguishable_example_numerics}
\entropyProduction=\frac{1}{Q}\sum_{q=1}^Q
\left[
\sum_{i_1=1}^N  
\Bigg\{
\frac{1}{\diffusion}
\bigg(
\vecDrift - \extPot'\big(\xvec^{(q)}_{i_1}\big) - 
\sum_{\substack{i_2=2\\ i_1\ne i_2}}
\pairPot'\big(\xvec^{(q)}_{i_1}-\xvec^{(q)}_{i_2}\big)
\bigg)^2
-\extPot''(\xvec^{(q)}_{i_2})
- \sum_{\substack{i_2=2\\ i_1\ne i_2}} 
\pairPot''\big(\xvec^{(q)}_{i_1}-\xvec^{(q)}_{i_2}\big)
\Bigg\}
\right]
\ ,
\end{equation}
which allows for the notion of an instantaneous entropy production in the form of the expression in the square bracket. This can be further reduced to an instantaneous entropy production of each particle,
\begin{equation*}
\frac{1}{\diffusion} 
\bigg(
\vecDrift - \extPot'\big(\xvec^{(q)}_{i_1}\big) 
- \sum_{\substack{i_2=2\\ i_1\ne i_2}}
\pairPot'\big(\xvec^{(q)}_{i_1}-\xvec^{(q)}_{i_2}\big)
\bigg)^2
-\extPot''(\xvec^{(q)}_{i_2})
- \sum_{\substack{i_2=2\\ i_1\ne i_2}}
\pairPot''\big(\xvec^{(q)}_{i_1}-\xvec^{(q)}_{i_2}\big)
\ .
\end{equation*}
\Eref{entropyProduction_interacting_indistinguishable_example_numerics} simplifies further if all interactions are harmonic or vanish, in which case the highest order correlations to be estimated are $\ave{\xvec_i\cdot\xvec_j}$.

\toasubsubsubsection{No entropy production of pair-interacting, indistinguishable, diffusive particles without drift}\seclabel{no_EPR_without_drift}
As a sanity check of
\Eref{entropyProduction_interacting_indistinguishable_example}, we calculate the entropy production of $N$ pair-interacting, indistinguishable particles, which are subject to diffusion but not to drift, assuming stationarity. In this case, the $N$-point density is Boltzmann,
\begin{equation}\elabel{density_indistinguishable_example}
\HMdensity{N}{N}(\xvec_1,\xvec_2,\ldots,\xvec_N) = \NC^{-1}
\exp{-\HC/\diffusion}
\quad\text{ with }\quad
\HC=\sum_{i=1}^N \extPot(\xvec_i) + \sum_{i=1}^N\sum_{j=i+1}^N \pairPot(\xvec_i-\xvec_j)
\end{equation}
and suitable normalisation $\NC^{-1}$, such that 
\Eref{marginalisation_densityN_consequences} holds. 
Without drift 
the entropy production should vanish.
The Hamiltionian written in the form
\Eref{density_indistinguishable_example} assumes an even pair-potential $\pairPot$, 
but this does not amount to a loss of generality,
because odd contributions can be shown to cancel in a Hamiltonian 
invariant under permutations of indeces, as is the case for
indistinguishable particles.
To show that the entropy production with \Eref{density_indistinguishable_example} vanishes, we use that the integral of $\nabla_{\xvec_1}^2 \HMdensity{N}{N}$ over all space vanishes by Gauss' theorem, and calculate it explicitly from \Eref{density_indistinguishable_example},
\begin{subequations}
\begin{align}\elabel{derivative_density_N_example_equilibrium_identity}
0&=
\frac{\diffusion}{ (N-1)!}
\int \ddint{x_{1,\ldots,N}}
\nabla_{\xvec_1}^2 \HMdensity{N}{N}(\xvec_1,\xvec_2,\ldots,\xvec_N) 
\\
\nonumber
&=\int \ddint{x_1} \HMdensity{N}{1}(\xvec_1)
\left\{
- \extPot''(\xvec_1) + \frac{\extPot'^2(\xvec_1)}{\diffusion}
\right\}
\\ \nonumber &\quad
+
\int \ddint{x_{1,2}} \HMdensity{N}{2}(\xvec_1,\xvec_2)
\left\{
- \pairPot''(\xvec_1-\xvec_2) +
\frac{\pairPot'^2(\xvec_1-\xvec_2)}{\diffusion} +
\frac{2 \extPot'(\xvec_1)}{\diffusion} \cdot \pairPot'(\xvec_1-\xvec_2)
\right\}
\\  &\quad
+
\int \ddint{x_{1,2,3}} \HMdensity{N}{3}(\xvec_1,\xvec_2,\xvec_3)
\frac{\pairPot'(\xvec_1-\xvec_2)}{\diffusion}\cdot\pairPot'(\xvec_1-\xvec_3) \ ,
\elabel{entropyProduction_indistinguishable_equilibrium_example_identity}
\end{align}
\end{subequations}
where we have used \Erefs{marginalisation_densityN}, \eref{marginalisation_densityN_i_indistinguishable} and \eref{def_rho_N_n_indistinguishable}
and the symmetry of the density under permutation of the arguments.
By inspection we find that \Eref{entropyProduction_indistinguishable_equilibrium_example_identity} is \Eref{entropyProduction_interacting_indistinguishable_example} at $\vecDrift=\nullvec$. In other words, the stationary entropy production of $N$ identical particles, subject to a pair- and an external potential, vanishes in the absence of drift, provided the particles are Boltzmann-distributed. Of course, this is what we expect from simple physical reasoning, but the present calculation offers an important sanity check in particular for the somewhat unusual $3$-point term.


\section*{Supplementary material}

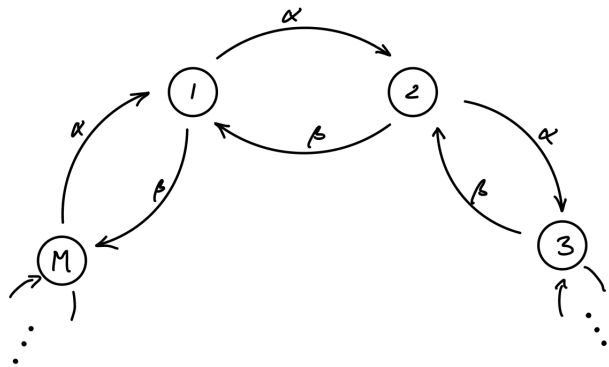

**Figure S-VI.1:** Cartoon of an $M$-state Markov process. Periodic states $i = 1, 2, \ldots, M$ are reached by transitions with rate $\alpha$ from $i-1$ and with rate $\beta$ from $i+1$.

a simple, exactly solvable system under the same approximation: $N$ particles hopping on a ring-lattice of $M$ states. Our results show that, although spatial correlations are captured correctly, the framework devised in [21] produces an unphysical entropy production, as it is not extensive in the particle number $N$, but instead extensive in the number of states $M$, consistent with those being the degrees of freedom of the description in terms of $\phi(x,t)$.

The outline of our derivation is as follows: In S-VI.1 we define the model and state its basic properties, such as average particle number, variance and entropy production, in S-VI.2 we introduce its continuum particle number description with additive noise, which finally produces the Onsager-Machlup functional Eq. (S-VI.26), in S-VI.3 and S-VI.4. We derive the correlation function of the Fourier modes of the density field description in S-VI.5 and validate it, before deriving the entropy production under this approximation as outlined in [21], Eq. (S-VI.37) in S-VI.6.

### S-VI.1 Biased hopping on a ring

In the following we consider $N$ independent particles subject to an $M$-state Markov process. The basic setup is sketched in Figure S-VI.1. States $i = 1, 2, \ldots, M$ are connected periodically, so that $i = 1$ may be thought of $i = M+1$ and $i = M$ as $i = 0$. Transitions from $i$ to $i+1$, clockwise moves, happen with rate $\alpha$ and transitions from $i$ to $i-1$, anti-clockwise moves, with rate $\beta$, implying $M > 2$ to render the setup and the notion of clockwise and anti-clockwise meaningful. No other transitions are allowed. The Markovian degree of freedom is $\phi_i(t)$, the number of particles on site $i$ at time $t$. As a count per site, we may refer to $\phi_i(t)$ as a density.

The density and its variance at stationarity are

$$\overline{\phi}_i := \lim_{t \to \infty} \langle \phi_i(t) \rangle = \frac{N}{M}, \quad \text{and} \quad \lim_{t \to \infty} \langle \phi_i^2(t) \rangle - \langle \phi_i(t) \rangle^2 = \frac{N(M-1)}{M^2}, \tag{S-VI.1}$$

by considering the stationary occupation of any site as a $N$-times repeated Bernoulli process with success probability $1/M$.

The entropy production of a single particle can be determined by elementary considerations [16] to be $(\alpha-\beta)\ln(\alpha/\beta)$ so that for $N$ particles,

$$\dot{S}_{\text{int}} = N(\alpha - \beta) \ln\left(\frac{\alpha}{\beta}\right) \quad \text{for} \quad M > 2 \tag{S-VI.2}$$

at stationarity, distinguishable or not. In the framework discussed in the present work, this is immediately confirmed by Eqs. (26) and Suppl. S-I.4.

### S-VI.2 Continuum particle number description

Following the approach in [21, 46, 59], we consider the particle density field $\phi_i(t)$ of state $i$ as a function of time $t$ as a continuum, $\phi_i \in \mathbb{R}$. Like Eq. (1) in [21], the density field $\phi_i(t)$ can be considered as being governed by a

conservative Langevin equation with *additive* noise such as

$$\dot{\boldsymbol{\phi}}(t) = (\alpha S + \beta S^\mathsf{T})\boldsymbol{\phi}(t) + \sqrt{f\alpha\frac{N}{M}}S\boldsymbol{\xi}_\alpha(t) + \sqrt{f\beta\frac{N}{M}}S^\mathsf{T}\boldsymbol{\xi}_\beta(t) \tag{S-VI.3}$$

where the column vector $\boldsymbol{\phi}(t)$ has components $\phi_i(t)$. The matrices

$$S = \begin{pmatrix} -1 & 0 & 0 & \cdots & 0 & 1 \\ 1 & -1 & 0 & \cdots & 0 & 0 \\ 0 & 1 & -1 & \cdots & 0 & 0 \\ \vdots & \vdots & \vdots & & \vdots & \vdots \\ 0 & 0 & 0 & \cdots & 1 & -1 \end{pmatrix} \tag{S-VI.4}$$

and its transpose $S^\mathsf{T}$, denoted by $\mathsf{T}$, result in a conservative movement of particles from site $i$ to $i+1$ and from $i$ to $i-1$ respectively. These matrices possess a number of very useful algebraic properties, discussed in Suppl. S-VI.3. To ease notation, we further introduce

$$\sigma \coloneqq \alpha S + \beta S^\mathsf{T} , \tag{S-VI.5}$$

for the deterministic part of the Langevin Eq. (S-VI.3). Without the noise the Langevin equation is the exact master equation of a single particle. Eq. (S-VI.3) may thus be seen as an attempt to, somewhat *ad hoc*, express the master equation on the particle count, by adding to the single particle dynamics a noisy "bath".

We have introduced a fudge-factor $f$ in (S-VI.3) as a way to trace the noise amplitude through the calculation. It also provides a mechanism to adjust the noise amplitude *a posteriori*. The amplitude $\propto \sqrt{N/M}$ is otherwise chosen to reflect that the variance of the noise is proportional to the mean, as for a Poisson distribution. We will verify below that this choice results in the density field $\boldsymbol{\phi}$ governed by Eq. (S-VI.3) reproducing the single site variance Eq. (S-VI.1) for $f = 1$. The two independent noise vectors $\boldsymbol{\xi}_\alpha(t)$ and $\boldsymbol{\xi}_\beta(t)$ have vanishing mean and correlation matrices

$$\langle \boldsymbol{\xi}_\alpha(t)\boldsymbol{\xi}_\alpha^\mathsf{T}(t') \rangle = \delta(t-t')\mathbf{1} \qquad \text{and} \qquad \langle \boldsymbol{\xi}_\beta(t)\boldsymbol{\xi}_\beta^\mathsf{T}(t') \rangle = \delta(t-t')\mathbf{1} \tag{S-VI.6}$$

with $\boldsymbol{\xi}_\alpha$ and $\boldsymbol{\xi}_\beta$ column vectors and $\mathbf{1}$ an $M \times M$ identity matrix. The noise vectors can, of course, be summed into a single noise term,

$$\boldsymbol{\zeta}(t;f) = \sqrt{f\alpha\frac{N}{M}}S\boldsymbol{\xi}_\alpha(t) + \sqrt{f\beta\frac{N}{M}}S^\mathsf{T}\boldsymbol{\xi}_\beta(t) \tag{S-VI.7}$$

with vanishing mean, $\langle \boldsymbol{\zeta}(t;f) \rangle = \mathbf{0}$, and correlator

$$\langle \boldsymbol{\zeta}(t;f)\boldsymbol{\zeta}^\mathsf{T}(t';f) \rangle = f\frac{N}{M}(\alpha+\beta)SS^\mathsf{T}\delta(t-t') . \tag{S-VI.8}$$

As shown in Suppl. S-VI.3, one eigenvalue of $S$, say $\lambda_0$, vanishes, which means that the corresponding 0-mode of $\zeta$ has no variance. To write a path-density $\mathcal{P}[\boldsymbol{\zeta}(t;f)]$ for the noise in the current form, we would need to regularise this correlation matrix, as well as the deterministic part of Eq. (S-VI.3). This is straight-forwardly doable, however, somewhat messy. To avoid this distraction, we change to a basis which allows the removal of the 0-mode from the path-densities altogether.

With the noise defined above and the deterministic part of the original Langevin Eq. (S-VI.3) effectively captured by $\sigma$, Eq. (S-VI.5), it may finally be written as

$$\dot{\boldsymbol{\phi}}(t) = \sigma\boldsymbol{\phi}(t) + \boldsymbol{\zeta}(t;f) . \tag{S-VI.9}$$

Eq. (S-VI.3) thus contains two approximations: Firstly, $\boldsymbol{\phi} \in \mathbb{R}^M$ is a *continuous degree of freedom* that evolves without any constraints other than the conservation imposed by $S$ and $S^\mathsf{T}$, even though it is introduced as a *local, instantaneous particle count*, which requires $\phi_i(t) \in \mathbb{N}_0$. Secondly, the additive noise has a (squared) amplitude proportional to the *steady state* $N/M$, rather than the *instantaneous* occupation number $\phi_i(t)$. There is no easy remedy for either of these two approximations, which ultimately will produce the wrong entropy production.

**S-VI.3 Properties of the lattice derivative matrix $S$**

The matrix $S$ as defined in Eq. (S-VI.4) plays the rôle of a spatial derivative and, unsurprisingly, has corresponding Fourier-mode eigenvectors, which we define as

$$\mathbf{e}_\mu = \begin{pmatrix} \exp(\mathring{\imath} k_\mu \cdot 1) \\ \exp(\mathring{\imath} k_\mu \cdot 2) \\ \exp(\mathring{\imath} k_\mu \cdot 3) \\ \vdots \\ \exp(\mathring{\imath} k_\mu \cdot M) \end{pmatrix} \tag{S-VI.10}$$

where the $\cdot$ is there to emphasise the multiplication in the exponent. The coefficients $k_\mu = 2\pi\mu/M$ parameterise the $M$ distinct modes $\mu = 0, 1, 2, \ldots, M-1$ and, in fact $\mathbf{e}_\mu = \mathbf{e}_{M+\mu}$. By definition, the $j$th component of $\mathbf{e}_\mu$ is $(\mathbf{e}_\mu)_j = \exp(\mathring{\imath} k_\mu j)$. Writing $S_{ij} = -\delta_{i,j}^M + \delta_{i-1,j}^M$ with

$$\delta_{i,j}^M = \begin{cases} 1 & \text{for } i = j \mod M \\ 0 & \text{otherwise} \end{cases} \tag{S-VI.11}$$

the eigenvalues are easily determined, $S\mathbf{e}_\mu = \lambda_\mu \mathbf{e}_\mu$ with

$$\lambda_\mu = e^{-\mathring{\imath} k_\mu} - 1 \tag{S-VI.12}$$

and, correspondingly, $S^\mathsf{T} \mathbf{e}_\mu = \lambda_\mu^* \mathbf{e}_\mu$, so that the eigenvectors are orthogonal,

$$\mathbf{e}_\mu \cdot \mathbf{e}_\nu = \mathbf{e}_{-\mu}^\dagger \mathbf{e}_\nu = M\delta_{\mu+\nu,0}^M \;. \tag{S-VI.13}$$

The eigenvalue of the 0-mode vanishes, $\lambda_M = \lambda_0 = 0$, and we will henceforth consider only $\mu = 1, 2, \ldots, M-1$.

From Eq. (S-VI.4) it follows by elementary calculation that

$$SS^\mathsf{T} = -(S + S^\mathsf{T}) = S^\mathsf{T} S \;, \tag{S-VI.14}$$

*i.e.* $S$ and $S^\mathsf{T}$ commute. The $\mathbf{e}_\mu$ are also eigenvectors of $\sigma$ introduced in Eq. (S-VI.5), as $\sigma \mathbf{e}_\mu = p_\mu \mathbf{e}_\mu$ with

$$p_\mu = \alpha\lambda_\mu + \beta\lambda_\mu^* = -(\alpha+\beta)(1-\cos(k_\mu)) + \mathring{\imath}(\beta-\alpha)\sin(k_\mu) \;, \tag{S-VI.15}$$

which has negative realpart for all $k_\mu$ as $k_\mu \neq 0$ for $\mu \neq 0$. The eigenvectors $\mathbf{e}_\mu$ can be used to re-express $\boldsymbol{\zeta}(t;f)$ and $\boldsymbol{\phi}(t)$ in terms of a more suitable basis, omitting the undesired 0-mode, say, $\boldsymbol{\zeta}(t;f) = \frac{1}{M}\sum_{\mu=1}^{M-1} \mathbf{e}_\mu z_\mu(t;f)$, or simply

$$\boldsymbol{\zeta}(t;f) = \begin{pmatrix} \zeta_1(t;f) \\ \zeta_2(t;f) \\ \zeta_3(t;f) \\ \vdots \\ \zeta_M(t;f) \end{pmatrix} = \frac{1}{M}\mathcal{E}\begin{pmatrix} z_1(t;f) \\ z_2(t;f) \\ z_3(t;f) \\ \vdots \\ z_{M-1}(t;f) \end{pmatrix} \quad \text{with} \quad \mathcal{E} = \begin{pmatrix} \mathbf{e}_1 & \mathbf{e}_2 & \mathbf{e}_3 & \ldots & \mathbf{e}_{M-1} \end{pmatrix} \tag{S-VI.16}$$

so that the $M \times (M-1)$ matrix $\mathcal{E}$ is composed from the column vectors $\mathbf{e}_1, \ldots, \mathbf{e}_{M-1}$, and is "essentially unitary", because

$$\mathcal{E}^\dagger = \mathcal{E}^{*\mathsf{T}} = \begin{pmatrix} \mathbf{e}_1^\dagger \\ \mathbf{e}_2^\dagger \\ \mathbf{e}_3^\dagger \\ \vdots \\ \mathbf{e}_{M-1}^\dagger \end{pmatrix} \quad \text{obeys} \quad \mathcal{E}^\dagger \mathcal{E} = M\mathbf{1} \;, \tag{S-VI.17}$$

with $\mathbf{1}$ an $(M-1) \times (M-1)$ identity matrix, as $\mathbf{e}_\mu^* = \mathbf{e}_{-\mu}$ by construction. Of course, not being square $\mathcal{E}$ cannot be

unitary and indeed $\mathcal{E}\mathcal{E}^\dagger$ looks rather grim, Eq. (S-VI.33). It follows that

$$\mathbf{z}(t;f) = \mathcal{E}^\dagger \boldsymbol{\zeta}(t;f) \ . \tag{S-VI.18}$$

By definition, the elements $(\mathcal{E})_{i\mu} = (\mathbf{e}_\mu)_i = \exp(\mathring{\imath} k_\mu i)$ depend only on the product $i\mu$ and therefore $\mathcal{E}$ is symmetric, $(\mathcal{E})_{i\mu} = (\mathcal{E})_{\mu i}$, for $1 \le i, \mu \le M-1$. One can further show that

$$S\mathcal{E} = \mathcal{E}\Lambda \quad \text{with} \quad \Lambda = \begin{pmatrix} \lambda_1 & & & & 0 \\ & \lambda_2 & & & \\ & & \lambda_3 & & \\ & & & \ddots & \\ 0 & & & & \lambda_{M-1} \end{pmatrix} \quad \text{so that} \quad \mathcal{E}^\dagger S^\mathsf{T} = \Lambda^* \mathcal{E}^\dagger \tag{S-VI.19}$$

and similarly

$$S^\mathsf{T} \mathcal{E} = \mathcal{E}\Lambda^* \quad \text{so that} \quad \mathcal{E}^\dagger S = \Lambda \mathcal{E}^\dagger \quad \text{and} \quad \mathcal{E}^\dagger \sigma = \left(\alpha\Lambda + \beta\Lambda^*\right)\mathcal{E}^\dagger \ . \tag{S-VI.20}$$

The noise correlator of $\mathbf{z}$ is determined by Eqs. (S-VI.8) and (S-VI.18) by direct computation

$$\langle \mathbf{z}(t;f)\mathbf{z}^\dagger(t';f) \rangle = fN(\alpha+\beta)\delta(t-t')\Lambda\Lambda^* \tag{S-VI.21}$$

using $\mathcal{E}^\dagger S S^\mathsf{T} \mathcal{E} = M\Lambda\Lambda^*$. As opposed to the correlator of $\boldsymbol{\zeta}$, Eq. (S-VI.8), the matrix on the right can be inverted and thus be used as the basis of an Onsager-Machlup functional.

While the noise $\boldsymbol{\zeta}$ is fully determined by the noise $\mathbf{z}$, as neither has a 0-mode, the density field $\boldsymbol{\phi}$ can have any 0-mode, only that it does not evolve in time and therefore is not to be considered a degree of freedom. We therefore introduce the modes $a_\mu(t)$ of $\boldsymbol{\phi}$ with an extra parameter $\overline{\boldsymbol{\phi}} = \overline{\phi}\mathbf{e}_0$ that is constant in time and has identical components, so that

$$\partial_t \overline{\boldsymbol{\phi}} = \mathbf{0} = \sigma \overline{\boldsymbol{\phi}} \ , \tag{S-VI.22}$$

and $\overline{\phi} = N/M$, Eq. (S-VI.1). Any $M$-component density field may then be written in terms of the background $\overline{\boldsymbol{\phi}}$ and an $(M-1)$-component vector $\mathbf{a}(t)$ of modes,

$$\boldsymbol{\phi}(t) = \overline{\boldsymbol{\phi}} + \frac{1}{M}\mathcal{E}\mathbf{a}(t) \ , \quad \text{so that} \quad \mathbf{a}(t) = \mathcal{E}^\dagger \boldsymbol{\phi}(t) \ . \tag{S-VI.23}$$

With this in place, we can write the Langevin Eq. (S-VI.9) in diagonal form, by acting from the left with $\mathcal{E}^\dagger$,

$$\dot{\mathbf{a}}(t) = \left(\alpha\Lambda + \beta\Lambda^*\right)\mathbf{a}(t) + \mathbf{z}(t;f) \ . \tag{S-VI.24}$$

Via Eq. (S-VI.23), Eq. (S-VI.24) provides an alternative equation of motion of the present Langevin dynamics that is more easily analysed.

### S-VI.4 The Onsager-Machlup functional

The path density of the noise $\mathbf{z}$ follows from its correlator Eq. (S-VI.21) as

$$\mathcal{P}[\mathbf{z}(t;f)] \propto \exp\left(-\frac{1}{2fN(\alpha+\beta)} \int \mathrm{d}t\, \mathbf{z}^\dagger(t;f)(\Lambda\Lambda^*)^{-1}\mathbf{z}(t;f)\right) \tag{S-VI.25}$$

and since a path $\mathbf{z}(t)$ is uniquely determined from that of $\mathbf{a}$ and vice versa via Eq. (S-VI.24), $\mathbf{z} = \dot{\mathbf{a}} - (\alpha\Lambda + \beta\Lambda^*)\mathbf{a}$, the Onsager-Machlup functional follows immediately, $\mathcal{P}[\mathbf{a}(t)] \propto \exp(\mathcal{G}[\mathbf{a}(t)])$ with

$$\mathcal{G}[\mathbf{a}(t)] = -\frac{1}{2fN(\alpha+\beta)} \int \mathrm{d}t\, \left(\dot{\mathbf{a}} - (\alpha\Lambda + \beta\Lambda^*)\mathbf{a}\right)^\dagger (\Lambda\Lambda^*)^{-1} \left(\dot{\mathbf{a}} - (\alpha\Lambda + \beta\Lambda^*)\mathbf{a}\right) \ , \tag{S-VI.26}$$

as the Jacobian is constant. By Fourier transforming the modes,

$$\mathbf{a}(\omega) = \int \mathrm{d}t\, e^{i\omega t}\mathbf{a}(t) \quad \text{and} \quad \mathbf{a}(t) = \int \mathrm{d}\omega\, e^{-i\omega t}\mathbf{a}(\omega) \tag{S-VI.27}$$

with $\mathrm{d}\omega = d\omega/(2\pi)$, the Onsager-Machlup functional becomes

$$\mathcal{G}[\mathbf{a}(\omega)] = -\frac{1}{2fN(\alpha+\beta)} \int \mathrm{d}\omega\, \mathbf{a}^\dagger(\omega)\bigl(i\omega\mathbf{1} + \alpha\Lambda + \beta\Lambda^*\bigr)^\dagger (\Lambda\Lambda^*)^{-1}\bigl(i\omega\mathbf{1} + \alpha\Lambda + \beta\Lambda^*\bigr)\mathbf{a}(\omega)\ . \tag{S-VI.28}$$

Because all matrices in Eq. (S-VI.28) are diagonal the following calculations are greatly simplified. For now, we will proceed with an infinite time domain, even when the calculation of the entropy production in Section S-VI.6 will require a finite time interval to be taken to infinity systematically.

### S-VI.5 Correlator of the density's Fourier modes $\mathbf{a}(t)$

Above, it was argued that the distribution $\mathcal{P}[\mathbf{z}]$ in Eq. (S-VI.25) produces correlator Eq. (S-VI.21). Correspondingly, the distribution $\mathcal{P}[\mathbf{a}]$ with functional in Eq. (S-VI.26) produces the correlation matrix

$$\langle \mathbf{a}(\omega)\mathbf{a}^\dagger(\omega')\rangle = fN(\alpha+\beta)\left[\bigl(i\omega\mathbf{1} + \alpha\Lambda + \beta\Lambda^*\bigr)^\dagger (\Lambda\Lambda^*)^{-1}\bigl(i\omega\mathbf{1} + \alpha\Lambda + \beta\Lambda^*\bigr)\right]^{-1} \delta(\omega-\omega')\ . \tag{S-VI.29}$$

To invert the matrix in square brackets we use that $\Lambda$ is diagonal, so

$$\left[\bigl(i\omega\mathbf{1} + \alpha\Lambda + \beta\Lambda^*\bigr)^\dagger (\Lambda\Lambda^*)^{-1}\bigl(i\omega\mathbf{1} + \alpha\Lambda + \beta\Lambda^*\bigr)\right]_{\mu\nu} = \delta_{\mu,\nu}(-i\omega + \alpha\lambda_\mu^* + \beta\lambda_\mu)(\lambda_\mu^*)^{-1}\lambda_\mu^{-1}(i\omega + \alpha\lambda_\mu + \beta\lambda_\mu^*)\ . \tag{S-VI.30}$$

Using Eq. (S-VI.15) to write $\lambda_\mu$ in terms of the poles $p_\mu$, and $a_\mu^*(\omega) = a_{M-\mu}(-\omega)$ via Eqs. (S-VI.23) and (S-VI.27), the correlators become

$$\langle a_\mu(\omega)a_\nu(\omega')\rangle = \frac{fN(\alpha+\beta)\lambda_\mu^*\lambda_\mu\delta_{\mu+\nu,0}^M\delta(\omega+\omega')}{(-i\omega-p_\mu)(i\omega-p_\mu^*)} \tag{S-VI.31}$$

bound to be real, as every factor is multiplied by its complex conjugate. Both brackets in the denominator are of the form $-i\omega$ plus a number that has a strictly positive real part, as $\Re(p_\mu) < 0$, Eq. (S-VI.15).

The inverse Fourier transform of Eq. (S-VI.31) is

$$\begin{aligned}\langle a_\mu(t)a_\nu(t')\rangle &= \int \mathrm{d}\omega \mathrm{d}\omega'\, e^{-i(\omega t + \omega' t')}\langle a_\mu(\omega)a_\nu(\omega')\rangle \\ &= fN\delta_{\mu+\nu,0}^M \exp\left(\frac{1}{2}(\alpha+\beta)(\lambda_\mu + \lambda_\mu^*)|t-t'|\right)\exp\left(\frac{1}{2}(\alpha-\beta)(\lambda_\mu - \lambda_\mu^*)(t-t')\right)\ ,\end{aligned} \tag{S-VI.32}$$

using $\lambda_\mu\lambda_\mu^* = -(\lambda_\mu + \lambda_\mu^*)$ with $\Re(\lambda_i + \lambda_i^*) = -2(1-\cos(k_\mu)) < 0$.

To validate this result, we may calculate the equal-time correlation matrix

$$\left\langle (\boldsymbol{\phi}(t) - \overline{\boldsymbol{\phi}})(\boldsymbol{\phi}(t) - \overline{\boldsymbol{\phi}})^\mathsf{T} \right\rangle = M^{-2}\mathcal{E}\left\langle \mathbf{a}(t)\mathbf{a}^\dagger(t)\right\rangle \mathcal{E}^\dagger = \frac{fN}{M^2}\begin{pmatrix} M-1 & & & -1 \\ & M-1 & & \\ & & \ddots & \\ -1 & & & M-1 \end{pmatrix} \tag{S-VI.33}$$

via Eq. (S-VI.23) and the matrix on the far right being $\mathcal{E}\mathcal{E}^\dagger$, with the diagonal confirming the variance Eq. (S-VI.1) with $f=1$. Closer inspection, for example using a Doi-Peliti field theory, shows that the correlator in Eq. (S-VI.32), is exact if $f=1$. In other words, the setup Eq. (S-VI.3) captures two-point correlations exactly.

The entropy production below draws on the symmetric equal-time derivative, which we write here simply as

$$\langle a_\mu(t)\dot{a}_\nu(t)\rangle = \frac{1}{2}\lim_{t'\downarrow t}\frac{d}{dt'}\langle a_\mu(t)a_\nu(t')\rangle + \frac{1}{2}\lim_{t'\uparrow t}\frac{d}{dt'}\langle a_\mu(t)a_\nu(t')\rangle$$
$$= -\frac{1}{2}fN\delta^M_{\mu+\nu,0}(\alpha-\beta)(\lambda_\mu - \lambda_\mu^*) \ . \tag{S-VI.34}$$

### S-VI.6 Entropy production

Using Seifert's scheme [60], the entropy production of the $M$-state Markov process with path density $\mathcal{P}([\boldsymbol{\phi}];T)$, Eq. (S-VI.26), for a path of duration $T$ is

$$\dot{S}_{\text{int}} = \lim_{T\to\infty}\frac{1}{T}\left\langle \ln\left(\frac{\mathcal{P}([\boldsymbol{\phi}];T)}{\mathcal{P}([\boldsymbol{\phi}^R];T)}\right)\right\rangle = \lim_{T\to\infty}\frac{1}{T}\langle \mathcal{G}([\mathbf{a}];T) - \mathcal{G}([\mathbf{a}^R];T)\rangle \ , \tag{S-VI.35}$$

where the constant Jacobian to transform $\boldsymbol{\phi}$ to $\mathbf{a}$ cancels in the fraction inside the logarithm and $\mathcal{G}([\mathbf{a}];T)$ is the Onsager-Machlup functional in Eq. (S-VI.26) with integration limits $0$ and $T$. In Eq. (S-VI.35), $\mathbf{a}^R(t)$ denotes the reverse path, $\mathbf{a}^R(t) = \mathbf{a}(T-t)$ so that the Onsager-Machlup functional $\mathcal{G}([\mathbf{a}^R];T)$ can be evaluated by replacing $\dot{\mathbf{a}}^R(t)$ by $-\dot{\mathbf{a}}(T-t)$. The ensemble average in Eq. (S-VI.35) needs to be taken over all *allowed, accessible* field configurations, but with $\boldsymbol{\phi}$ being continuous and $\overline{\boldsymbol{\phi}}$ fixed this poses no restriction. In constructing the Onsager-Machlup functional, a decision had been made implicitly or explicitly about the Itô *vs.* Stratonovich nature of the derivative $\dot{\mathbf{a}}$. Avoiding ambiguity, we use in the following the symmetrised version of the derivative, Eq. (S-VI.34), so that

$$\dot{S}_{\text{int}} = \lim_{T\to\infty}\frac{1}{T}\int_0^T dt\, \frac{1}{fN(\alpha+\beta)}\left(\langle \dot{\mathbf{a}}(t)^\dagger (\Lambda\Lambda^*)^{-1}(\alpha\Lambda+\beta\Lambda^*)\mathbf{a}(t)\rangle + \langle ((\alpha\Lambda+\beta\Lambda^*)\mathbf{a}(t))^\dagger (\Lambda\Lambda^*)^{-1}\dot{\mathbf{a}}(t)\rangle\right)$$
$$= \lim_{T\to\infty}\frac{1}{T}\int_0^T dt\,\frac{(\alpha-\beta)^2}{2(\alpha+\beta)}\sum_{\mu=1}^{M-1}\left(2 - \frac{\lambda_\mu}{\lambda_\mu^*} - \frac{\lambda_\mu^*}{\lambda_\mu}\right) \ . \tag{S-VI.36}$$

The fudge-factor $f$ has cancelled because it is the amplitude of the correlator and thus appears with its inverse in the action functional. It is obvious that this type of cancellation will occur in any bilinear action. Gone with the fudge factor is also the particle number $N$.

Otherwise, Eq. (S-VI.36) shows all the characteristics of the entropy production rate of drift-diffusion: It is quadratic in the hopping bias, $\alpha-\beta$, inversely proportional in the total hopping rate $\alpha+\beta$, which plays the rôle of a diffusion constant, and the integrand is independent of $t$, as the system is in the stationary state, so that the integral simply cancels the normalisation $1/T$. With some algebraic manipulation, $\lambda_\mu/\lambda_\mu^* + \lambda_\mu^*/\lambda_\mu = -2\cos k_\mu$ and $\sum_{\mu=1}^{M-1}\cos k_\mu = -1$ for $M \geq 2$ as $\cos k_0 = 1$, Eq. (S-VI.36) becomes

$$\dot{S}_{\text{int}} = \frac{(\alpha-\beta)^2}{\alpha+\beta}(M-2) \quad \text{for} \quad M \geq 2 \ . \tag{S-VI.37}$$

This is the final result for the entropy production on the basis of the path probability from the Langevin equation (S-VI.3) of the continuum particle number description. Taking the continuum limit $M\to\infty$ while maintaining finite drift and diffusion results in Eq. (S-VI.37) diverging.

Comparing Eq. (S-VI.37) to the exact expression Eq. (S-VI.2), immediately reveals some problems: While the logarithm might be recovered by making $\alpha-\beta$ small, the linearity in the particle number $N$ of the exact expression is replaced by a linearity in the number of states $M$ in Eq. (S-VI.37). It generally bears all the hallmarks of $M$ rather than $N$ being the number of degrees of freedom entering into this expression of the entropy production. One cannot argue that this $M$ is a proxy for the particle number $N$, as it is not multiplied by the expected particle number per site $\overline{\phi}$. As a result $\dot{S}_{\text{int}}$ of Eq. (S-VI.37) diverges as $M\to\infty$, irrespective of whether $\overline{\phi}$ is held constant or not in the limit. In short, Eq. (S-VI.37) produces the wrong result, consistent with this expression capturing the *states* as the degrees of freedom, rather than the *particles*.

It may not come as a surprise that an approximation scheme that changes the phase space from countable and discrete to uncountable and continuous shows a very different entropy production. In this light, it appears to be anything but a coarse-graining, to turn the $N$ individual degrees of freedom of positions $i \in \{1, 2, \ldots, M\}$, that evolve stochastically in time, to the vastly larger phase space of a density $\phi_i : \{1, 2, \ldots, M\} \to \mathbb{R}$, similarly evolving stochastically in time.

In principle, the path probabilities and thus the entropy production in the present framework can be derived exactly on the basis of Dean's multiplicative noise [32]. However, this results in the Onsager-Machlup functional carrying an inverse of the field, which generally poses a challenge, certainly a difficult one in the presence of interaction. Even when that is overcome, the expectation in Eq. (S-VI.35) would need to be taken over the set of allowed field configurations, which now would be sums of $\delta$-functions.

This concludes the present derivation. Apparently, in this simple setup, the Langevin Eq. (S-VI.3) and the subsequent Onsager-Machlup functional Eq. (S-VI.26) capture the fluctuations correctly, but cannot be used as the starting point to construct the observable that determines the entropy production from the path probabilities, because they consider $\boldsymbol{\phi}$ as continuous degrees of freedom subject to additive noise, whereas in the original process the components of $\boldsymbol{\phi}$ are non-negative integers, $\phi_i \in \mathbb{N}_0$. The relationship between Langevin equation and entropy production as exploited in [21] is in principle exact. But by assuming that the Langevin equation that approximates the dynamics can also be used to approximate the entropy production, it seems the wrong degree of freedom is subsequently considered as the one generating entropy.


\end{document}